\documentclass[11pt,preprint]{aastex}
\usepackage{lscape}
\newcommand{\etal}{{et al.\ }}
\newcommand{\Kepler}{{\sl Kepler}\ }

\newcommand{\be}{\begin{equation}}
\newcommand{\ee}{\end{equation}}
\newcommand{\lp}{\left(}
\newcommand{\rp}{\right)}

\slugcomment{accepted to ApJS for \Kepler special issue}
\shorttitle{Transit Timing Observations from \Kepler}
\shortauthors{Ford et al.}

\begin{document}
\title{
Transit Timing Observations from \Kepler:  
I. Statistical Analysis of the First Four Months
}
\author{
Eric B. Ford\altaffilmark{1}, 
Jason F. Rowe\altaffilmark{2,3}, 
Daniel C. Fabrycky\altaffilmark{4}, 
Joshua A. Carter\altaffilmark{6}, 
Matthew J. Holman\altaffilmark{6}, 
Jack J. Lissauer\altaffilmark{3}, 
Darin Ragozzine\altaffilmark{6}, 
Jason H. Steffen\altaffilmark{7},  
Natalie M. Batalha\altaffilmark{8},
William J. Borucki\altaffilmark{3}, 
Steve Bryson\altaffilmark{3}, 
Douglas A. Caldwell\altaffilmark{2,3}, 
Thomas N. Gautier III\altaffilmark{10},
Jon M. Jenkins\altaffilmark{2,3},
David G. Koch\altaffilmark{3},
Jie Li\altaffilmark{2,3},  
Philip Lucas\altaffilmark{11},
Geoffrey W. Marcy\altaffilmark{12},
Sean McCauliff\altaffilmark{9}, 
Fergal R. Mullally\altaffilmark{2,3},
Elisa Quintana\altaffilmark{2,3}, 
Martin Still\altaffilmark{13,3}, 
Peter Tenenbaum\altaffilmark{2,3}, 
Susan E. Thompson\altaffilmark{2,3}, 
Joseph D. Twicken\altaffilmark{2,3}
}
\altaffiltext{1}{Astronomy Department, University of Florida, 211 Bryant Space Sciences Center, Gainesville, FL 32111, USA}
\altaffiltext{2}{SETI Institute, Mountain View, CA, 94043, USA}
\altaffiltext{3}{NASA Ames Research Center, Moffett Field, CA, 94035, USA}
\altaffiltext{4}{UCO/Lick Observatory, University of California, Santa Cruz, CA 95064, USA}
\altaffiltext{5}{Hubble Fellow}
\altaffiltext{6}{Harvard-Smithsonian Center for Astrophysics, 60 Garden Street, Cambridge, MA 02138, USA}
\altaffiltext{7}{Fermilab Center for Particle Astrophysics, P.O. Box 500, MS 127, Batavia, IL 60510}
\altaffiltext{8}{San Jose State University, San Jose, CA 95192, USA}
\altaffiltext{9}{Orbital Sciences Corporation/NASA Ames Research Center, Moffett Field, CA 94035, USA}
\altaffiltext{10}{Jet Propulsion Laboratory/California Institute of Technology, Pasadena, CA 91109, USA}
\altaffiltext{11}{Centre for Astrophysics Research, University of Hertfordshire, College Lane, Hatfield, AL10 9AB, England}
\altaffiltext{12}{University of California, Berkeley, Berkeley, CA 94720}
\altaffiltext{13}{Bay Area Environmental Research Institute/NASA Ames Research Center, Moffett Field, CA 94035, USA}
\email{eford@astro.ufl.edu}
\begin{abstract}
The architectures of multiple planet systems can provide valuable constraints on models of planet formation, including orbital migration, and excitation of orbital eccentricities and inclinations.  NASA's \Kepler mission has identified 1235 transiting planet candidates (Borucki \etal 2011).  The method of transit timing variations (TTVs) has already confirmed 7 planets in two planetary systems (Holman et al.\ 2010; Lissauer et al.\ 2011a).  
We perform a transit timing analysis of the \Kepler planet candidates.  We find that at least $\sim$11\% of planet candidates currently suitable for TTV analysis show evidence suggestive of TTVs, representing at least $\sim$65 TTV candidates.  In all cases, the time span of observations must increase for TTVs to provide strong constraints on planet masses and/or orbits, as expected based on n-body integrations of multiple transiting planet candidate systems (assuming circular and coplanar orbits).  We find the fraction of planet candidates showing TTVs in this data set does not vary significantly with the number of transiting planet candidates per star, suggesting significant mutual inclinations and that many stars with a single transiting planet should host additional non-transiting planets.  
We anticipate that \Kepler could confirm (or reject) at least $\sim$12 systems with multiple transiting planet candidates via TTVs.   
Thus, TTVs will provide a powerful tool for confirming transiting planets and characterizing the orbital dynamics of low-mass planets.  
If \Kepler observations were extended to at least seven years, then TTVs would provide much more precise constraints on the dynamics of systems with multiple transiting planets and would become sensitive to planets with orbital periods extending into the habitable zone of solar-type stars. 
\end{abstract}


\section{Introduction} 

%
NASA launched the \Kepler space mission on March 6, 2009, to measure the frequency of small exoplanets.  In its nominal mission \Kepler will observe over 100 square degrees nearly continuously for three and a half years, so it can detect multiple transits of planets in the habitable zones of solar-type stars.  The spacecraft carries consumables that could support an extended mission which would improve sensitivity for detecting small planets and would dramatically improve the constraints from transit timing studies.  
%
%
\Kepler began collected engineering data (``quarter'' 0; Q0) for stars brighter than \Kepler magnitude ($K_p$) 13.6 on May 2, 2009, and science data for over 150,000 stars on May 13, 2009.  The first ``quarter'' (Q1) of \Kepler data extends through June 15, 2009 and the second quarter (Q2) runs from June 20 to September 16, 2009.  On February 1, 2011, the \Kepler team released light curves during Q0, Q1 and Q2 for all planet search targets via the Multi-Mission Archive at the Space Telescope Science Institute (MAST; \url{http://archive.stsci.edu/kepler/}). 
%
%
The \Kepler team has performed an initial transiting planet search to identify Kepler Objects of Interest (KOIs) that show transit-like events during Q0-2 (Borucki \etal 2011; hereafter B11).  B11 lists 1235 KOIs as active planet candidates.  
Other KOIs are recognized as likely astrophysical false positives (e.g., blends with background eclipsing binaries) and are reported in B11 Table 4.  As the team performs tests of KOIs, a ``vetting flag'' is used to indicate which KOIs are the strongest and weakest planet candidates with the expected reliability ranging from $\ge$98\% for confirmed planets to $\ge$60\% for those which have yet to be fully vetted.  Table 2 of B11, lists the putative orbital period, transit epoch, transit duration, planet size, and vetting flag for KOIs which are active planet candidates and were observed to transit in Q0-2.  
%
%
The current sample of planet candidates is incomplete due to selection effects.  In particular, planets with long orbital periods ($P>125$d), small planets ($R_p\le~2R_\oplus$), and planets around faint, active and/or large stars are affected (B11).   Further candidates will likely be identified as additional data are analyzed and the \Kepler pipeline is refined.

%
%
The light curve of two eclipsing objects can be very sensitive to perturbations that result in non-Keplerian motion. The variability in the times of eclipsing binaries have been studied for decades.  Typically, an ``O-C'' diagram, highlighting the difference between the observed ephemeris and the ephemeris calculated from a constant period, has been used to detect additional bodies, apsidal motion, and other effects (Bozkurt \& De{\u g}irmenci 2007 and reference therein; Slawson \etal 2011).  The application of this technique to planetary transits is known as Transit Timing Variations (TTVs) which have been studied extensively both theoretically and observationally over the past several years.  Astrophysically interesting deviations from a linear transit ephemeris that are potentially observable by \Kepler can be caused most readily by a perturbing planet (e.g., Miralda-Escud\'{e} 2002; Agol \etal 2005; Holman \& Murray 2005; Ford \& Holman 2007) or moon (e.g., Simon \etal 2007; Kipping \etal 2009), though perturbations by a stellar companion or higher-order gravitational effects can occasionally be significant (e.g., Carter \& Winn 2010).  In principle, variations in transit duration, depth, or overall shape can also be used to study various astrophysical properties (e.g., Miralda-Escud\'{e} 2002; Ragozzine \& Wolf 2009; Carter \& Winn 2010).  
The physics of TTV is very similar to eclipse timing variations (ETVs) for binary stars, which are not uncommon in the Kepler Binary Star Catalog (Prsa \etal 2011; Slawson \etal 2011; Orosz \etal 2011).  
The symmetry of transits (even for large eccentricities) means that transit times are much less susceptible to degeneracies than other transit parameters (e.g., duration, impact parameter and limb darkening; Col{\'o}n \& Ford 2009).  Thus, transit times can be measured with the highest precision and accuracy, and are generally expected to be the first recognizable signs of dynamical perturbations.  
In cases where TTVs are detected, a follow-up investigations of transit duration variations would be warranted.  

%
%
In this paper, we analyze putative transit times (TTs) by \Kepler planet candidates that show at least three transits in Q0-2.
%
%
We describe our methods for measuring transit times and constructing ephemerides in \S\ref{secMethods}.  
We discuss the results of several statistical analyses applied to each of the planet candidates under consideration in \S\ref{secTests}.  We present a list of planet candidates with early indication of TTVs in \S\ref{secCandidates}.  We compare the expected and observed TTVs for candidate multiple transiting planet systems in \S\ref{secMultis}.  Finally, we discuss the implications of these early results for planet formation and the future of the TTV method in \S\ref{secDiscuss}.

\section{Measurement of Transit Times from Kepler Data}
\label{secMethods}
\subsection{Bulk Transit Time Measurements} 
\label{secMeasure}
A combination of pipelines is used to identify KOIs as described in Latham \etal (2011).  We measure transit times based on the long cadence (LC), optimal aperture photometry performed by the \Kepler SOC pipeline version 6.2 (Jenkins \etal 2010).  The pipeline produces both ``calibrated'' light curves (PA data) for individual analysis and ``corrected'' light curves (PDC) which are used to search for transits.  This paper presents a bulk set of transit times measured from PDC data.  For a detailed analysis of an individual system, we strongly recommend that users inspect both PA and PDC data to determine which is best suited for their target star and applications.  For example, the TTV detections of Kepler-9b \& c and Kepler-11 b-f were based on PA data (Holman \etal 2010; Lissauer \etal 2011a).  

We fold light curves at the orbital period reported in B11.  Next, we fit a limb-darkened transit model to the folded light curve, allowing the epoch, planet-star radius ratio, transit duration, and impact parameter to vary.  We use the best-fit transit model as a fixed template to measure the transit time of each individual transit (while holding other parameters fixed).  The best-fit transit times are determined by Levenberg-Marquardt minimization of $\chi^2$ (Press et al.\ 1992).  In a few cases, we iterate the procedure, aligning transits based on the measured period in order to generate an acceptable light curve model.  For a small fraction of candidates (KOI 2.01, 403.01, 496.01, 508.01, 559.01, 617.01, 625.01, 678.01, 687.01, 777.01), we measure TTs during Q1 only, due to complications in the Q2 light curves (e.g., stellar noise, and/or grazing short-period events).  Two other types of complications merit some explanation.  
For bright stars that saturate the CCD, causing electrons to bleed into neighboring pixels.  When the spacecraft rotates each quarter, a target moves from one CCD to another and the aperture mask specifying which pixels are downloaded changes.  If the aperture mask does not capture all the electrons, this can cause the transit light curve to vary from quarter to quarter, breaking the code used for measuring transit times.  Another complication that interfered with measuring TTs during Q2 arises since the Q2 photometry for most targets is not as high quality as the Q1 photometry, due to a variety of technical issues (e.g., replacing guide stars that were discovered to be variable or binary resulted in the use of even more problematic guide stars, causing high-frequency spacecraft motion).  For a few KOIs with small signal-to-noise per transit, this prevented accurate measurement of TTs during Q2.  The photometric quality in subsequent quarters has improved significantly, as demonstrated in publically availiable light curves for Kepler-9, 10 and 11.

\begin{figure*}
\epsscale{1.0}
\plotone{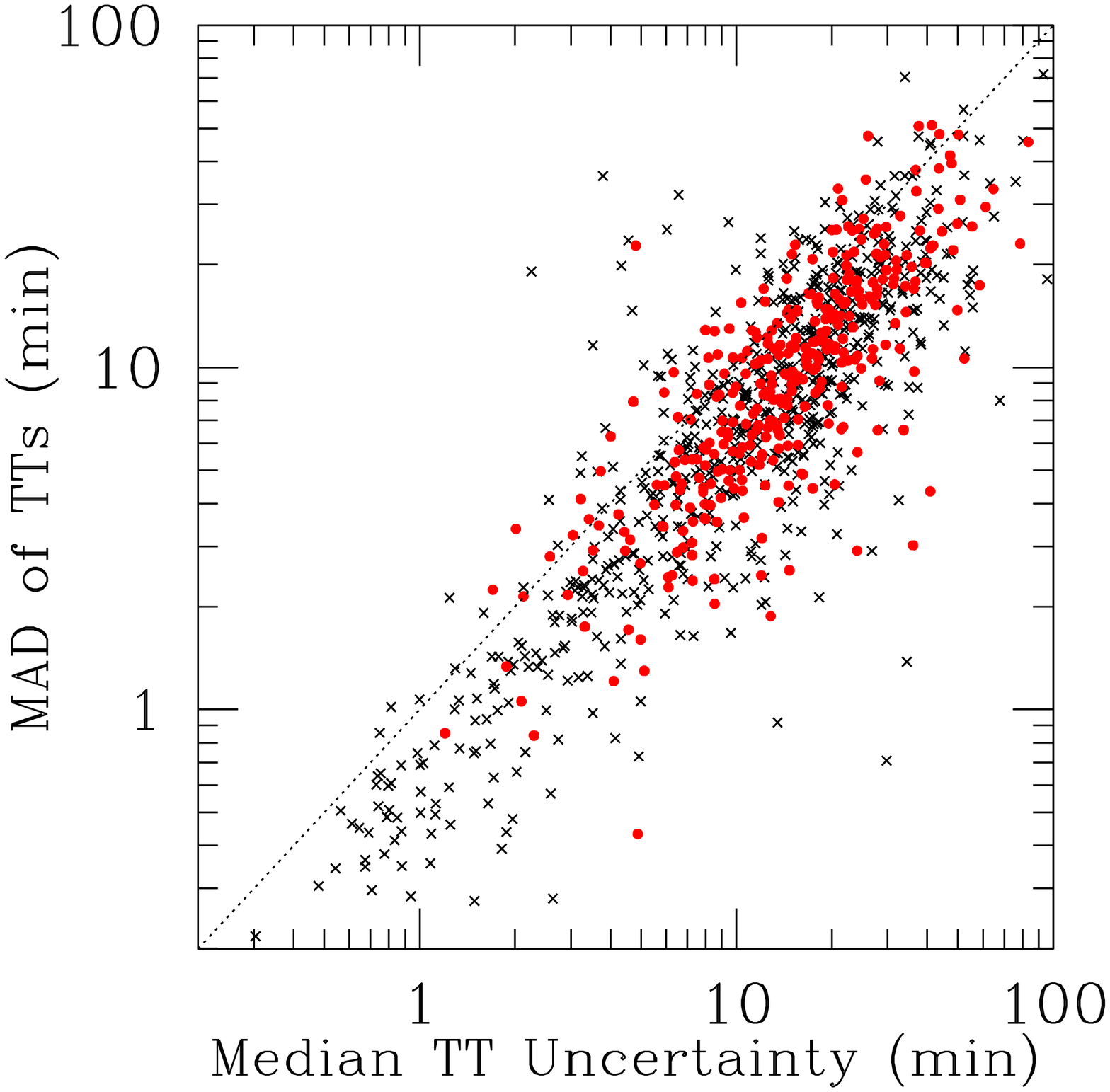}
\caption{Median absolution deviation of transit times from the ephemeris of B11 versus the median uncertainty in transit time observations during Q0-2.  Systems with one planet candidate are marked with an X, and multiple planet candidate systems are marked with a (red) disk.}
\label{figMadVsSigma}
\end{figure*}

We estimate the uncertainty in each transit time from the covariance matrix. 
For most planet candidates, typical scatter in its TT relative to a linear ephemeris is comparable to the median timing uncertainty (see Fig.\ \ref{figMadVsSigma}).  
Even when the transit model used for transit time measurements is not ideal, good TTVs and errors can be extracted using this technique, since the transit time is the only free parameter and only a roughly correct shape is needed to identify accurate times and errors. Subsets of the observed times reported in Table \ref{TabTTs} were tested with different transit time estimation codes, which employ different techniques.  While a more thorough analysis can sometimes reduce the timing uncertainty or minimize the number of apparent TTV outliers, the results are consistent across different algorithms.

%
For target stars where multiple transiting planet candidates have been identified, we sequentially fit each transit separately.  Prior to fitting for TTVs, we remove the best-fit transit-models for additional candidates present in the light curve.  For example, if there are 3 planet candidates in the system, then the fitting procedure would first fit for the template light curves in the following order:  1) fit candidate .01, 2) remove .01 from the light curve, 3) fit candidate .02 from residuals, 4) remove .02 from light curve, 5) fit candidate .03 from residuals, and 6) remove .03 from light curve.  In most cases, the sequence .01, .02, ... is from the highest to the lowest signal to noise (integrated over all transits).  Next, we would measure the transit times according to the following plan:  1) remove .02 and .03 from original light curve, 2) fit for .01 template, 3) measure TTs for candidate .01, 4) remove .01 and .03 from original light curve, 5) fit for .02 template, 6) measure TTs for candidate .02, 7) remove .01 and .02 from original light curve, 8) fit for .03 template, and 9) measure TTs for candidate .03.  In some cases, we repeat the measurement of TTs by aligning the transits using the first set of TTs before generating the template.

\subsection{Transit Timing Models}
\label{secModels}
Once each set of individual transit times (TTs) has been measured, we calculate multiple sets of TTVs by comparing them to multiple ephemerides.  The TTVs reported in Table \ref{TabTTs} are measured relative to the linear ephemeris published in B11 (E$_L$5).  A postive TTV corresponds to a transit occuring later than the ephemeris.  Second, we considered the TTVs relative to the best-fit linear ephemeris calculated from TTs in Q0-2 (E$_L$2; Table \ref{TabEphems}).  
The E$_L$5 ephemerides differ from the E$_L$2 ephemerides in that epoch ($\hat{E}_0$) and period ($\hat{P}$) were determined using \Kepler data up to and including Q5, rather than Q2.  
Finally, we also considered TTVs relative to the best-fit quadratic ephemeris calculated from TTs in Q0-2 (E$_Q$2).  The quadratic ephemeris is given by
\be
\label{eqnQuad}
\hat{t}_n = \hat{E}_0 + n \hat{P} (1 + n \hat{c}),
\ee
where $\hat{E}_0$ is the best-fit time of the zeroth transit, $\hat{P}$ is the best-fit orbital period, and $\hat{c}$ is the best-fit value of the curvature.  For a linear ephemeris, $\hat{c}=0$.  
%

\label{secResults}
\section{Assessing the Significance of Transit Timing Variations}
\label{secTests}
Transit times for each planet candidate considered are provided in the electronic version of Table \ref{TabTTs}.  
Times are measured relative to the E$_L$5 ephemerides given in Borucki \etal 2011b.
We perform several tests to determine if TTVs are statistically significant.  These address three questions: 1) ``Is the observed scatter in TTs greater than expected?'', 2) ``Is there is a long-term trend in the TTs?'', and 3) ``Is there a simple periodic variation in the TTs?''.  

\subsection{Scatter in Transit Times}
\label{secTestScatter}
First, we calculated $X^2 \equiv \sum_{n\in~Q0-2} \lp \lp t_n - \hat{t}_n \rp / \sigma_n \rp^2$ for $\hat{t}_n$ calculated from both the E$_L$2 and E$_L$5 ephemerides.  As $X^2$ closely resembles a $\chi^2$ variable, we first apply a $\chi^2$ test, but based on $X^2$.  If we assume that $X^2$ follows a $\chi^2$ distribution, 
then a $\chi^2$ test based would result in a $p$-value less than 0.001 for 15-17\% of candidates.  As expected, the fraction is lower if we use the E$_L$2 ephemerides, since the same times are being used for the model fit and the calculation of $X^2$.  If the errors in the TT measurements, normalized by the estimated timing uncertainty, were accurately described by a standard normal distribution, then one would expect only a few false alarms.  However, we note that a disproportionate fraction (42\%) of the planet candidates with seemingly significant scatter in their transit times have an average signal to noise (S/N) in each transit of less than 3, while planet candidates with such a small S/N per transit represent only ($\sim$23\%) of planet candidates considered.  While it is possible that planets with low S/N transit are more likely to have significant TTVs, we opt for a more conservative interpretation that the distribution of errors in our transit times measurements for low S/N transits are not accurately described by a normal distribution with a dispersion given by the estimated timing uncertainties.  In particular, when measuring the times of low S/N transits, the $X^2$ surface as a function of $t_n$ can be bumpy due to noise in the observed light curve, resulting in a much greater probability of measuring a significantly discrepant transit time than assumed by our normal model.  

To avoid a high rate of false alarms due to random noise, we focus our attention on planet candidates where the typical S/N in a transit exceeds 3.  If we assume that $X^2$ follows a $\chi^2$ distribution, then results in $\sim$11-13\% (86 or 103/805) of the remaining planet candidates having a $p$-value less than 0.001.  We observe that for many of these cases, the contribution to $X^2$ is dominated by a small fraction of the TTs.  Based on the inspection of many light curves around these isolated discrepant transit times, we find that occasionally the measured TTs can be significantly skewed by a couple of deviant photometric measurements.  In principle, this could occur due to random noise, but for high S/N transits, stellar variability or improperly corrected systematic errors are the more likely culprits.  For example, systematic errors often appear after events such as reaction wheel desaturations, reaction wheel zero crossings, and data gaps (see data release notes at MAST for details; \url{http://archive.stsci.edu/kepler/data\_release.html}).

To further guard against a few discrepant TTs leading to a spurious detection of TTVs, we calculate a statistic that is similar to $X^2$, but one that is more robust to outliers,
\be
X^{\prime 2} \equiv \frac{\pi~N_{TT} \lp~\mathrm{MAD}\rp^2}{2\sigma_{TT}^2},
\ee
where $N_{TT}$ is the number of TTs measured in Q0-2, $\sigma_{TT}$ is the median TT measurement uncertainty.  Here MAD is the median absolute deviation of the measured TTs from an ephemeris, given by
\be
\mathrm{MAD} \equiv \mathrm{median}_{i=1}^{N_{TT}}\left(\left|t_i-\hat{t}_i\right|\right),
\ee
where $\hat{t}_i$ is the time predicted by a given ephemeris. 
The $X^{\prime 2}$ statistic can be viewed as the ratio of a robust variance of the transit times to the square of the typical measurement uncertainty for transit times.  We calculate alternative $p$-values for rejecting the null hypothesis that there are no deviations from a linear ephemeris by replacing $\chi^2$ with $X^{\prime 2}$.  Since the distribution of $X^{\prime 2}$ may deviate from a $\chi^2$ distribution, the $p$-values and resulting significance levels are not reliable.  Our $X^{\prime 2}$ test merely provides a means of filtering the list of KOIs to identify those with potentially significant TTVs.  The advantage of $X^{\prime 2}$ for our application is that $X^{\prime 2}$ is much less sensitive to the assumption that measurement errors are well characterized by a normal distribution.  A small fraction of significant outliers can greatly increase $X^2$ but has a minimal effect on $X^{\prime 2}$.  While this is more robust than a standard $\chi^2$ test, the statistical significance will not be accurately estimated if the measurement errors are severely non-Gaussian.  Of course, the increased robustness comes at a  price: $X^{\prime 2}$ is not useful for recognizing TTV signals where only a small fraction of TTs deviate from a standard linear ephemeris.  We expect this is a good trade, given the predicted TTV signatures (Veras et al.\ 2011) and the characteristics of our TT measurements.   

We identify planet candidates showing strong indications of a TTV signal based on an ad hoc threshold
which would correspond to a $p$-value of $p\le~10^{-3}$, if $X^{\prime 2}$ followed a $\chi^2$ distribution.  
Restricting our attention to planet candidates for which the individual transits are detected in the light curve with a S/N greater than 3, this test identifies roughly $\sim$4-5\% (29 or 39/805) as showing excess scatter relative to a linear ephemeris.  We find the larger fraction ($\sim$5\%) when comparing to the E$_L$2 ephemerides (rather than the E$_L$5 ephemerides).  We speculate that this may be due to the E$_L$5 ephemerides being more robust to a small number of outlying TTs in Q2 which would have a greater effect on the E$_L$2 ephemerides.  
We discuss these candidates further in \S\ref{secCandidates}.
A summary of these and other summary statistics for all planet candidates considered appears in Table \ref{TabMetrics}.  

We also considered calculating $X^2$, but based on ``clipped'' TTs from Q0-2, where we reject TTs with either a formal uncertainty greater than twice the median timing uncertainty or with an absolute deviation greater than three times the MAD of TTs. This results in $\sim$4\% of planet candidates with at least 3 transits during Q0-2 and a S/N per transit greater than 3 as showing excess scatter relative to the E$_L$2 ephemeris.  While the rate of TTV candidates is consistent with the results of the $X^{\prime 2}$ statistic described above, the exact systems flagged differ from those identified basedon $X^{\prime 2}$ test.   Roughly half of these are cases with less than 10 transits during Q0-2, in which case the threshold for clipping is poorly defined.  Therefore, we prefer the tests based on $X^{\prime 2}$ for the present dataset.  
%
%
%
%

\subsection{Long Term Trends in Transit Times}
\label{secTestTrend}
\subsubsection{Difference in Best-fit Orbital Periods}
\label{secTestPeriod}
Next, we search the TTs in Q0-2 for evidence of a long-term trend.  As a first test, we compare the orbital periods we measure from our E$_L$2 ephemeris to the period of the E$_L$5 ephemeris reported in B11 that is based on data from Q0-5.  We find period differences greater than three times the formal uncertainty in the orbital periods for roughly 14\% (111/805) of planet candidates with a S/N per transit greater than 3 and at least 3 transits during Q0-2.  Most of these have only three transits in Q0-2 or only slightly exceed the threshold, perhaps due to a slight underestimate of the uncertainty in the orbital period.  Of the planet candidates for which the periods disagree very significantly and there are at least four transits in Q0-2, most show an easily recognizable long-term trend and were identified as interesting based on $X^{\prime 2}$, as described in \S\ref{secTestScatter}.  

For planet candidates with an apparently discrepant period and no more than 5 transit times measured in Q0-2, often the E$_L$2 ephemeris could have been significantly affected by a single outlying TTV point during Q0-2.  KOI 1508.01 has 6 transits, and may also have been affected by an outlier.  Two planet candidates have large and nearly linear residuals suggesting an inaccurate ephemeris was given in B11.  Similarly, further analysis has revealed that the orbital period of KOI 730.03 is much more likely to be very nearly twice that of the value given in B11.  Based on data through Q0-2 only, we provide alternative ephemerides of:
$(107.5984\pm0.0090) + n (19.7198\pm0.0044)$ for KOI 730.03,
$(118.05868\pm0.00011) + n (2.8153306\pm0.0000084)$ for KOI 767.01 and
$(147.41414\pm0.00085) + n (1.2094482\pm0.000038)$ for KOI 1540.01.
After discarding the cases above, the most compelling candidates with at least five transit times observed in Q0-2 that were identified by this method and not based on $X^{\prime 2}$ test are {\bf KOI-524.01} and {\bf 662.01}.  KOI-961.01 is formally highly significant, but we caution that the short transit duration may affect the TT measurements.  {\bf KOIs 226.01, 238.01, 248.01, 564.01, 700.02, 818.01} and {\bf 954.01} are also identified based on an apparent discrepancy ($\ge~4-\sigma$) between the periods of the E$_L$2 and E$_L$5 ephemerides.  KOIs 295.01, 339.02 and 834.03 also meet this criteria, but are even less secure since they have a S/N per transit of less than three.  If the putative signals are real, then they should become obvious with a longer time span of TT observations.  

\subsubsection{Difference in Best-fit Epochs}
\label{secTestEpoch}
We can perform a test similar to \S\ref{secTestPeriod}, but comparing the epochs we measure from our E$_L$2 ephemeris to the epoch from the E$_L$5 ephemeris reported in B11.
This test identifies 30 planet candidates for which the best-epoch for Q0-2 differs from that reported for Q0-5 from B11 at the $\ge~4\sigma$ level (excluding some with known issues regarding the TT measurements).  Of these, 16 had not been identified based on $X^{\prime 2}$ or the test for a difference in orbital periods between Q0-2 and Q0-5: 
{\bf 10.01, 
94.02, 
137.02, 
148.03, 
217.01, 
279.01, 
377.02, 
388.01, 
417.01, 
443.01, 
658.01, 
679.01} and 
{\bf 1366.01}.
In addition, the test based on $X^{\prime 2}$ flagged KOIs 1169.01 and 360.01 
that have poor TT measurements due to low S/N, 800.02 that may be affected by an outlier.  While 279.01 and 679.01 have only three TTs during Q0-2, that is sufficient for testing whether the epoch matches the B11 ephemeris.  
Particularly interesting are KOIs 279.01, and 658.01 and 663.02 that are in systems with two planet candidates and
94.02, 137.02, 148.03, 377.02, 884.02 and 961.01 that are in systems with three planet candidates (see \S\ref{secStrongCandidates} \& \ref{secMultis}).

\subsubsection{$\mathcal{F}$-test for Comparing Quadratic and Linear Models}
While TTV signatures can be quite complex, the time scale for resonant dynamical interactions among planetary systems is typically orders of magnitude longer than the orbital period.  Thus, we expect that the dominant TTV signature of many resonant planetary systems can be well approximated by a gradual change in the orbital period over timescales of months to years.  Indeed, such a pattern led to the confirmation of Kepler-9 b\&c (Holman \etal 2010).  (There is also a pattern of alternating transit arriving earlier and later than the quadratic ephemeris.  However this ``chopping'' signature occurs on an orbital timescale, but a much smaller amplitude.)  
Thus, we fit both linear and quadratic ephemerides (see \S\ref{secModels}) to the Q0-2 data, calculate $X_{E_L2}^2$ and $X_{E_Q2}^2$ for the two models, respectively.  We perform a variant of the $F$-test to assess whether including a curvature term significantly improves the quality of the fit.  We use a test statistic as $\mathcal{F} \equiv (N_{TT}-3) X_{E_L2} / \left[ (N_{TT}-2) X_{E_Q2} \right]$.  This is similar to an $F$-statistic, but we use $\mathcal{F}$ since it is based on a ratio of $X$ statistics that do not necessarily follow $\chi^2$ distributions.  This method has the advantage of being less sensitive to the accuracy of the TT uncertainty estimates, as they affect both the numerator and denominator of the $\mathcal{F}$ statistic similarly.  
If we assumed that $\mathcal{F}$ followed an $F$ distribution, then an $F$-test would not result in a $p$-value less than $10^{-3}$ for any of the KOIs that are still viable planet candidates.  For our application, the $\mathcal{F}$-test has limited power, since we fit only TTs measured in Q0-2, resulting in a shorter time span for a gradual change in the period to accumulate.  We note that the $\mathcal{F}$-test does strongly favor the quadratic model for two KOIs previously identified as being due to stellar binary or triple systems (KOI-646.01, Fabrycky \etal in prep; 1153.01), where the curvature is sufficiently large that the period changes appreciably during Q0-2. 

Next, we consider systems for which $\mathcal{F}$ would imply a $p$-value of 0.05 or less, if $\mathcal{F}$ were to follow an $F$ distribution.  Quadratic ephemerides for these KOIs are given in Table \ref{tabQuad}.  
Approximately 1\% of KOIs considered are identified.  Half of these correspond to candidates with only 4 or 5 transits in Q0-2, so it is impractical to assess the quality of a three-parameter model accurately.  Of the remaining systems, KOI 142.01 and 227.01 were already identified by the $X^{\prime 2}$ test, while KOI-528.01 and 1310.01 were not.  Upon visual inspection, {\bf 528.01} appears to be a plausible detection of TTVs, but 1310.01 does not, as the TTs and uncertainties are also consistent with a linear ephemeris.  
These candidates are discussed further in \S\ref{secCandidates}.

\subsection{Periodic Variations in Transit Times}
\label{secTestHarmonic}
Over long time scales, dynamical interactions in two-planet systems can result in complex TTV signatures characterized by many frequencies.  However, on short time scales, TTV signatures can often be well described by a single periodic term.  For example, for closely packed, but non-resonant, planetary systems, the TTV signature is often dominated by the reflex motion of the star due to other planets.  In this case, we would expect a typically small TTV signal on an orbital time scale.   

\begin{figure*}
\epsscale{1.0}
\plottwo{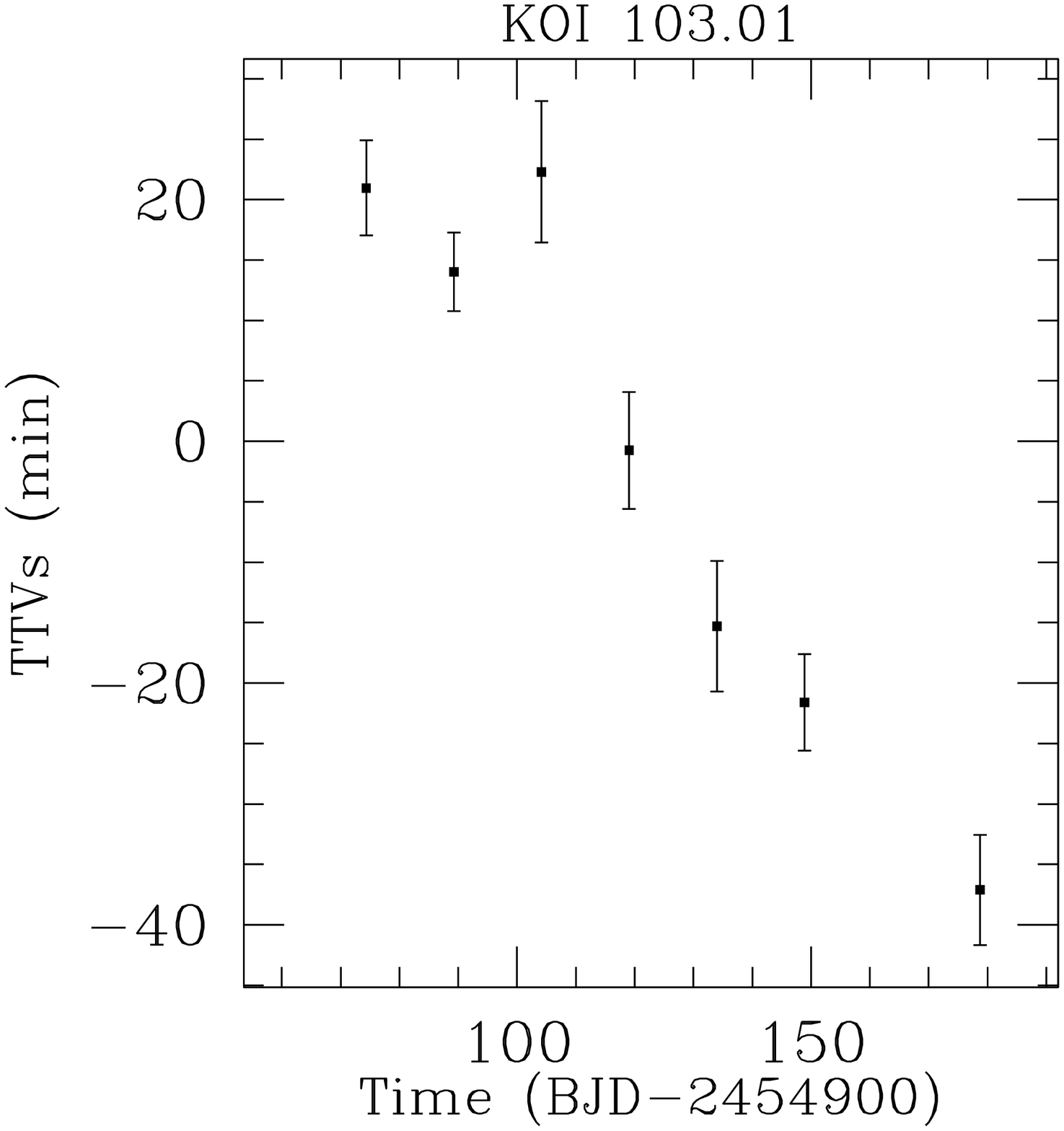}{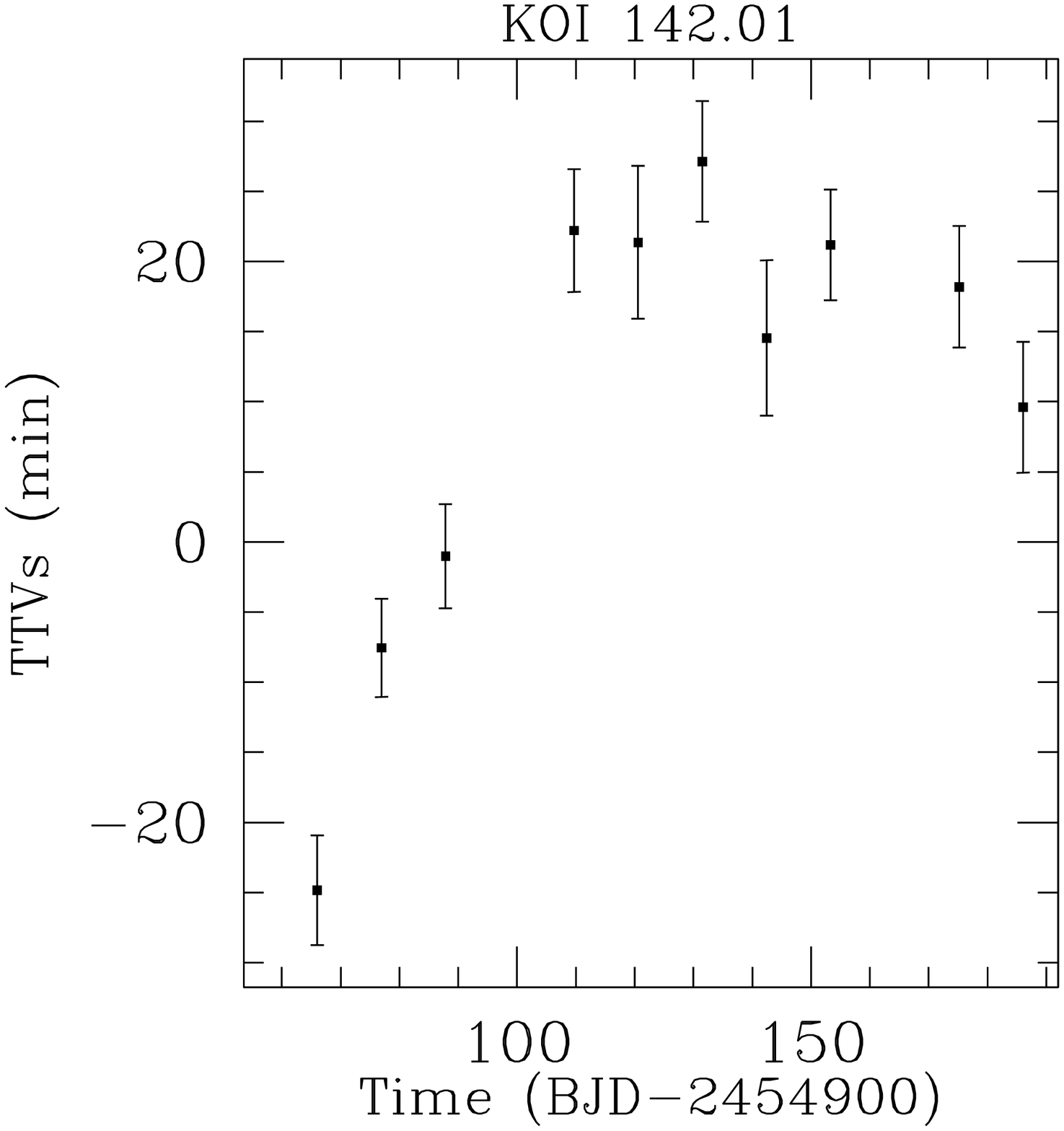} \\
\plottwo{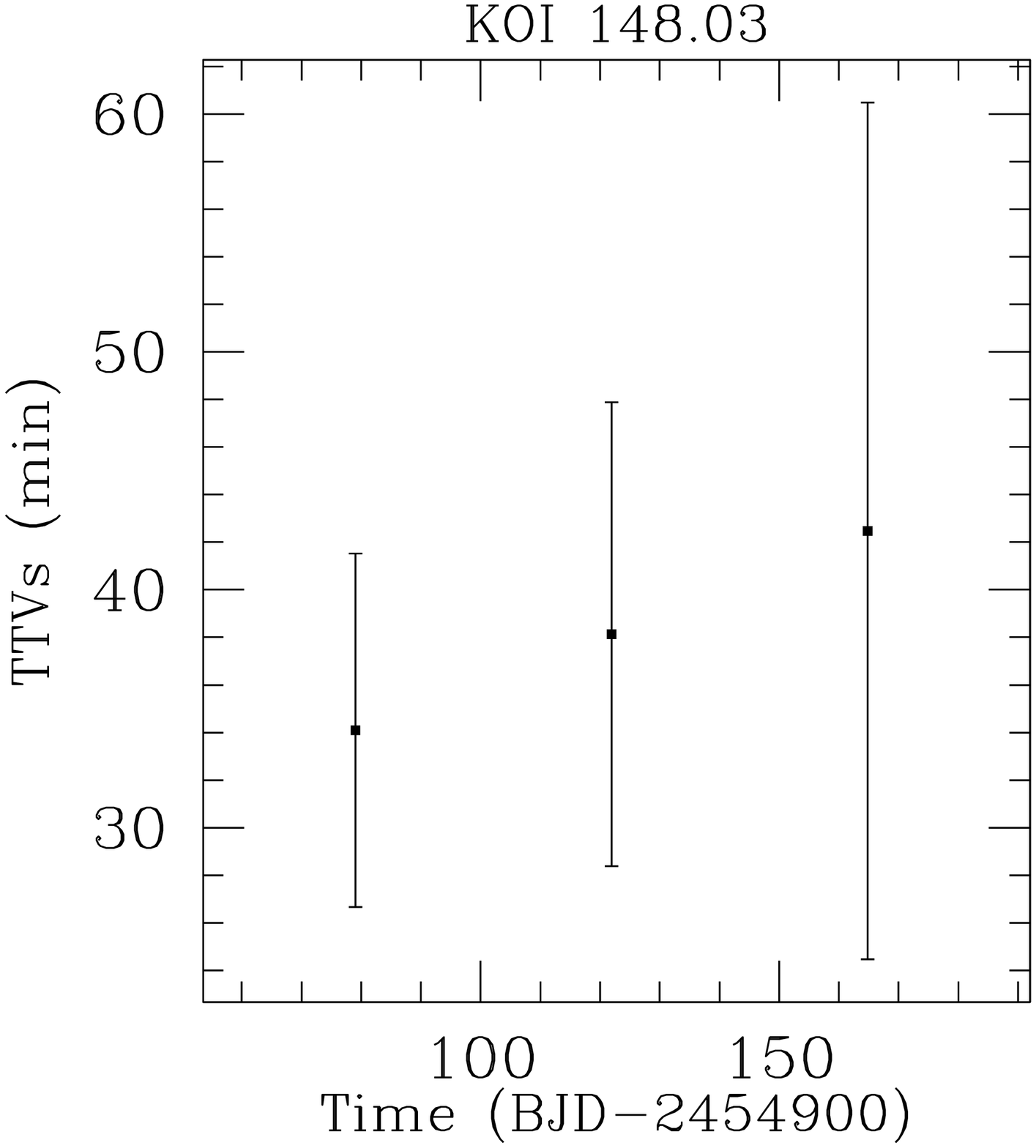}{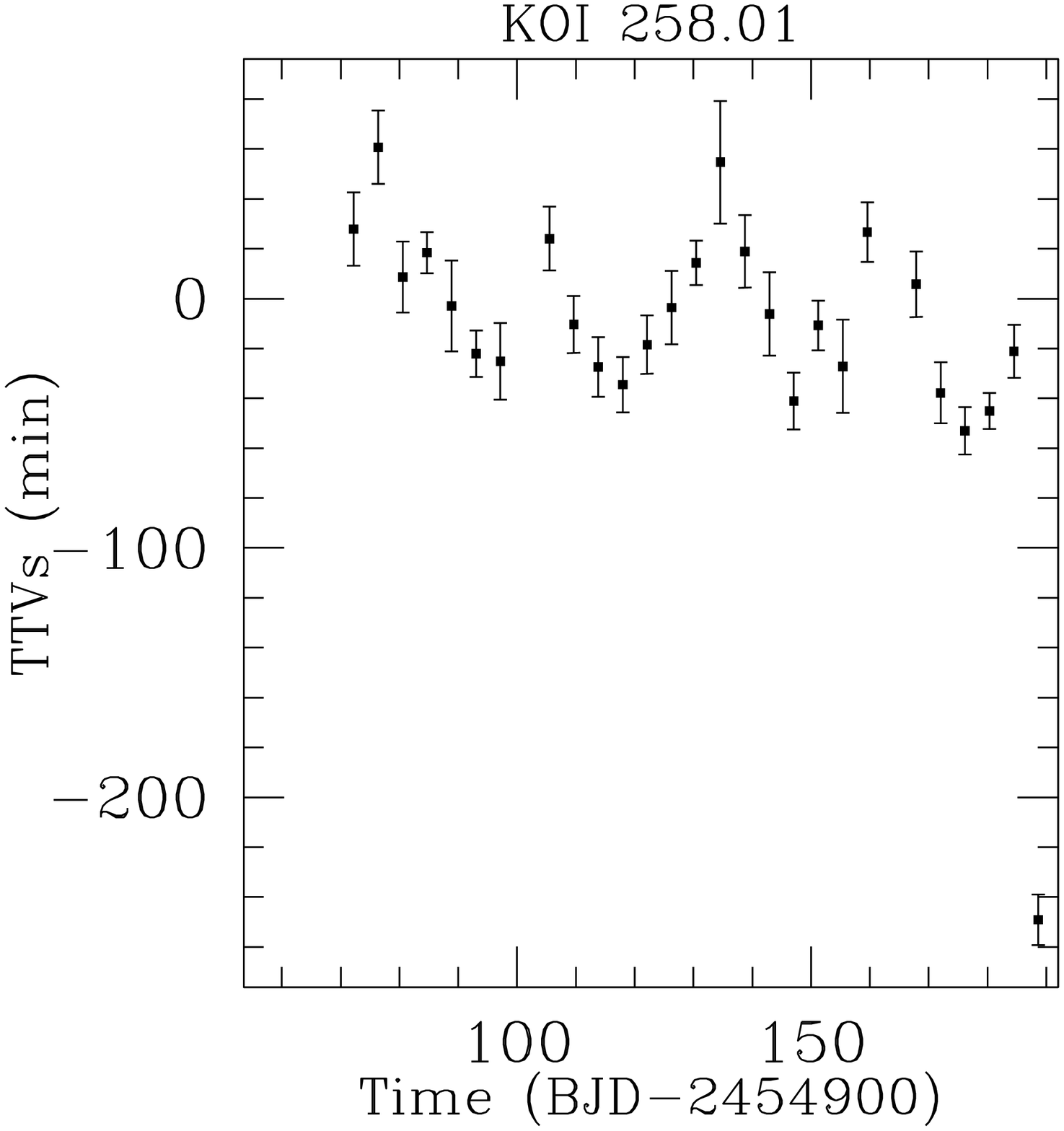}
\caption{Transit timing measurements for four examples of strong TTV candidates: KOI 103.01, 142.01, 148.03, 258.01.  
Note that the TTs are measured relative to the EL5 ephemerides given in Borucki et al. 2011b and this is based on transit times measured through Q5.
KOI 103.01 shows TTVs indicative of a long-term change in the orbital period.  KOI 142.01 already shows significant curvature during Q0-2, suggesting an orbital period or libration timescale not much longer than the timespan of observations.  While the TTs of KOI 148.03 appear consistent with a constant orbital period, they are significantly offset relative to the ephemeris of B11, suggesting a long-term change in the orbital period.  KOI 258.01 appears to show periodic TTVs on a relatively short timescale.  There are preliminary indications that KOI 258.01 may show an occultation or secondary eclipse.}
\label{figKoiGoodCandidates}
\end{figure*}

Given the relatively short time span of the observations, we search for planet candidates with a TTV signature that can be approximated by a single sinusoid,
\be
\hat{t}_{n} = \hat{E}_0 + n \hat{P} + \hat{A} \sin (2\pi~ n\hat{P} / \hat{P}_{TTV}) + \hat{B} \cos (2\pi~n\hat{P} /\hat{P}_{TTV}),
\ee
$\hat{A}$ and $\hat{B}$ determine the amplitude and phase of the TTV signal, while $2\pi/\hat{P}_{TTV}$ gives the frequency of the model TTV perturbation.  For a given $\hat{P}_{TTV}$, finding the best-fit (i.e., minimum $X^2$) model is a linear minimization problem.  Thus, we perform a brute force search over $\hat{P}_{TTV}$ to identify the best-fit simple harmonic model (Ford et al.\ 2011).  Assessing whether the best-fit harmonic model is significantly better than a standard linear ephemeris is notoriously difficult (e.g., Ford \& Gregory 2007), even when the distribution and magnitude of measurement uncertainties are well understood.  Therefore, we plotted the cumulative distribution of summary statistics and recognized a break in the distribution corresponding to a tail that included approximately 2\% of the planet candidates considered, which also included several KOIs which have since been moved to the false positives list.  One quarter of the KOIs in this tail had eight or fewer TTs measured in Q0-2, so it was not practical to assess the quality of a five-parameter fit accurately.  Of the remaining KOIs with a possible periodic signal, 90\% have some other evidence suggesting that they are a stellar binary.  Three KOIs remain:  KOI-258.01 (most significant), 1465.01, and 1204.01 (least significant).  The first two were also identified by the $X^{\prime 2}$ analysis described in \S\ref{secTestScatter}.  KOI-258.01 appears to show a periodic pattern with amplitude $\sim$40 minutes and time scale of either $\sim$28 or 58 days (see Fig.\ \ref{figKoiGoodCandidates}), however there is also a hint of a secondary eclipse for KOI-258.01. 
For KOI-1465.01 the combination of a short transit duration and long cadence observations might have resulted in the PDC data having artifacts that render the TTs unreliable.  For KOI-1204.01, there is only a hint of a periodicity and additional observations will be necessary to assess the significance of the putative signal.  

\section{Planet Candidates of Particular Interest}
\label{secIndiv}
In this section, we investigate KOIs of particular interest, including those which were identified as potentially having transit timing variations based on one of the statistical tests described in \S\ref{secTests} and several in multiple transiting planet candidate systems.

\subsection{Planet Candidates with Potential Transit Timing Variations}
\label{secCandidates}
We provide a summary of the planet candidates identified by our statistical tests in Table \ref{tabNotes}.  
For completeness sake, the table includes all planet candidates from B11, with some indication of TTVs and at least three transits in Q0-2.  However, we regard many as weak candidates.  For a strong candidate, we typically require: 1) a S/N per transit of at least 4, 2) at least 5 TTs observed in Q0-2 for most tests, and 3) no indications of potential difficulties measuring the transit times (e.g., near data gap, short duration, PDC artifacts, heavily spotted star).  When comparing the best-fit epochs, we require only three transits during Q0-2.  
Many of these candidates were identified by the $X^{\prime 2}$ statistic which appears to provide a balance of sensitivity and robustness when searching for excess scatter in the present data set.  Those detected by other tests have already been discussed in \S\ref{secTestTrend} \& \S\ref{secTestHarmonic} or are indicated with a flag of 6 in Table \ref{tabNotes}.  As the number of transits observed by \Kepler increases, it is expected that model fitting will eventually provide superior results (Ford \& Holman 2007).  For example, Kepler 9b\&c, and Kepler 11b-f were confirmed by transit timing variations, but only Kepler-9c shows TTVs in the Q0-2 data.  For Kepler-9b, the TTVs are not detectable during Q0-2, as the libration timescale ($\sim$10 years) is much greater than the time span of observations (Holman \etal 2009).  The TTVs of planets in Kepler-11 are much smaller and demonstrate the benefits of fitting a dynamical model to TTs of all planets simultaneously (Lissauer \etal 2011a).  

The analysis presented in this manuscript considers each candidate individually.  For multi-candidate systems, it is possible that there exists a significant anti-correlation in TTV signals (Ford et al.\ in prep; Steffen et al.\ in prep), even though the individual signals do not reach the various significance thresholds given in Section 3. If the relationship between TTVs of multi-candidate systems can be proven to be non-random, it significantly weakens the probability that the candidate signals are due to false positives (Ragozzine \& Holman 2010), even if the properties of the planets cannot be directly inferred. This method of confirming planets will become much stronger with additional data.

\subsubsection{False Positives}
\label{secFalsePos}
KOI 928.01 attracted the early attention of the \Kepler Transit Timing Working Group, despite its low S/N per transit.  A detailed analysis suggesting that this is likely a stellar triple system will be presented in Steffen \etal (2011).  Several other single TTV candidates were identified as false positives and removed from the B11 planet candidate list.

\subsubsection{Weak TTV Candidates}
\label{secWeakCandidates}
KOIs 156.02, 260.01, 346.01, 579.01, 
751.01, 756.02, 786.01, 1019.01, 1111.01, 1236.02, 1241.02, 1396.02, 1508.01 and 1512.01 appear to have excess scatter based on the $X^{\prime 2}$ test.  However, the small S/N in each transit makes the TT measurements and uncertainties unreliable.  
Similarly, KOI 295.01, 339.02, 700.02, 800.02, and 834.03 appear to have a change in orbital period and/or epoch between Q0-2 and Q0-5, but may be affected by the small S/N in each transit.  In particular, KOI 260.01, 346.01, 800.02, 1111.01, 1508.01 and 1512.01 appear to be affected by an outlier.  

Several relatively weak TTV candidates will receive further attention once more TTs are available.  For example,
KOIs 124.02, 148.03 (see Fig.\ \ref{figKoiGoodCandidates}), 209.01 and 707.03 also appear to have excess scatter and have host stars with multiple transiting planet candidates.  These planet candidates are less likely to be affected by random noise, given their higher S/N transits, but it is difficult to assess whether the scatter is significant, since they have only three or four TTs measured in Q0-2.  

Of the TTV candidates with only 3 or 4 transits observed in Q0-2, {\bf KOI 148.03, 279.01} and {\bf 377.02} stand out, as all three have host stars with multiple transiting planet candidates and a significant offset in epoch relative to the B11 ephemeris.  This is suggestive of long-term trends emerging during Q0-5.  Indeed, KOI 377.02 has already been confirmed as Kepler-9b (Holman \etal 2011).  
In the case of KOI 148.03, the scatter in the TTs of the other planet candidates is small, but there may be a feature in common near 160 days, perhaps due to dynamics.  Clearly, a more detailed analysis incorporating TTs beyond Q0-2 is merited. 

\subsubsection{Strong TTV Candidates}
\label{secStrongCandidates}
%
%
Based on Q0-2 data, the largest timing variations for active planet candidates are {\bf KOIs 227.01, 277.01, 1465.01, 884.02, 103.01, 142.01,} and {\bf 248.01} (starting with largest magnitude of TTVs).  Each of these has well-measured transit times due to a high S/N in each transit, a transit duration of over three hours (minimizing complications due to long cadence and PDC corrections), and a star with limited variability.  Each shows a clear trend of TTVs, indicating at least a period change between Q0-2 and Q0-5.  
Fig.\ \ref{figKoiGoodCandidates} (top row) shows the TTVs of two examples, KOI 103.01 and 142.01.  
For KOIs 142.02 and 227.01, the period derivative can be measured from Q0-2 data alone (see Table \ref{tabQuad}).  Each of these candidates has been tested with centroid motion tests, high-resolution imaging (except 884 and 1465), and at least some spectroscopic observations.  Only KOI 142 shows any indications of the KOI being due to a blend with an eclipsing binary star.  Given the large magnitude of the period derivatives, the transiting body must be strongly perturbed, by a companion that is massive, close and/or in a mean-motion resonance (MMR).  Despite the clear timing variations, we do not consider these confirmed planets, out of an abundance of caution.  
Confirming them would require 
excluding nearly all known possible false positives, as was done for Kepler-9d (Torres et al.\ 2011) and Kepler-10c (Fressin et al.\ 2011).  

For example, it is possible that some (or all) of these may be examples of physically bound triple systems consisting of the bright target star and a low-mass eclipsing binary.  
Eclipse timing variations are not uncommon in the Kepler Binary Star Catalog (Prsa \etal 2011; Slawson \etal 2011; Orosz \etal 2011).  If this is the case, then it may eventually be possible to detect eclipses of additional bodies due to orbital precession, as observed in the triply eclipsing triple system, KOI-126 (Carter \etal 2011).  
Indeed, some of the first KOIs investigated by the \Kepler Transit Timing Working Group have timing variations of a similar magnitude and time scale but turned out to be triple systems (KOIs 646, 928).  Since these systems were selected to have the largest amplitude TT variations during Q0-2, it would not be surprising if they were atypical of multiple transiting planet candidate systems.  The \Kepler Science Team will investigate these systems further to determine which are indeed planets and which are multiple star systems. If they are planets, then the planet radii are $\sim2-3 R_\oplus$, assuming stellar radii from the Kepler Input Catalog (Brown \etal 2011).  Even if all were to turn out to be false positives, the timing variations will play a critical role in understanding the nature of these KOIs.

Several other planet candidates were identified as having significant scatter in their TTs by the $X^{\prime 2}$ test or a difference between our E$_L$2 ephemeris and the E$_L$5 ephemeris of B11.  Further information for these is given in Table \ref{tabNotes}.  A few, such as {\bf KOI 151.01} and {\bf 270.01}, appear to have hints of a pattern in the TTs, but large timing uncertainties make interpretation difficult at this time.  We expect that further TT observations will clarify which are real signals and enable confirmation and/or complete dynamical model modeling.  
One potential source of astrophysical false positives that could masquerade as planets with dynamical TTVs is an isolated star+planet (or eclipsing binary) combined with stellar activity and/or spots.  This is a particular concern for planet candidates that appear to have excessive scatter of their TTVs, rather than a long-term trend or periodic pattern.  The \Kepler Follow-up Observation Program will be conducting a variety of follow-up measurements to help eliminate potential false positives and likely confirming many of the multiple transiting planet candidates.

\subsubsection{Strong TTV Candidates in Multiple Transiting Planet Candidate Systems}
\label{secStrongMultis}
Here we identify only planet candidates with a S/N per transit of at least 4, five or more TTs measured in Q0-2 and a host star with multiple transiting planet candidates:  {\bf KOIs 137.02, 153.01, 244.02, 248.01, 270.01, 528.01, 528.03, 564.01, 658.01, 663.02, 693.02, 884.02, 935.01} and {\bf 954.01}.  For most of these, there is no obvious pattern in the TTVs, perhaps due to measurement errors, or perhaps due to a complex TTV signature.  Visual inspection does reveal several planet candidates with tantalizing patterns in the observed TTs.  

{\bf KOI 884.02} shows the most pronounced pattern, with a min-to-max variation of over an hour within Q0-2.  Any TTVs of KOI 884.01 are only suggestive at this time.  The period ratio with KOI 884.01 is $P_{884.02}/P_{884.01}=2.17$, near, but well beyond the nominal 1:2 MMR.  

{\bf KOI 137.02} shows a significant shift in epoch between the E$_L$2 and E$_L$5 ephemerides.  The variations in KOI 137.01 are not statistically significant on their own, but are suggestive.  The period ratio with KOI 137.01 is $P_{137.02}/P_{137.01}=1.94$, slightly inside the nominal 1:2 MMR. A detailed analysis of this system will be presented in Cochran \etal (2011).  

{\bf KOI 244.02} shows a significant shift in epoch between the E$_L$2 and E$_L$5 ephemerides, but KOI 244.01 does not.  The period ratio with KOI 244.01 is $P_{244.02}/P_{244.01}=2.04$, near the nominal 1:2 MMR.

{\bf KOI 663.02} and {\bf KOI 248.01} show significant shifts in both period and epoch between the E$_L$2 and E$_L$5 ephemerides.  KOI 663.02 is not near a low-order period commensurability with another transiting planet candidate.  KOI 248.01 is near the 3:2 MMR withi 248.02, with a period ratio $P_{248.02}/P_{248.01}=1.52$.

{\bf KOI 528.01} appears to have a long-term trend and/or a periodic pattern of TTVs with a $\sim$20-30 minute min-to-max in Q0-2.  There are not sufficient transits of the other candidates in Q0-2 for a comparison of their TTVs.  KOI 528.01 is slightly beyond the nominal 1:2 MMR with KOI 528.03 with a period ratio $P_{528.03}/P_{528.01}=2.15$.

{\bf KOI 270.01} could have a sizable TTV amplitude ($\sim$40 minutes), but relatively large TT measurement uncertainties prevent such a pattern from being clearly recognized in the present data.  The period ratio with KOI-270.02 is $P_{270.02}/P_{270.01}=2.68$.

\subsection{Candidate Multiple Transiting Planet Systems}
\label{secMultis}
We discuss a few particularly interesting systems with multiple transiting planet candidates.  In order to form an order-of-magnitude estimate of the magnitude of TTVs that are likely to arise in systems with multiple transiting planets, we performed an n-body integration for each of these systems.  We assume coplanar and circular initial conditions and assign masses according to 
\be
M_p/M_\oplus = (R_p/R_\oplus)^{2.06},
\label{eqnMassRadius}
\ee
as described in Lissauer \etal (2011b).  
As an example, the TTs predicted by the baseline model for the four of the planet candidates of KOI 500 are presented in Fig.\ \ref{fig500}.  
The magnitude of TTVs predicted by this baseline model for all stars with multiple transiting planet candidates are reported in Table \ref{TabPredictTTVs}.
As the TTV signature can be very sensitive to masses and initial conditions (Veras \etal 2011), we do not expect that these simulations will accurately model the TTV observations.  
However, they can help us develop intuition for interpreting TT observations.  For example, we can often predict a timescale and associated amplitude of TTVs to within a factor of $\sim~2$, but the phase of the associated TTVs is more sensitive to the detailed initial conditions.  For systems with at least one eccentric planet, there can be additional TTV frequencies with much larger amplitudes.  
the exact phases

\subsubsection{Non-detection of TTVs in Multiple Transiting Planet Candidate Systems}
\label{secNoTtvs}
Our n-body simulations of multiple transiting planet candidate systems indicate that we should not be surprised that many multiple transiting systems have not yet been detected by TTVs.  For our assumed mass-radius relation and circular, coplanar orbits, only KOI 137.01 and 250.02 would have TTVs more than twice the median TT uncertainty and at least three transits during Q0-2.  The only indication of TTVs in KOI 137.01 is from the $X^2$ statistics calculated using the clipped TTs relative to the E$_2$L ephemeris.  Interestingly, we do not detect significant TTVs for 250.02.  One possible explanation is that the planets have smaller masses, resulting in TTVs with a smaller amplitude and/or longer timescale.  Alternatively, the planets may have eccentricities that significantly affect the TTV signal during Q0-2.  Or, there may be additional non-transiting planets that significantly affect the orbital dynamics.  Yet another possibility is that one or both of these KOIs are actually a blend of two planetary systems (or one planetary system and one background eclipsing binary), rather than multiple planets in a single planetary system.  Since we chose these two systems based on the predicted TTV signature being among the largest during Q0-2, it would not be surprising if they were atypical of the multiple transiting planet candidate systems.  Fortunately, simulations predict that the RMS amplitude will grow to over 17 (137.01) and 5 (250.02) times the median timing uncertainty, so further observations are very likely to resolve the nature of these systems.

Looking at the predicted TTVs over the 3.5 year nominal mission lifetime for all the multiple transiting planet candidate systems, we find that at least 25 transiting planet candidates in 12 systems would be expected to have detectable TTVs.  The number of multiple transiting planet systems with detectable TTVs could increase considerably if significant eccentricities are common (Steffen \etal 2010; Veras \etal 2011; Moorhead \etal 2011).

\subsubsection{Dynamical Instability \& Multiple Transiting Planet Candidate Systems}
\label{secInstability}
Our n-body integrations show that the nominal circular, coplanar models of only three KOIs (191, 284 and 730) are violently unstable (Lissauer \etal 2011b).  As it is unlikely that a (presumably old) planetary system would go unstable on a timescale much less than the age of the system, we presume that this is an artifact of our choice of masses and/or orbital parameters.  This also demonstrates the power of dynamical studies to help constrain the masses and orbits of systems of transiting planet candidates. Alternatively, these KOIs could be a blend of two stars, each with one transiting planet.  This would be the natural conclusion, if a future, more thorough investigation were to find that all plausible masses and orbits would quickly result in instability.  
Indeed, we note that KOI 191.02, 284.02 and 284.03 were assigned a vetting flag of 3 in B11, indicating that there is a increased probability of confusion for these candidates.  However, the close commensurabilities of the orbital periods of the planet candidates in KOI 191 (5:4) and 730 (8:6:4:3) make it extremely unlikely for these to be blend scenarios.  The dynamics of systems like KOI-730 is discussed in Fabrycky \etal (in prep).
Assuming KOI 191, 284 and 730 are stable multiple transiting planet systems, they could have very large TTVs due to their strong mutual gravitational interactions that place them near the edge of instability.  As the time span of \Kepler TT observations increases, TTVs become increasingly sensitive to the masses and orbits of such planetary systems.  

\subsubsection{Specific Systems}
%
%
We do not analyze KOI 377 (Kepler-9; Holman \etal 2010), KOI 72 (Kepler-10; Batalha \etal 2011), KOI 157 (Kepler-11; Lissauer \etal 2011a), KOI 137 (Cochran \etal in prep) or KOI 730 (Fabrycky \etal in prep) further, as more thorough analyses have already been published using data beyond Q0-2 or are in preparation.

%
{\em KOI-500} hosts five transiting planet candidates.  If confirmed, they would be even more tightly packed than the planets of Kepler-11 and would indicate that planetary systems like Kepler-11 are not extremely rare (Lissauer \etal 2011b).  The TTs during Q0-2 are shown in Fig.\ \ref{fig500}.  KOI 500.01 was identified as a TTV candidate due to the difference in the best-fit epoch during Q0-2 from that during Q0-5, as reported in B11.  While the apparent discrepancy is suggestive, the Q0-2 data is not sufficient to confirm the planetary nature of the KOI.  This is not surprising as confirming 5 of the 6 planets orbiting Kepler-11 with TTVs required data extending over Q0-6 (Lissauer \etal 2011a).  Fortunately, the nominal model predicts that much larger TTVs ($\sim$hour) will become apparent for four of the 5 planet candidates in future \Kepler data.  The large amplitude is very likely related to the near period commensurabilities of the orbital periods of the outer four planet candidates (4:6:9:12).  Each neighboring pair of planets has a period ratio slightly greater than the nominal MMR, as is typical for near-resonant systems identified by \Kepler (Lissauer \etal 2011b).  Further, each pair deviates from the nominal resonance by a similar amount, strongly suggesting that the four bodies are dynamically interacting.  Unfortunately, \Kepler's nominal mission lifetime is shorter than the dominant timescale of TTVs predicted for KOI-500.  In order to establish that a TTV signal is periodic, one needs to observe $\ge~2$ cycles.  Hopefully, the \Kepler mission can be extended, as the increased time baseline would dramatically enhance the sensitivity of TTV observations to the planet masses and orbital parameters. 

\section{Discussion}
\label{secDiscuss}

\subsection{Precision of Transit Times from Kepler}
Our analysis of TT measurements during Q0-Q2 demonstrate that \Kepler is capable of providing precise transit times which can be expected to enable the dynamical confirmation of transiting planet candidates and detection of non-transiting planets.  After discarding those planet candidates with S/N per transit of less than 4 or just a few transits in Q0-2, the median absolute deviation (MAD) of transit times from a linear ephemeris during Q0-2 is as small as $\sim$20 seconds for Jupiter-size planets, and 1.5 minutes for Neptune and super-Earth-size planets.  The median (taken over planet candidates) MAD of TTs is approximately $\sim$11 minutes for super-Earth-size planets, $\sim$6.5 minutes for Neptune-size planets, and $\sim$1.5 minutes for Jupiter-size planets.  Since large planets can be detected around fainter stars at low signal-to-noise, each class includes some candidates with relatively poor timing precision (up to an hour).  

All the TT measurements presented in this paper were based on LC data only and were performed in a semi-automated fashion.  Experience with systems subjected to detailed TTV studies suggests that TT precision could often be significantly improved, if individual attention to devoted to mitigating systematic effects and choosing which transits yield the best templates.  This work can serve as a broad, but shallow survey of \Kepler planet candidates.  Our results can help scientists choose the most interesting systems for follow-up work to perform more detailed light curve modeling (as well as other types of follow-up such as observations from other observatories, theoretical investigations and observations using short cadence mode).

In addition to LC observations, \Kepler collects photometry in short cadence mode ($\approx$ 1 minute samples) for a small fraction of its targets.  In some cases, short cadence data may further improve timing precision (e.g., Batalha \etal 2011; Carter \etal 2011).   It is expected that the differences in the times and errors inferred from the two data types are small for purely white, Gaussian noise and adequate transit phase coverage.  However, when either precondition is not met, the accuracy of TTs can depend on whether LC or SC data are used.

Short cadence times show improved accuracy and smaller errors for situations where the ingress/egress phases are short in duration and largely unresolved in the long cadence photometry.  In the case of KOI-137, this improvement is significant, with timing errors differing by nearly a factor of two at some epochs. In a similar vein, the short cadence times are more robust against deterministic (as opposed to stochastic) trends that have characteristic timescales less than the long cadence integration time.  In particular, brightening anomalies occurring during transit associated with the occultation of a cool spot on the star by the planet will be unresolved with long cadence photometry and will likely result in a timing bias.

For several planet candidates, the orbital period is a near multiple of the long
cadence integration time.  The result is sparse transit phase coverage that is not improved as the number of observed epochs increases.  In these cases, the short cadence photometry provides a more complete phase coverage and, subsequently, times with smaller error bars.

For low signal-to-noise transit events, stochastic temporally-correlated noise may dominate on short cadence timescales.  As a result, the short cadence timing uncertainties may be unrealistically optimistic if they were inferred assuming a ``white'' noise model.
Here, the long cadence photometry will give more reliable errors when assessed with the same noise model by effectively averaging over high frequency noise.  Short cadence times estimated for Kepler-10b seem to be affected by correlated noise, under the expectation of a linear ephemeris, given the relatively small timing errors and the relatively large deviates (compared to the long cadence times).

\subsection{Frequency of TTV Signals \& Multiple Planet Systems}
\label{secTtvMultiCor}
While TTVs have been used to confirm transiting planet candidates, TTVs have yet to provide a solid detection of a non-transiting planet.  In principle, one strength of the TTV method is that it is sensitive to planets which do not transit the host star (Agol et al.\ 2005; Holman \& Murray 2005).  A longer time baseline will be required before TTVs yield strong detections.  Yet, we can already use the multiple transiting planet candidate systems and the number of preliminary TTV signals to estimate the frequency of multiple planet systems.  

Doppler planet searches find that systems with multiple giant planets are common ($\ge~28\%$; Wright \etal 2009).  \Kepler is probing new regimes of planet mass and orbital separation, so it will be interesting to compare the frequency of multiple planet systems among systems surveyed by Doppler and \Kepler observations.  B11 estimates the false alarm rate of \Kepler planet candidates to range from $\le~2\%$ for the confirmed planets (``vetting flag''=1) to $\le~20\%$ for well-vetted candidates (vetting flag=2) to $\le~40\%$ for those that are yet to be fully vetted (vettting flag=3 or 4).  Morton \& Johnson (2011) use the specifications for \Kepler to predict that even incomplete vetting could result in a false positive rate of less than $\sim$10\% for planet candidates larger than $1.3 R_{\oplus}$ and a typical host star.  The most common mode of false positive involves a blend of multiple stars which are closely superimposed on the sky (Torres \etal 2011).  Contriving such scenarios for KOIs with multiple planet candidates is more difficult.  The odds of three or more physically unassociated stars being blended together is extremely small.  In many cases, requiring dynamical stability excludes the possibility of multiple transit-like events being due to a multiply eclipsing star systems.  The most common types of false positives for a candidate multiple transiting systems are expected to be either a blend of two stars each with one transiting planet or a blend of one star with a transiting planet and one eclipsing binary (Torres \etal 2011).  Even these scenarios become highly improbable for pairs of planets that are close to a MMR.  Thus, candidate multiple transiting planet systems, and especially near-resonant candidate multiple planet systems, are expected to have very few false positives (Holman \etal 2010; Latham \etal 2011; Lissauer \etal 2011b).   

Given the low rate of expected false positives, we can use the frequency of stars with multiple transiting planet candidates to estimate a lower bound on the frequency of systems with multiple planets (with sizes and orbits detectable by \Kepler using the present data), assuming that all systems are coplanar.  Table \ref{tabFreq} lists the number of stars with at least one transiting planet candidate ($N_{st}$) and the number of transiting planet candidates ($N_{tr}$), separated by the number of candidates per star ($N_{cps}$).  The probability that at least two coplanar  candidates transit for a randomly positioned observer is simply $R_\star/a_{(2)}$, where $a_{(2)}$ is the semi-major axis of the planet with the second smallest orbital period.  Among the planet candidates in B11, the average $a/R_\star$ for single planet candidates is $\sim$30.  For stars with two (multiple) transiting planet candidates, the average $a_{(2)}/R_\star$ is 44 (39).  Thus, the difference in the detection rates due to purely geometric considerations would be modest if the two planets are stricly coplanar.  The fraction of \Kepler planet candidate host stars with at least two (exactly two) transiting planets (with sizes and orbits detectable by \Kepler using the present data) is at least $\sim$23\% (17\%).  For stars with at least three transiting planet candidates, the average $a_{(3)}/R_\star$ is 50, so purely geometric considerations are more significant even if the orbits are strictly coplanar.  While only $\sim5.6\%$ of \Kepler planet candidate host stars have at least three candidates, adopting a minimum geometric correction factor of 50/30, yields a fraction of \Kepler planet candidate host stars with at least three similar transiting planets of at least $\sim9.4\%$.  
Such a high occurance rate of multiple candidate systems requires that there be large numbers of systems with one transiting planet where additional (more distant) planets are not seen to transit, even in the most conservative coplanar case. 
Of course, the true rates of multiplicity could be much higher, if systems have significant mutual inclinations.  Lissauer \etal (2011b) provide more detailed analysis of the observed rate of multiple transiting planet systems and its implications for their inclination distribution and multiplicity rate.

Next, we make use of the fact that TTV observations are sensitive to non-transiting planets.  Table \ref{tabFreq} includes multiple values of the number of planet candidates ($N_{\rm TTV}$) that were identified as likely having TTVs by various sets of tests described in \S\ref{secTests} and the number of planet candidates which these tests were applied to ($N_{\rm tr}$).  We evaluate the robustness of our results by applying various sets of tests for TTVs to different samples of planet candidates.  In Table \ref{tabFreq} the columns labeled sSnN refer only to planet candidates with a single transit S/N of at least S and at least N TTs measured during Q0-2.  Some planet candidates in the s4n3 sample have too few transits to apply the $X^{\prime 2}$ test, $\mathcal{F}$-test and period comparison test.  Similarly, for the epoch and period comparison tests, we required a S/N per transit of at least 4, so these tests were not applied to all the planet candidates in the s3n5.  The s4n5 sample is the smallest, but is the least likely to result in false alarms when searching for TTV signals.  To further reduce the risk of false alarms, we do not include the $X^{\prime 2}$ test for excess scatter when analyzing the s4n5 sample.  
The columns labeled ``\% TTV'' give the fraction of planet candidates that were identified by the tests for TTVs that were applied to the given sample. 

We find that $\sim$11-20\% of planet candidates suitable for TTV analysis show some evidence for TTVs, depending on the tests applied (see Table \ref{tabFreq}).  We obtain similar rates when we consider only systems with $N_{cps}=1-3$ transiting planet candidates.  For $N_{cps}\ge~4$, the accuracy of the resulting rates are limited by small number statistics.   

Of planet candidates which are near a 1:2 MMR with another planet candidate, 25\% show some evidence for a long-term trend.  This could be due to planets near the 1:2 MMR being more likely to have large TTVs, but we caution that this results is based on a small sample size and an early estimate for the frequency of TTVs.

Regardless of which sample and tests are chosen, we do not find significant differences in the fraction of planet candidates which show TTV signals as a function of $N_{cps}$.  This suggests stars with a single transiting planet are nearly as likely to have additional planets that cause TTVs, as stars with multiple transiting planets are to have masses and orbits that result in detectable TTVs.  Since large TTVs most naturally arise for systems that are densely packed and/or have pairs of planets in or near a MMR, a system with a single transiting planet that shows TTVs is likely to have a significant dispersion of orbital inclinations.  A dispersion of inclinations has the effect of increasing the probability that a randomly located observer will observe a single planet to transit and decreasing the probability of observing multiple planets to transit.  Thus, a dispersion of inclinations may help explain the relatively small frequency of systems with two transiting planet candidates relative to the frequency of systems with one transiting planet candidate.
However, simply increasing the inclination dispersion to match the ratio of two transiting planet candidate systems to one transiting planet candidate systems fails to produce the observed rate of systems with three or more transiting planet candidates.  This suggests that a single population model is insufficient to explain the observed multiplicity frequencies  (Lissauer \etal 2011b).

\subsection{Frequency of False Positives and Planets in Close Binaries}
The small fraction of systems with very large TTVs is consistent with the notion that the \Kepler planet candidate list has a small rate of false positives.  In particular, physically bound triple systems are one of the most difficult types of astrophysical false positives to completely eliminate (i.e., an eclipsing binary that is diluted by light from a third star).  In many cases, wide triple systems would be recognized based on centroid motion during the transit (B11).  For KOIs that are not near the threshold of detection there is a relatively narrow range of orbital periods that would escape detection by the centroid motion test and be dynamically stable (for stellar masses).  In many cases, such a triple system would exhibit eclipse timing variations, as are often seen in the \Kepler binary star catalog (Prsa \etal 2011; Slawson \etal 2011; Orosz \etal 2011).  We identify only a handful of systems with large period derivatives that are consistent with a stellar triple system.  This suggests that the \Kepler planet list contains few physical triple stars with eclipsing timing variations and that the current \Kepler planet candidates are rarely in a tight binary systems.

\subsection{Future Prospects for TTVs} 
\label{secFuture}
We identify over 60 transiting planet candidates that show significant evidence of TTVs, even on relatively short timescales.  Even for the early TTV candidates identified here, an increased number of transits and time span of \Kepler observations will be necessary before TTVs can provide secure detections of non-transiting planets.  Additionally, follow-up observations to determine the stellar properties and reject possible astrophysical false positives will also be important for confirming planets to be discovered by TTVs.  

We expect the number of TTV candidates to increase considerably as the number and timespan of \Kepler observations increase.  
For non-resonant systems (e.g., Kepler-11), TTVs typically have timescales of order the orbital period, but relatively small amplitudes (Nesvorn\'{y} 2009; Veras \etal 2011).  In this case, \Kepler will be most sensitive when the planets are closely spaced (e.g., Kepler-11).  Our analysis of TTs during Q0-2 identified no periodic signals at a confidence level of $\le 0.01$.  When searching for a simple periodic signal in time series, the minimum detectable signal decreases dramatically as the number of observations increases beyond $\sim$12 observations and continues to decrease faster than classical $\sim~N^{-1/2}$ scaling even as the number of observations grows to $\sim$40 observations.  Thus, over the 3.5 year nominal lifetime of \Kepler the increased number of TT observations will significantly improve {\em Kepler's} sensitivity to closely spaced, non-resonant systems.  In this regime, \Kepler will be most sensitive to TTVs of short-period planets, since they will provide enough transits during the mission lifetime to detect a periodic signal.  Some TTV candidates identified by \Kepler will become targets for ground-based follow-up in the post-\Kepler era.

For planetary systems near a MMR (e.g., Kepler-9), both the amplitude and timescale of the TTVs can be quite large.  Fortunately, continued observations provide the double benefit of increased number of observations and an increasing signal size.  To illustrate this point, we used n-body integrations to predict the RMS TTV of multiple transiting planet candidate systems identified in B11 for a nominal circular, coplanar model (see Fig. \ref{TtvPredictCum}).  During Q0-2 less than 2\% had a TTV signature with an RMS more than 10 minutes, but the fraction grows to over 10\% over the 3.5 year mission lifetime (see Table \ref{TabPredictTTVs}).  Both this result and the $\sim$12\% of suitable planet candidates showing evidence for a long-term drift in TTs suggest that the TTV method will become a powerful tool for detecting non-transiting planets as well as confirming transiting planets.  
For systems with multiple transiting planets, the additional information makes the interpretation of TTVs even more powerful for confirming their planetary nature and that they orbit the same host star (e.g., Holman \etal 2010; Lissauer \etal 2011a).  With continued observations, TTVs become very sensitive to the planet mass and orbital parameters.  

Based on the distribution of orbital periods of \Kepler transiting planet candidates in B11, at least $\sim$16\% of multiple transiting planet candidate systems contain at least one pair of transiting planets close to a 2:1 period commensurability ($1.83\le~P_{\rm out}/P_{\rm in}\le 2.18$).  If we assume that planets near the 1:2 MMR are nearly coplanar, then true rate of detectable planets near the 1:2 MMR (if both had a favorable inclination) is at least $\sim$~25\%.  If a significant fraction of these systems are not in the low inclination regime, then the true rate of pairs of planets near the 1:2 MMR would be even larger (Lissauer \etal 2011b).  This rate is approximately double the rate of systems that show TTVs basd on our initial analysis of early \Kepler data, implying that the number of systems with TTVs could double over the course of the mission.  

In conclusion, transit timing is extremely complementary to Doppler observations form confirming planets.  On one hand, transit timing only works for planets with detectable TTVs.  On the other hand, TTVs can be quite sensitive to low-mass planets that are extremely challenging for Doppler confirmation.  Additionally, TTVs are likely to be particularly useful for confirming some of the \Kepler planet candidates with host stars that are problematic for confirmation via Doppler observations (e.g., faint stars, active stars, hot and/or rapidly rotating stars). 

Of particular interest for the \Kepler mission is whether the transit timing method might be able to discover or confirm rocky planets in the habitable zone.  Due to the TT uncertinaties for Earth-size planets, a detection of TTVs of the rocky planet itself would require a large signal which is only likely if the small planet is near a resonance with another planet.  Further complicating matters, planets in the habitable zone will have only a few transits for solar-mass stars.   Fortunately, \Kepler is detecting many systems with multiple transiting planets, which would open the door to TT measurements of both the planet in the habitable zone and an interior planet (see Fig.\ \ref{figTtvHz}).  Indeed, \Kepler has identified over a dozen planet candidates that are in or near the habitable zone and are associated with stars that have multiple transiting planet candidates.  In most cases, the period ratio between the transiting planet candidates is large, so the TTVs could be small (unless there are additional non-transiting planets).  However, N-body integrations suggest that continued \Kepler observations of several pairs could prove very useful for confirming (or rejecting) these planet candidates based on TT.  Of course, many other stars with a planet candidate in or near the habitable zone may harbor additional non-transiting planets.  The distribution of period ratios of \Kepler transiting planet candidates shows that planets near the 1:2, 2:3 and 1:3 MMRs are not uncommon.  For reference, Kepler-9 b \& c were confirmed on the basis on 9(b)+6(c)=15 TT observations.  Obtaining 15 TTs for two planets in a 1:2 MMR requires observing for 5-6 times the orbital period of the outer planet.  Thus, it is feasible that a transiting planet in the habitable zone identified by \Kepler could be confirmed using TTVs, provided that there is another transiting planet near an interior MMR and that the \Kepler mission were extended to $\ge~6$ years.  The prospects improve significantly for stars less massive than the sun, thanks to the shorter orbital period at the habitable zone.

\acknowledgements  Funding for this mission is provided by NASA's Science Mission Directorate.  We thank the entire Kepler team for the many years of work that is proving so successful.  
We thank David Latham for a careful reading of the manuscript.  
E.B.F acknowledges support by the National Aeronautics and Space Administration under grant NNX08AR04G issued through the Kepler Participating Scientist Program.  This material is based upon work supported by the National Science Foundation under Grant No. 0707203.
D. C. F. and J. A. C. acknowledge support for this work was provided by NASA through Hubble Fellowship grants \#HF-51272.01-A and \#HF-51267.01-A awarded by the Space Telescope Science Institute, which is operated by the Association of Universities for Research in Astronomy, Inc., for NASA, under contract NAS 5-26555.

{\it Facilities:} \facility{Kepler}.

\clearpage

\begin{figure*}
\epsscale{1.0}
\plottwo{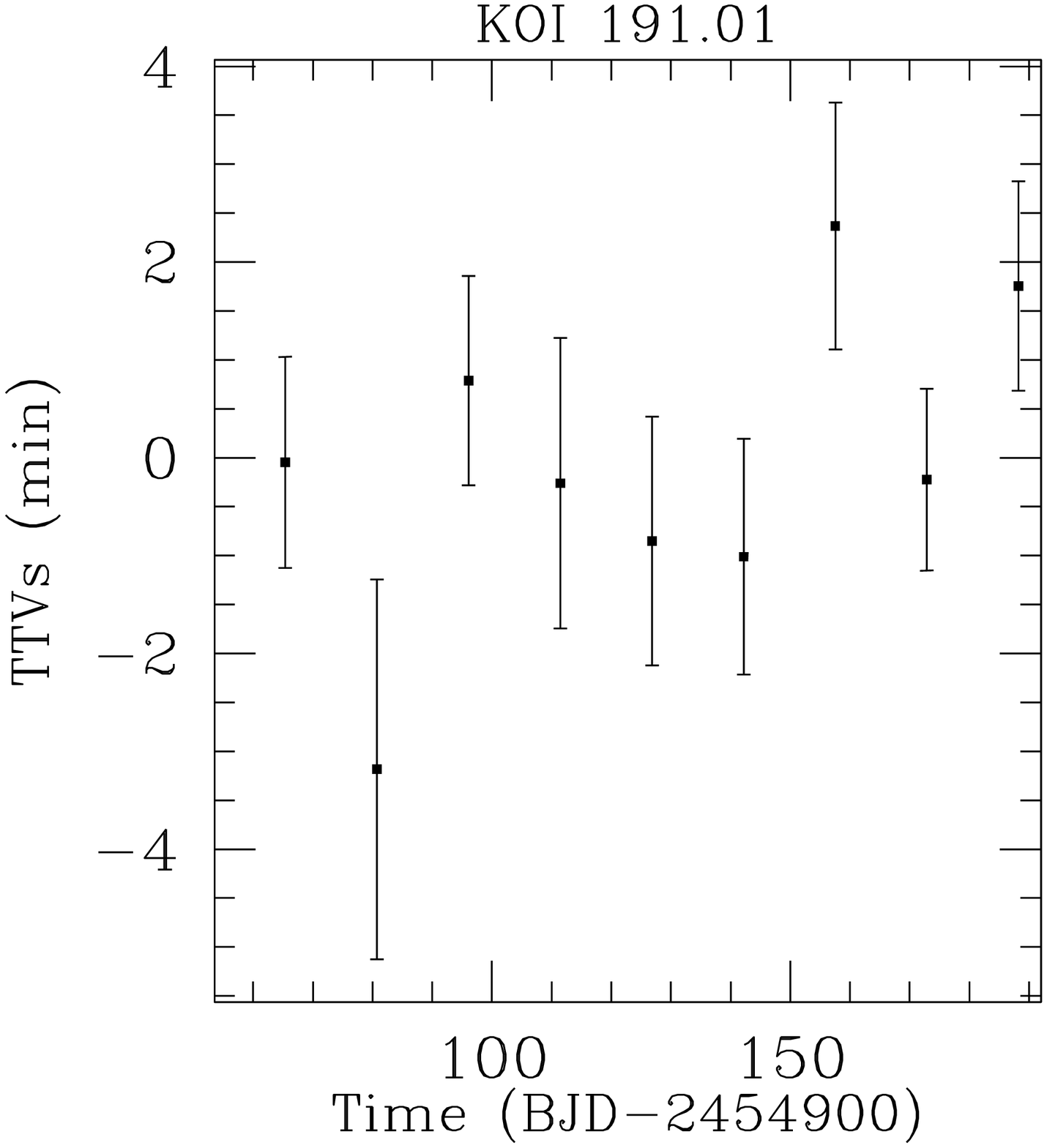}{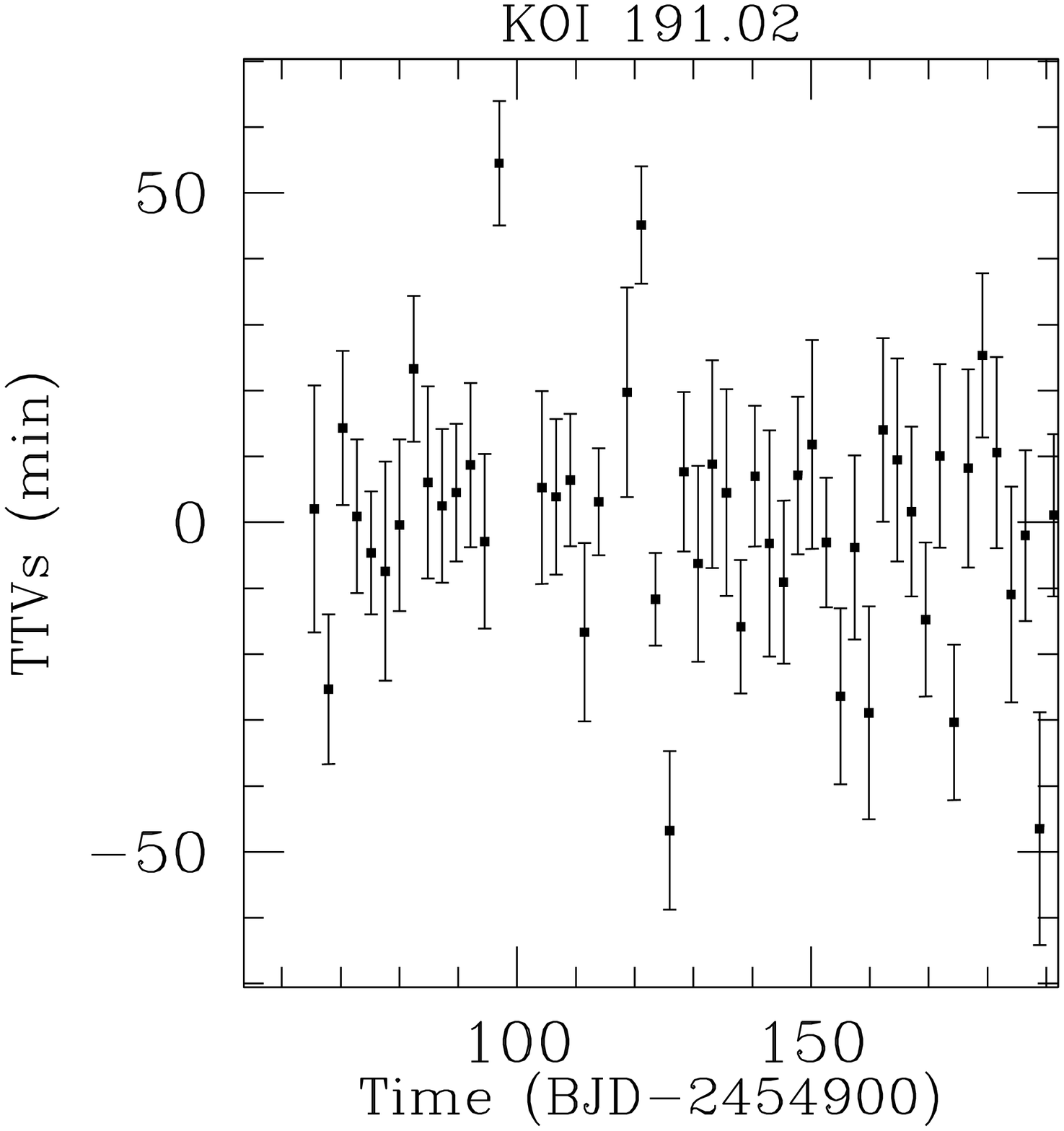}
\plottwo{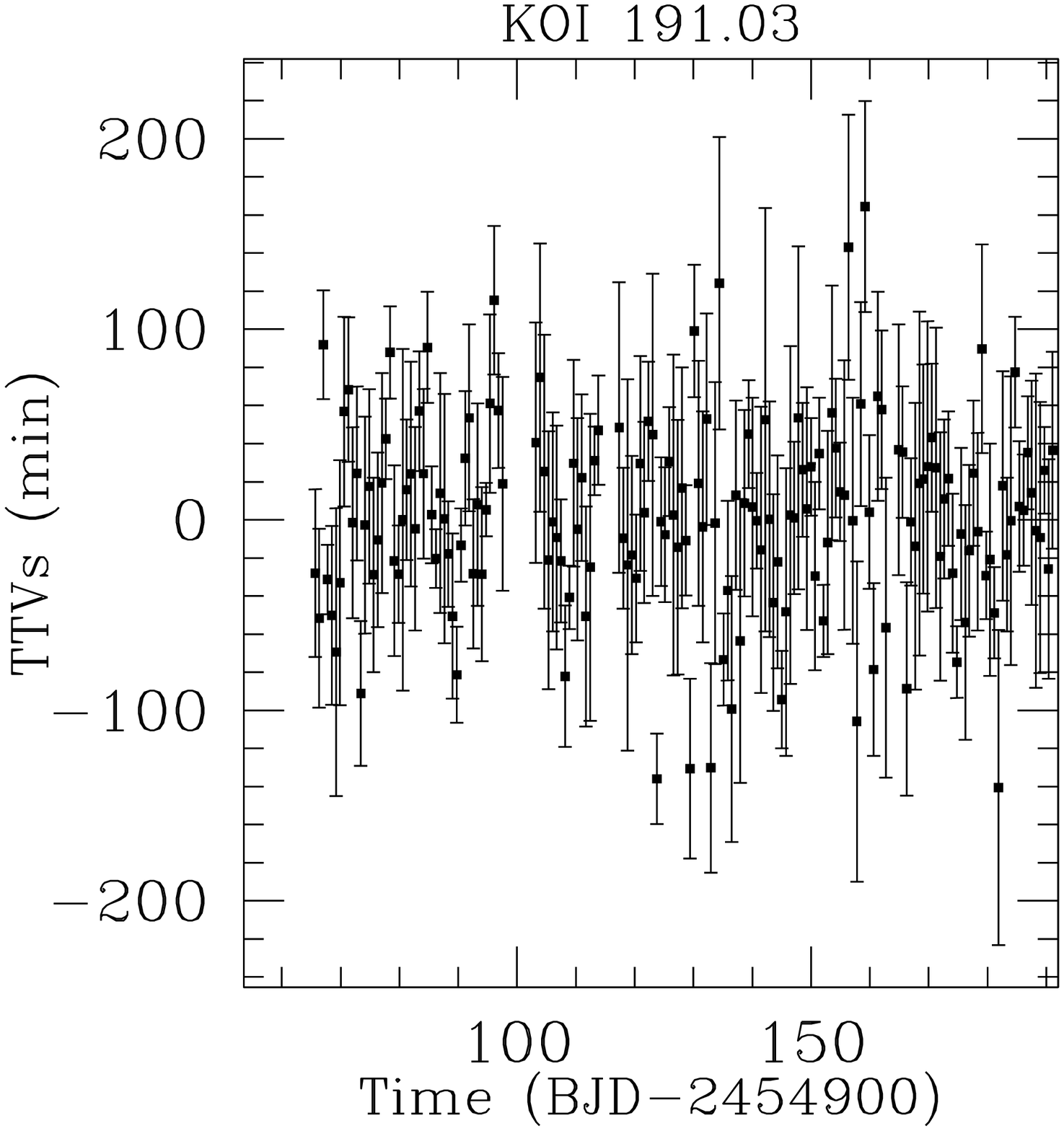}{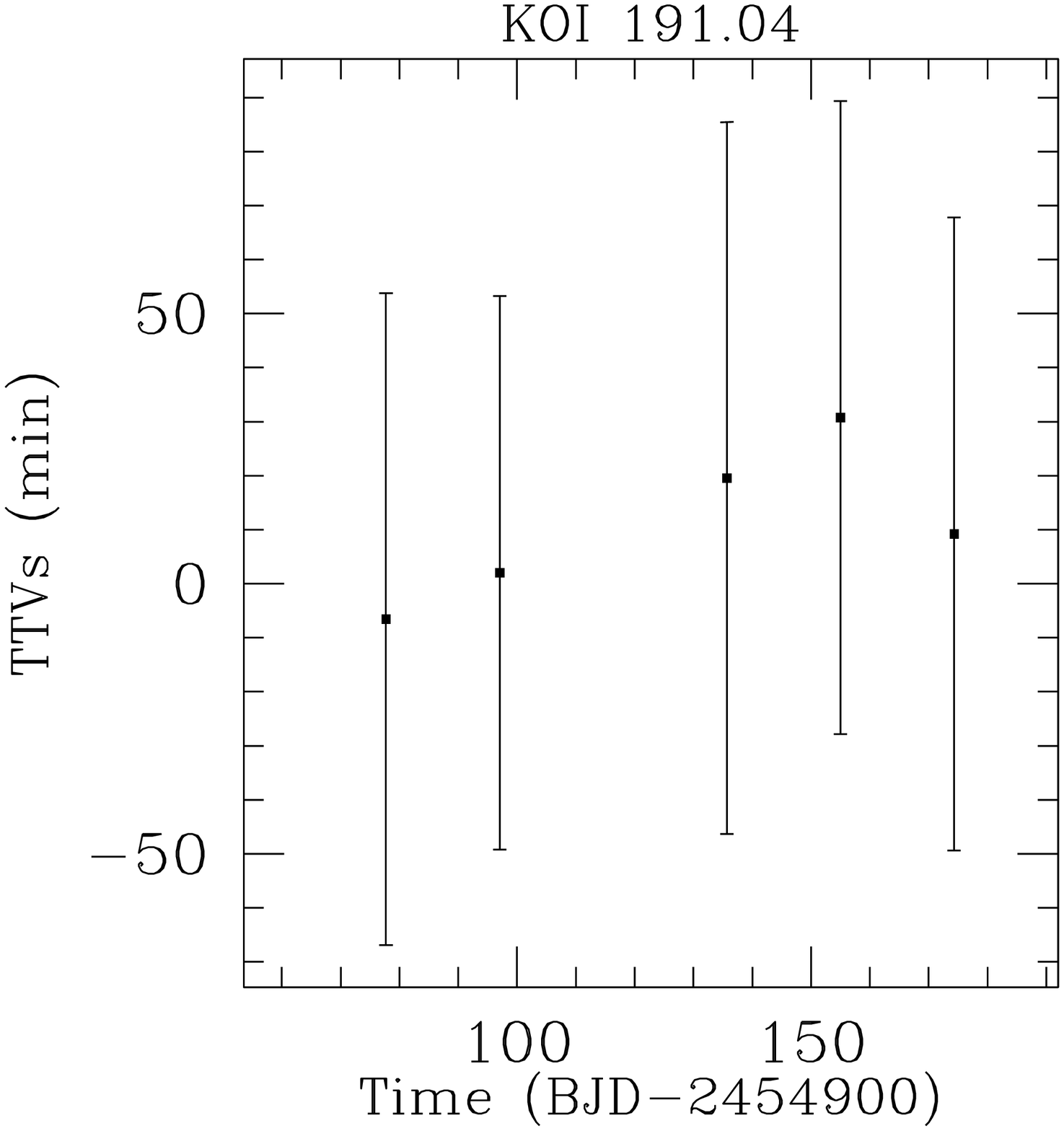}
\caption{Transit timing measurements for four transiting planet candidates associated with KOI 191.  These provide examples of datasets for which we do {\em not} find significant evidence of TTVs.}
\label{figKoiNonDetections}
\end{figure*}

\begin{figure*}
\epsscale{1.0}
\plotone{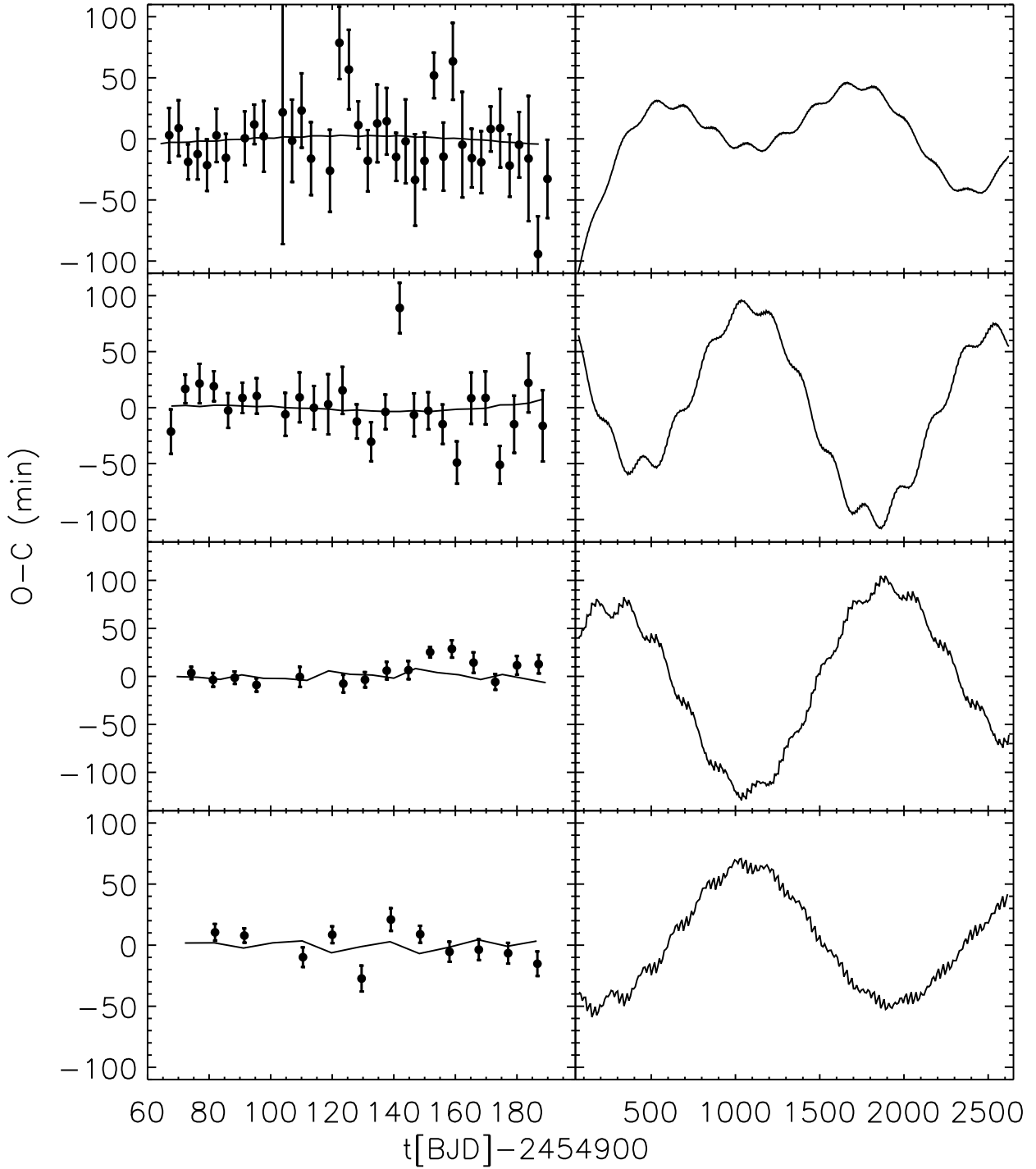}
\caption{Transit timing observations (points, left column only) and the TTVs predicted by n-body integrations (lines). This is not a fit, but rather the output for a nominal circular orbital model (Lissauer \etal 2011).  The right-hand column shows the predictions over 7-years, while the left-hand column zooms in on the first two quarters reported here.  Rows are for KOI 500.03 (top), 500.04 (upper middle), 500.01 (lower middle) and 500.02 (bottom).  KOI 500.05 is not shown, as the TT error bars are $\sim$hour and the model TTVs are less than a second.
\label{fig500}}
\end{figure*}

\begin{figure*}
\epsscale{1.0}
\plotone{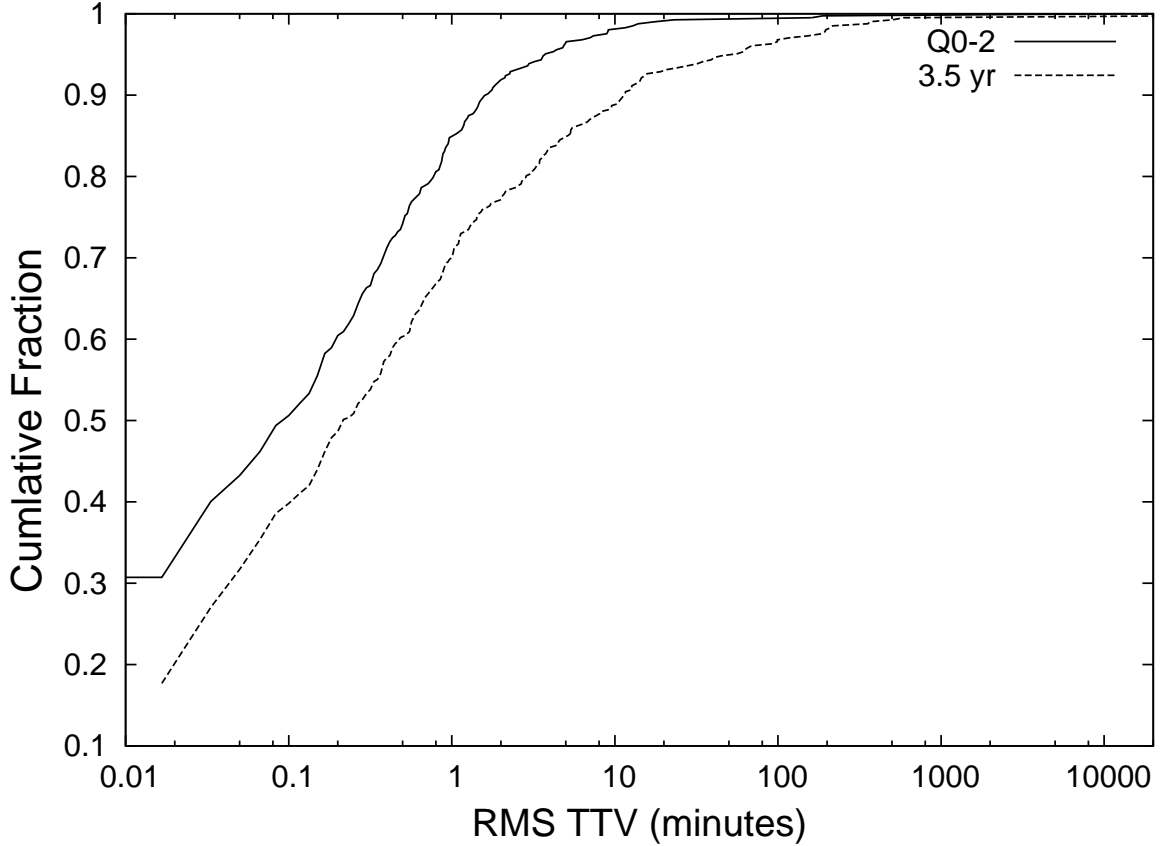}
\caption{Cumulative distribution of the predicted RMS TTVs for systems of multiple transiting planets over the four months of Q0-2 (solid) and the 3.5 years mission lifetime (dashed).  The predictions are based on n-body integrations starting from coplanar and circular orbits.  Actual TTV amplitudes could be much higher for even modest eccentricities.  Even for the case of all circular orbits, over half have a TTV amplitude that is measurable with ground-based follow-up.  At least 10\% of current multiple transiting planet candidates are expected to have amplitudes of $\sim$10 minutes or more, allowing for detailed dynamical modeling based on TTV observations.
\label{TtvPredictCum}}
\end{figure*}

\begin{figure*}
\epsscale{0.7}
\plotone{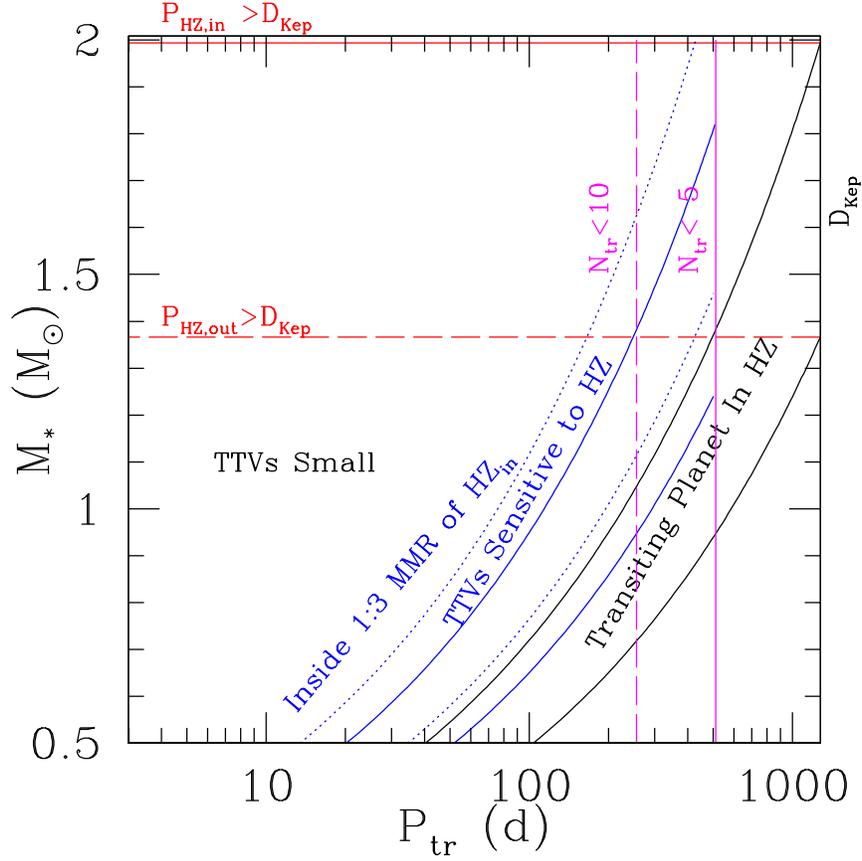}
\caption{Orbital periods and stellar masses for which confirmation of a transiting planet in the habitable zone is practical.  The x-axis is the orbital period of a transiting planet ($P_{\rm tr}$), limited by the nominal mission lifetime ($D_{\rm Kep}=$3.5 years).  The black curve approximate the orbital periods corresponding to the inner and outer edge of the habitable zone.  The solid (dotted) magenta lines indicate the orbital period beyond which \Kepler would observe no more than 5 (10) transits during $D_{\rm Kep}$.  It would be extremely difficult to interpret TTVs for planets to the right of these curves.  Therefore, it is unlikely that the nominal 3.5 year \Kepler mission would measure the masses of planets in the HZ of stars more massive than the sun based on their TTVs.  The solid (dotted) blue curves indicate the orbital period of a planet near the 1:2 (1:3) MMR with the inner and outer edges of the habitable zone.  A second planet significantly to the left of the blue curves will not typically result in detectable TTV signature due to interactions with a planet in the habitable zone.  The solid (dotted) line indicates the stellar mass ($M_\star$) above which the orbital period at the inner (outer) edge of the habitable zone exceeds $D_{\rm Kep}$.  The most promising prospects for TTVs confirming a planet in the habitable zone involve a system with one transiting planet in the habitable zone (between black curves) and a second transiting planet that is between the blue and black curves.  
\label{figTtvHz}}
\end{figure*}

\clearpage

\begin{deluxetable}{lcccc} 
\tabletypesize{\scriptsize}
\tablecaption{Transit Times for \Kepler Transiting Planet Candidates during Q0-2}
\tablewidth{0pt}
\tablehead{
\colhead{KOI} & \colhead{n} & \colhead{$t_n$}        & \colhead{TTV$_n$} & \colhead{$\sigma_n$} \\
\colhead{}    & \colhead{}      & \colhead{BJD-2454900} & \colhead{(d)}  & \colhead{(d)} 
   }
\startdata
 1.01 & 0	& 65.645036	& -0.000204	& 0.000073 \\
 1.01 & 1 	& 68.115649	& -0.000058	& 0.000051 \\
 1.01 & 2 	& 70.586262	& -0.000001 & 0.000079 \\
 1.01 & 3 	& 73.056875	& -0.000042	& 0.000186 \\
 1.01 & 4 	& 75.527488	& 0.000045	& 0.000098 \\
 1.01 & 5 	& 77.998101	& -0.000020	& 0.000103 \\
 1.01 & 6 	& 80.468714	& -0.000041	& 0.000071 \\
 1.01 & 7 	& 82.939327	& -0.000091	& 0.000080 \\
 1.01 & 8 	& 85.409940	& 0.000004	& 0.000074 \\
 1.01 & 9 	& 87.880553	& -0.000123	& 0.000054 \\
\enddata
\tablecomments{
Note that the TTs are measured relative to the EL5 ephemerides given in Borucki et al. 2011b and this is based on transit times measured through Q5. In some cases, a long-term trend manifests itself as all the reported values of TTs having the same sign.  Table \ref{TabTTs} is published in its entirety in the electronic edition of the {\it Astrophysical Journal Supplement}.  A portion is shown here for guidance regarding its form and content.  An electronic version is availiable online at: \url{http://astro.ufl.edu/$sim$eford/data/kepler/}} 
\label{TabTTs}
\end{deluxetable}

\begin{deluxetable}{lcccc}
\tabletypesize{\scriptsize}
\tablecaption{Linear Ephemerides Used for TTV analysis}
\tablewidth{0pt}
\tablehead{
\colhead{KOI} & \colhead{Epoch\tablenotemark{a}} & \colhead{Period} \\ 
\colhead{}    & \colhead{(d)}   & \colhead{(d)} 
}
\startdata
1.01 & 55.762580 & 2.47061310 \\ 
2.01 & 54.357810 & 2.20473550 \\ 
3.01 & 57.812270 & 4.8878177 \\ 
4.01 & 90.526100 & 3.849370 \\ 
5.01 & 65.973500 & 4.7803247 \\ 
7.01 & 56.611260 & 3.213682 \\ 
10.01 & 54.118090 & 3.522297 \\ 
12.01 & 79.597720 & 17.855038 \\ 
13.01 & 53.564980 & 1.7635892 \\ 
17.01 & 54.485750 & 3.2347003 \\ 
\enddata
\tablenotetext{a}{BJD-2454900}
\tablecomments{Table \ref{tabNotes} is published in its entirety in the electronic edition of the {\it Astrophysical Journal Supplement}.  A portion is shown here for guidance regarding its form and content.}
\label{TabEphem}
\end{deluxetable}

 \begin{deluxetable}{lc|cccc|cccc|cc} 
\tabletypesize{\scriptsize}
\rotate
\tablecaption{Metrics for Transit Timing Variations of \Kepler Planet Candidates.}
\tablewidth{0pt}
\tablehead{
\colhead{} & 
\colhead{} & 
\multicolumn{4}{c}{(E$_L$5,T2)} & 
\multicolumn{4}{c}{(E$_L$2,T2)} 
 \\
\colhead{KOI} & 
\colhead{$\sigma_{TT}$} & 
\colhead{MAD\tablenotemark{a}} & \colhead{WRMS\tablenotemark{b}} & \colhead{MAX\tablenotemark{c}} & \colhead{$p_{X^{\prime 2}}$\tablenotemark{d}} & 
\colhead{MAD\tablenotemark{a}} & \colhead{WRMS\tablenotemark{b}} & \colhead{MAX\tablenotemark{c}} & \colhead{$p_{X^{\prime 2}}$\tablenotemark{d}} & 
\colhead{$\frac{\left|\Delta~P\right|}{\sigma_P}$\tablenotemark{e}} & \colhead{$\frac{\left|\Delta~E\right|}{\sigma_E}$\tablenotemark{f}}
\\
\colhead{} & 
\colhead{(min)} & 
\colhead{(min)} & \colhead{(min)} & \colhead{(min)} & \colhead{} & 
\colhead{(min)} & \colhead{(min)} & \colhead{(min)} & \colhead{} &
& \\
}
\startdata
1.01 &  0.2  &  0.1  &  0.1  &  2.4  & 0.99999727 &  0.1  &  0.1  &  2.5  & 0.99989265 &  1.3  &  1.1  \\ 
2.01 &  0.3  &  0.2  &  0.3  &  0.7  & 0.66745223 &  0.1  &  0.3  &  0.8  & 0.99935567 &  1.4  &  1.4  \\ 
3.01 &  0.7  &  0.3  &  0.4  &  3.2  & 0.99982414 &  0.3  &  0.4  &  3.1  & 0.99929283 &  1.2  &  0.6  \\ 
4.01 &  3.5  &  2.4  &  3.5  &  8.6  & 0.84167045 &  2.8  &  3.4  &  9.4  & 0.35265519 &  0.0  &  2.4  \\ 
5.01 &  1.9  &  1.4  &  1.6  &  3.1  & 0.711099 &  1.7  &  1.5  &  3.0  & 0.11421086 &  2.2  &  0.6  \\
\enddata
\tablecomments{Table \ref{TabMetrics} is published in its entirety in the electronic edition of the {\it Astrophysical Journal Supplement}.  A portion is shown here for guidance regarding its form and content.}
\tablenotetext{a}{Median Absolute Deviation of transit times in Q0-2 from ephemeris}
\tablenotetext{b}{Weighted Root Mean Square deviation of transit times in Q0-2 from ephemeris}
\tablenotetext{c}{MAXimum absolute deviation of transit times in Q0-2 from ephemeris}
\tablenotetext{d}{$p$-value for a $\chi^2$-like-test assuming $X^{\prime 2}$ follows a $\chi^2$ distribution, as described in \S\ref{secTestScatter}}
\tablenotetext{e}{Absolute value of difference of best-fit periods for L$_2$,T2 and L$_5$,T2 ephemerides normalized by formal uncertainty}
\tablenotetext{f}{Absolute value of difference of best-fit transit epochs for L$_2$,T2 and L$_5$,T2 ephemerides normalized by formal uncertainty}
\label{TabMetrics}
\end{deluxetable}

\begin{deluxetable}{rccccccccccc} 
\tabletypesize{\scriptsize}
\rotate
\tablecaption{Quadratic Ephemerides for Kepler Objects of Interest based on transit times during Q2.}
\tablewidth{0pt}
\tablehead{
\colhead{KOI} & 
\colhead{$E$\tablenotemark{a}} & 
\colhead{$\sigma_E$} & 
\colhead{$P$} & 
\colhead{$\sigma_P$} & 
\colhead{$c$\tablenotemark{b}} & 
\colhead{$\sigma_c$} &
\colhead{MAD\tablenotemark{c}} & 
\colhead{WRMS\tablenotemark{d}} & 
\colhead{MAX\tablenotemark{e}} & 
\colhead{$p_{X^{\prime 2}}$\tablenotemark{f}} & 
\colhead{$p_{\mathcal{F}}$\tablenotemark{g}} \\
\colhead{} & \colhead{(d)} & \colhead{(d)} & \colhead{(d)} & \colhead{(d)} & \colhead{} & \colhead{} & \colhead{(d)} & \colhead{(d)} & \colhead{(d)} & \colhead{} & \colhead{} }
\startdata
42.01 & 149.89207 & 0.002317258 & 17.835806 & 0.000999096 & -7.93E-05 & 4.83E-05 & 0.000224292 & 0.000465821 & 0.00087804 & 0.98156568 & 0.053661725 \\
124.02 & 107.5387 & 0.004856891 & 31.713798 & 0.003447988 & 0.000155781 & 9.53E-05 & 0.000101866 & 8.70E-05 & 0.000177141 & 0.96675223 & 0.018503953 \\
142.01 & 120.58229 & 0.0014794 & 10.912549 & 0.000260135 & 6.28E-05 & 7.86E-06 & 0.001685495 & 0.002572359 & 0.00727168 & 0.67069648 & 0.005875349 \\
227.01 & 122.54014 & 0.001542628 & 17.678654 & 0.000469767 & -7.68E-05 & 1.55E-05 & 0.000969901 & 0.001511303 & 0.003312192 & 0.82842292 & 0.028645506 \\
314.01 & 124.63071 & 0.001690889 & 13.780802 & 0.000429775 & 3.71E-05 & 1.52E-05 & 0.002035933 & 0.001290098 & 0.003126117 & 0.35704363 & 0.078718429 \\
467.01 & 151.46194 & 0.001891538 & 18.007959 & 0.000842961 & -4.92E-05 & 2.47E-05 & 6.71E-05 & 6.55E-05 & 0.000102345 & 0.93992411 & 0.029226395 \\
528.01 & 138.41866 & 0.003407732 & 9.578232 & 0.000663483 & -5.97E-05 & 2.13E-05 & 0.002067159 & 0.003516547 & 0.007456802 & 0.99547272 & 0.035870139 \\
649.01 & 139.37393 & 0.010461445 & 23.447238 & 0.004043237 & -0.000165971 & 0.000148364 & 0.000121626 & 0.000120244 & 0.000228574 & 0.98082178 & 0.017380622 \\
\enddata
\tablenotetext{a}{Epoch}
\tablenotetext{b}{curvature; see Eqn.\ \ref{eqnQuad}}
\tablenotetext{c}{Median Absolute Deviation of TTs from quadratic ephemeris}
\tablenotetext{d}{Weighted Root Mean Square deviation of TTs from quadratic ephemeris}
\tablenotetext{e}{MAXimum absolute deviation of TTs from quadratic ephemeris}
\tablenotetext{f}{$p$-value for a $chi^2$-test based on $X^{\prime 2}$ relative to quadratic ephemeris, as described in \S\ref{secTestScatter}}
\tablenotetext{g}{$p$-value for an $F$-like-test that compares linear and quadratic ephemerides, assuming that $\mathcal{F}$ follows an $F$-distribution, as described in \S\ref{secTestScatter}}
\label{tabQuad}
\end{deluxetable}



\begin{deluxetable}{rccccrccl} 
\tabletypesize{\scriptsize}
\tablecaption{Notes for \Kepler Planet Candidates with Putative Transit Timing Variations.}
\tablewidth{0pt}
\tablehead{
\colhead{KOI} & 
\colhead{$P$} & 
\colhead{$R_p$\tablenotemark{a}} & 
\colhead{S/N\tablenotemark{b}} & 
\colhead{$T_{dur}$\tablenotemark{c}} & 
\colhead{nTT\tablenotemark{d}} &
\colhead{nPC\tablenotemark{e}} &
\colhead{TTV\tablenotemark{f}} & 
\colhead{Comment} \\
\colhead{} & 
\colhead{(d)} & \colhead{($R_\oplus$)} & \colhead{} & \colhead{(hr)} & \colhead{} & \colhead{ } & \colhead{Flag} }
\startdata
10.01&  3.52230 &  10.5 &  22.6 &  3.3 & 35& 1& 2& epoch offset \\ 
13.01&  1.76359 &  20.4 &  130.7 &  3.2 & 35& 1& 3& outlier \\ 
42.01&  17.83278 &  2.6 &  10.8 &  4.5 & 5& 1& 2& quadratic? \\ 
94.02&  10.42361 &  4.0 &  10.1 &  5.3 & 3& 3& 2& offset \\ 
103.01&  14.91155 &  2.3 &  14.5 &  3.4 & 7& 1& 1& Period \& epoch differ \\ 
\enddata
\tablecomments{Table \ref{tabNotes} is published in its entirety in the electronic edition of the {\it Astrophysical Journal Supplement}.  A portion is shown here for guidance regarding its form and content.}
\tablenotetext{a}{Putative radius of planet in Earth radii (from B11)}
\tablenotetext{b}{Typical Signal to Noise Ratio of an individual transit}
\tablenotetext{c}{Transit duration (from B11)}
\tablenotetext{d}{Number of transit times measured in Q0-2}
\tablenotetext{e}{Number of transiting planet candidates for host star}
\tablenotetext{f}{1=pattern to eye, 2=trend or periodicity, 3=excess scatter and no trend, 4=low S/N per transit and/or few transits, 5=note about difficulty measuring TTs, 6=excess scatter significant only after clipping}
\label{tabNotes}
\end{deluxetable}

\begin{deluxetable}{lcccccc} 
\tabletypesize{\scriptsize}
\tablecaption{Predicted Transit Time Variation Magnitude for \Kepler Transiting Planet Candidates}
\tablewidth{0pt}
\tablehead{
\colhead{KOI} & \colhead{RMS Q2} & \colhead{Min-to-Max Q2} & \colhead{RMS 3.5yr} & \colhead{Min-to-Max 3.5yr} & \colhead{RMS 7yr} & \colhead{Min-to-Max 7yr} \\
\colhead{   } & \colhead{(s)} & \colhead{(s)} & \colhead{(s)} & \colhead{(s)} & \colhead{(s)} & \colhead{(s)}  }
\startdata
70.01 & 1.2 & 3.5 & 1.3 & 5.1 & 1.3 & 5.2 \\ 
70.02 & 1.4 & 5 & 1.3 & 6 & 1.3 & 6 \\ 
70.03 & \nodata & \nodata & 3.4 & 9.7 & 3.4 & 10 \\ 
70.04 & 23 & 71 & 23 & 91 & 23 & 94 \\ 
72.01 & 0.0003 & 0.001 & 0.003 & 0.01 & 0.01 & 0.05 \\ 
72.02 & 0.01 & 0.03 & 0.2 & 0.5 & 0.2 & 0.5 \\ 
\enddata
\tablecomments{We report the magnitude (root mean square and min-to-max) of transit timing variations expected based on n-body integrations using estimated nominal masses and initially circular orbits (Lissauer et al.\ 2011b).  We assume that all members of multiple planet candidate systems are true planets and orbit the same star.  Integrations extend for the duration of the first two quarters of {\em Kepler} data, the nominal 3.5 year mission life time and 7.5 years, representative of a hypothetical extended mission.  Eccentric models can dramatically affect both the predicted TTV magnitude and timescale.
Table \ref{TabPredictTTVs} is published in its entirety in the electronic edition of the {\it Astrophysical Journal Supplement}.  A portion is shown here for guidance regarding its form and content.  } 
\label{TabPredictTTVs}
\end{deluxetable}

\begin{deluxetable}{cccccccccccc} 
\tabletypesize{\scriptsize}
\tablecaption{TTV Candidates \& Transiting Planet Candidates}
\tablewidth{0pt}
\tablehead{
\colhead{$N_{\mathrm cps}$\tablenotemark{a}} &  
\colhead{$N_{\mathrm st}$\tablenotemark{b}} & 
\colhead{$N_{\rm tr}$\tablenotemark{c}} & 
\colhead{$N_{\rm tr}$\tablenotemark{d}} & 
\colhead{$N_{\rm tr}$\tablenotemark{e}} & 
\colhead{$N_{\rm tr}$\tablenotemark{f}} & 
\colhead{$N_{\rm TTV}$\tablenotemark{g}} & 
\colhead{$N_{\rm TTV}$\tablenotemark{h}} & 
\colhead{$N_{\rm TTV}$\tablenotemark{i}} & 
\colhead{\% TTV\tablenotemark{j}} &
\colhead{\% TTV\tablenotemark{k}} &
\colhead{\% TTV\tablenotemark{l}} \\
\colhead{} &  
\colhead{} & 
\colhead{} & 
\colhead{s4n3} & 
\colhead{s3n5} & 
\colhead{s4n5} & 
\colhead{s4n3} & 
\colhead{s3n5} & 
\colhead{s4n5} & 
\colhead{s4n3} &
\colhead{s3n5} &
\colhead{s4n5} 
} 
\startdata
1   & 809 & 809 & 462 & 453 &371&  95 & 59 & 45 & 21\% & 13\% & 12\% \\
2   & 115 & 230 & 121 & 110 & 83&  24 & 16 & 13 & 20\% & 15\% & 16\% \\
3   &  45 & 135 &  71 &  67 & 57&  14 &  8 &  6 & 20\% & 12\% & 11\% \\
4   &   8 &  32 &  15 &  14 & 10&   2 &  1 &  1 & 13\% &  7\% & 10\% \\
5   &   1 &   5 &   2 &   3 &  2&   1 &  1 &  0 & 50\% & 33\% &  0\% \\
6   &   1 &   6 &   4 &   3 &  3&   0 &  0 &  0 &  0\% &  0\% &  0\% \\
All & 979 &1217 & 675 & 650 &526& 136 & 85 & 65 & 20\% & 13\% & 12\% \\
\enddata
\tablecomments{Number and fractions of systems with evidence showing for TTVs separated by the number of transiting planet candidates in the system}
\tablenotetext{a}{Number of candidates per star}
\tablenotetext{b}{Number of stars}
\tablenotetext{c}{Number of transiting planet candidates (only those used for this analysis)}
\tablenotetext{d}{Number of transiting planet candidates with a S/N per transit $\ge~4$ and at least 3 transits in Q0-2}
\tablenotetext{e}{Number of transiting planet candidates with a S/N per transit $\ge~3$ and at least 5 transits in Q0-2}
\tablenotetext{f}{Number of transiting planet candidates with a S/N per transit $\ge~4$ and at least 5 transits in Q0-2}
\tablenotetext{g}{Number of transiting planet candidates from s4n3 sample that show indications of TTVs}
\tablenotetext{h}{Number of transiting planet candidates from s3n5 sample that show indications of TTVs, excluding epoch offset test}
\tablenotetext{i}{Number of transiting planet candidates from s4n5 sample that show indications of TTVs, excluding $X^{\prime 2}$ test}
\tablenotetext{j}{\% of planets from s4n3 sample that show indications of TTVs}
\tablenotetext{k}{\% of planets from s3n5 sample that show indications of TTVs, excluding the epoch offset test}
\tablenotetext{l}{\% of planets from s4n5 sample that show indications of TTVs, ecluding $X^{\prime 2}$ test}
\label{tabFreq}
\end{deluxetable}

\clearpage

\setcounter{table}{3}
\begin{deluxetable}{lc|cccc|cccc|cc} 
\tabletypesize{\scriptsize}
\rotate
\tablecaption{Metrics for Transit Timing Variations of \Kepler Planet Candidates.}
\tablewidth{0pt}
\tablehead{
\colhead{} & 
\colhead{} & 
\multicolumn{4}{c}{(E$_L$5,T2)} & 
\multicolumn{4}{c}{(E$_L$2,T2)} 
 \\
\colhead{KOI} & 
\colhead{$\sigma_{TT}$} & 
\colhead{MAD\tablenotemark{a}} & \colhead{WRMS\tablenotemark{b}} & \colhead{MAX\tablenotemark{c}} & \colhead{$p_{X^{\prime 2}}$\tablenotemark{d}} & 
\colhead{MAD\tablenotemark{a}} & \colhead{WRMS\tablenotemark{b}} & \colhead{MAX\tablenotemark{c}} & \colhead{$p_{X^{\prime 2}}$\tablenotemark{d}} & 
\colhead{$\frac{\left|\Delta~P\right|}{\sigma_P}$\tablenotemark{e}} & \colhead{$\frac{\left|\Delta~E\right|}{\sigma_E}$\tablenotemark{f}}
\\
\colhead{} & 
\colhead{(min)} & 
\colhead{(min)} & \colhead{(min)} & \colhead{(min)} & \colhead{} & 
\colhead{(min)} & \colhead{(min)} & \colhead{(min)} & \colhead{} &
& \\
}
\startdata
1.01 &  0.2  &  0.1  &  0.1  &  2.4  & 0.99999727 &  0.1  &  0.1  &  2.5  & 0.99989265 &  1.3  &  1.1  \\ 
2.01 &  0.3  &  0.2  &  0.3  &  0.7  & 0.66745223 &  0.1  &  0.3  &  0.8  & 0.99935567 &  1.4  &  1.4  \\ 
3.01 &  0.7  &  0.3  &  0.4  &  3.2  & 0.99982414 &  0.3  &  0.4  &  3.1  & 0.99929283 &  1.2  &  0.6  \\ 
4.01 &  3.5  &  2.4  &  3.5  &  8.6  & 0.84167045 &  2.8  &  3.4  &  9.4  & 0.35265519 &  0.0  &  2.4  \\ 
5.01 &  1.9  &  1.4  &  1.6  &  3.1  & 0.711099 &  1.7  &  1.5  &  3.0  & 0.11421086 &  2.2  &  0.6  \\ 
7.01 &  3.5  &  2.1  &  3.4  &  18.3  & 0.98587454 &  2.1  &  3.2  &  15.8  & 0.97553341 &  1.5  &  1.4  \\ 
10.01 &  0.8  &	 18.3 &	 21.0 &	 35.2 &	0	 & 6.2 	&  7.1 &  12.7 &	0.015052582 &	 2.6 \\
12.01 &  1.0  &  0.7  &  1.3  &  2.1  & 0.60479879 &  0.7  &  0.7  &  1.0  & 0.28837559 &  0.8  &  1.3  \\
\enddata
\tablecomments{Table \ref{TabMetrics} is published in its entirety in the electronic edition of the {\it Astrophysical Journal Supplement}.  A portion is shown here for guidance regarding its form and content.}
\tablenotetext{a}{Median Absolute Deviation of transit times in Q0-2 from ephemeris}
\tablenotetext{b}{Weighted Root Mean Square deviation of transit times in Q0-2 from ephemeris}
\tablenotetext{c}{MAXimum absolute deviation of transit times in Q0-2 from ephemeris}
\tablenotetext{d}{$p$-value for a $\chi^2$-like-test assuming $X^{\prime 2}$ follows a $\chi^2$ distribution, as described in \S~3.1}
\tablenotetext{e}{Absolute value of difference of best-fit periods for L$_2$,T2 and L$_5$,T2 ephemerides normalized by formal uncertainty}
\tablenotetext{f}{Absolute value of difference of best-fit transit epochs for L$_2$,T2 and L$_5$,T2 ephemerides normalized by formal uncertainty}
\label{TabMetrics}
\end{deluxetable}

\setcounter{table}{5}
\begin{deluxetable}{rccccrccl} 
\tabletypesize{\scriptsize}
\tablecaption{Notes for \Kepler Planet Candidates with Putative Transit Timing Variations.}
\tablewidth{0pt}
\tablehead{
\colhead{KOI} & 
\colhead{$P$} & 
\colhead{$R_p$\tablenotemark{a}} & 
\colhead{S/N\tablenotemark{b}} & 
\colhead{$T_{dur}$\tablenotemark{c}} & 
\colhead{nTT\tablenotemark{d}} &
\colhead{nPC\tablenotemark{e}} &
\colhead{TTV\tablenotemark{f}} & 
\colhead{Comment} \\
\colhead{} & 
\colhead{(d)} & \colhead{($R_\oplus$)} & \colhead{} & \colhead{(hr)} & \colhead{} & \colhead{ } & \colhead{Flag} }
\startdata
10.01&  3.52230 &  10.5 &  22.6 &  3.3 & 35& 1& 2& epoch offset \\ 
13.01&  1.76359 &  20.4 &  130.7 &  3.2 & 35& 1& 3& outlier \\ 
42.01&  17.83278 &  2.6 &  10.8 &  4.5 & 5& 1& 2& quadratic? \\ 
94.02&  10.42361 &  4.0 &  10.1 &  5.3 & 3& 3& 2& offset \\ 
103.01&  14.91155 &  2.3 &  14.5 &  3.4 & 7& 1& 1& Period \& epoch differ \\ 
124.02&  31.71954 &  2.8 &  10.7 &  5.1 & 4& 2& 4& few TTs; quadratic? \\ 
131.01&  5.01418 &  9.0 &  78.6 &  4.7 & 16& 1& 3& plausible \\ 
137.01& 	7.6415774 &		6.0 &	22.0 &	3.5 &	16&	3&	5&  if clipped \\
\enddata
\tablenotetext{a}{Putative radius of planet in Earth radii (from B11)}
\tablenotetext{b}{Typical Signal to Noise Ratio of an individual transit}
\tablenotetext{c}{Transit duration (from B11)}
\tablenotetext{d}{Number of transit times measured in Q0-2}
\tablenotetext{e}{Number of transiting planet candidates for host star}
\tablenotetext{f}{1=pattern to eye, 2=trend or periodicity, 3=excess scatter and no trend, 4=low S/N per transit and/or few transits, 5=note about difficulty measuring TTs, 6=excess scatter significant only after clipping}
\label{tabNotes}
\end{deluxetable}

\setcounter{table}{6}
\begin{deluxetable}{lcccccc} 
\tabletypesize{\scriptsize}
\tablecaption{Predicted Transit Time Variation Magnitude for \Kepler Transiting Planet Candidates}
\tablewidth{0pt}
\tablehead{
\colhead{KOI} & \colhead{RMS Q2} & \colhead{Min-to-Max Q2} & \colhead{RMS 3.5yr} & \colhead{Min-to-Max 3.5yr} & \colhead{RMS 7yr} & \colhead{Min-to-Max 7yr} \\
\colhead{   } & \colhead{(s)} & \colhead{(s)} & \colhead{(s)} & \colhead{(s)} & \colhead{(s)} & \colhead{(s)}  }
\startdata
70.01 & 1.2 & 3.5 & 1.3 & 5.1 & 1.3 & 5.2 \\ 
70.02 & 1.4 & 5 & 1.3 & 6 & 1.3 & 6 \\ 
70.03 & \nodata & \nodata & 3.4 & 9.7 & 3.4 & 10 \\ 
70.04 & 23 & 71 & 23 & 91 & 23 & 94 \\ 
72.01 & 0.0003 & 0.001 & 0.003 & 0.01 & 0.01 & 0.05 \\ 
72.02 & 0.01 & 0.03 & 0.2 & 0.5 & 0.2 & 0.5 \\ 
82.01 & 161 & 484 & 183 & 612 & 185 & 609 \\ 
82.02 & 442 & 1303 & 484 & 1624 & 485 & 1639 \\
\enddata
\tablecomments{We report the magnitude (root mean square and min-to-max) of transit timing variations expected based on n-body integrations using estimated nominal masses and initially circular orbits (Lissauer et al.\ 2011b).  We assume that all members of multiple planet candidate systems are true planets and orbit the same star.  Integrations extend for the duration of the first two quarters of {\em Kepler} data, the nominal 3.5 year mission life time and 7.5 years, representative of a hypothetical extended mission.  Eccentric models can dramatically affect both the predicted TTV magnitude and timescale.
Table \ref{TabPredictTTVs} is published in its entirety in the electronic edition of the {\it Astrophysical Journal Supplement}.  A portion is shown here for guidance regarding its form and content.  } 
\label{TabPredictTTVs}
\end{deluxetable}

\clearpage
\clearpage
\section*{Appendix}

Here we provide an appendix (online only) of TTVs for several \Kepler planet candidates of particular interest.  The TT are plotted relative to the E$_L$5 ephemeris provided by B11.  
Using the Q0-Q2 ephemeris would cause the weighted average of TTVs was zero.  Thus, an offset of the weighted average of plotted TTVs (relative to zero) and/or a slope of the TTVs may indicate a gradual change in the orbital period between the ephemeris measured based on Q0-2 and the ephemeris based on Q0-5 data and provided in B11.

\section*{ TTV Candidates}
%
\begin{figure*}\plottwo{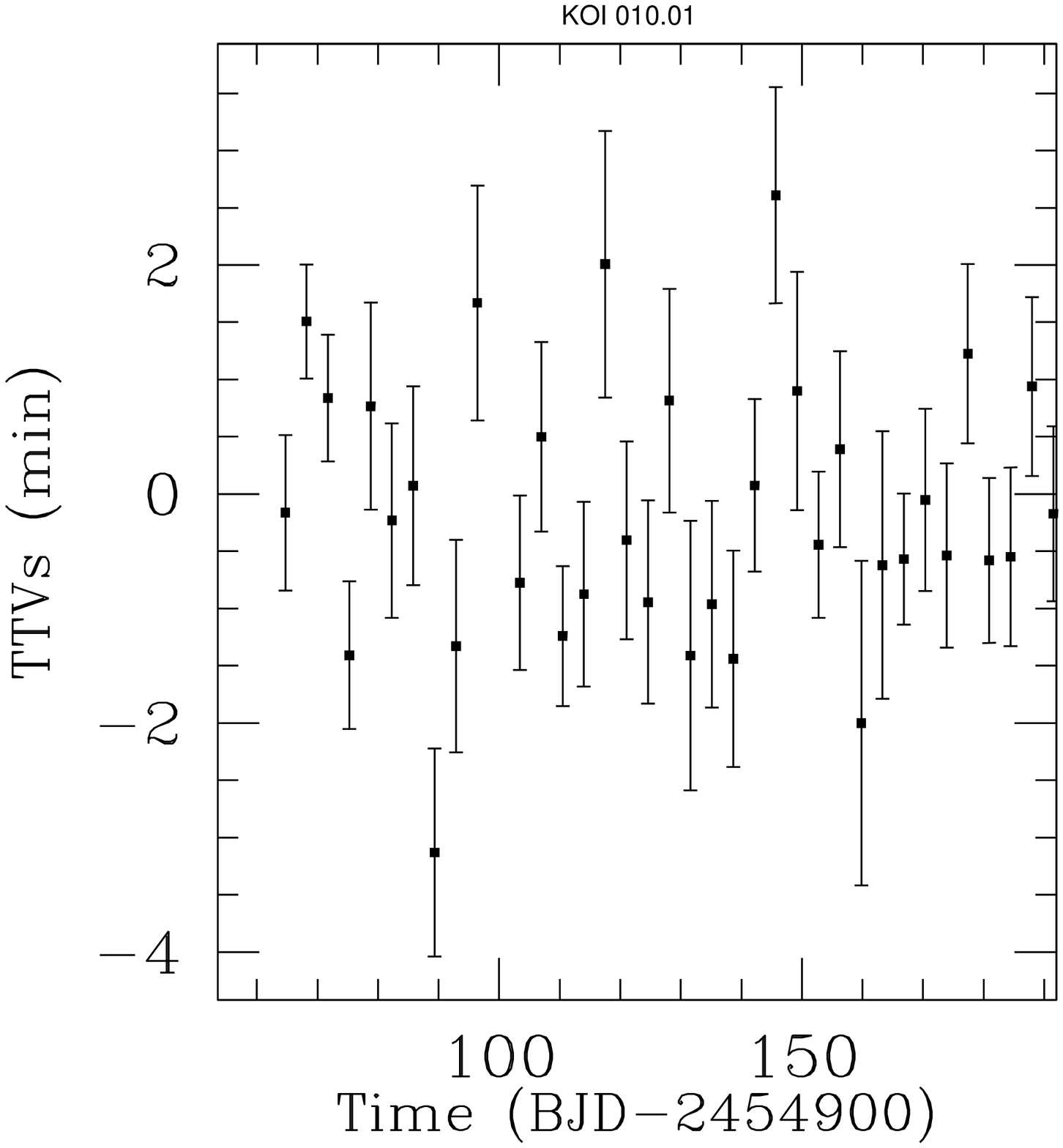}{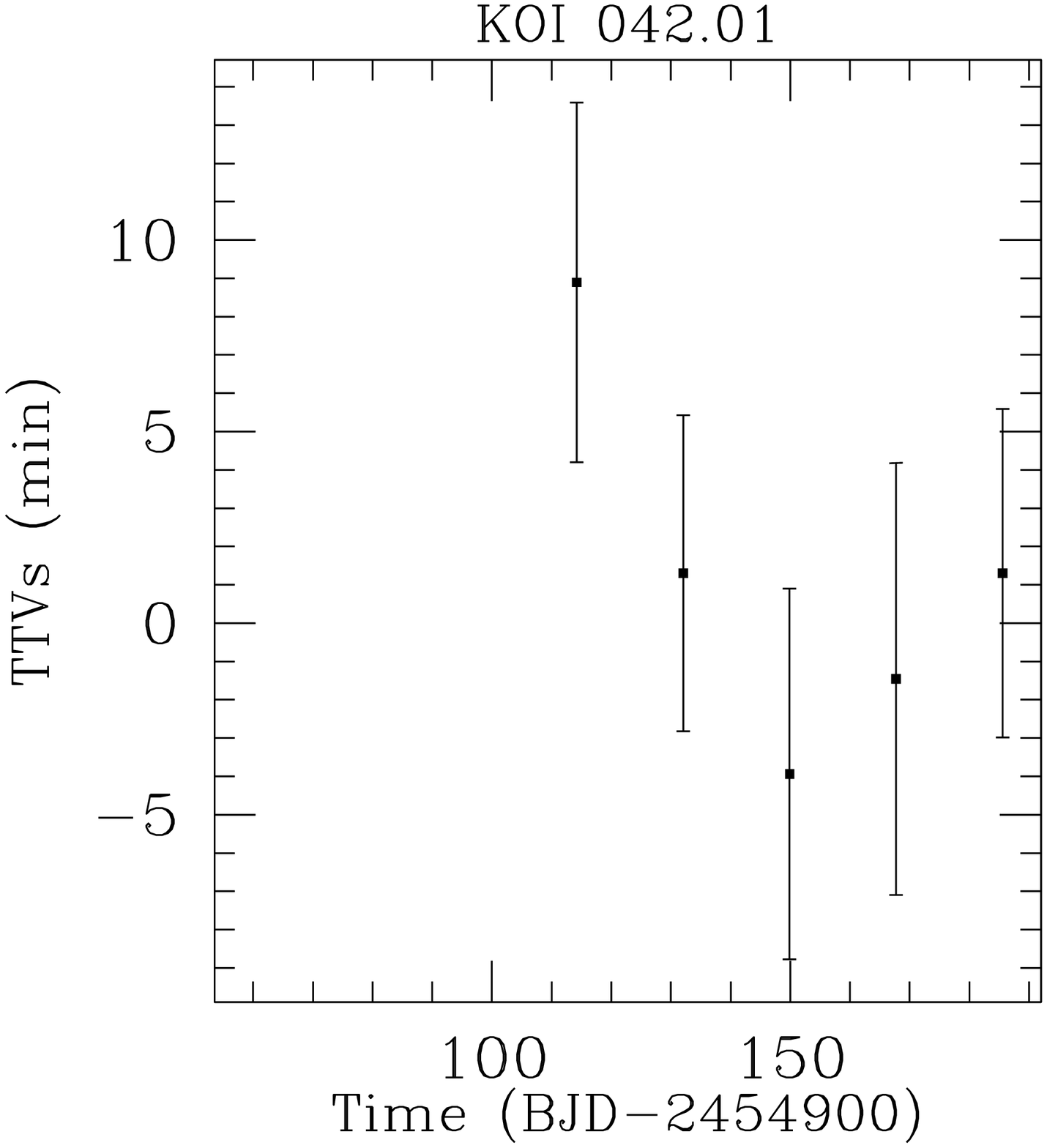}\end{figure*}
\begin{figure*}\plottwo{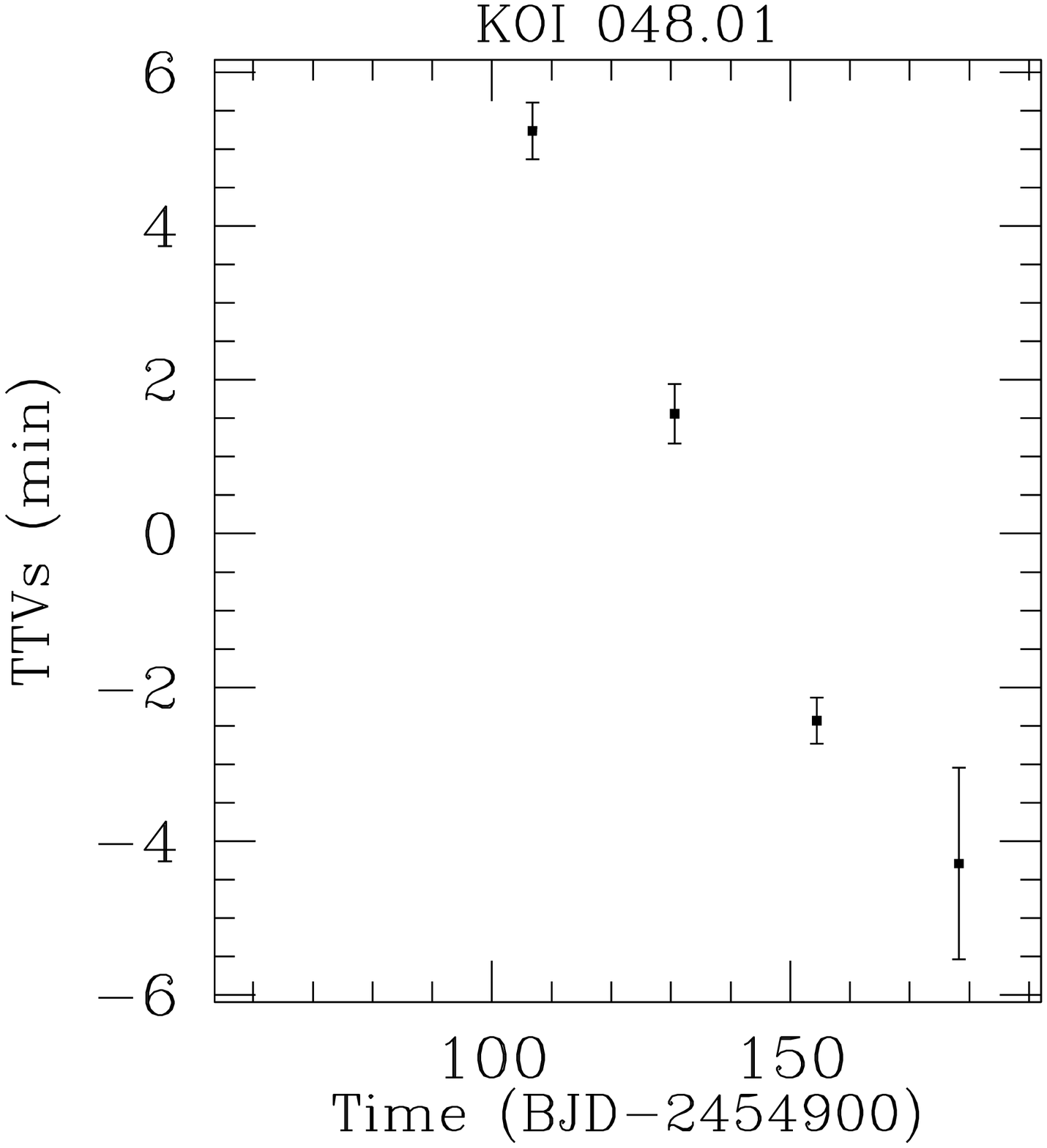}{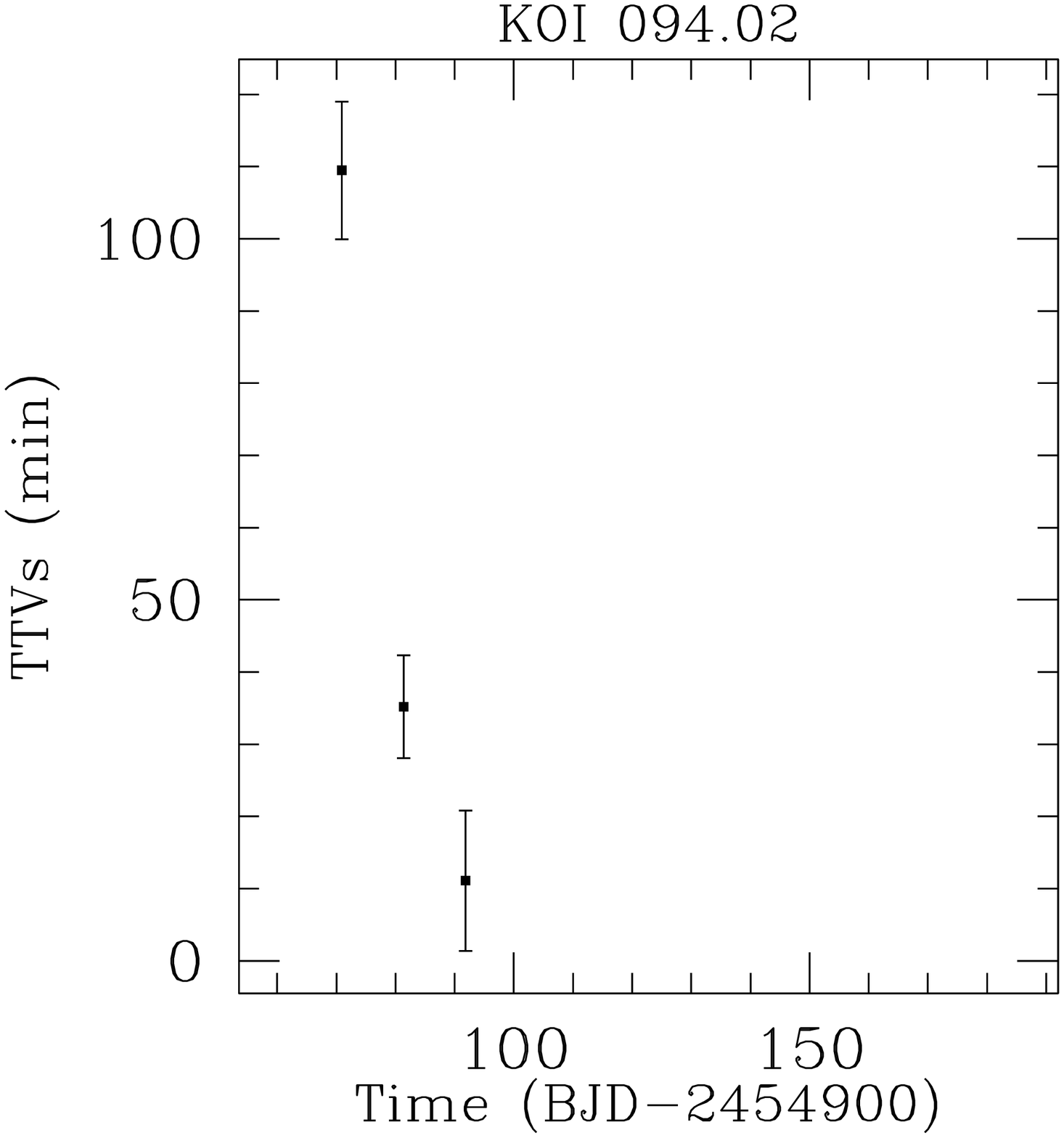}\end{figure*}\clearpage
\begin{figure*}\plottwo{koi103.01.eps}{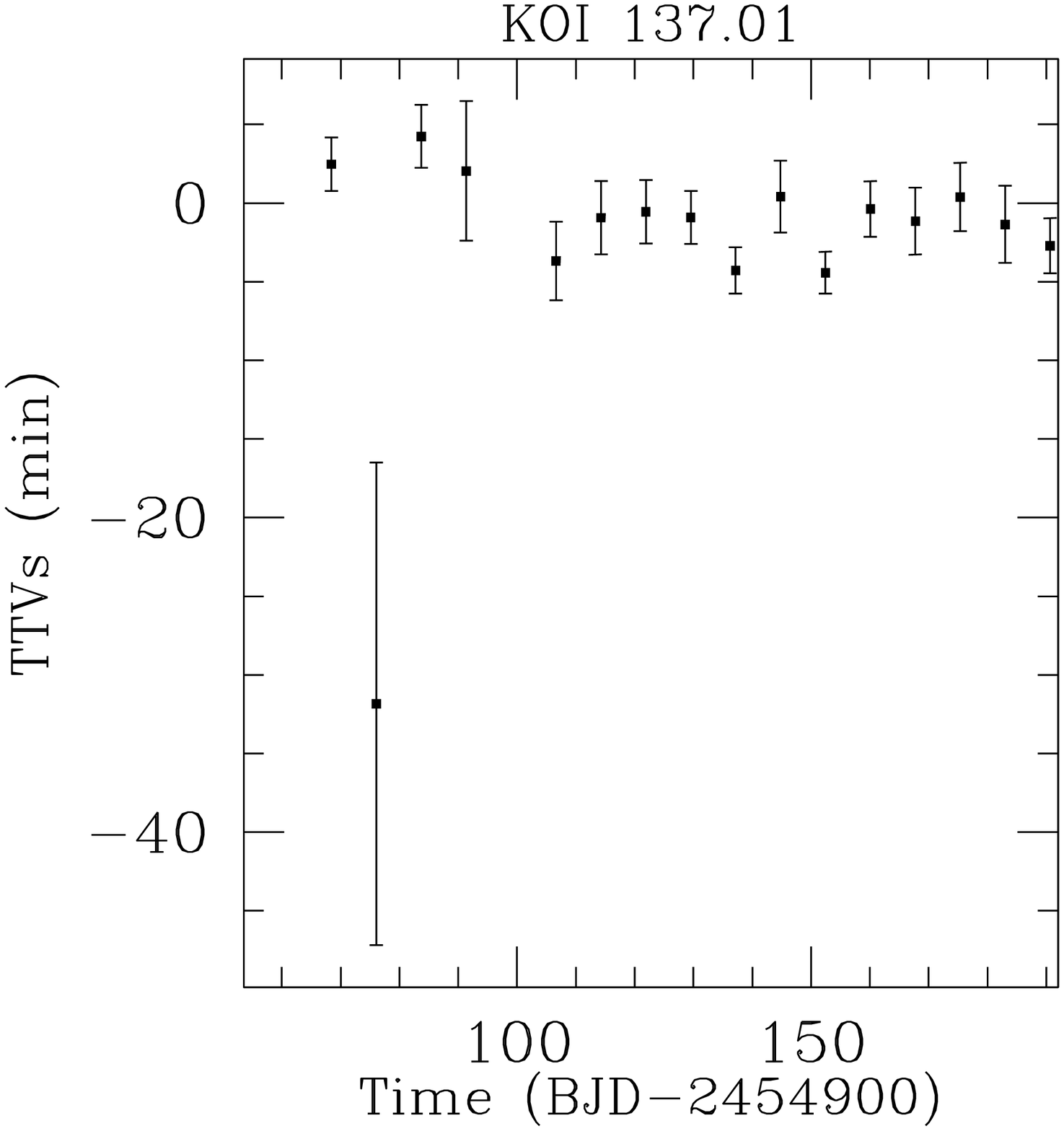}\end{figure*}
\begin{figure*}\plottwo{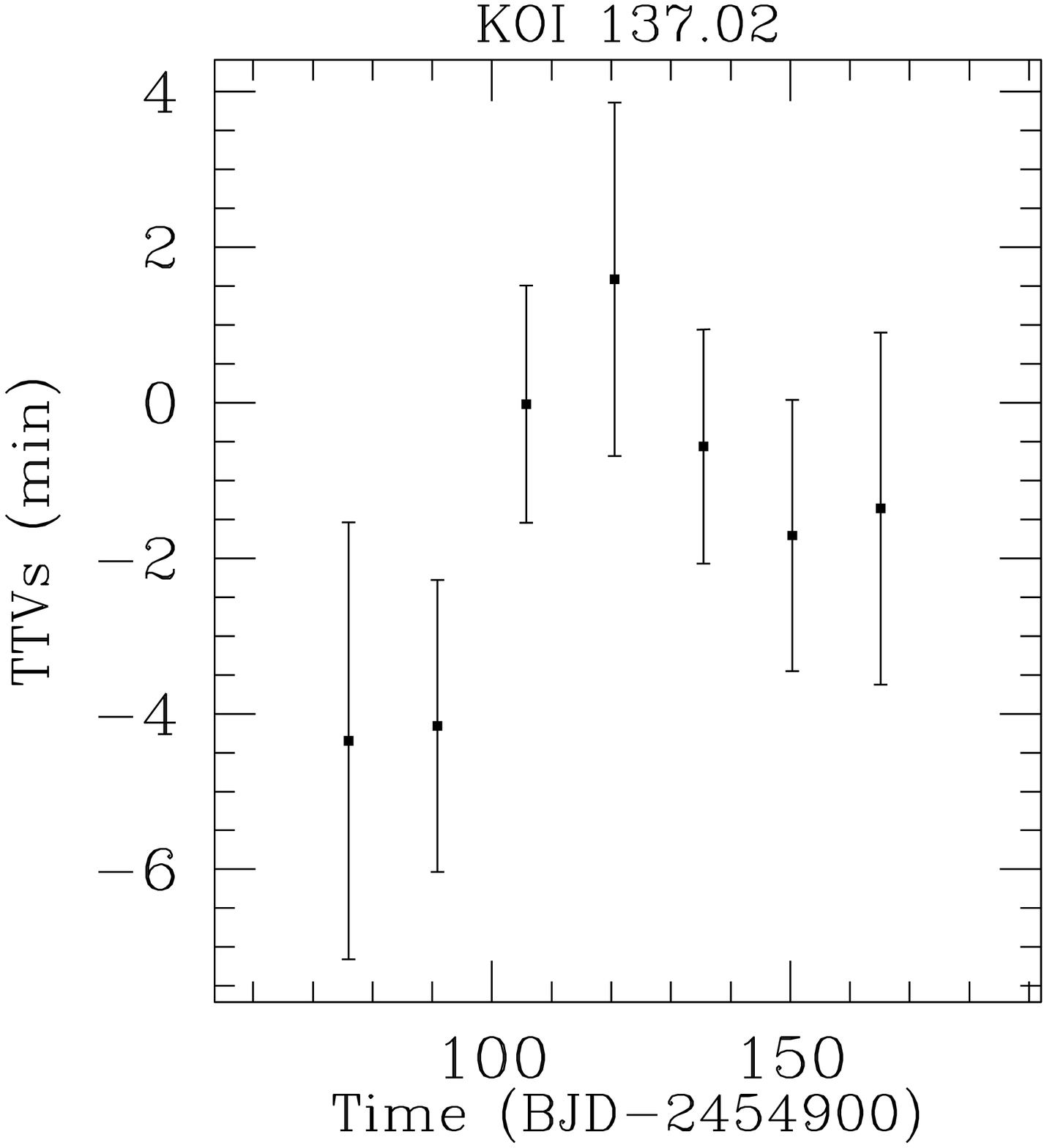}{koi142.01.eps}\end{figure*}\clearpage
\begin{figure*}\plottwo{koi148.03.eps}{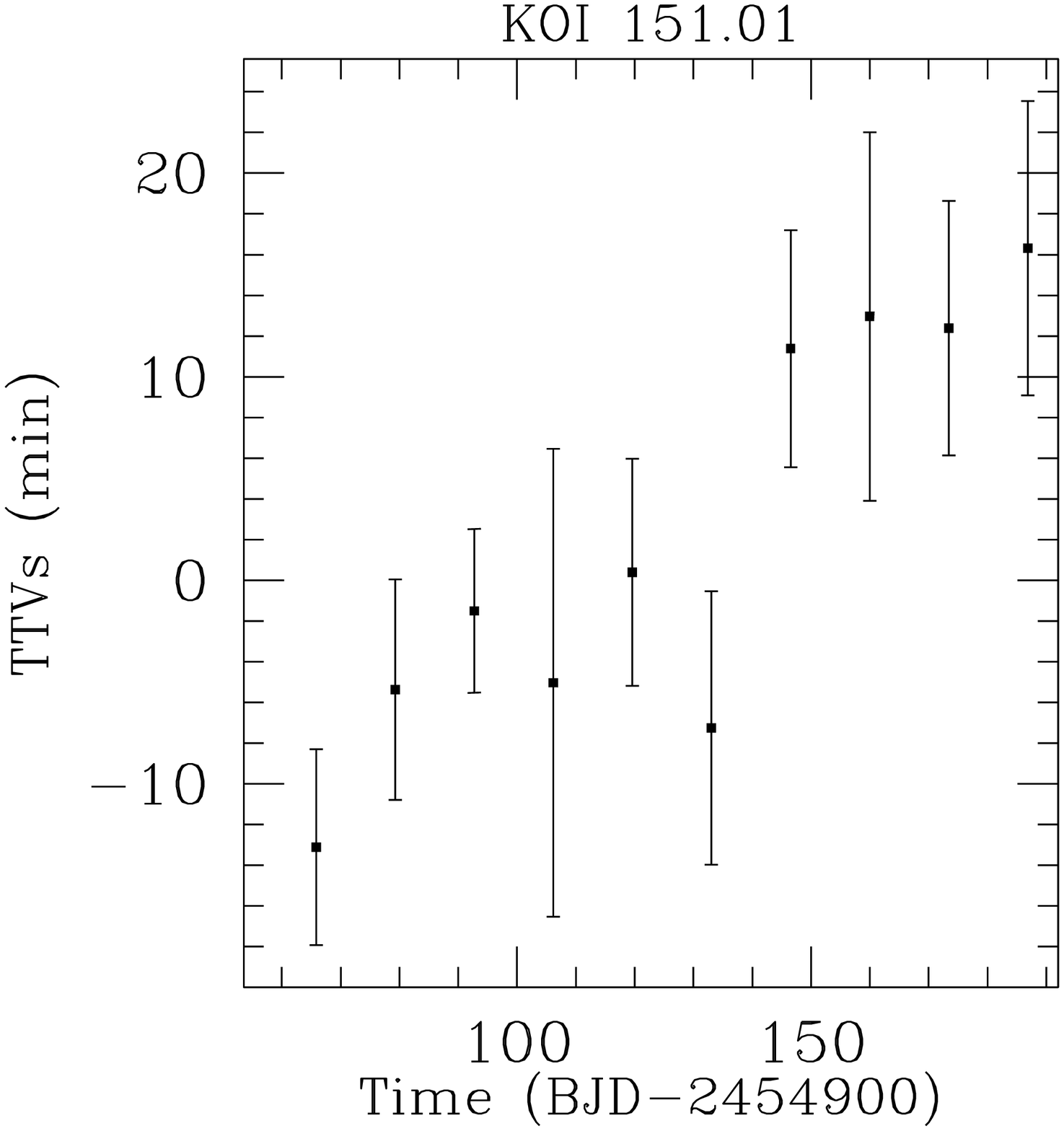}\end{figure*}
\begin{figure*}\plottwo{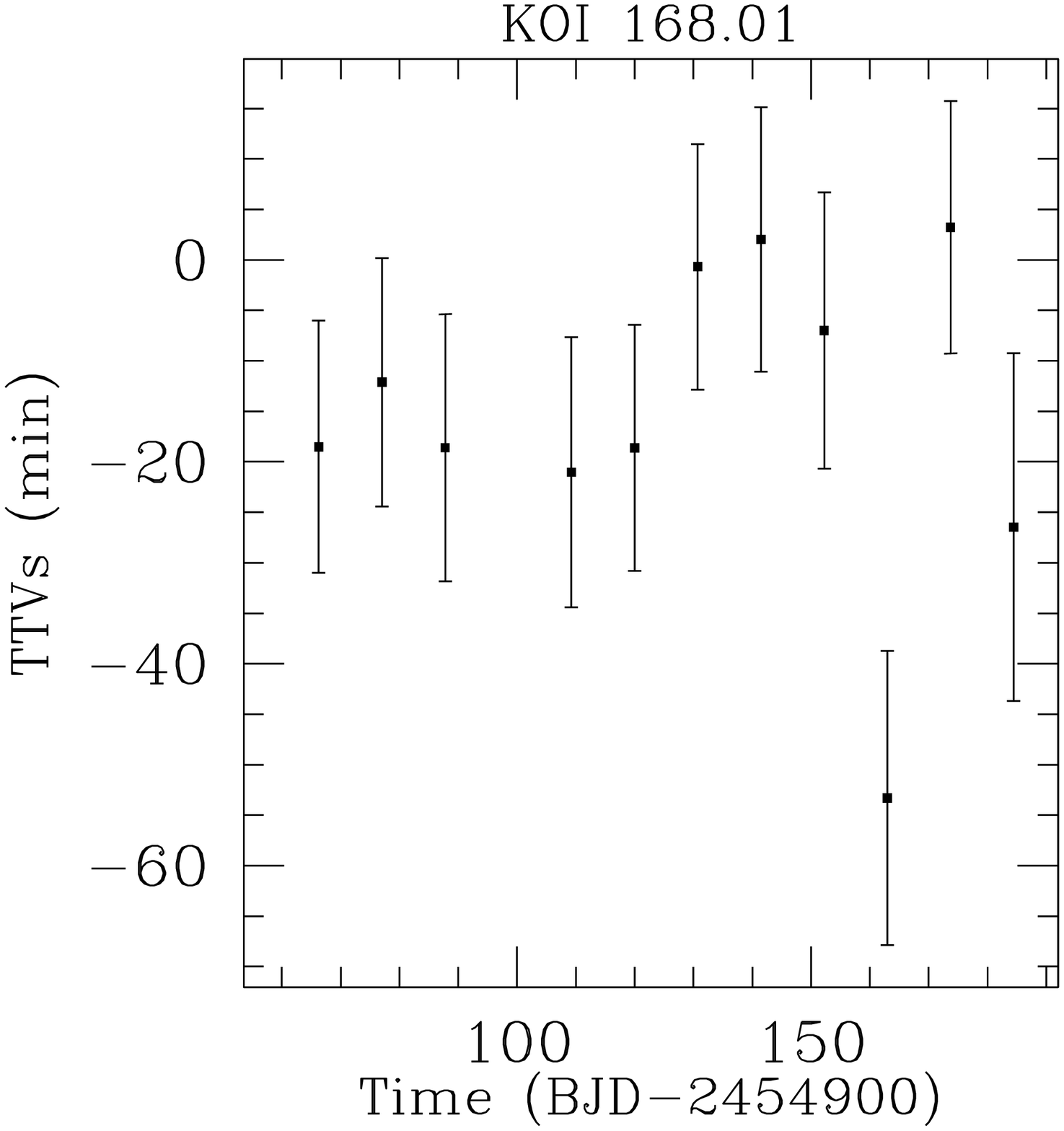}{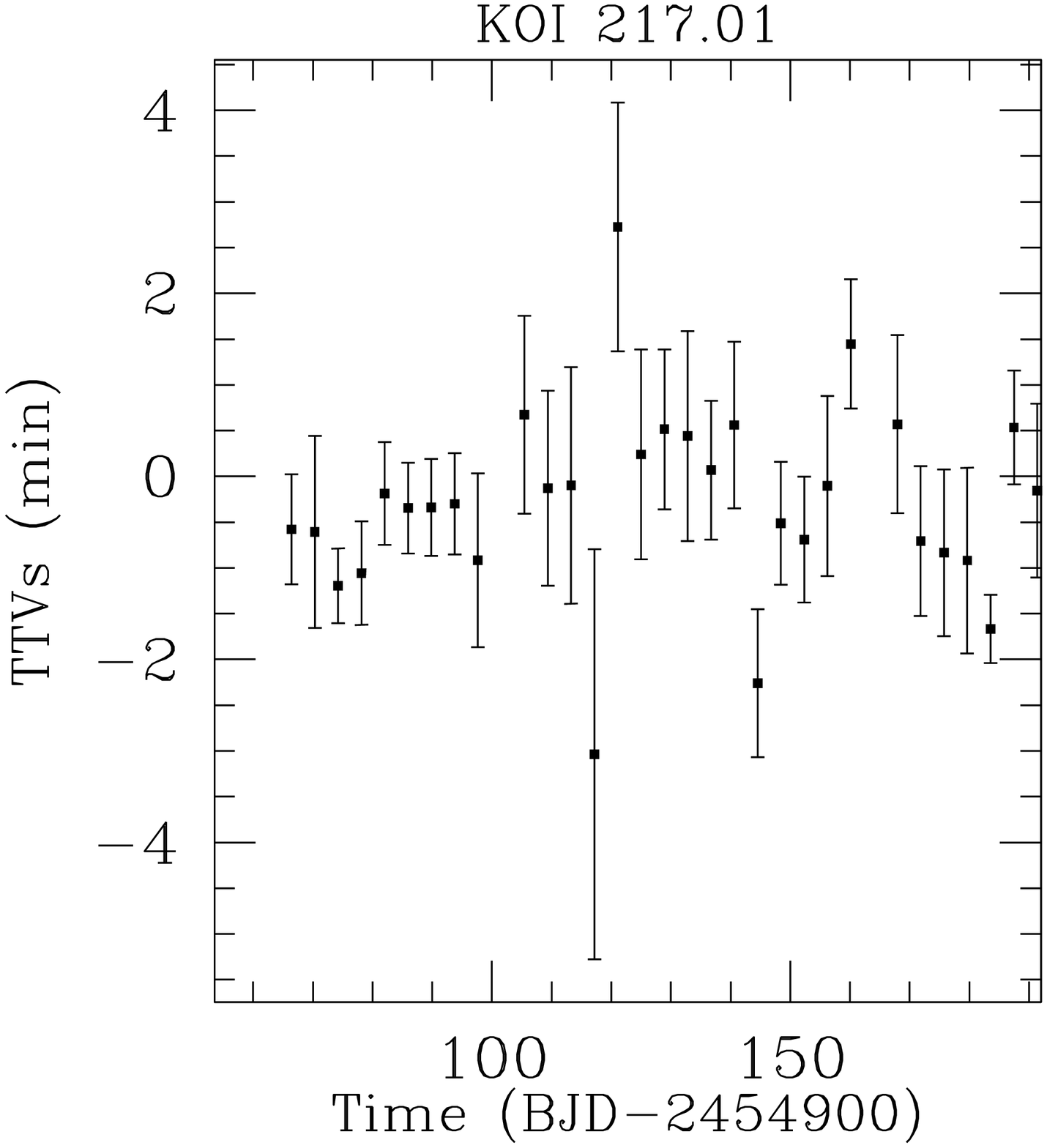}\end{figure*}\clearpage
\begin{figure*}\plottwo{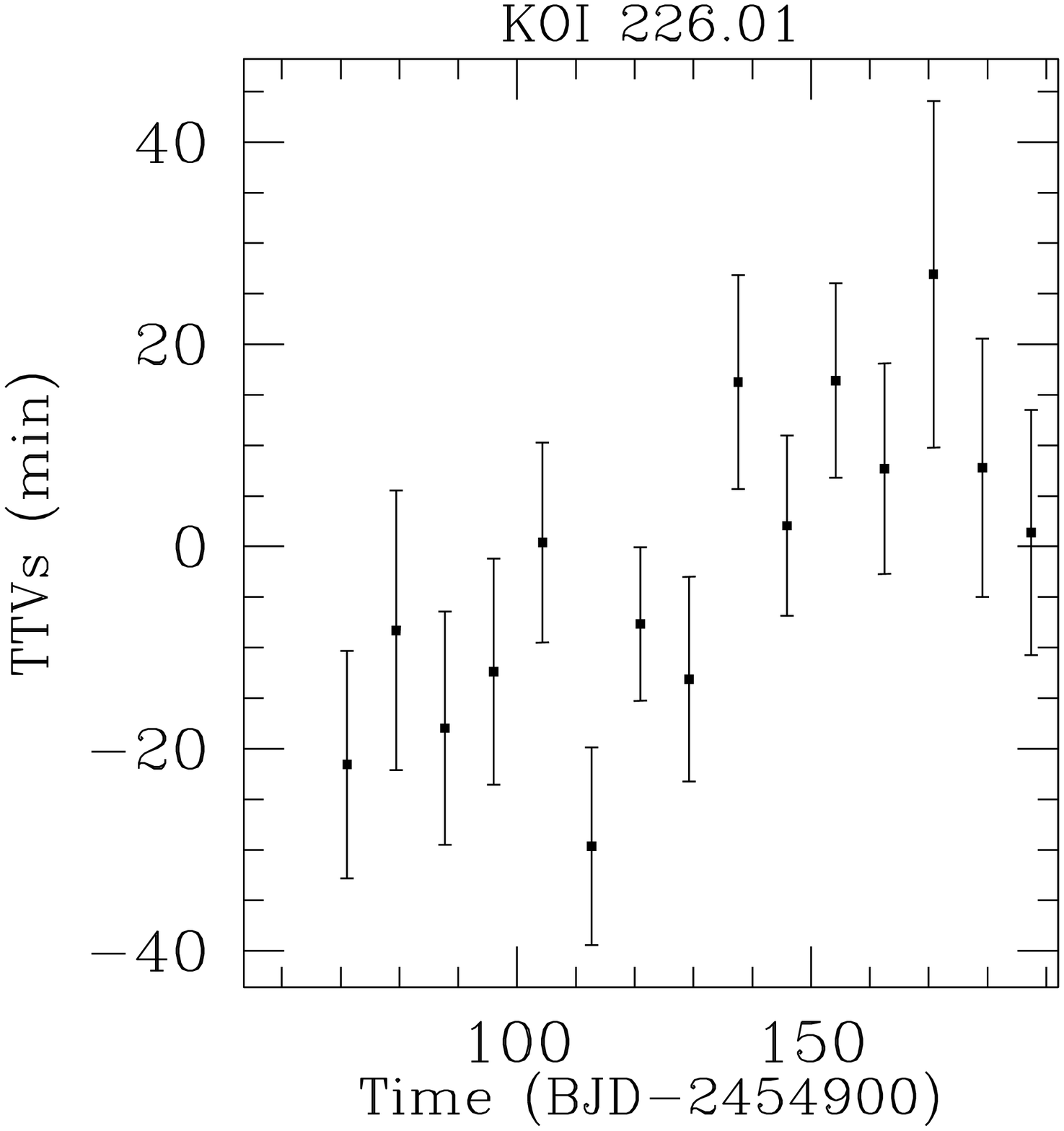}{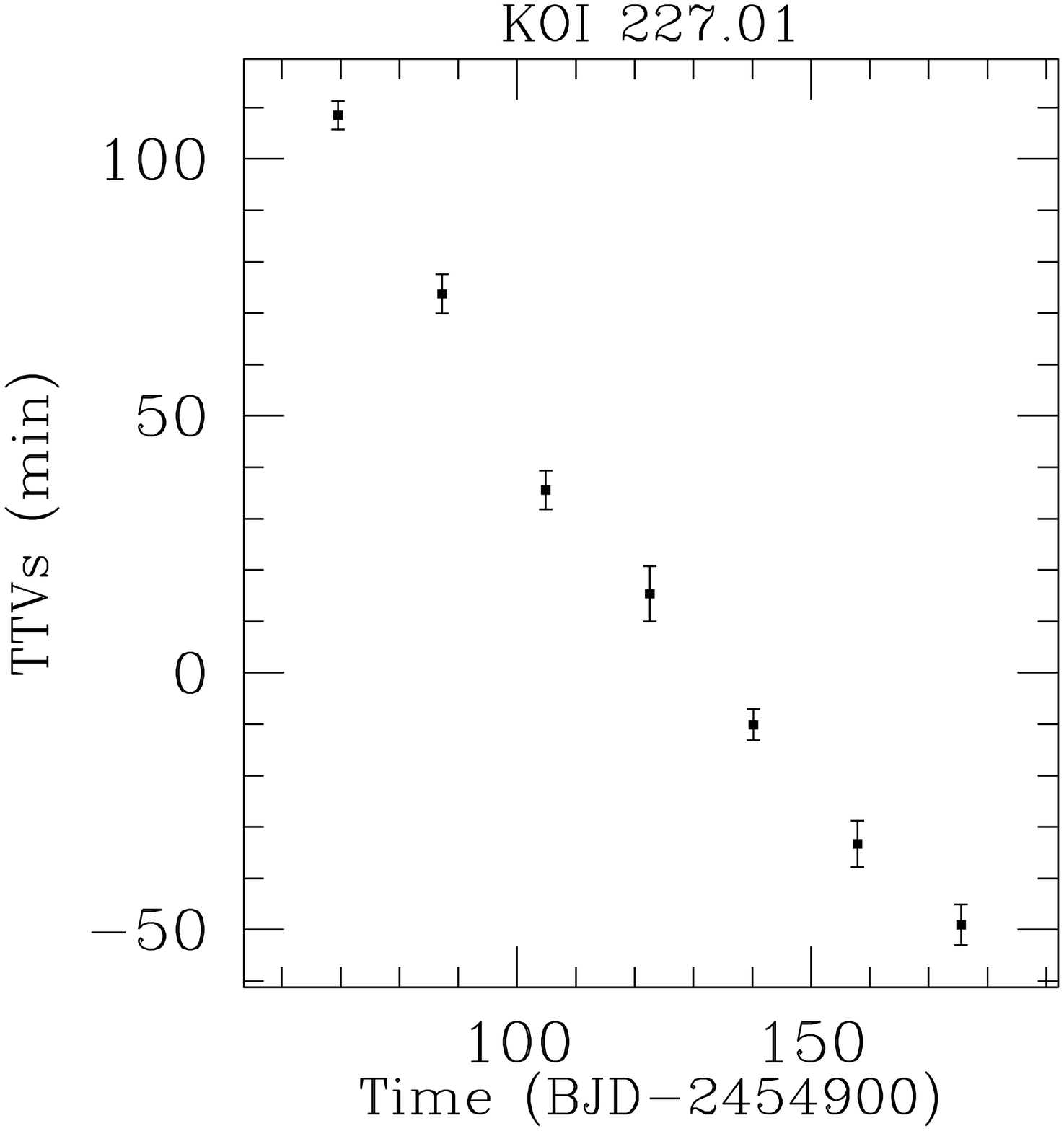}\end{figure*}
\begin{figure*}\plottwo{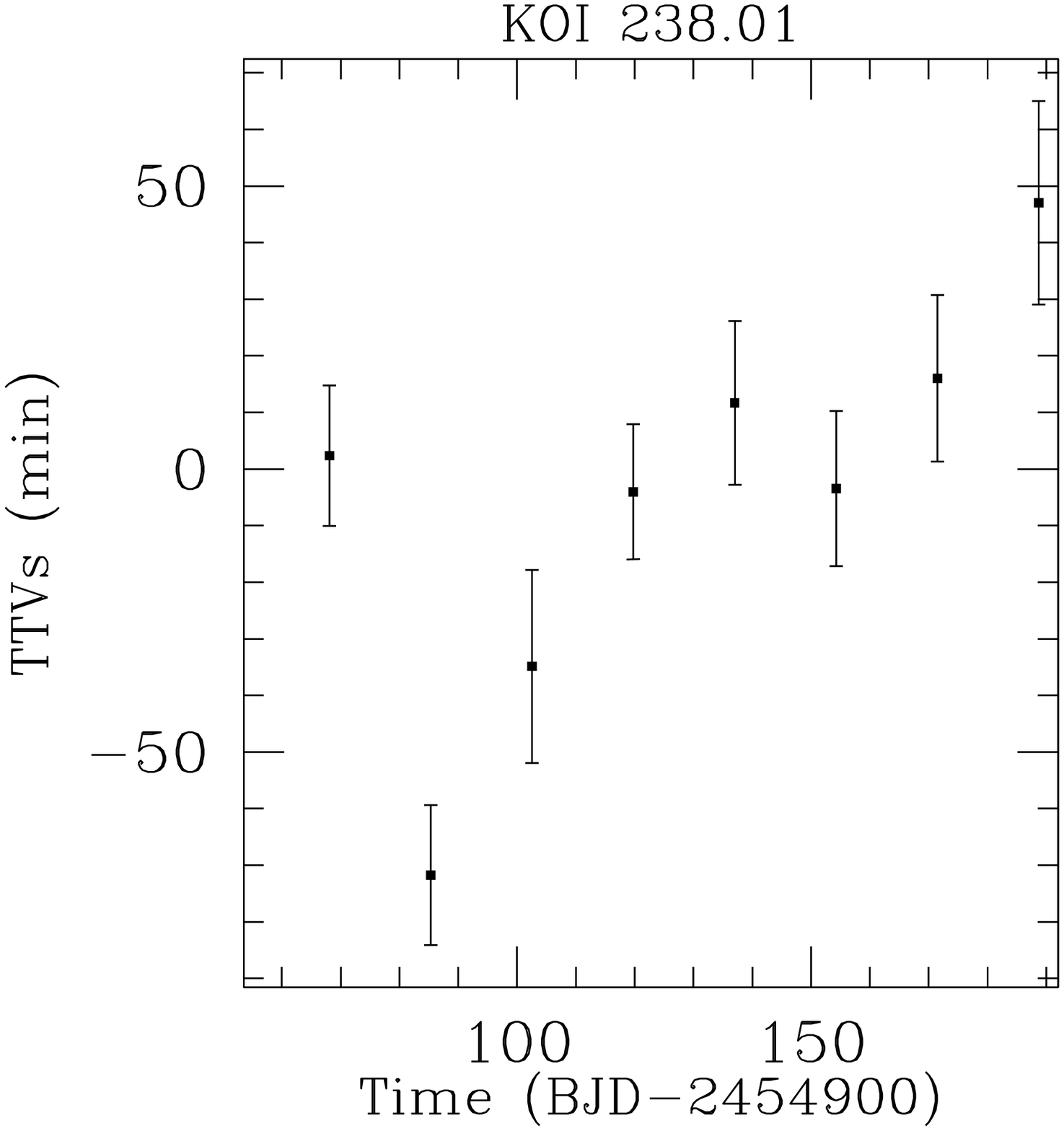}{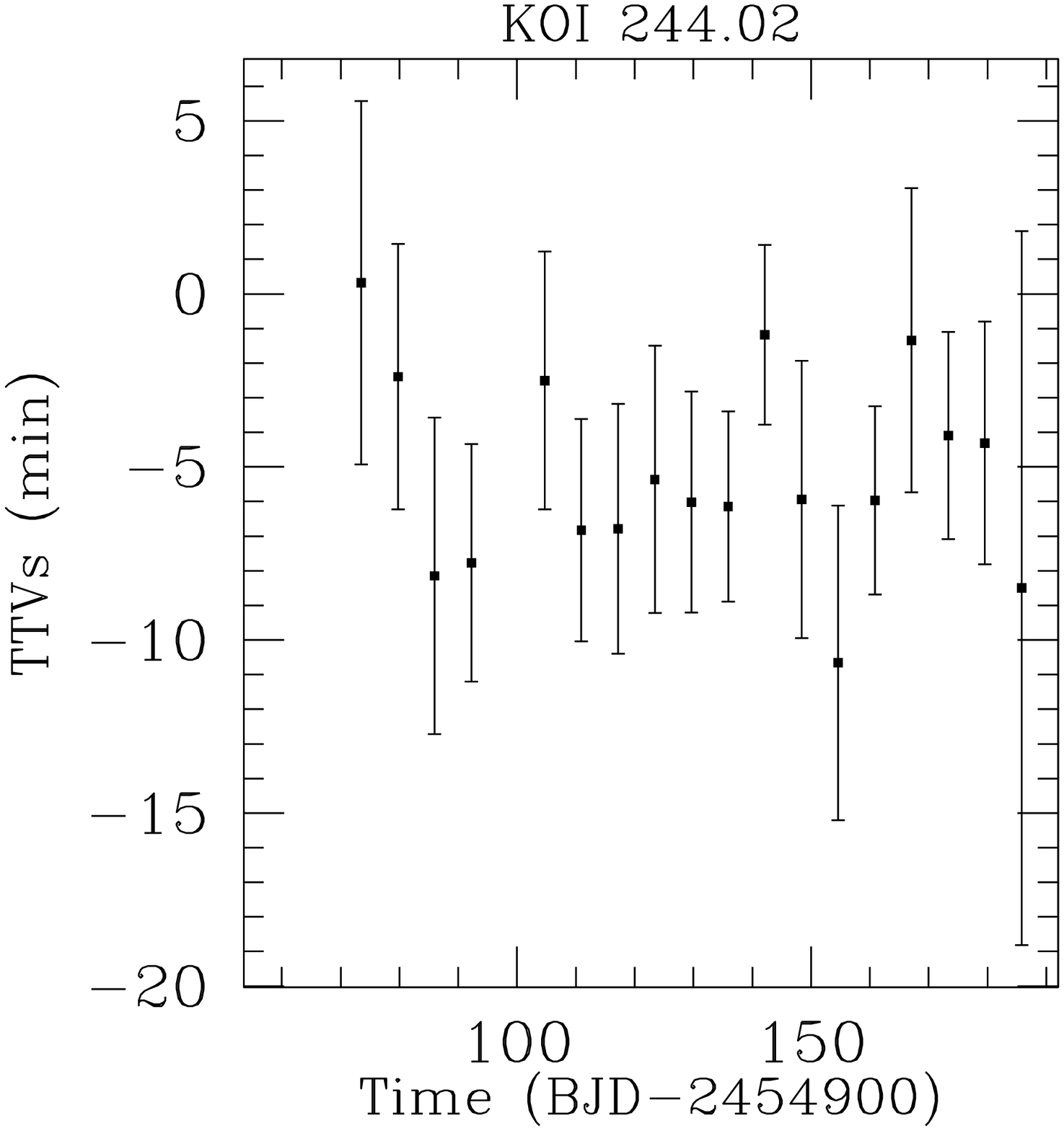}\end{figure*}\clearpage
\begin{figure*}\plottwo{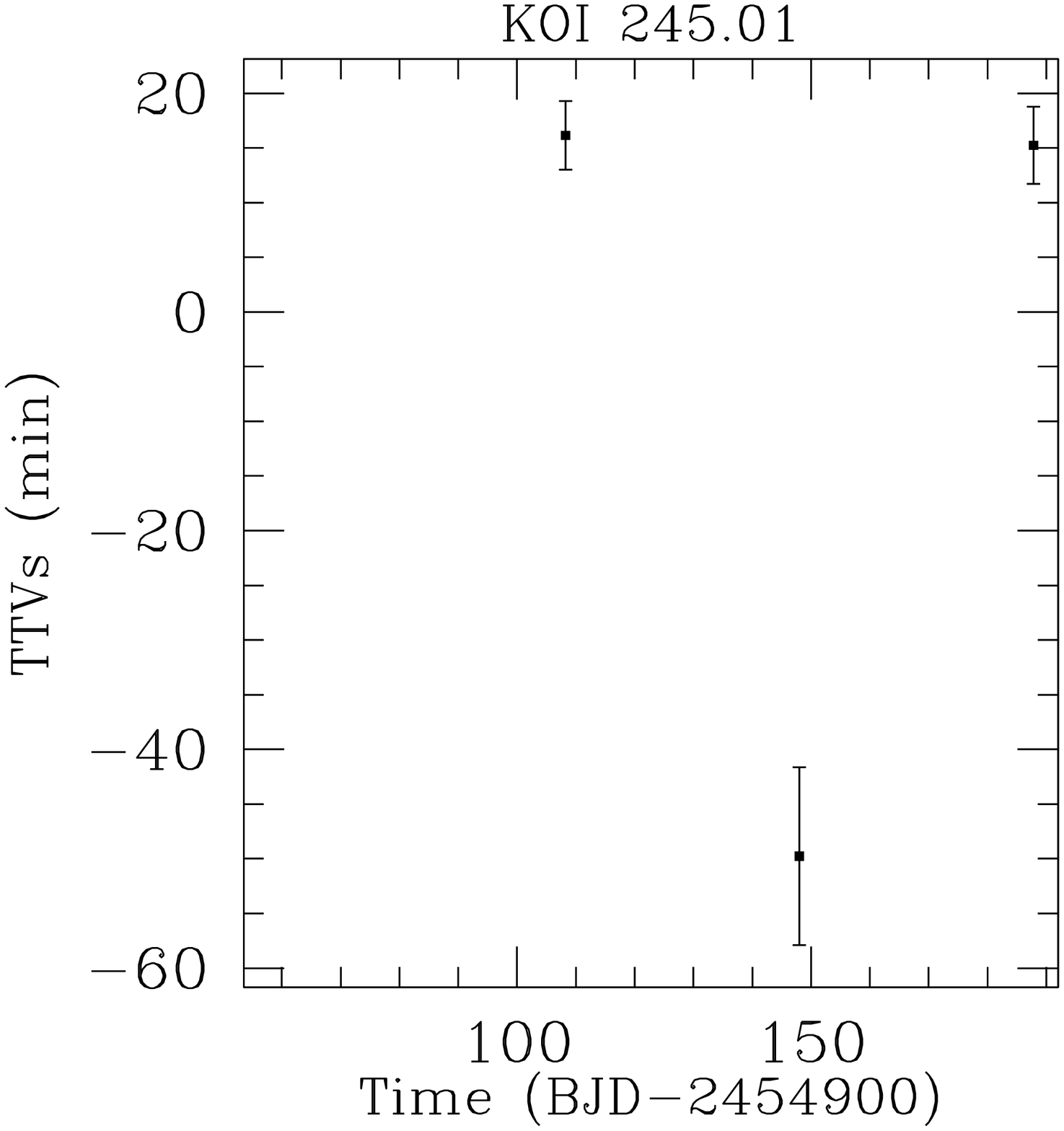}{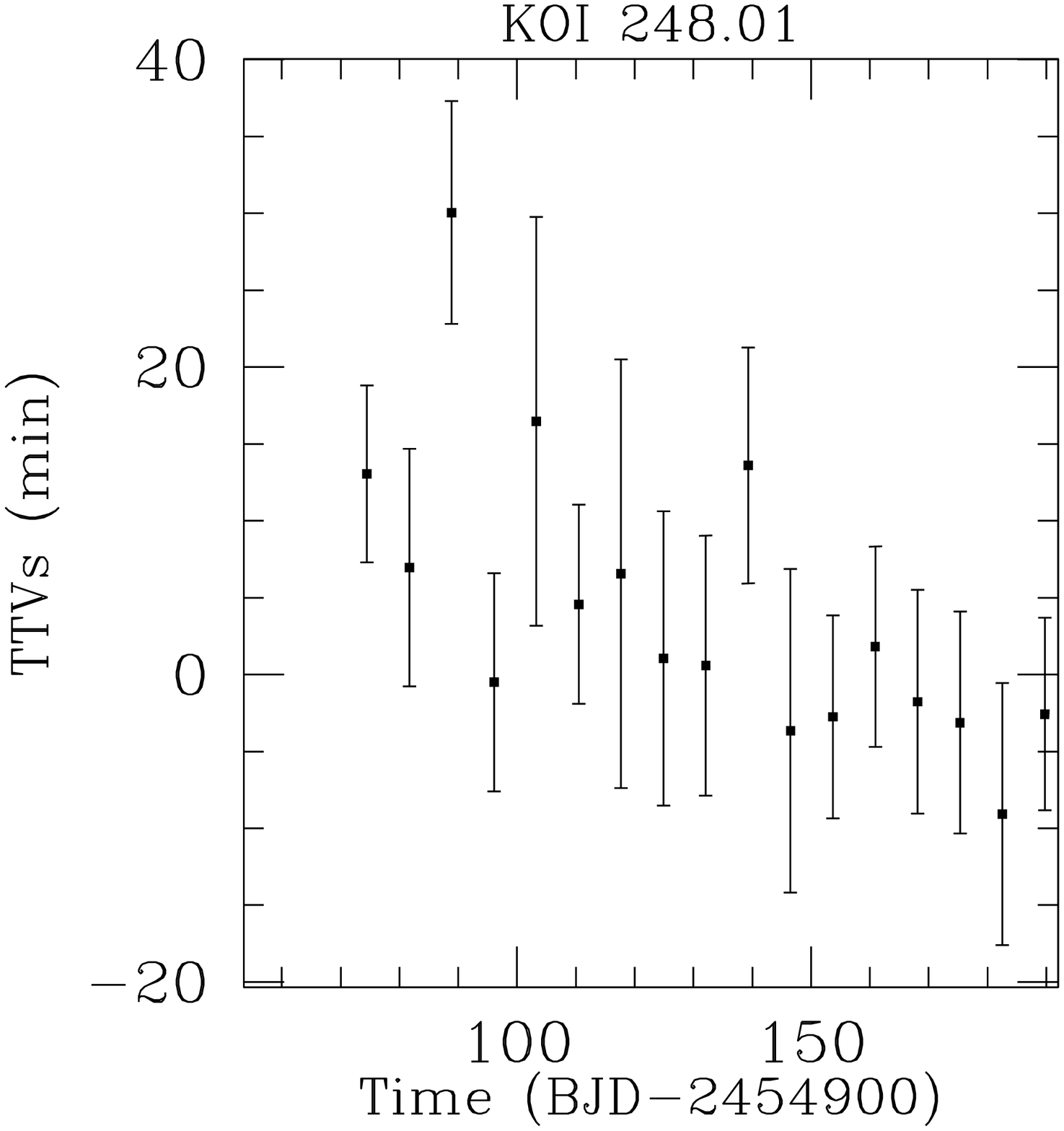}\end{figure*}
\begin{figure*}\plottwo{koi258.01.eps}{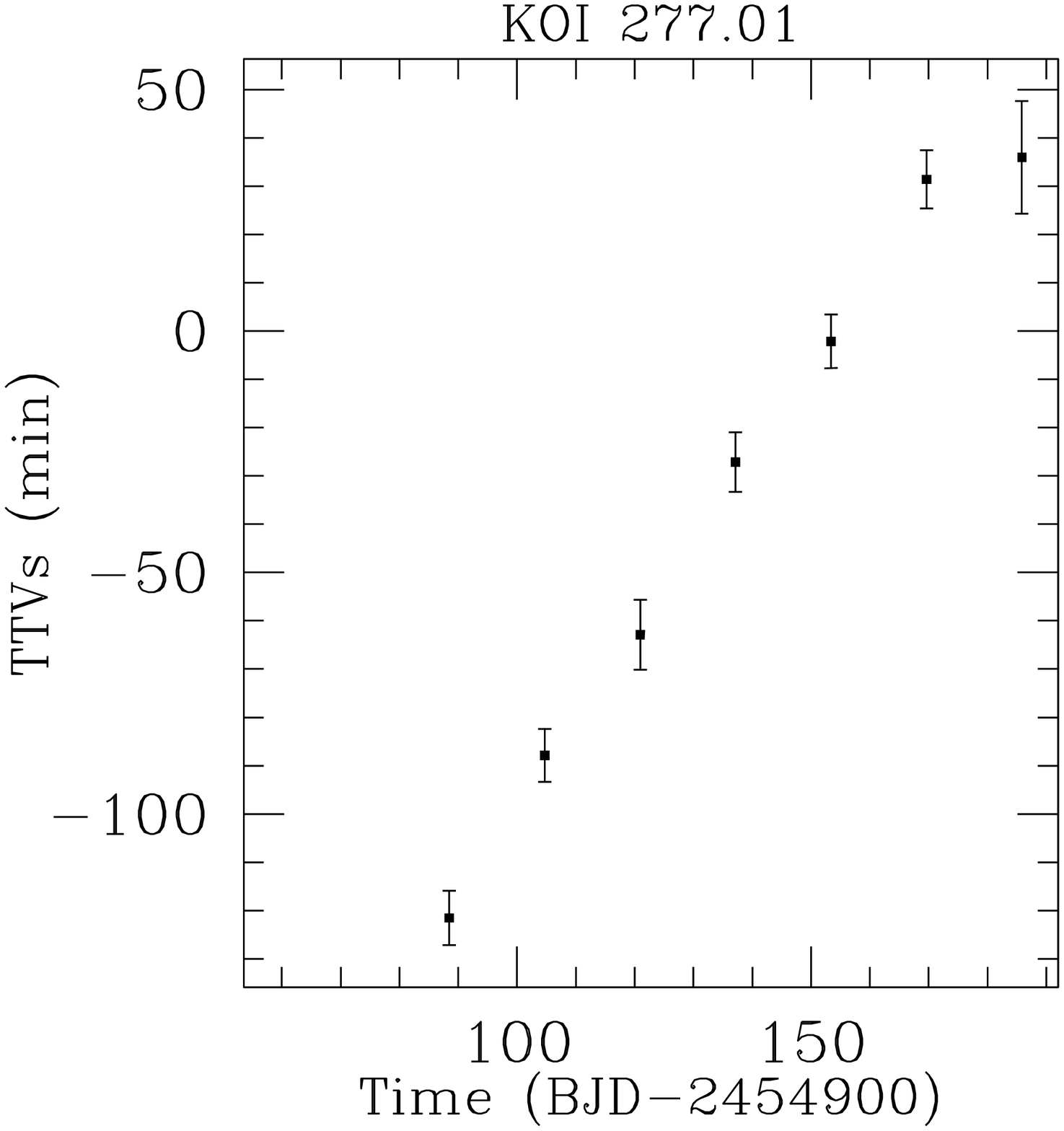}\end{figure*}\clearpage
\begin{figure*}\plottwo{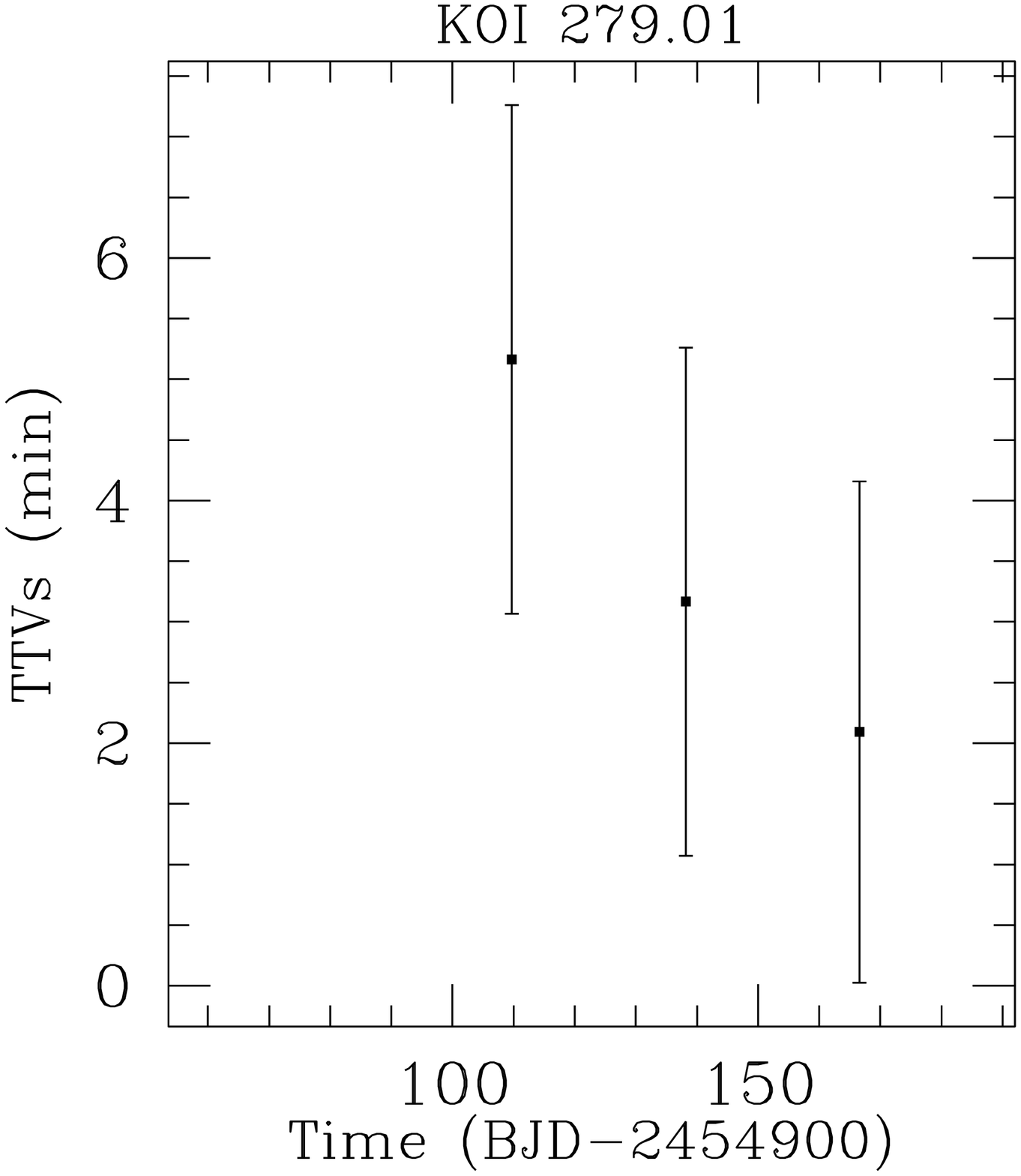}{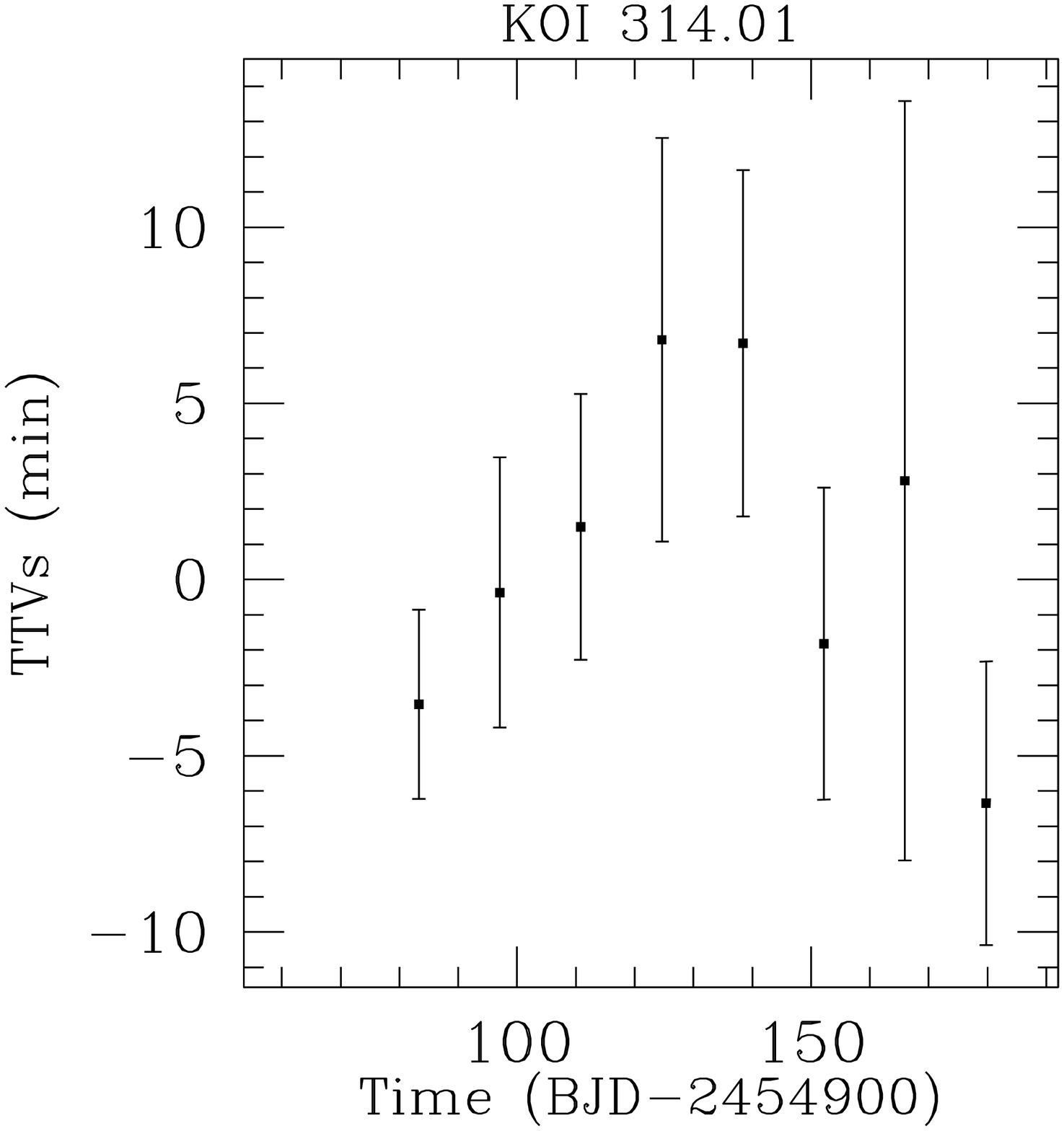}\end{figure*}
\begin{figure*}\plottwo{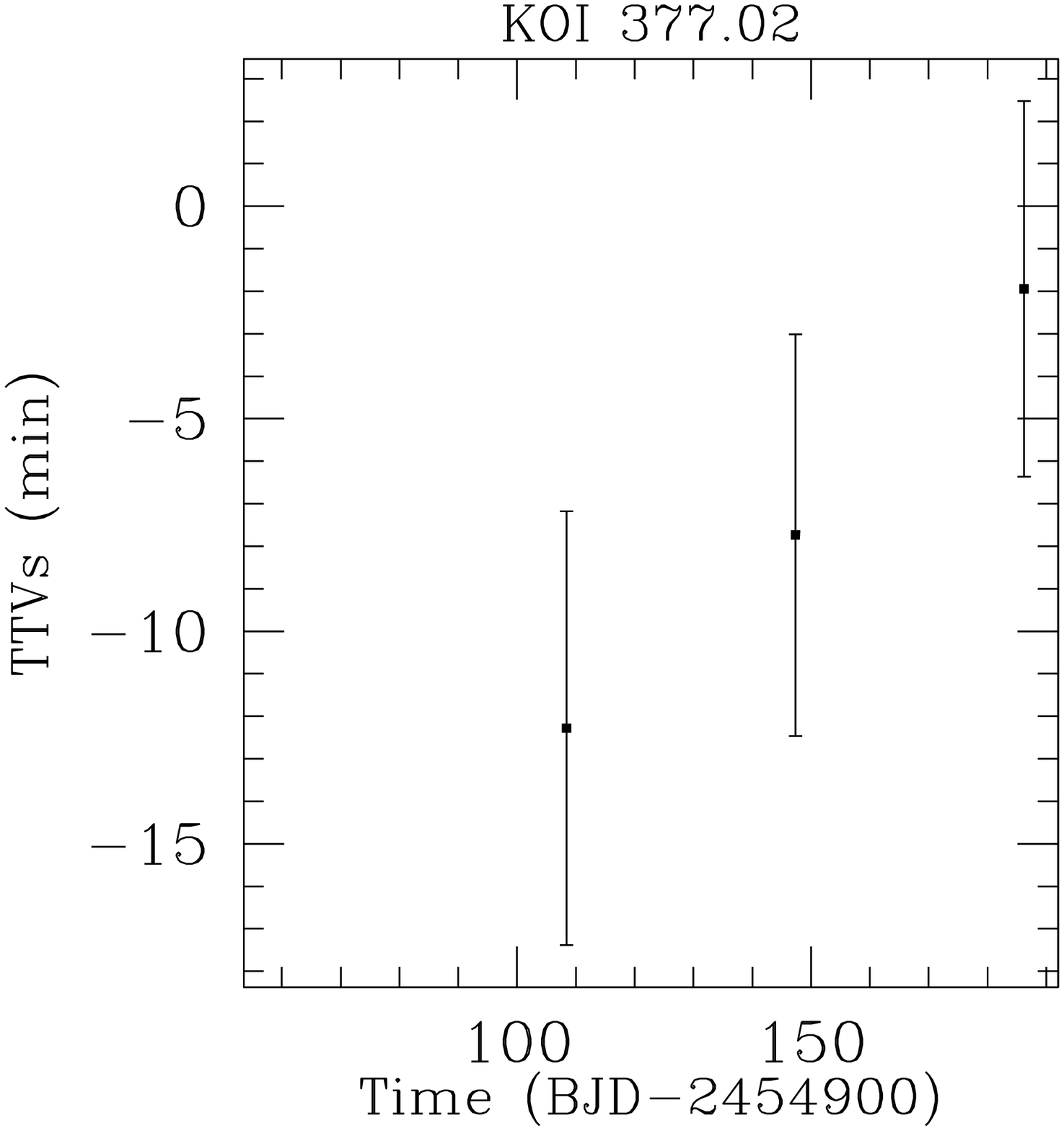}{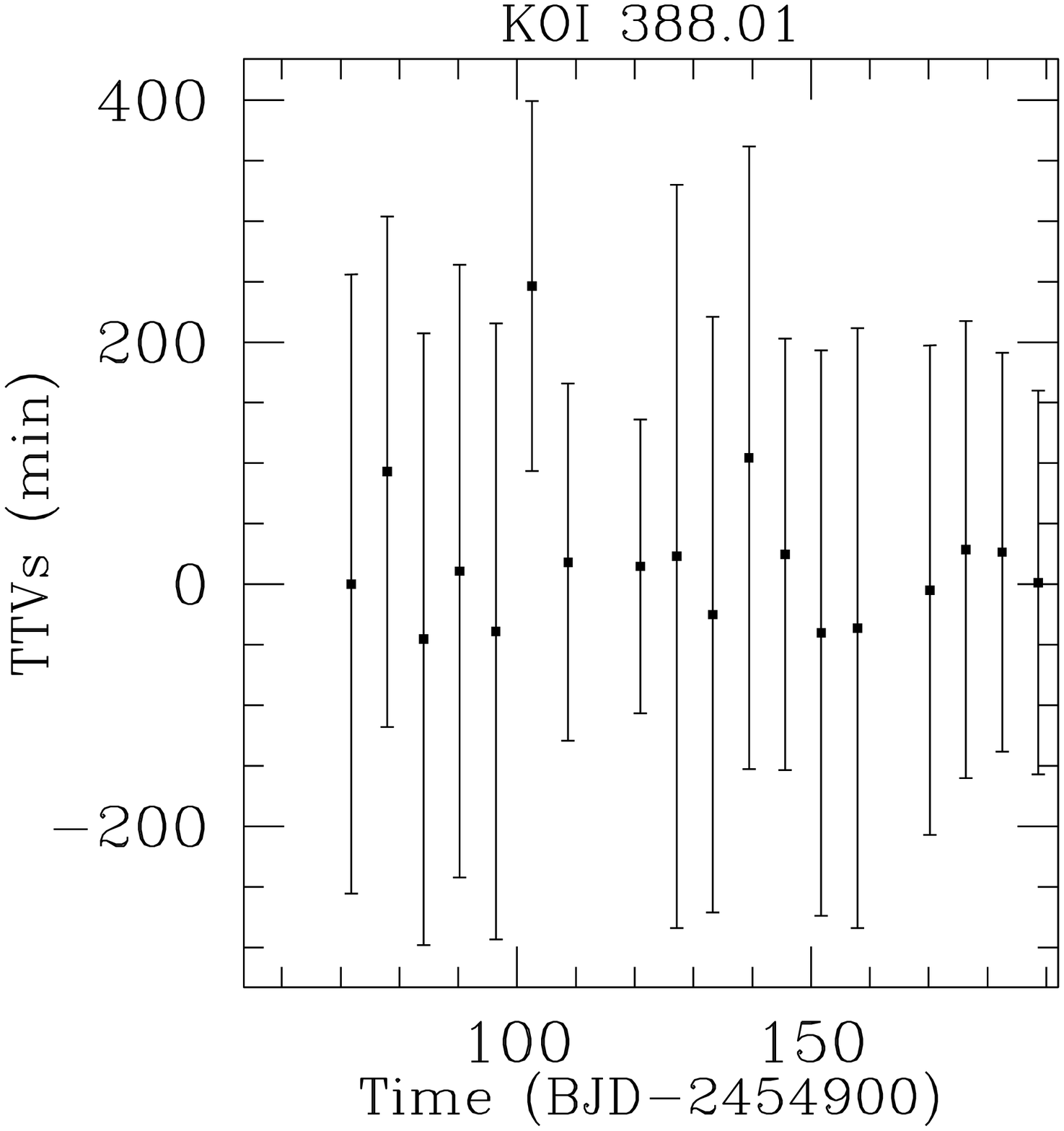}\end{figure*}\clearpage
\begin{figure*}\plottwo{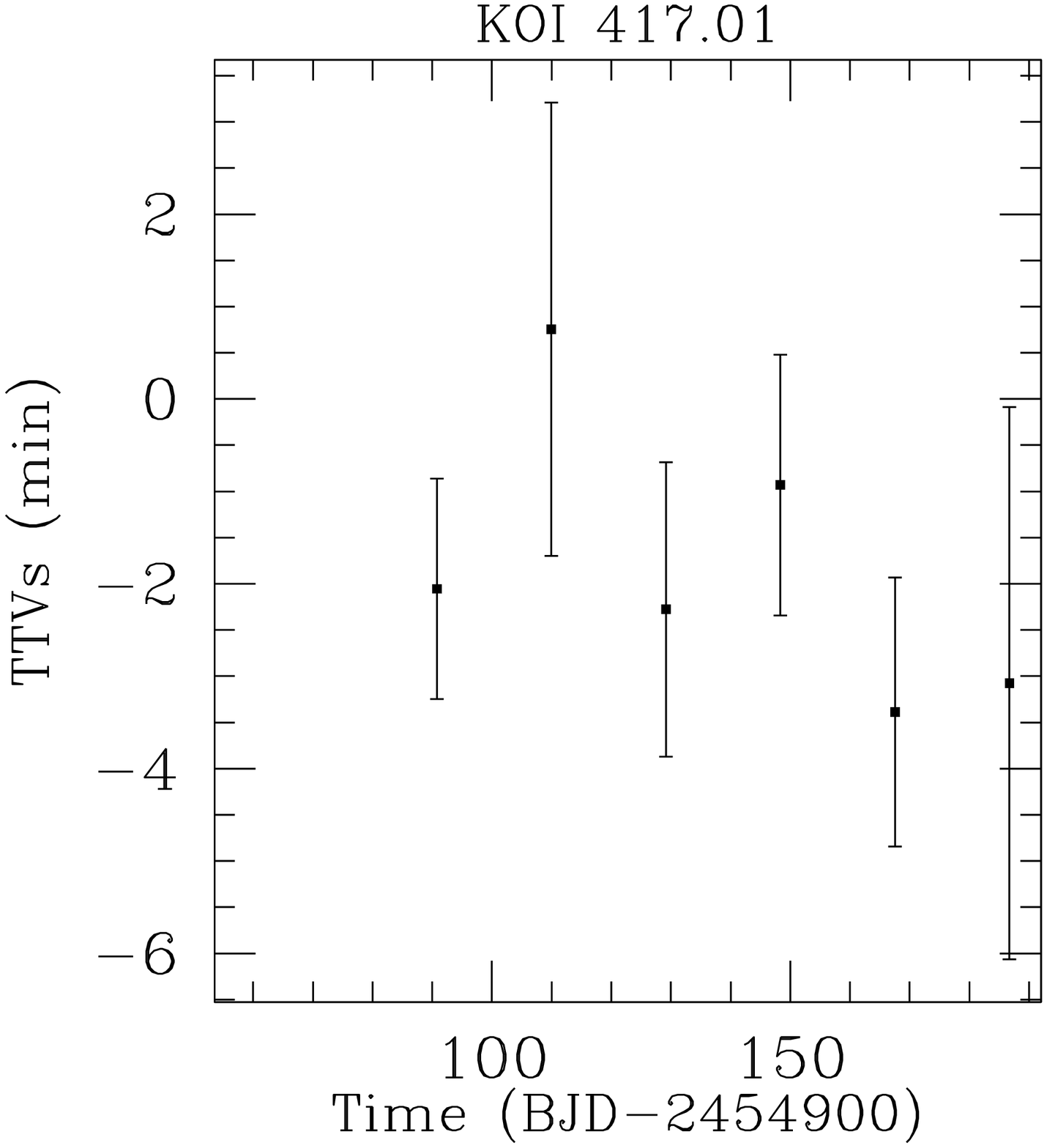}{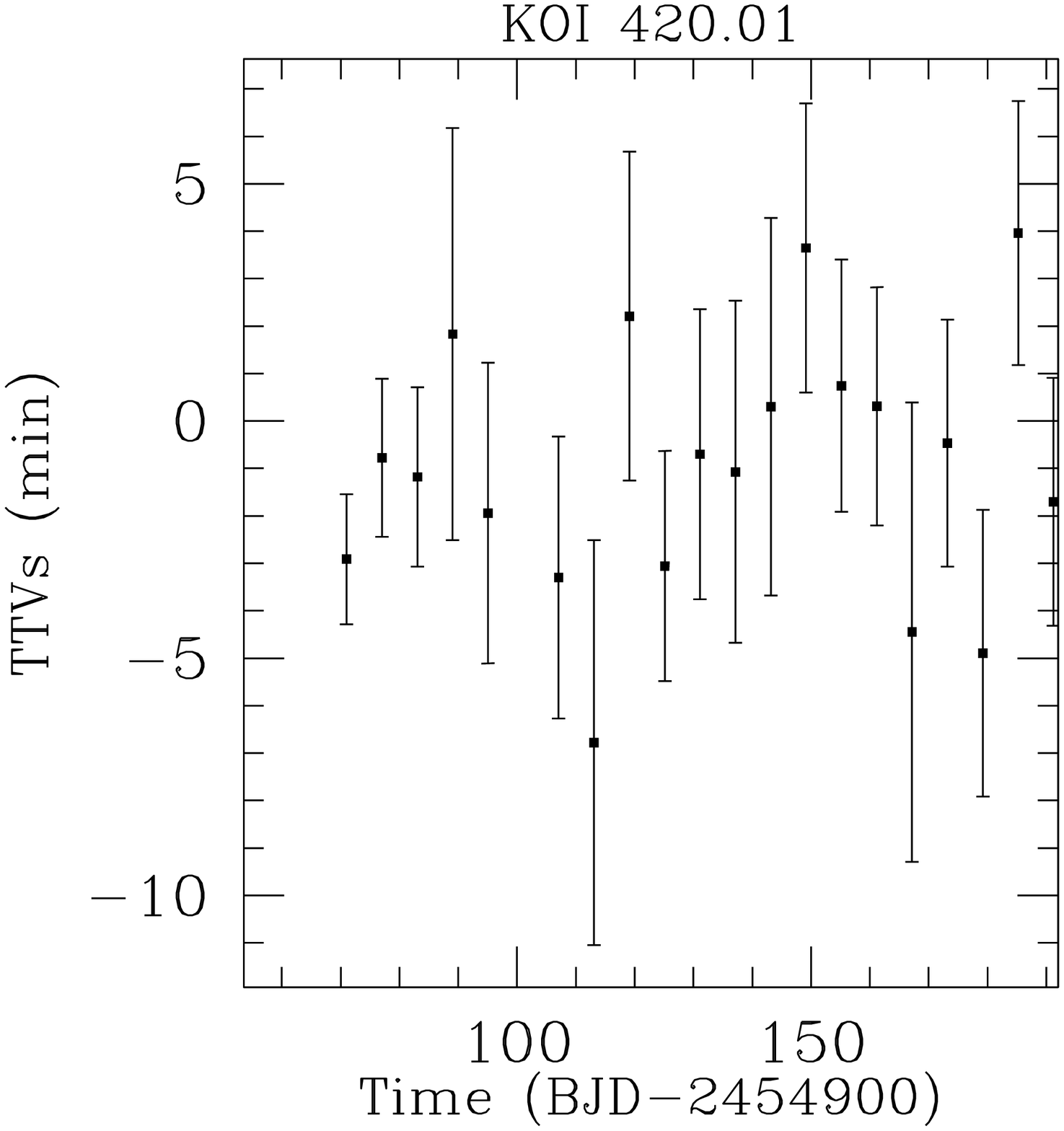}\end{figure*}
\begin{figure*}\plottwo{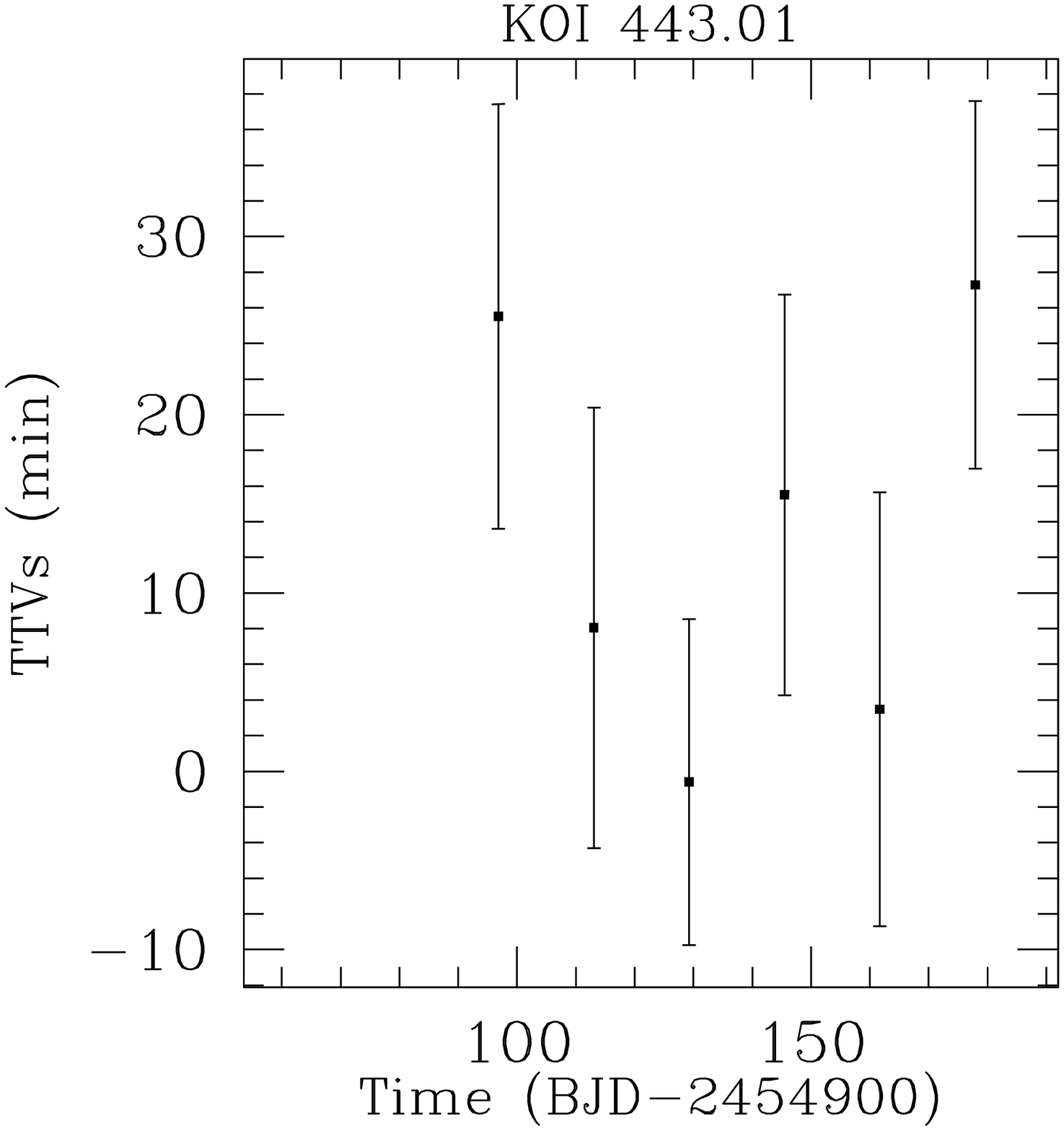}{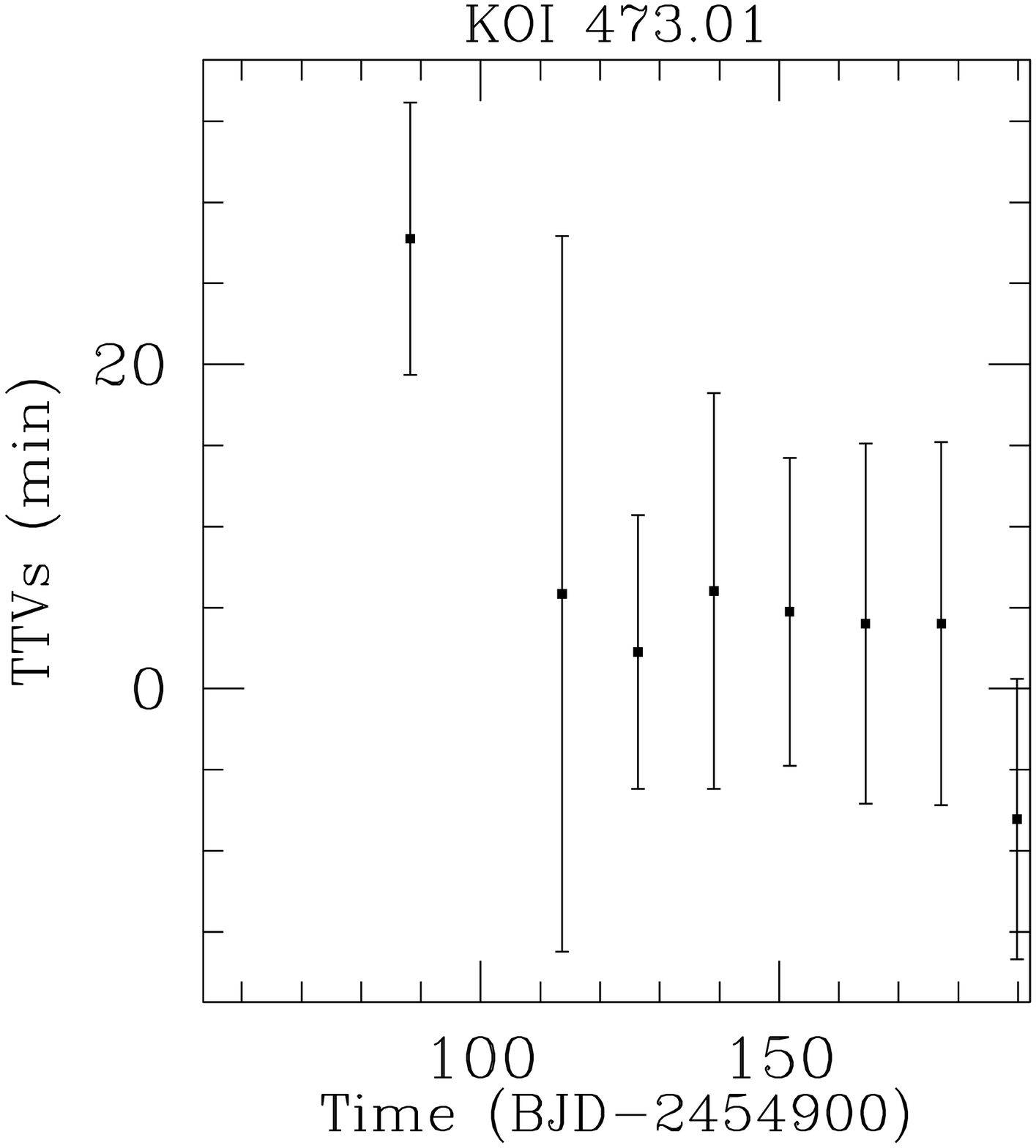}\end{figure*}\clearpage
\begin{figure*}\plottwo{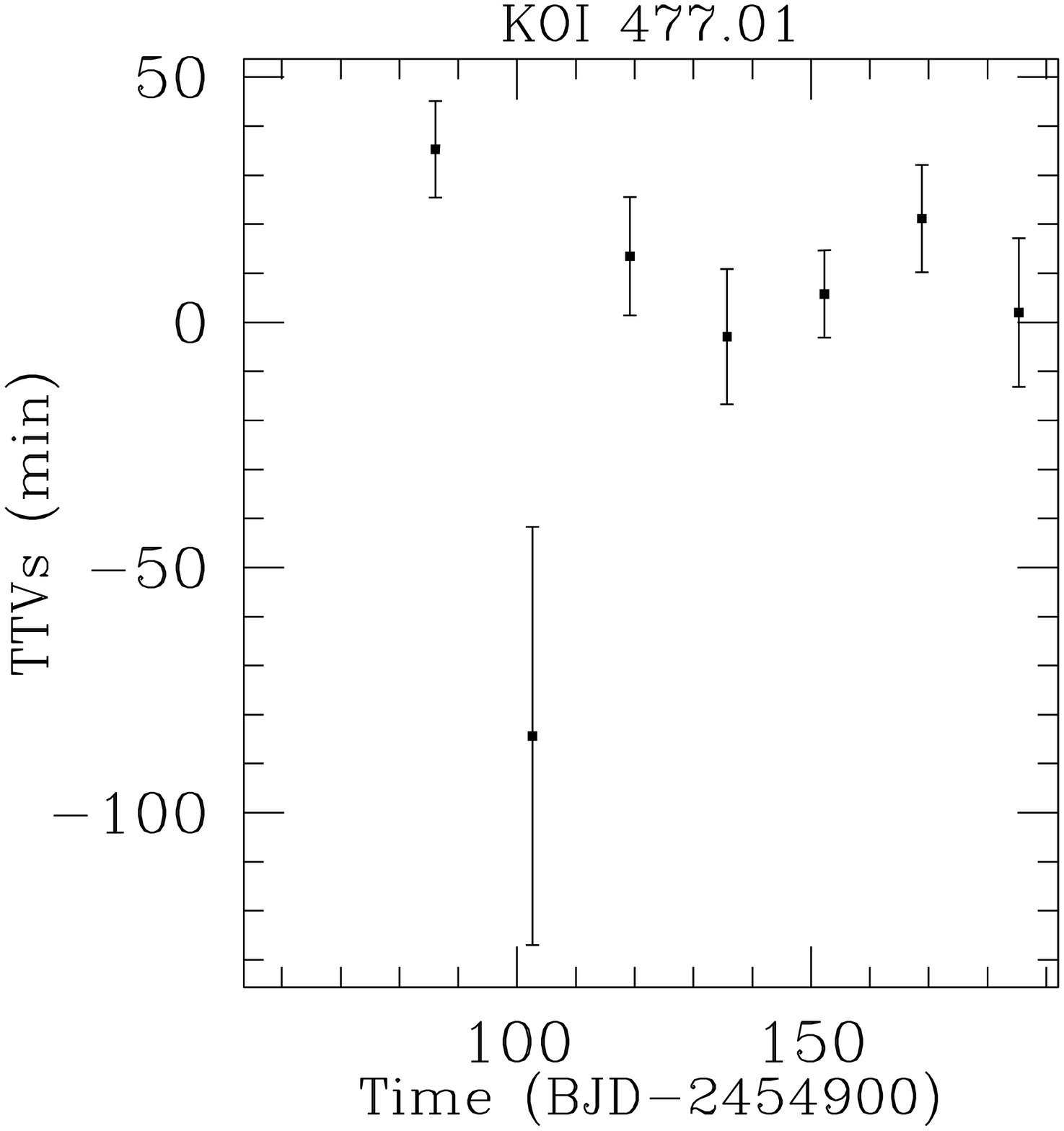}{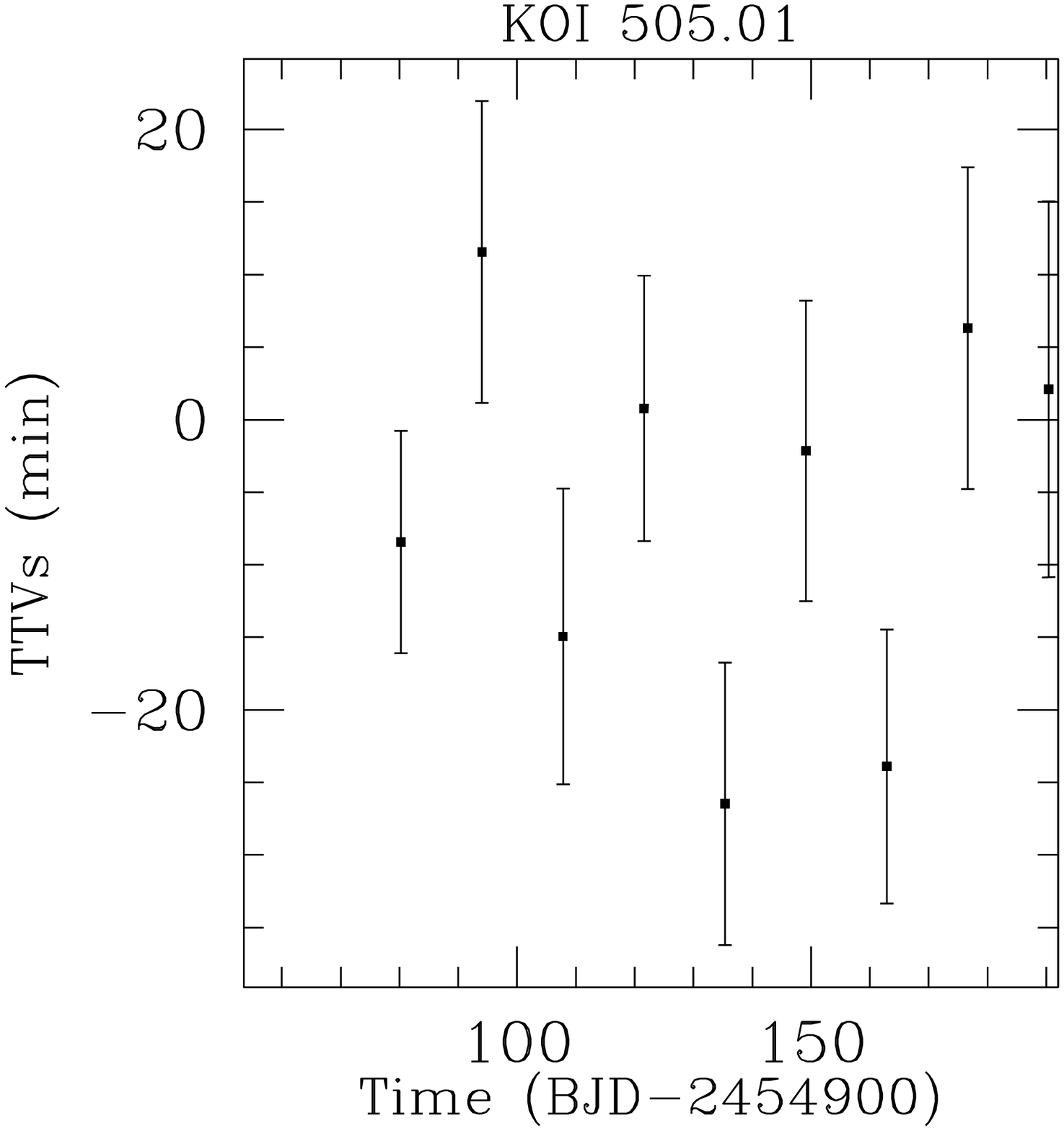}\end{figure*}
\begin{figure*}\plottwo{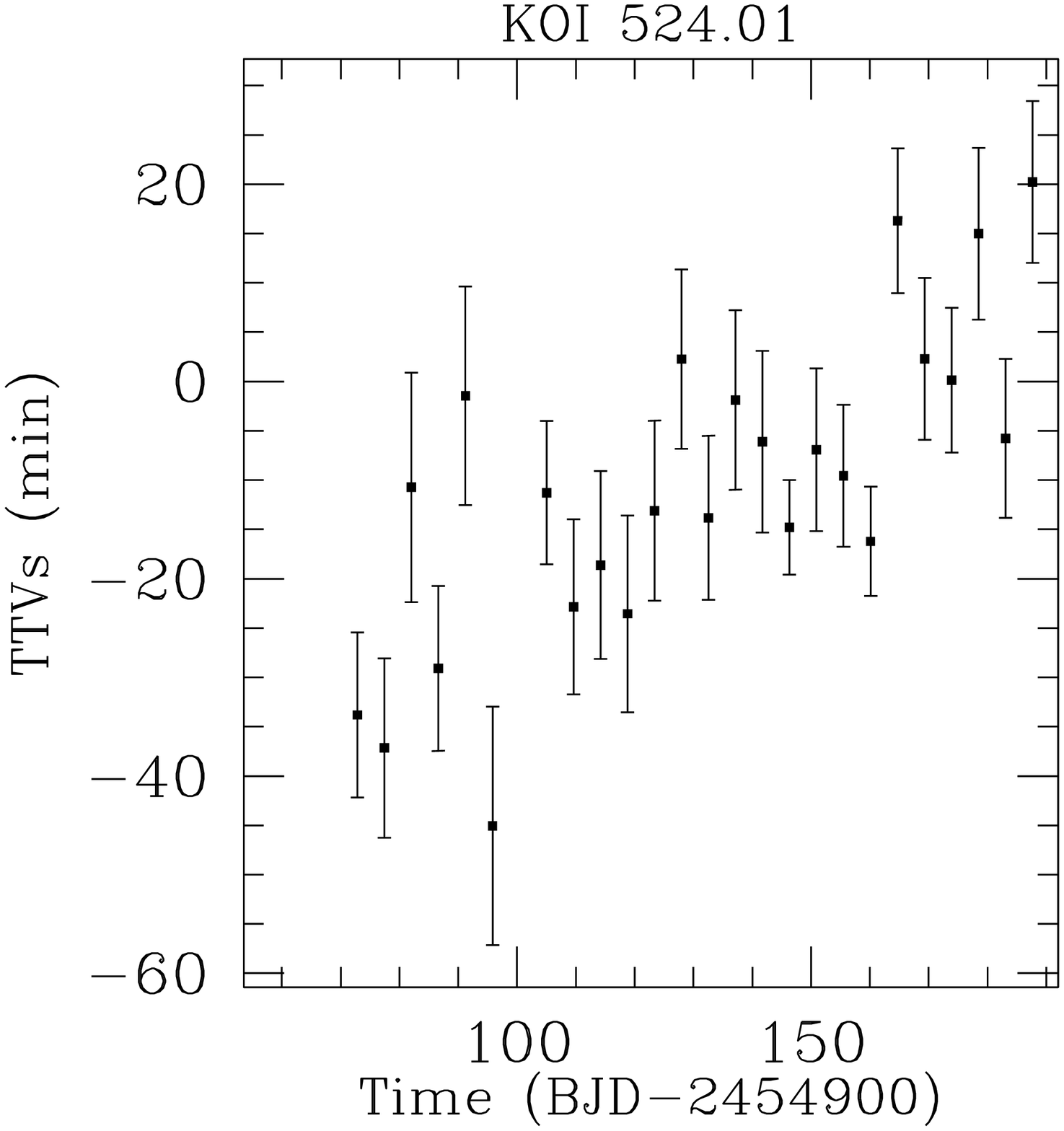}{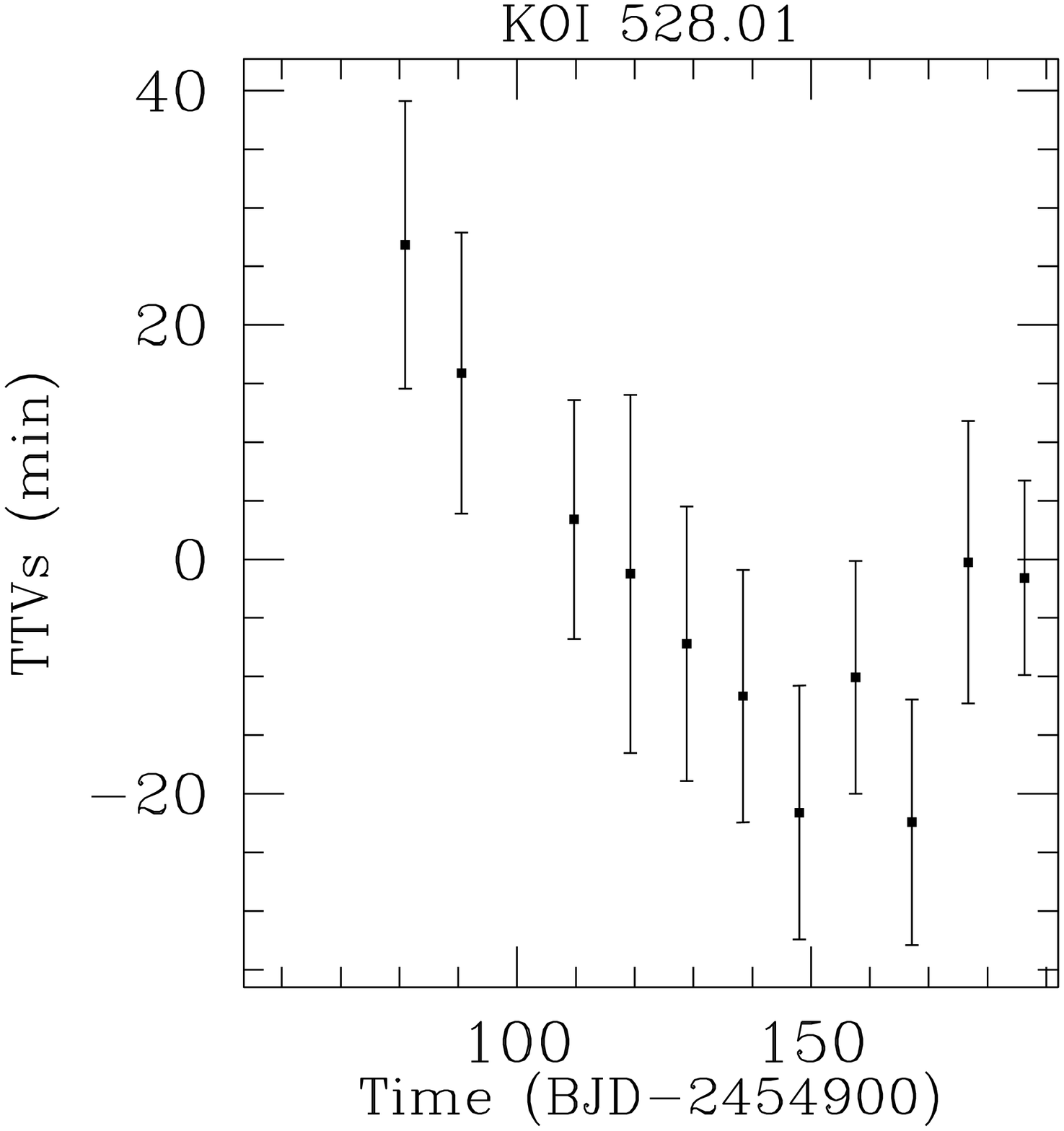}\end{figure*}\clearpage
\begin{figure*}\plottwo{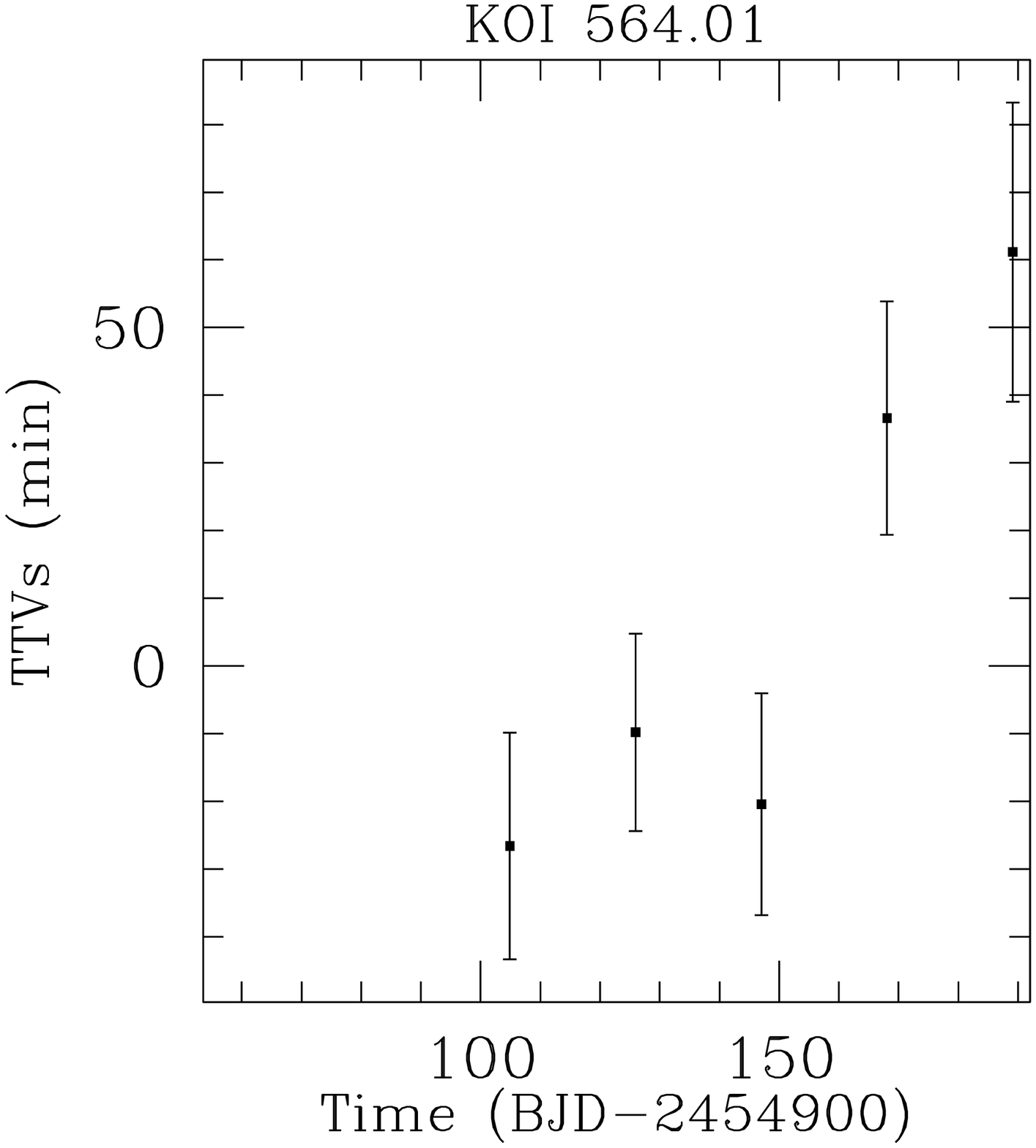}{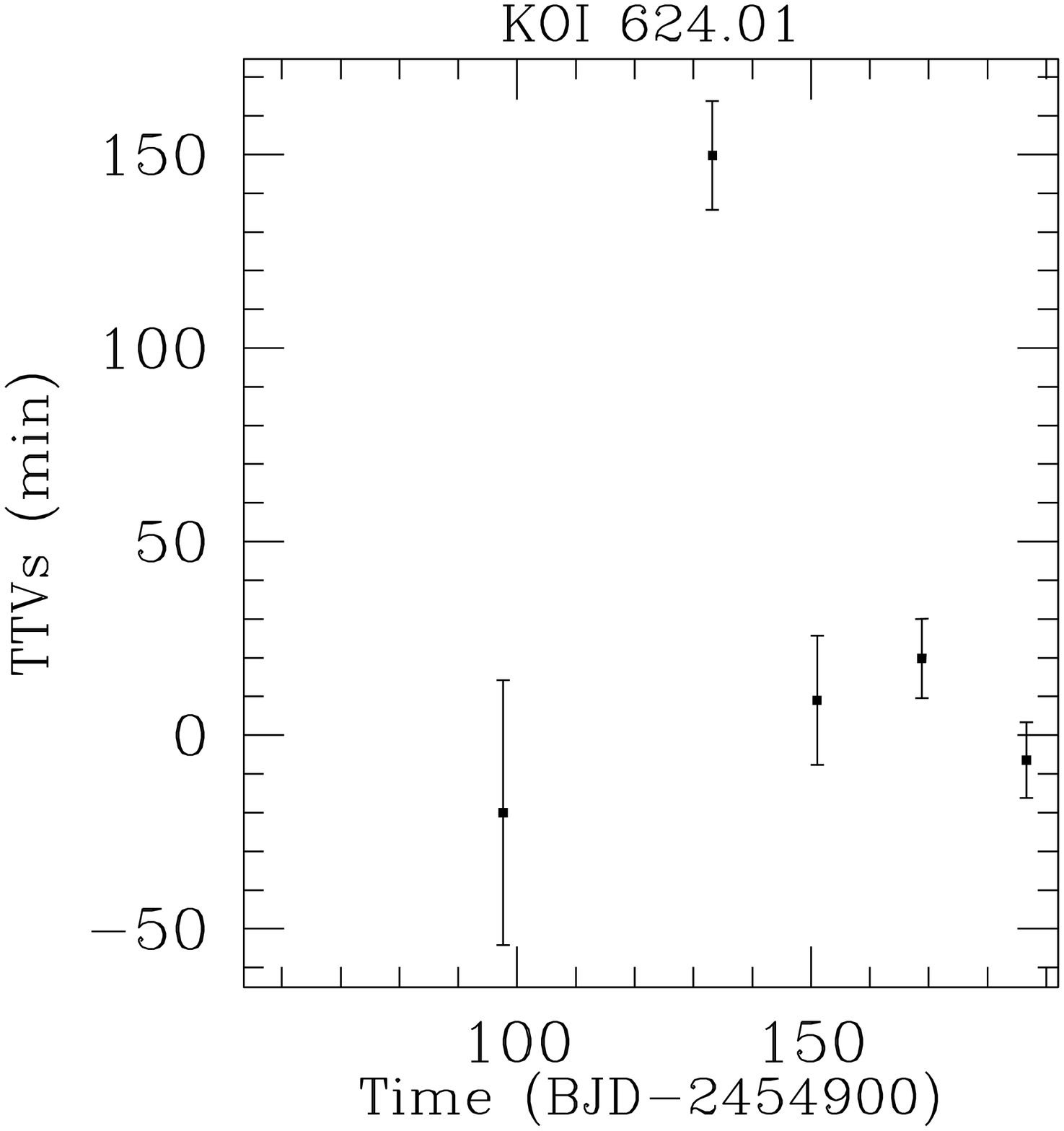}\end{figure*}
\begin{figure*}\plottwo{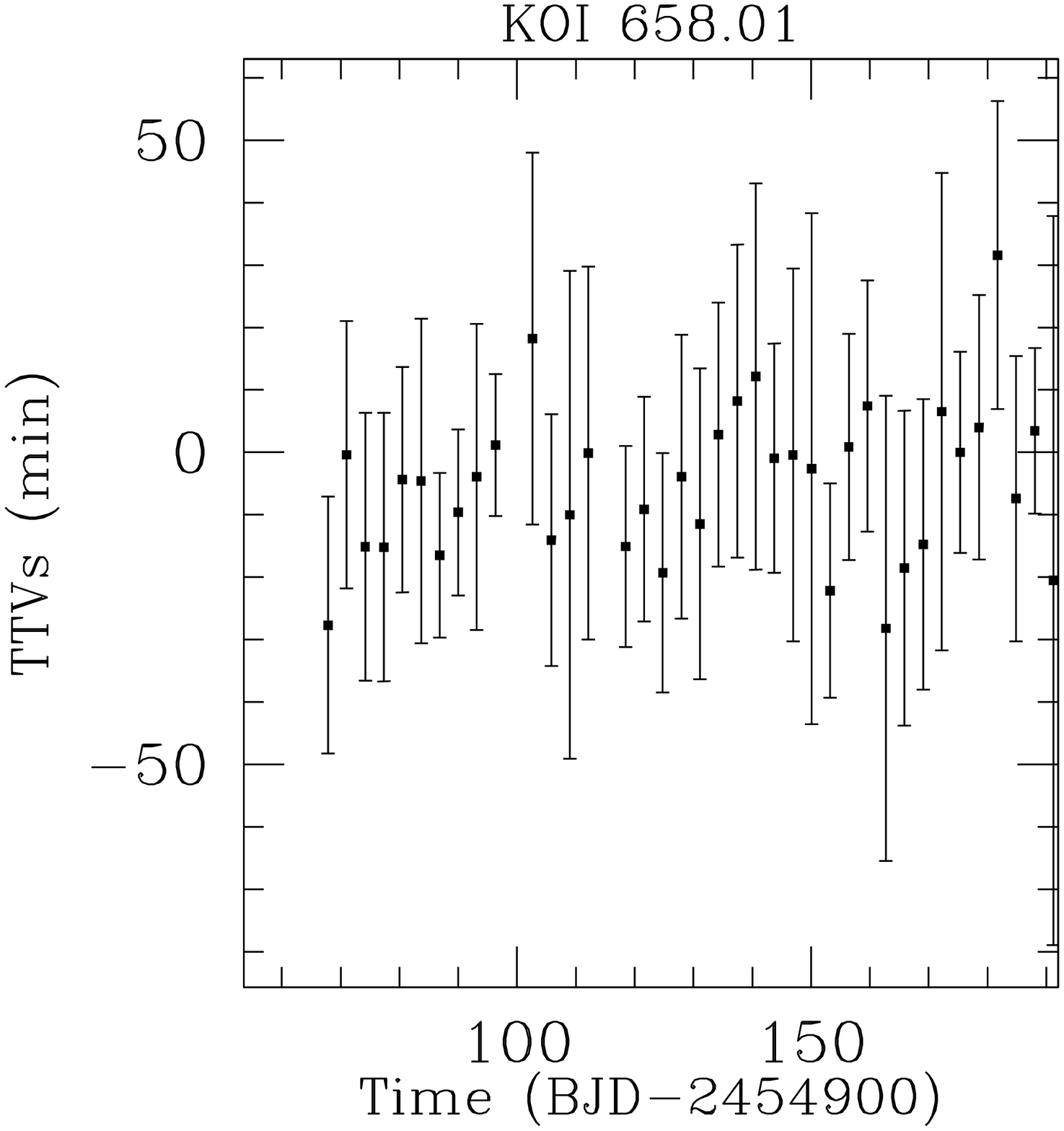}{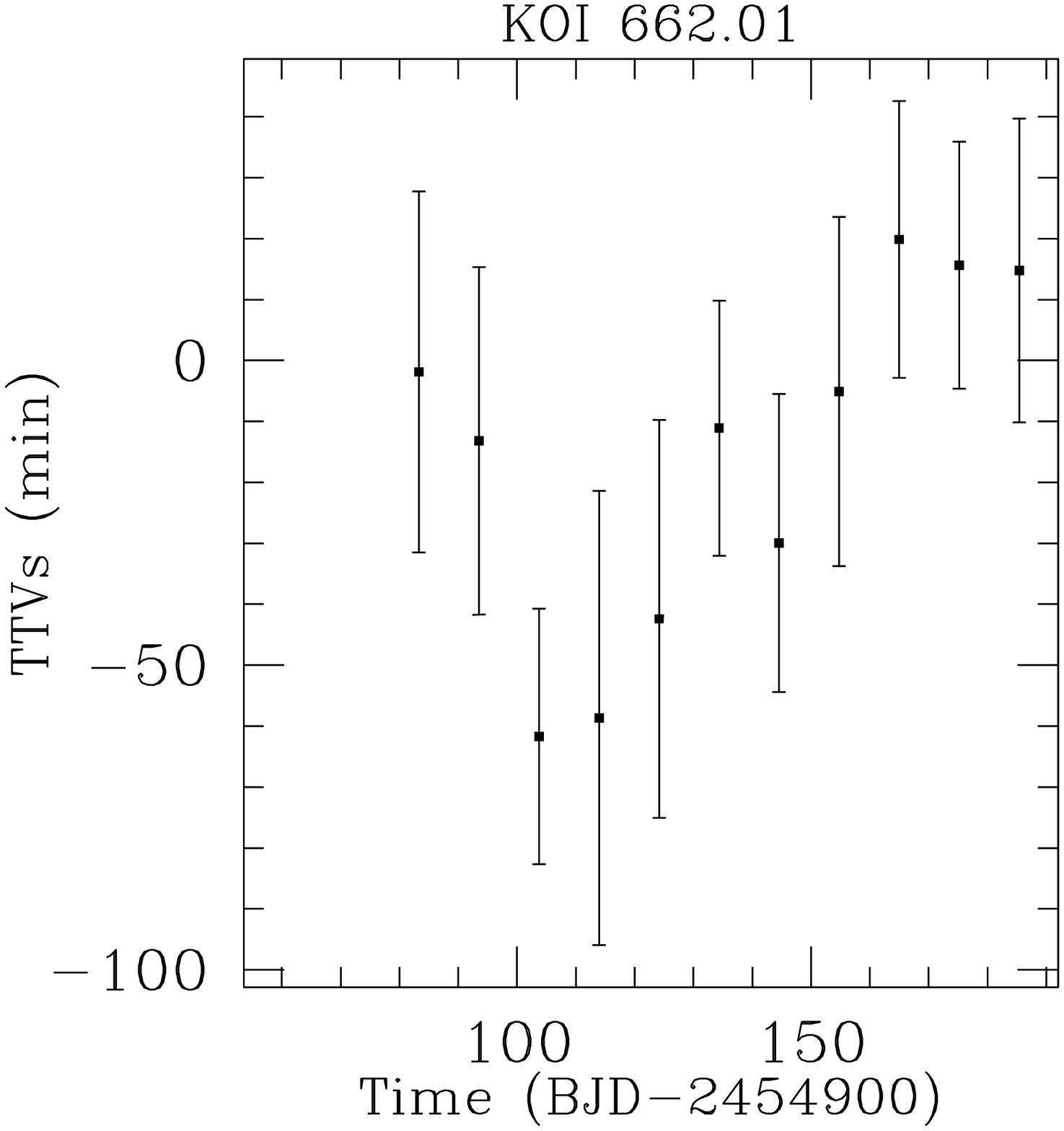}\end{figure*}\clearpage
\begin{figure*}\plottwo{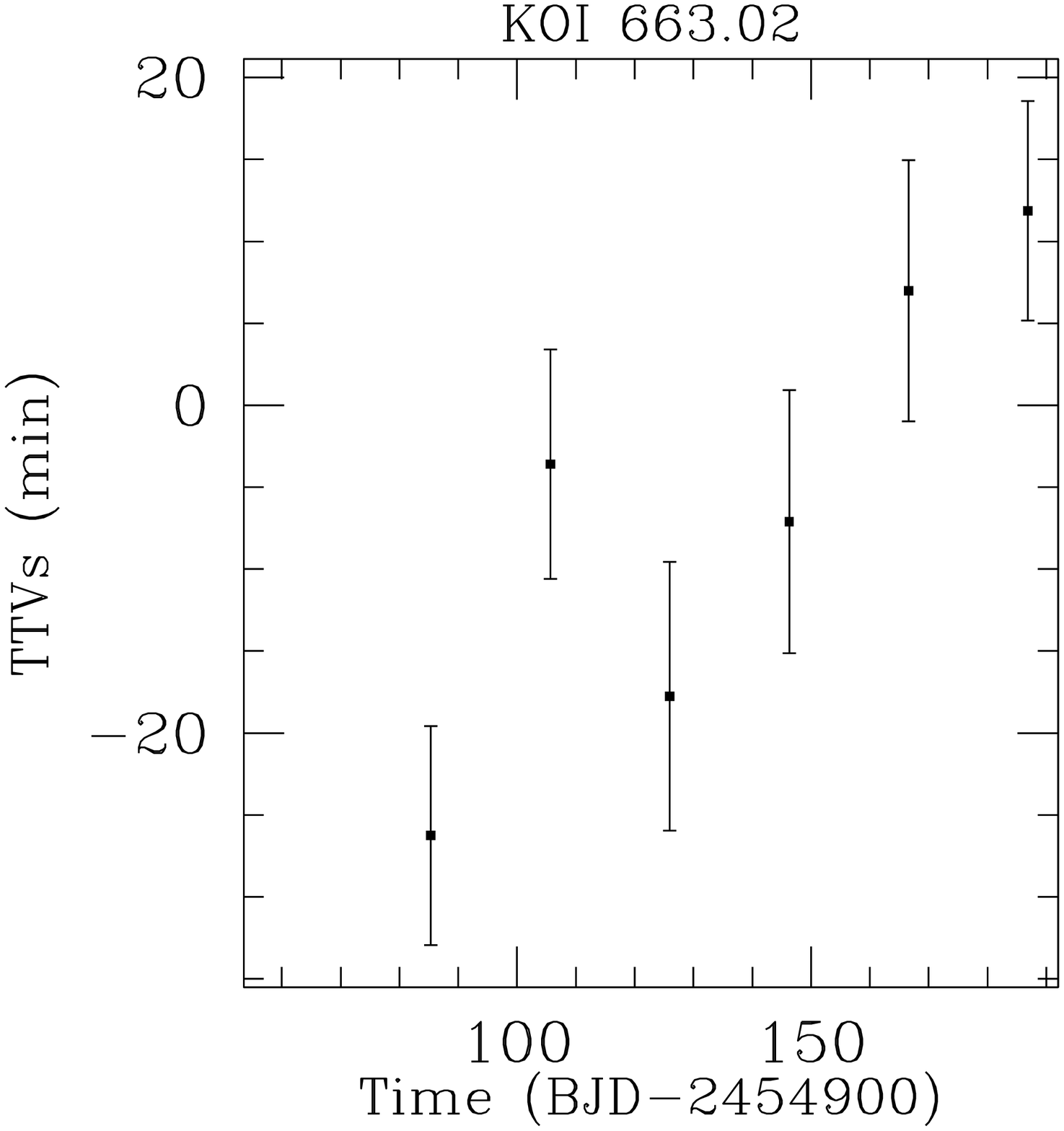}{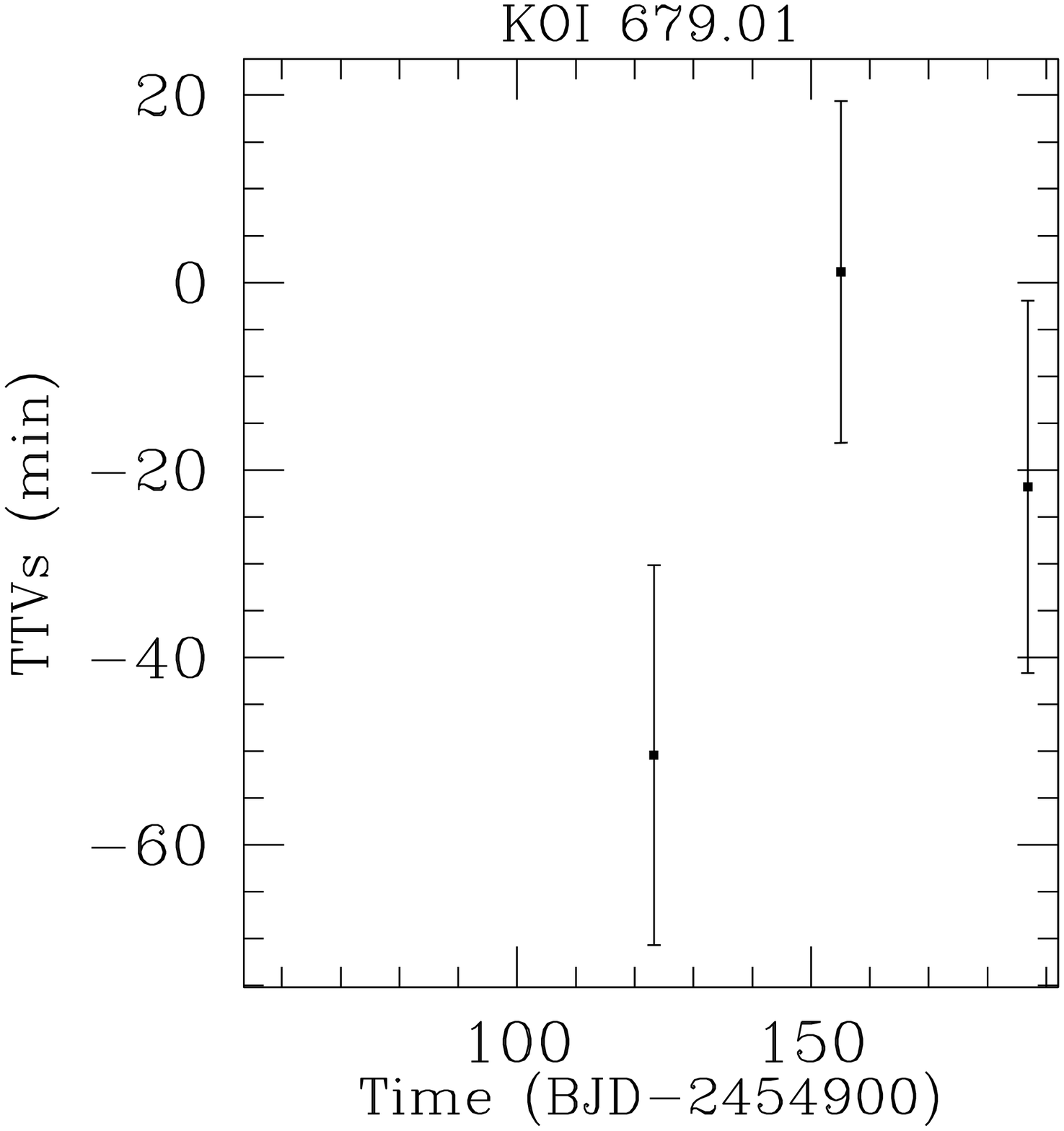}\end{figure*}
\begin{figure*}\plottwo{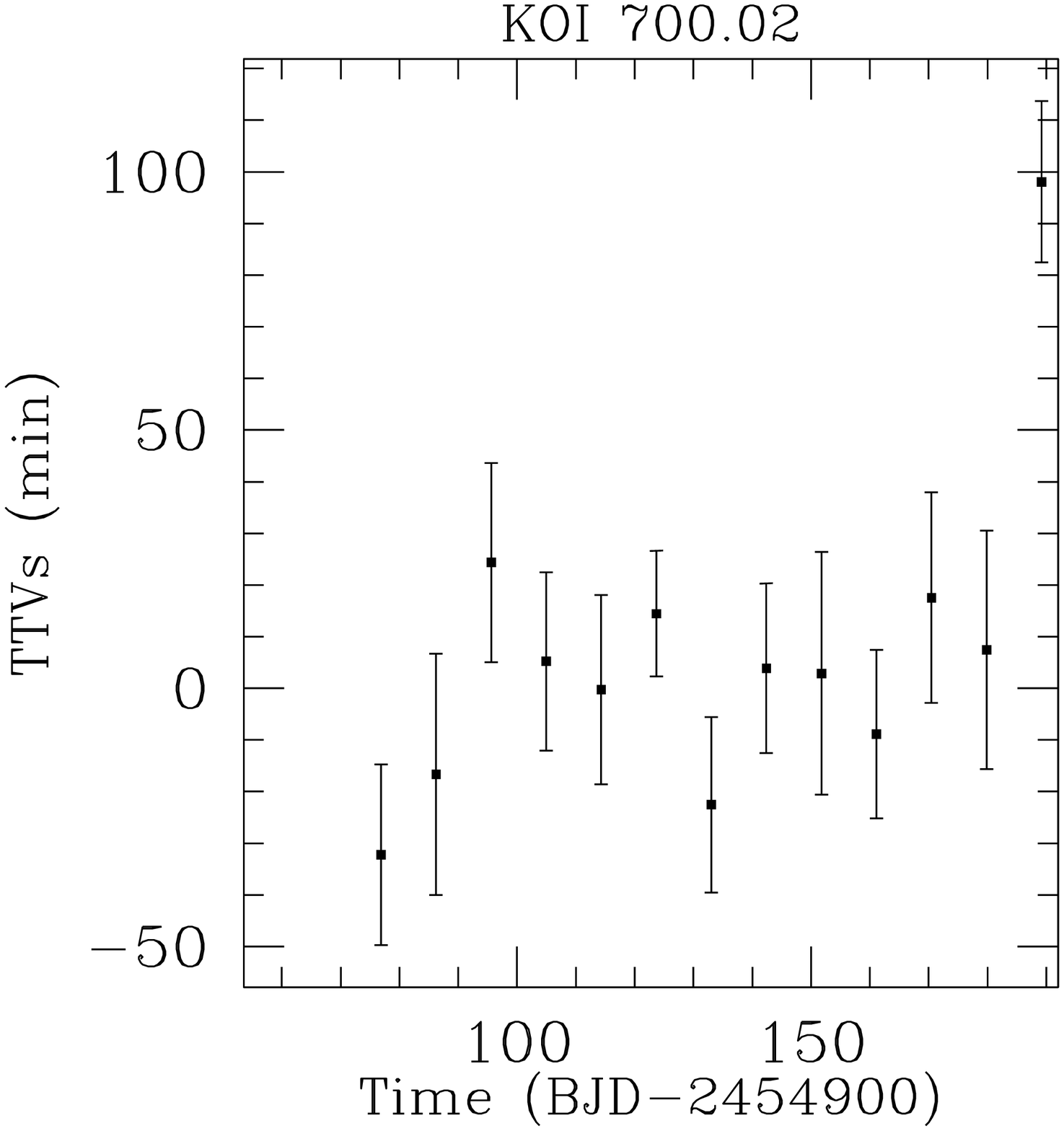}{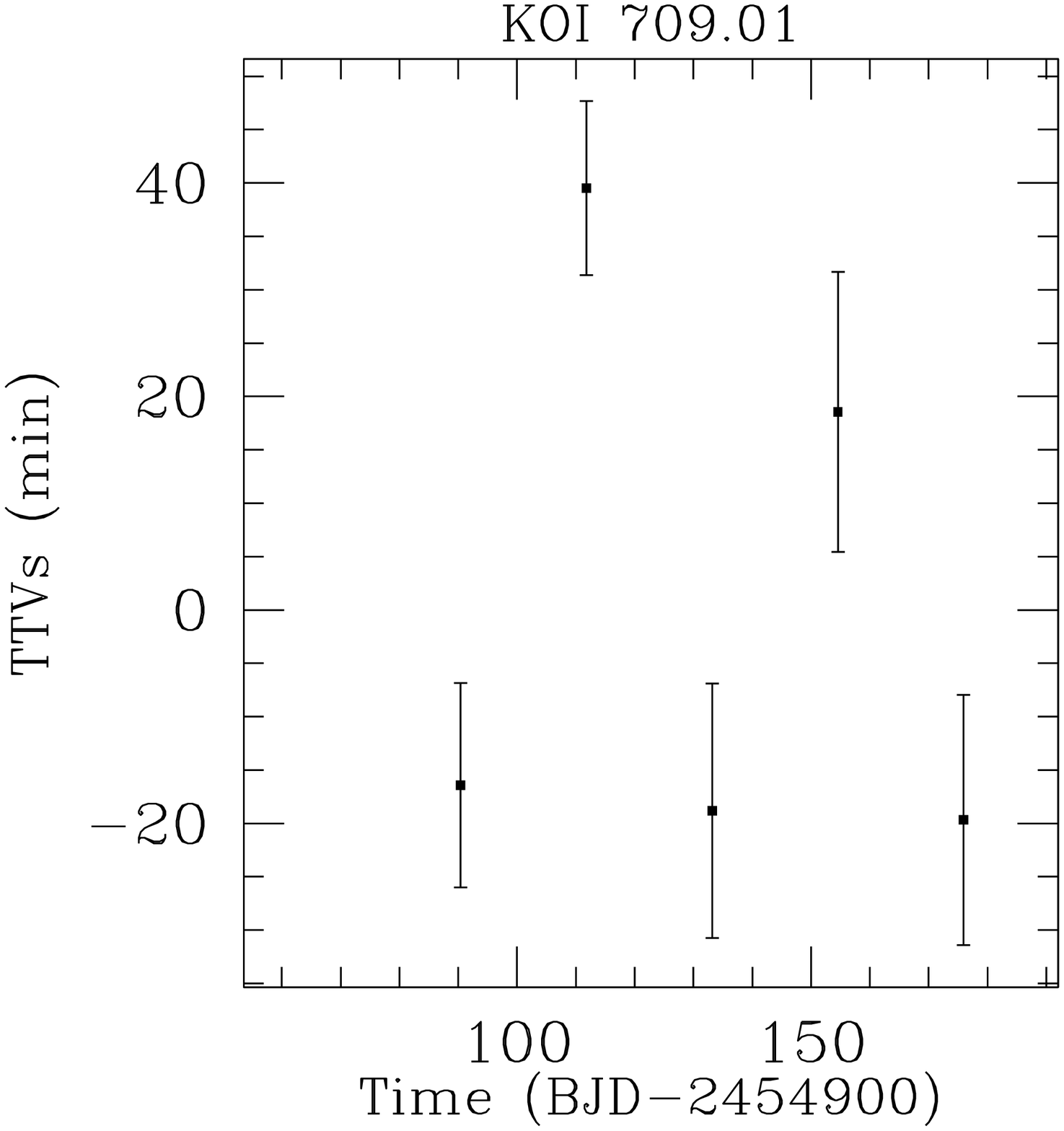}\end{figure*}\clearpage
\begin{figure*}\plottwo{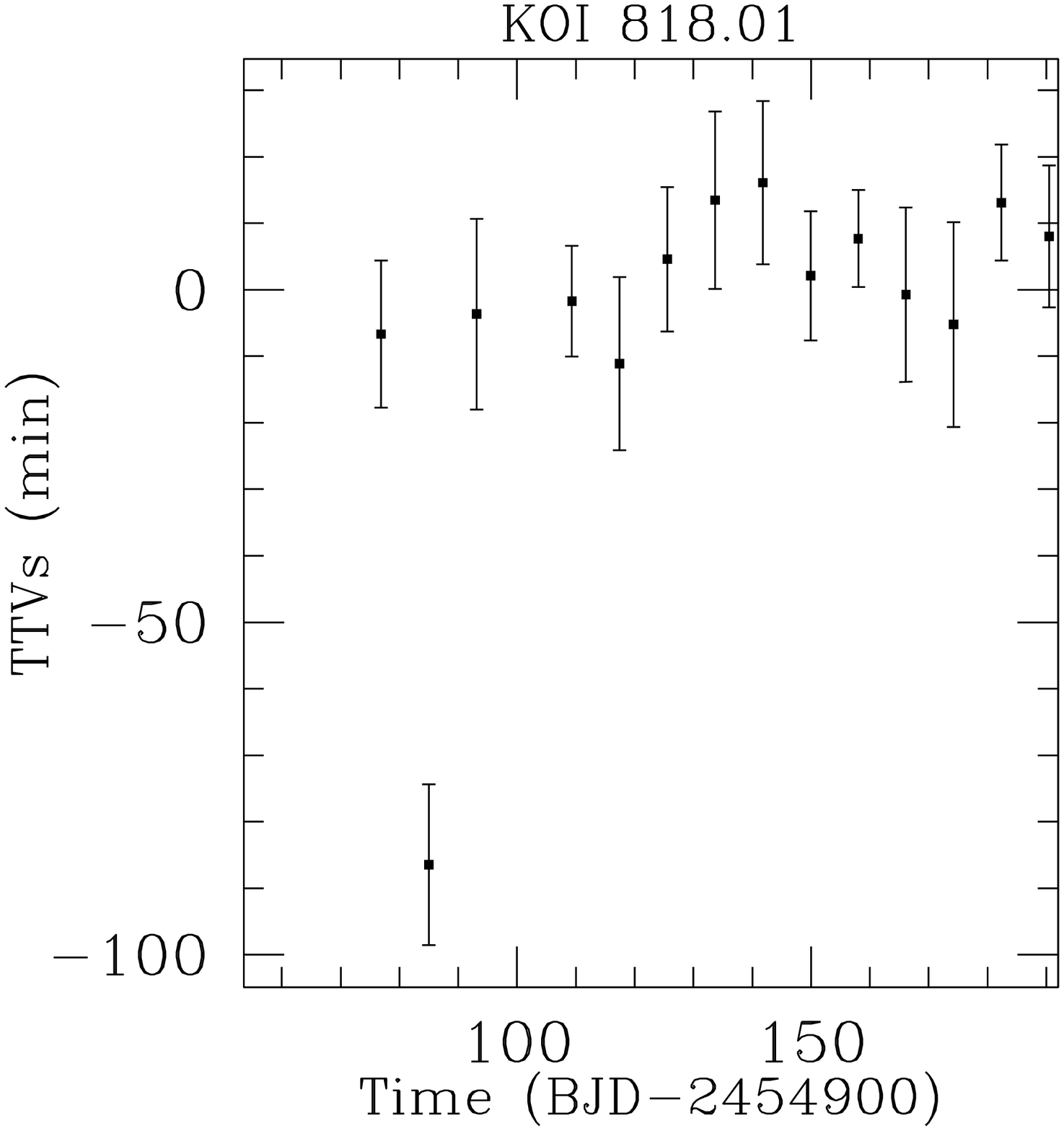}{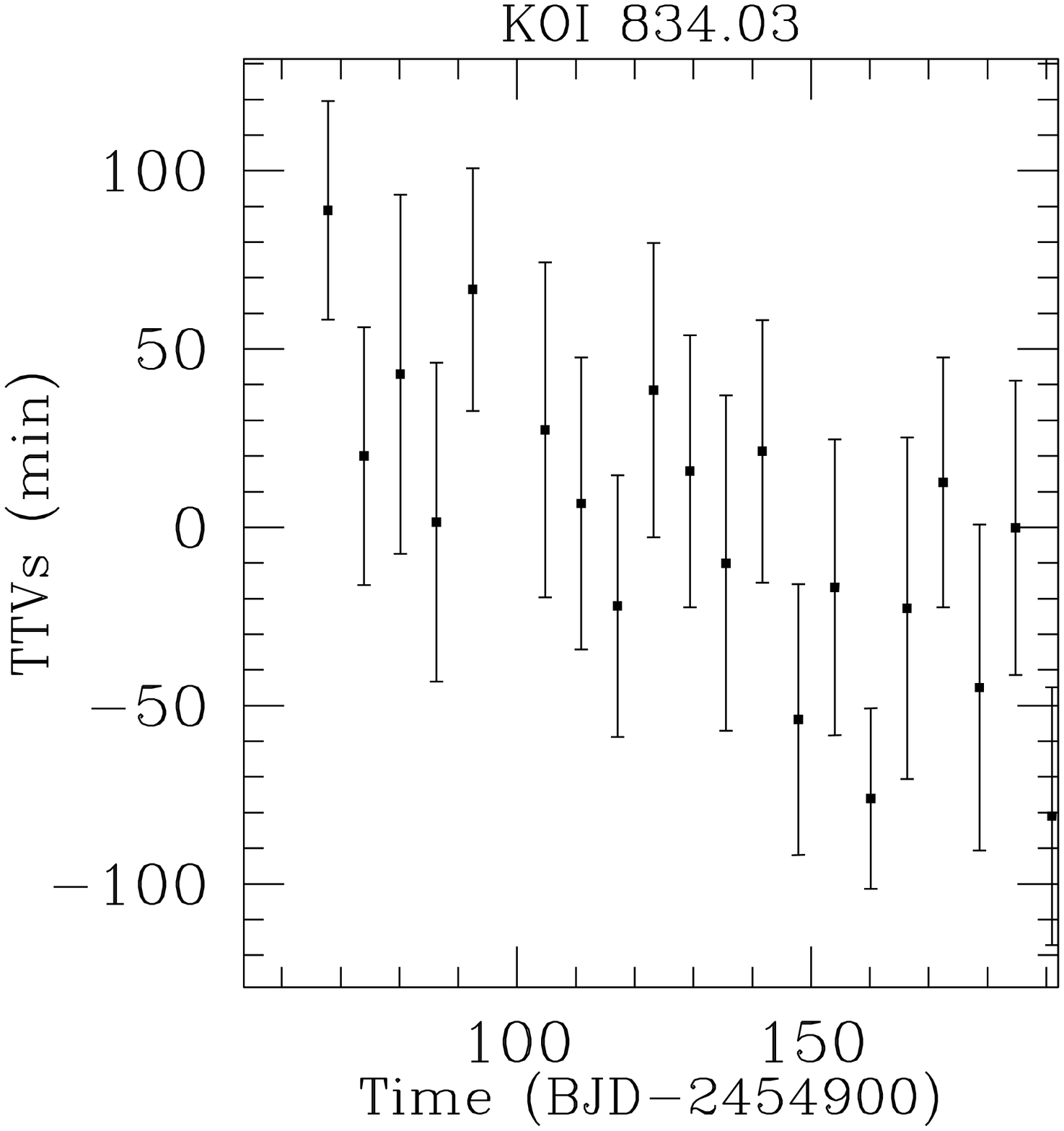}\end{figure*}
\begin{figure*}\plottwo{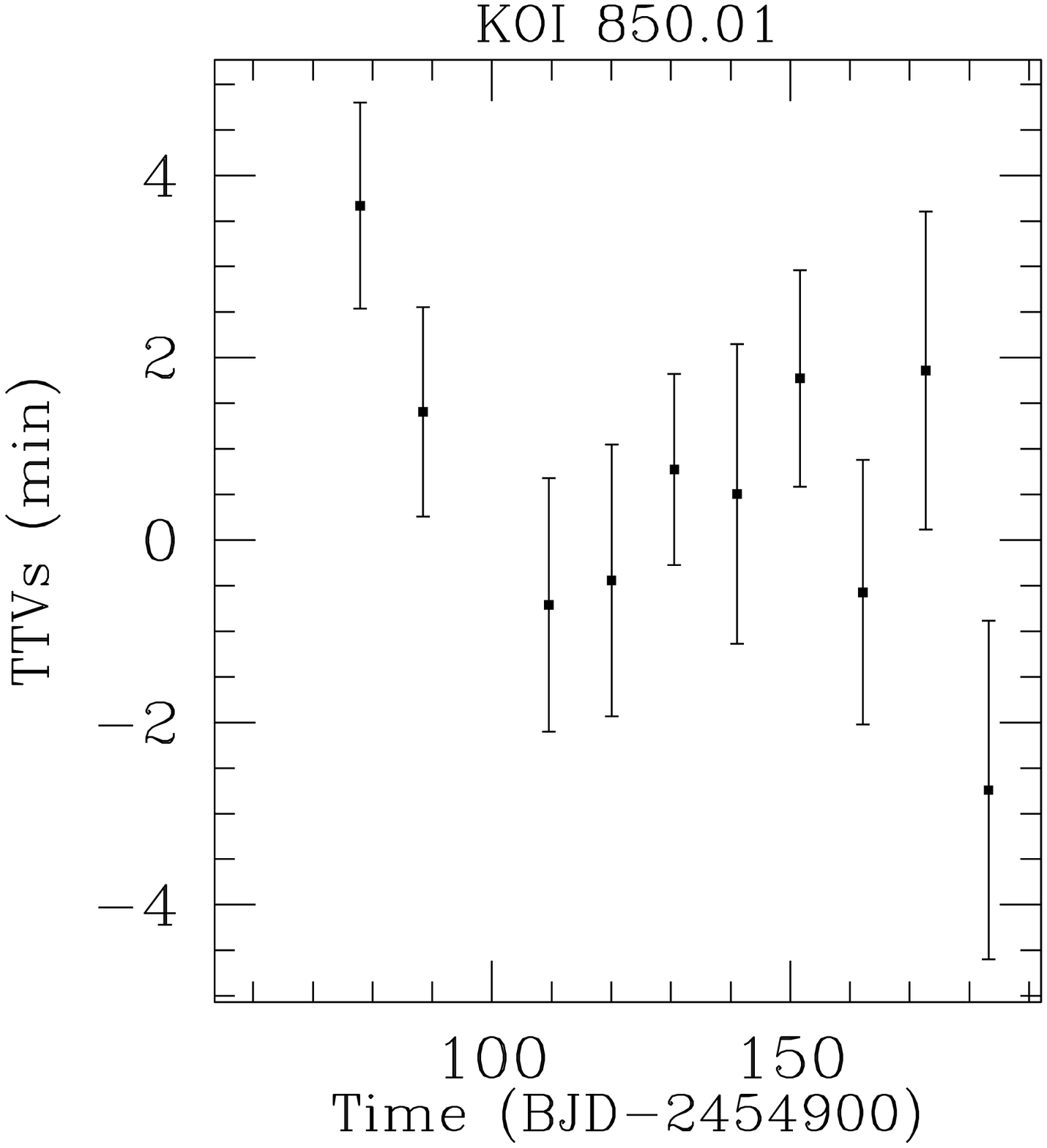}{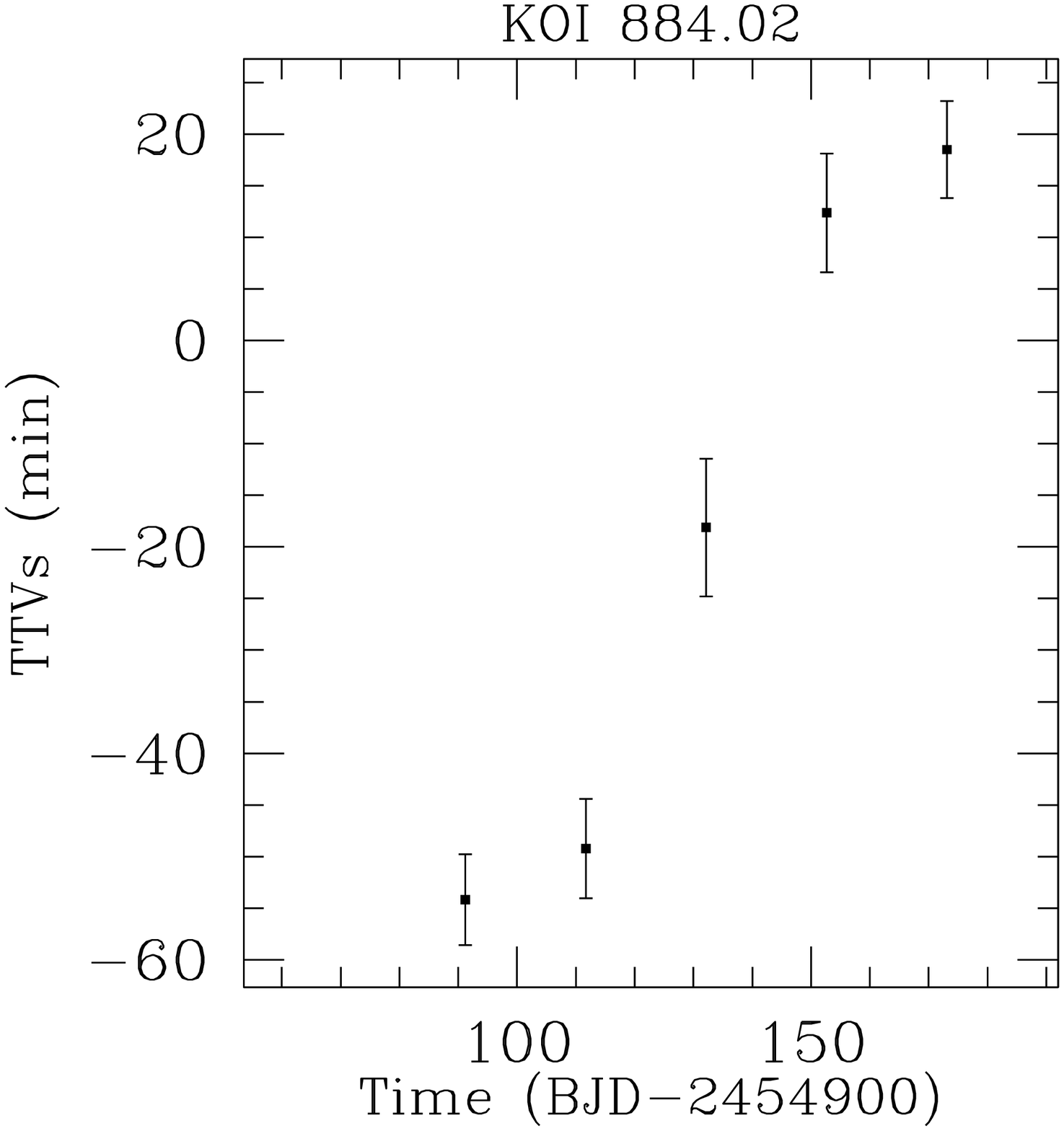}\end{figure*}\clearpage
\begin{figure*}\plottwo{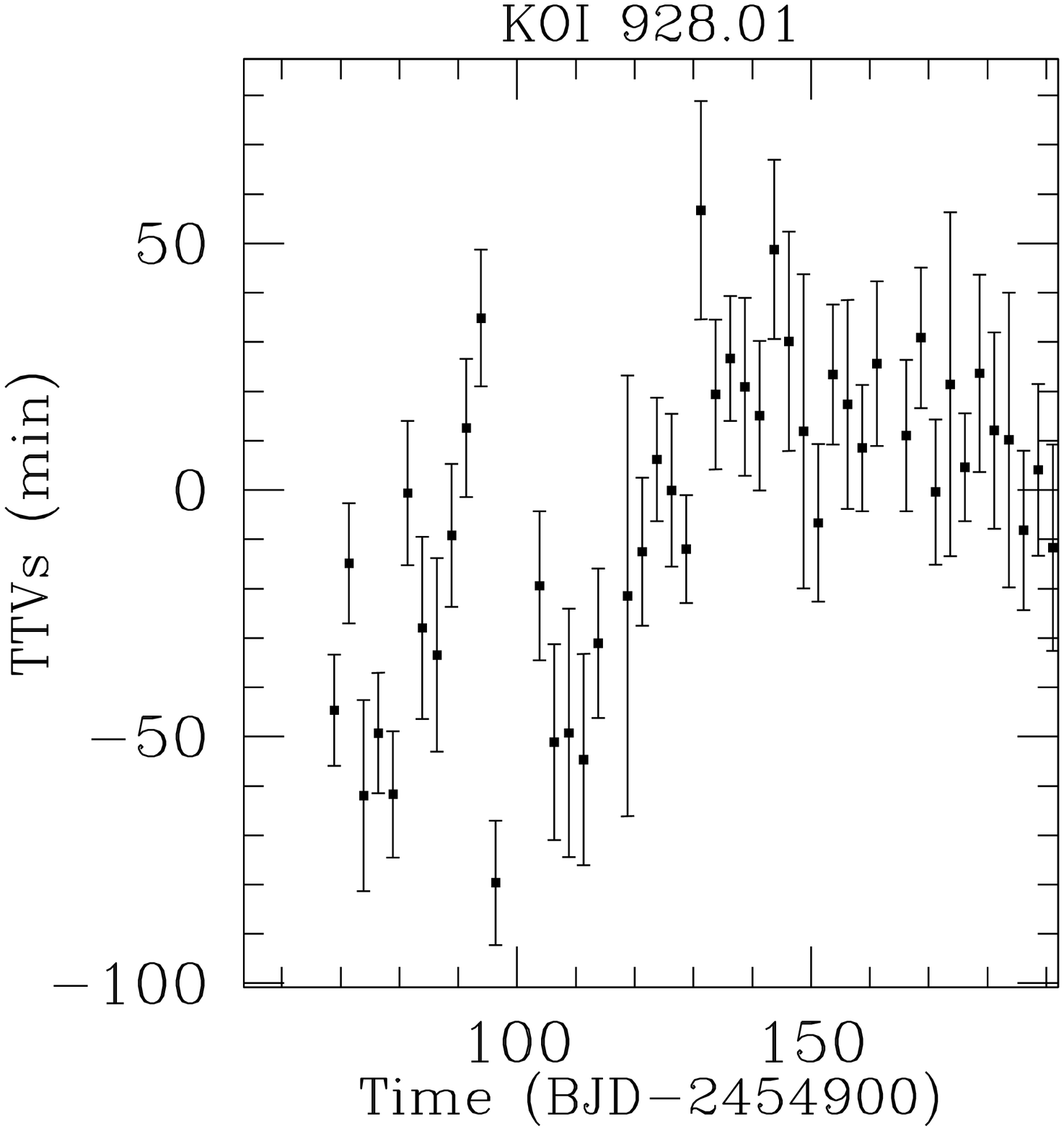}{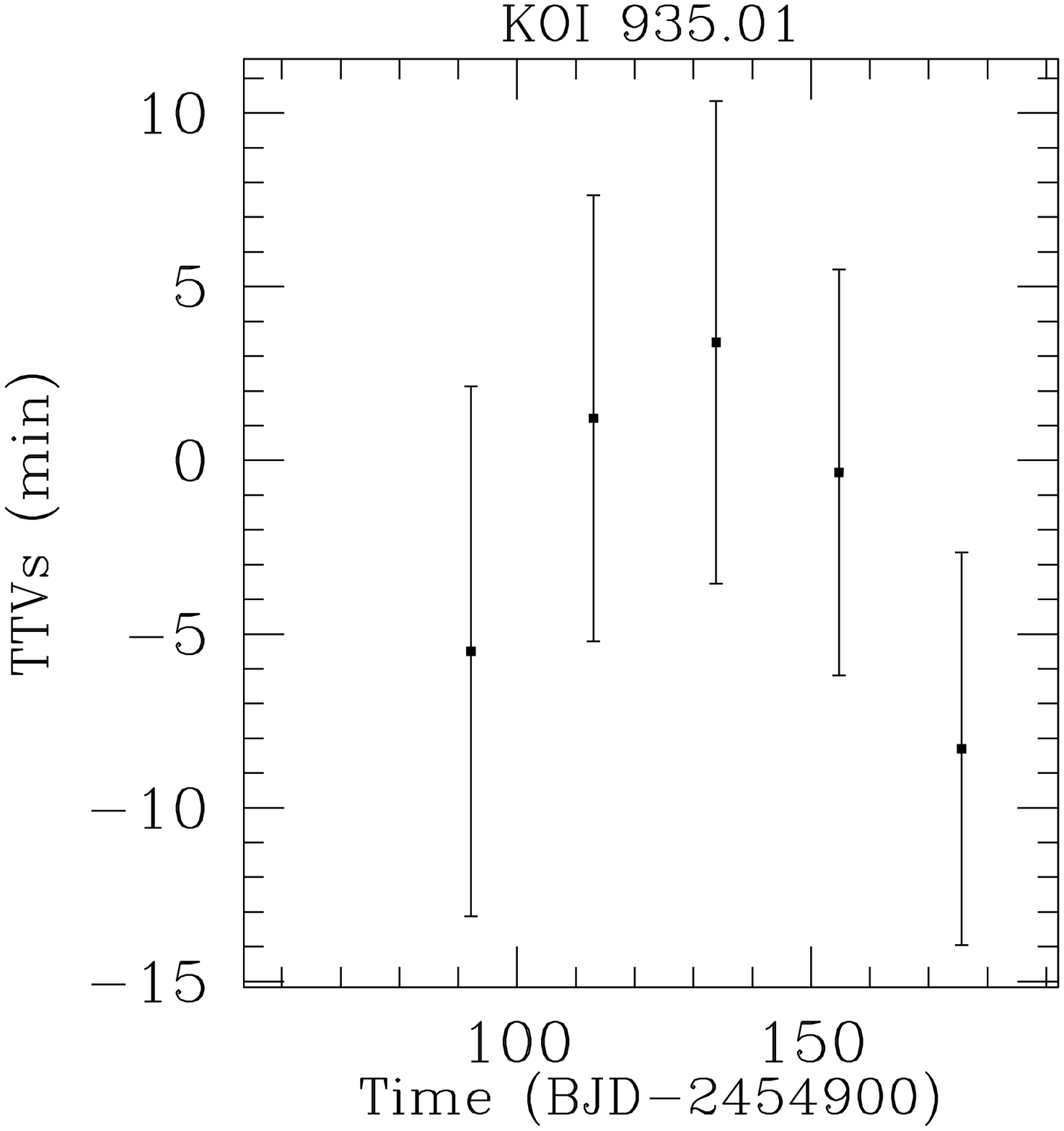}\end{figure*}
\begin{figure*}\plottwo{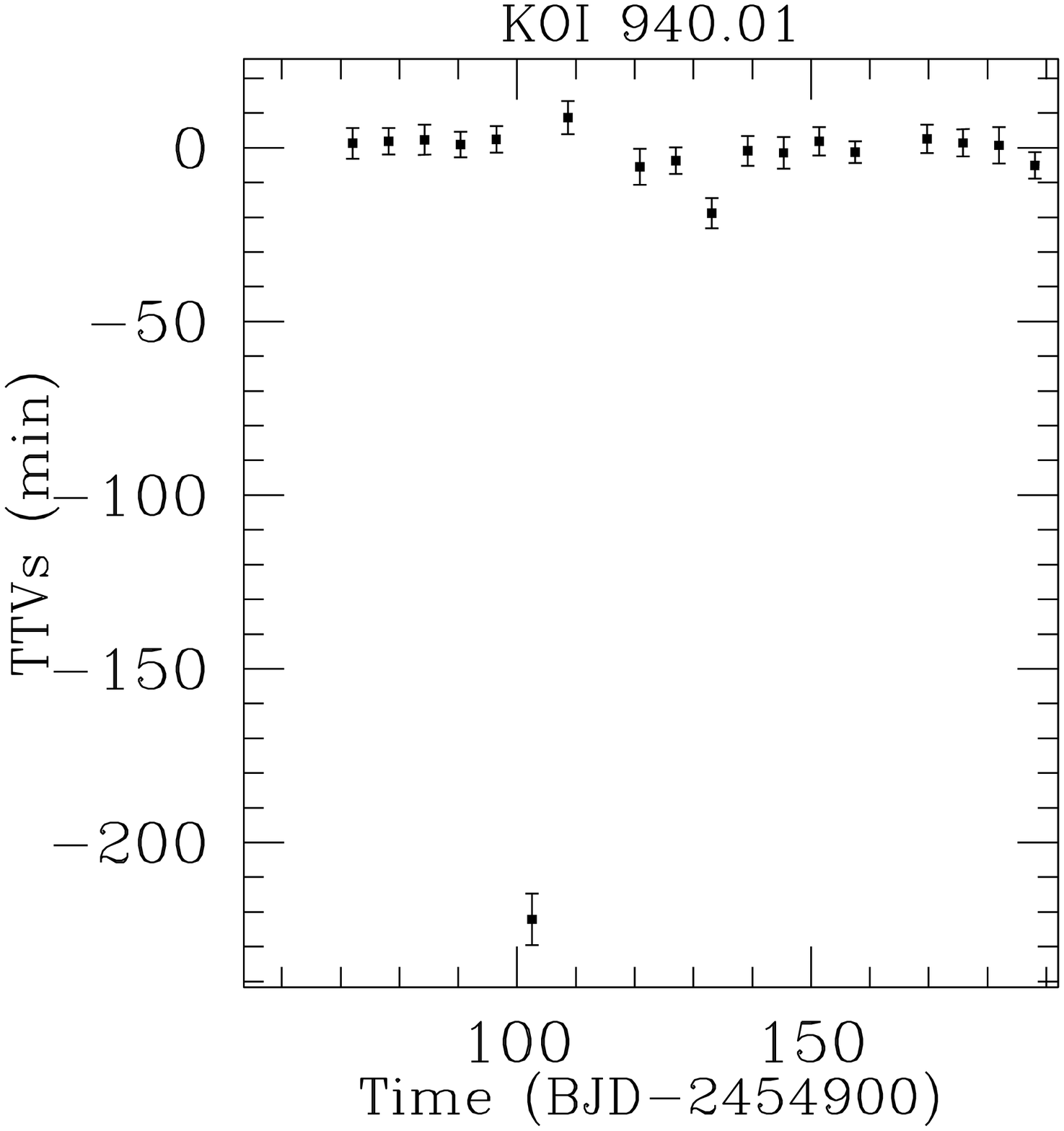}{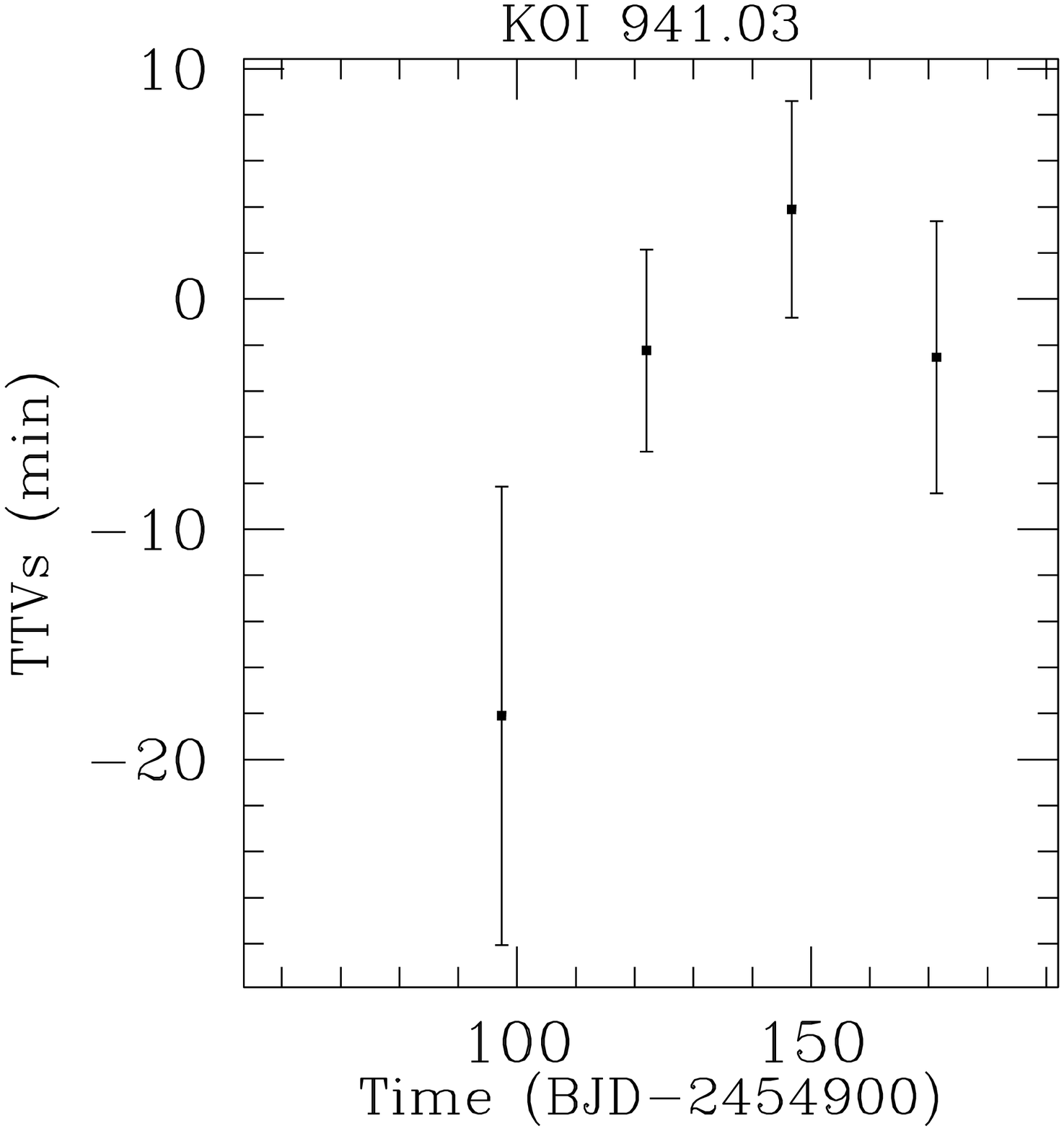}\end{figure*}\clearpage
\begin{figure*}\plottwo{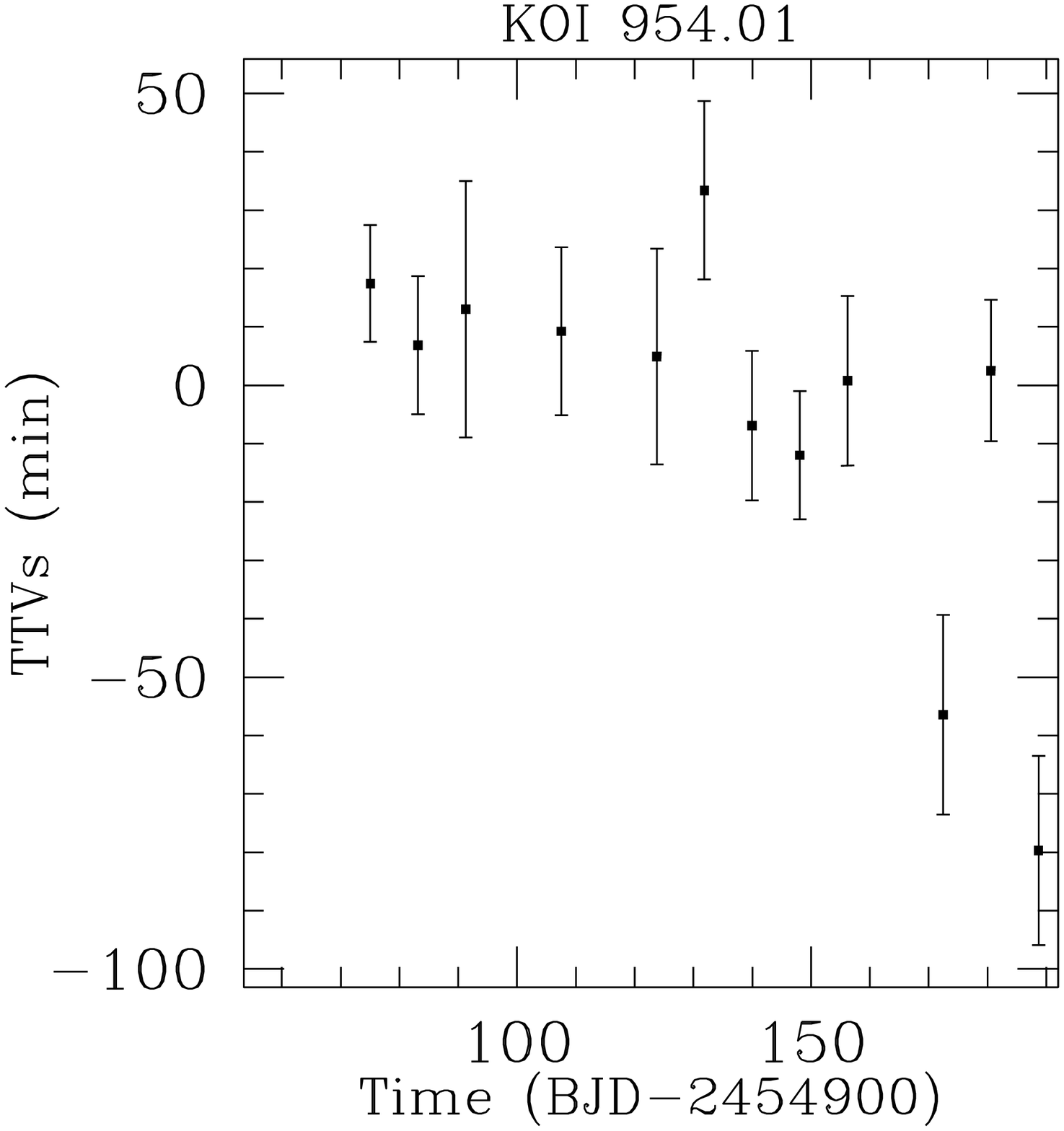}{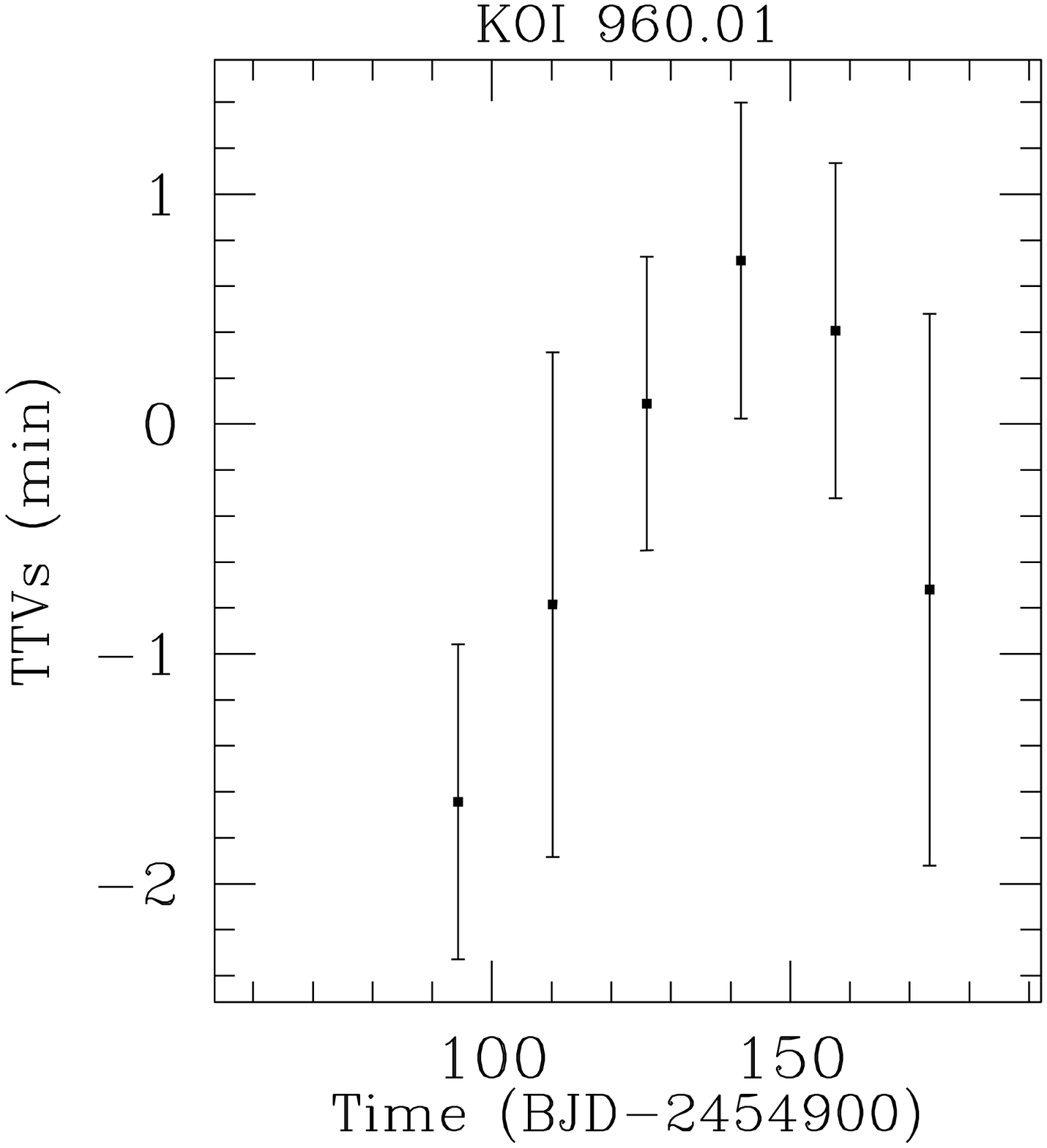}\end{figure*}
\begin{figure*}\plottwo{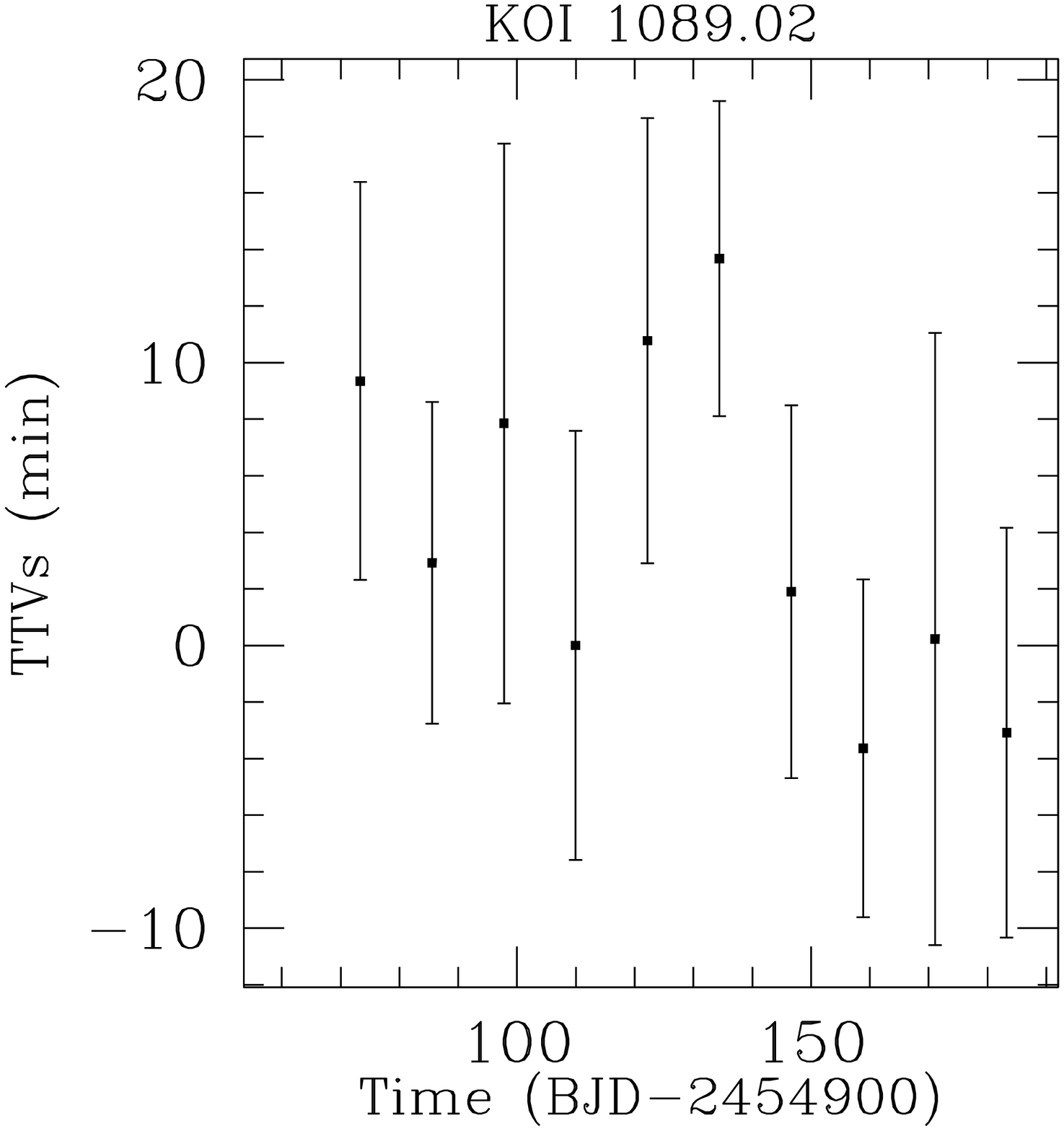}{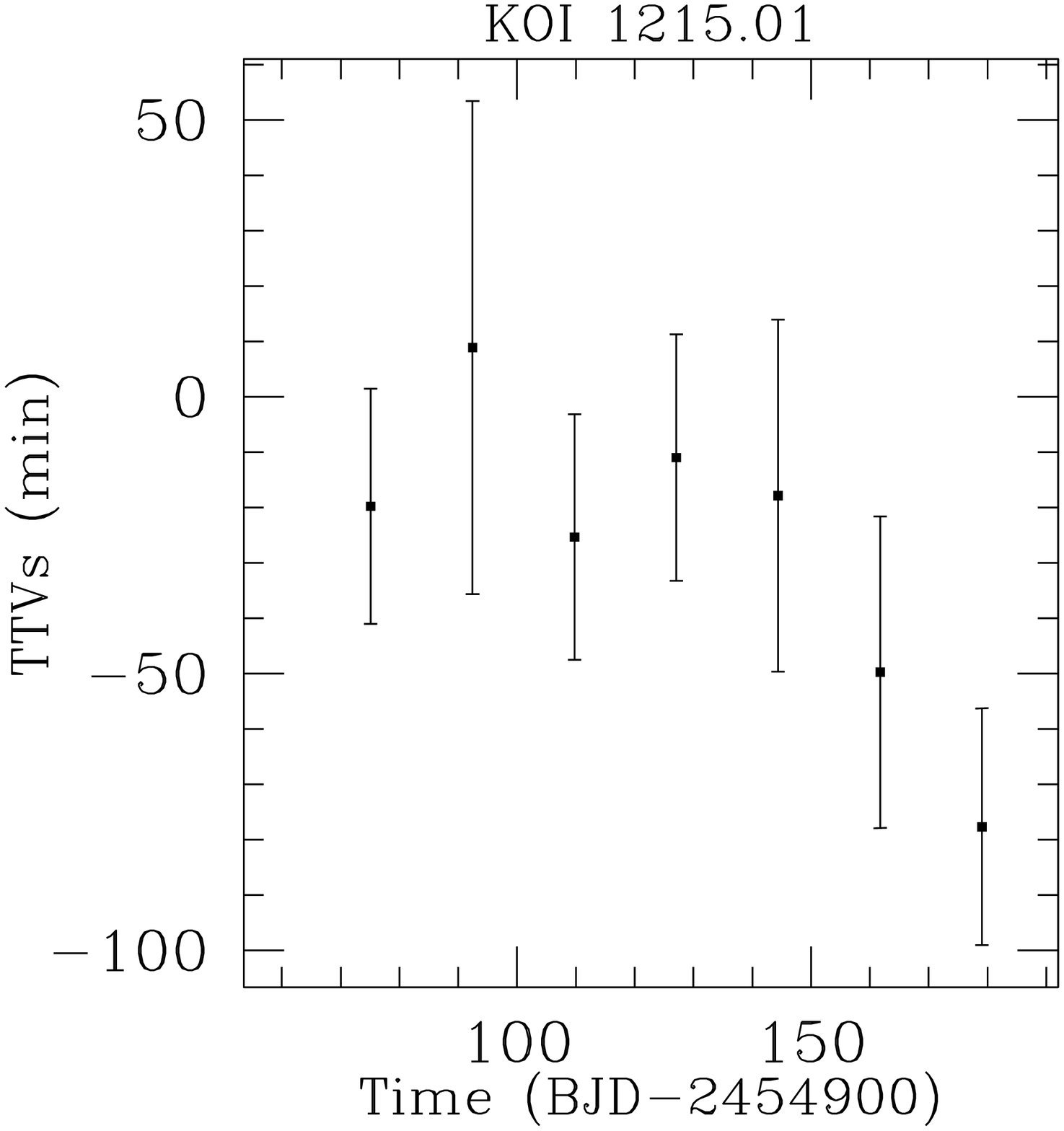}\end{figure*}\clearpage
\begin{figure*}\plottwo{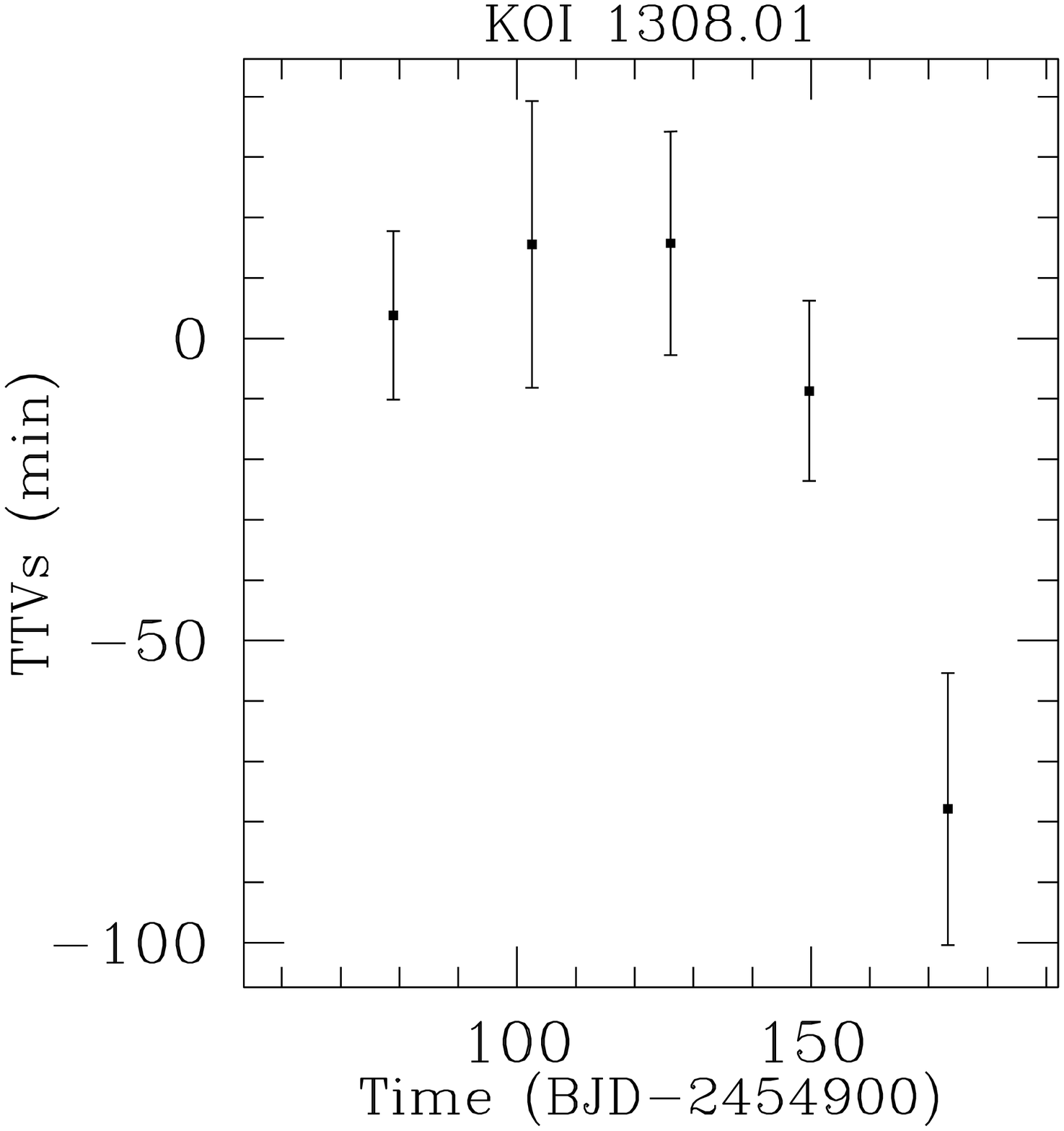}{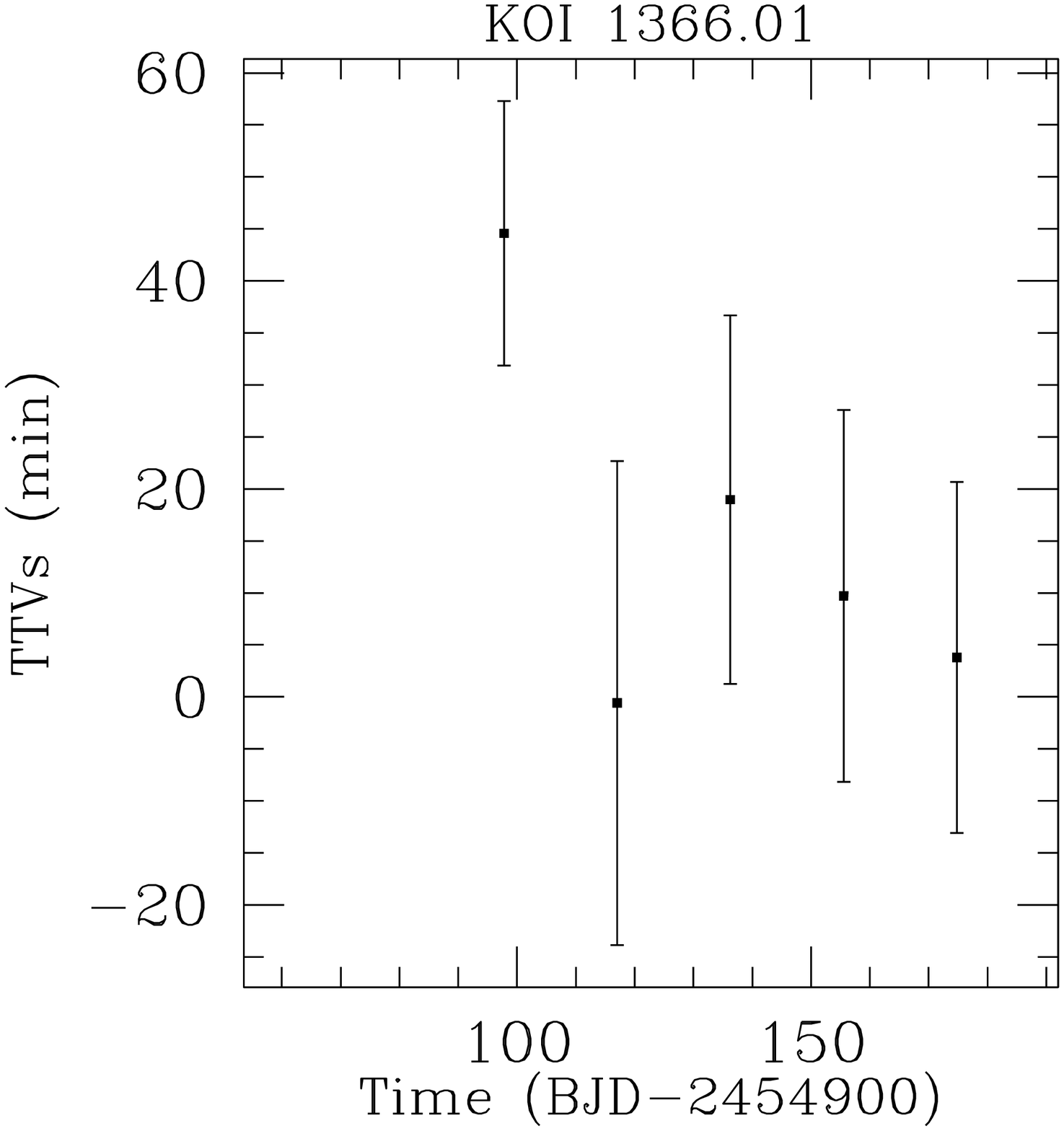}\end{figure*}
\begin{figure*}\plottwo{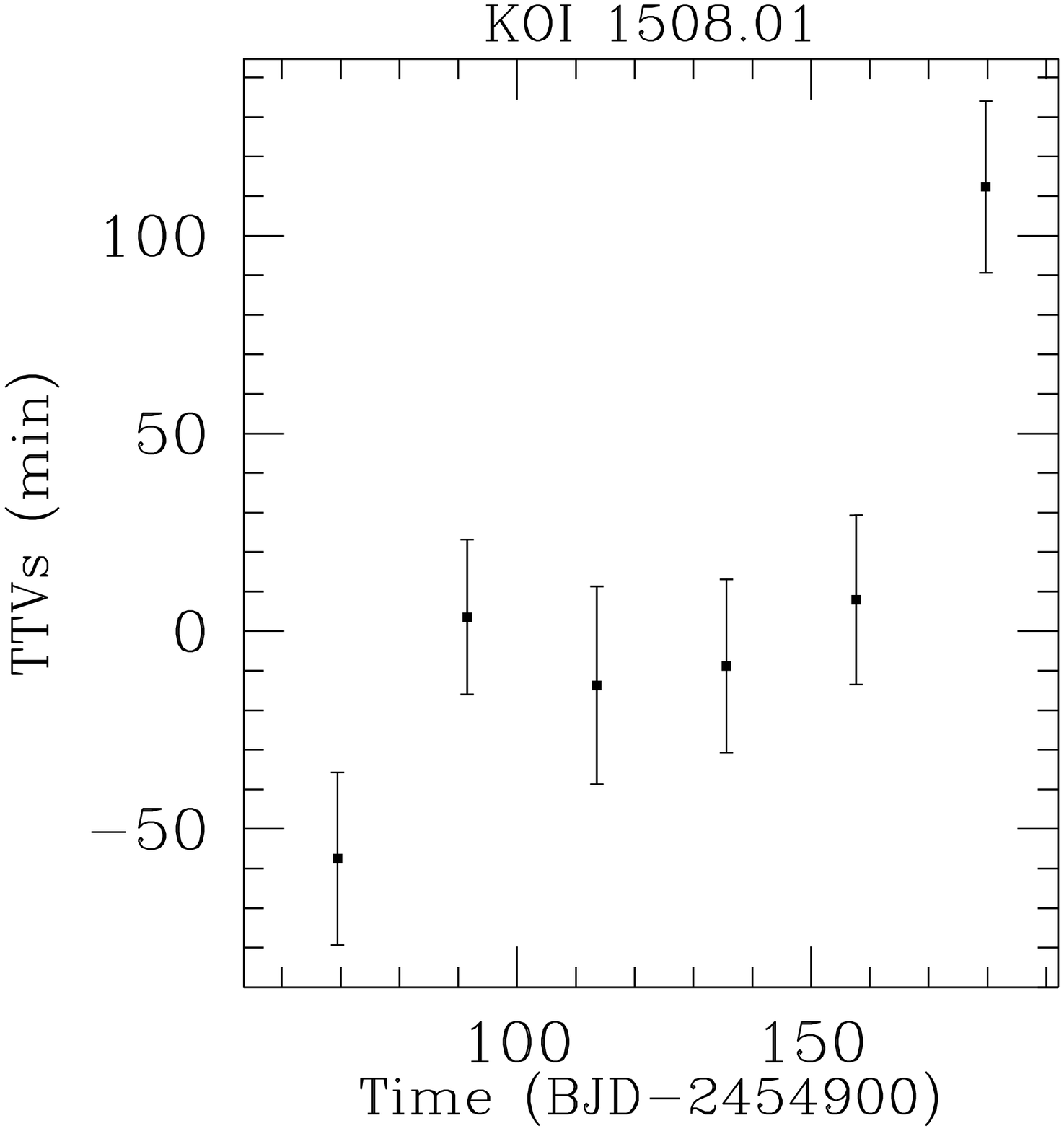}{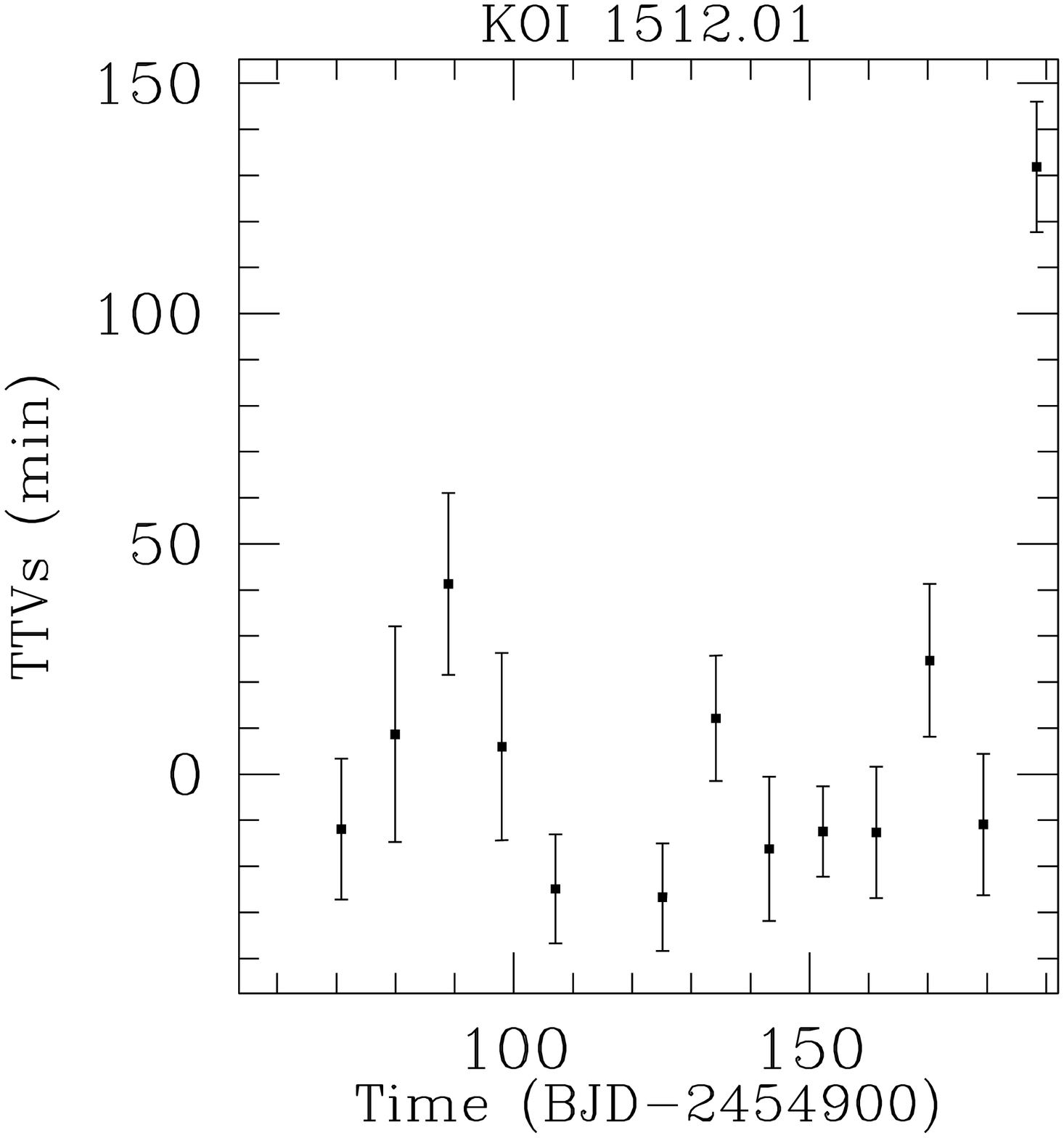}\end{figure*}

\clearpage

\section*{Weak TTV Candidates}

\begin{figure*}\plottwo{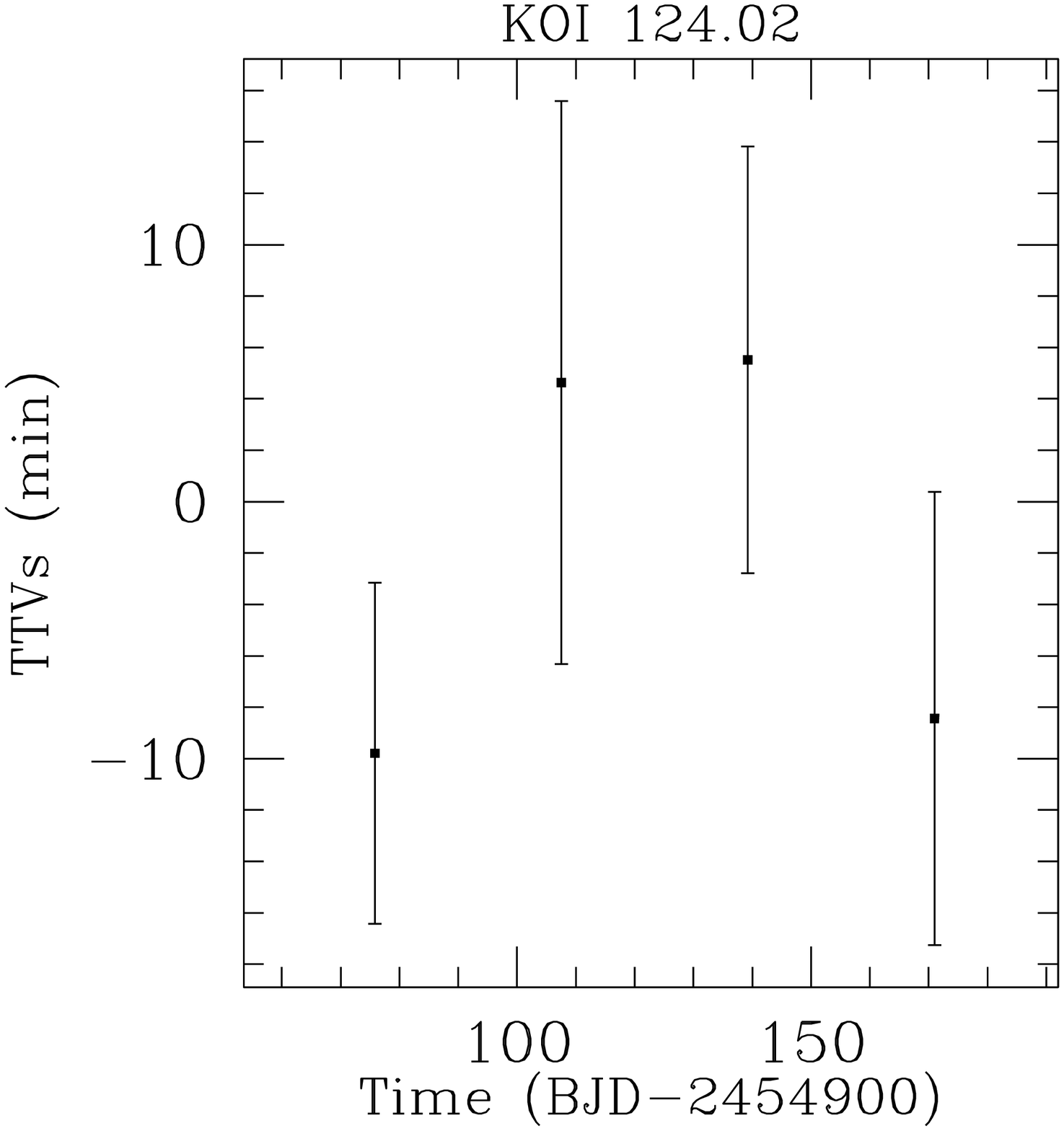}{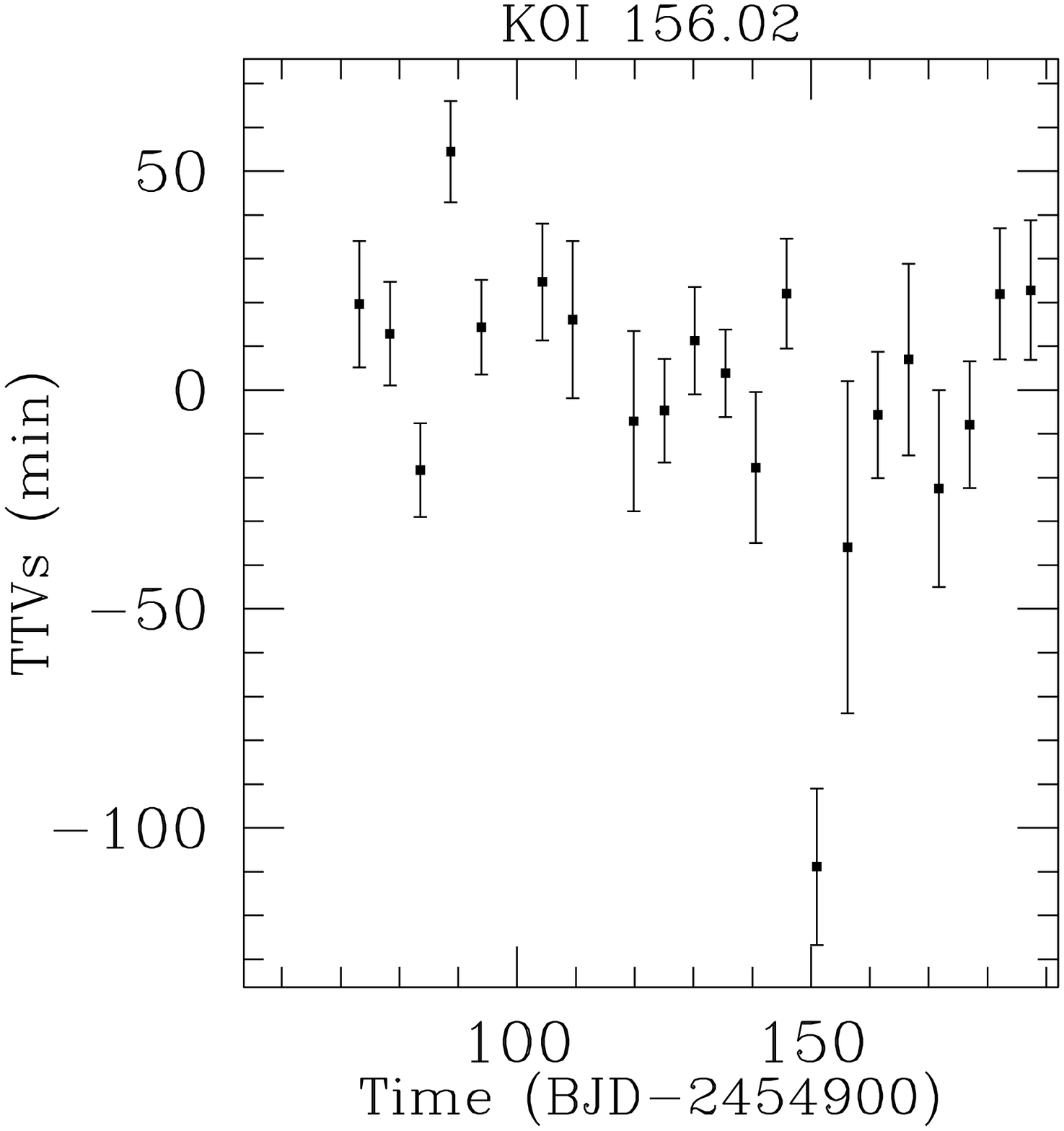}\end{figure*}
\begin{figure*}\plottwo{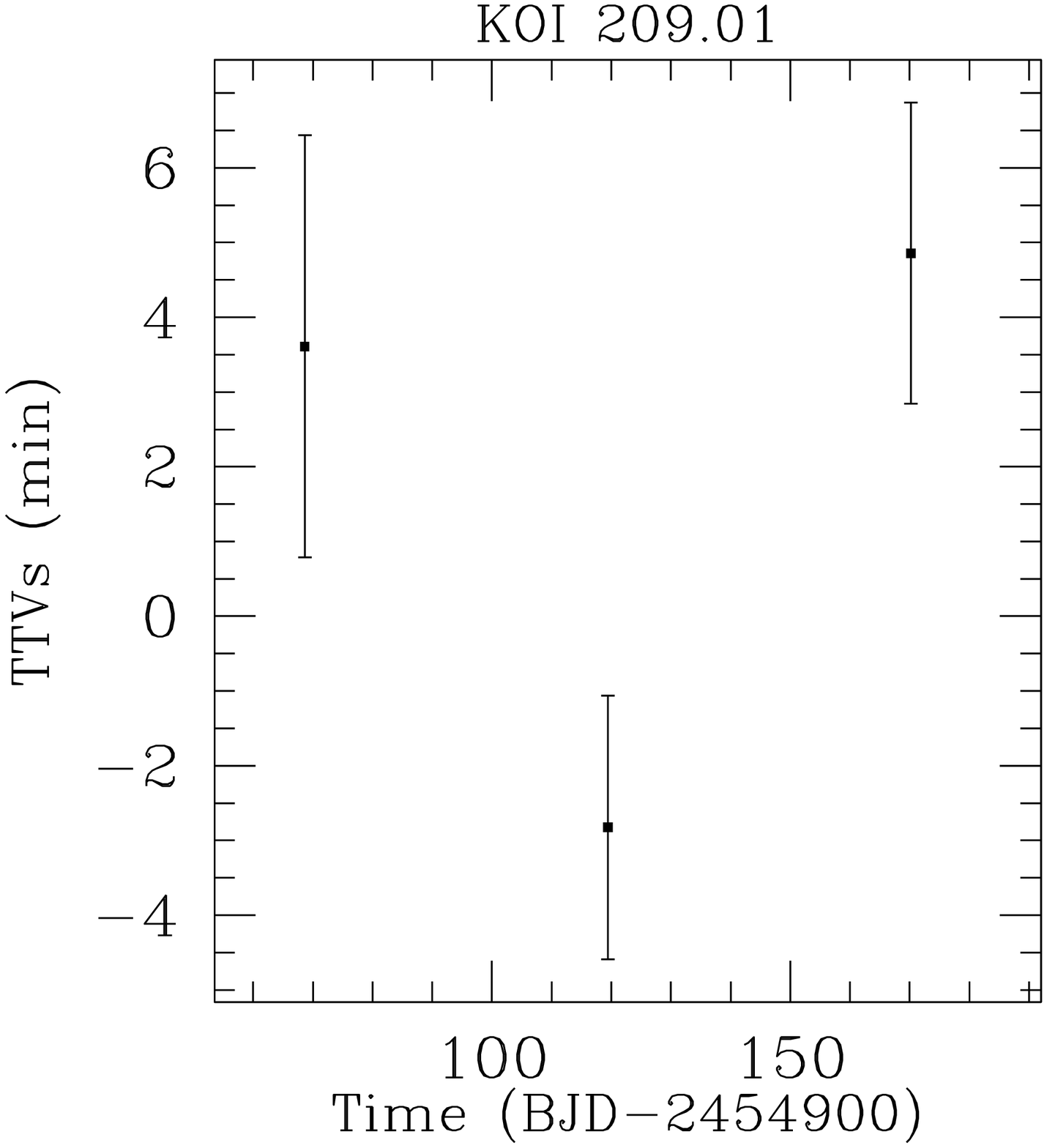}{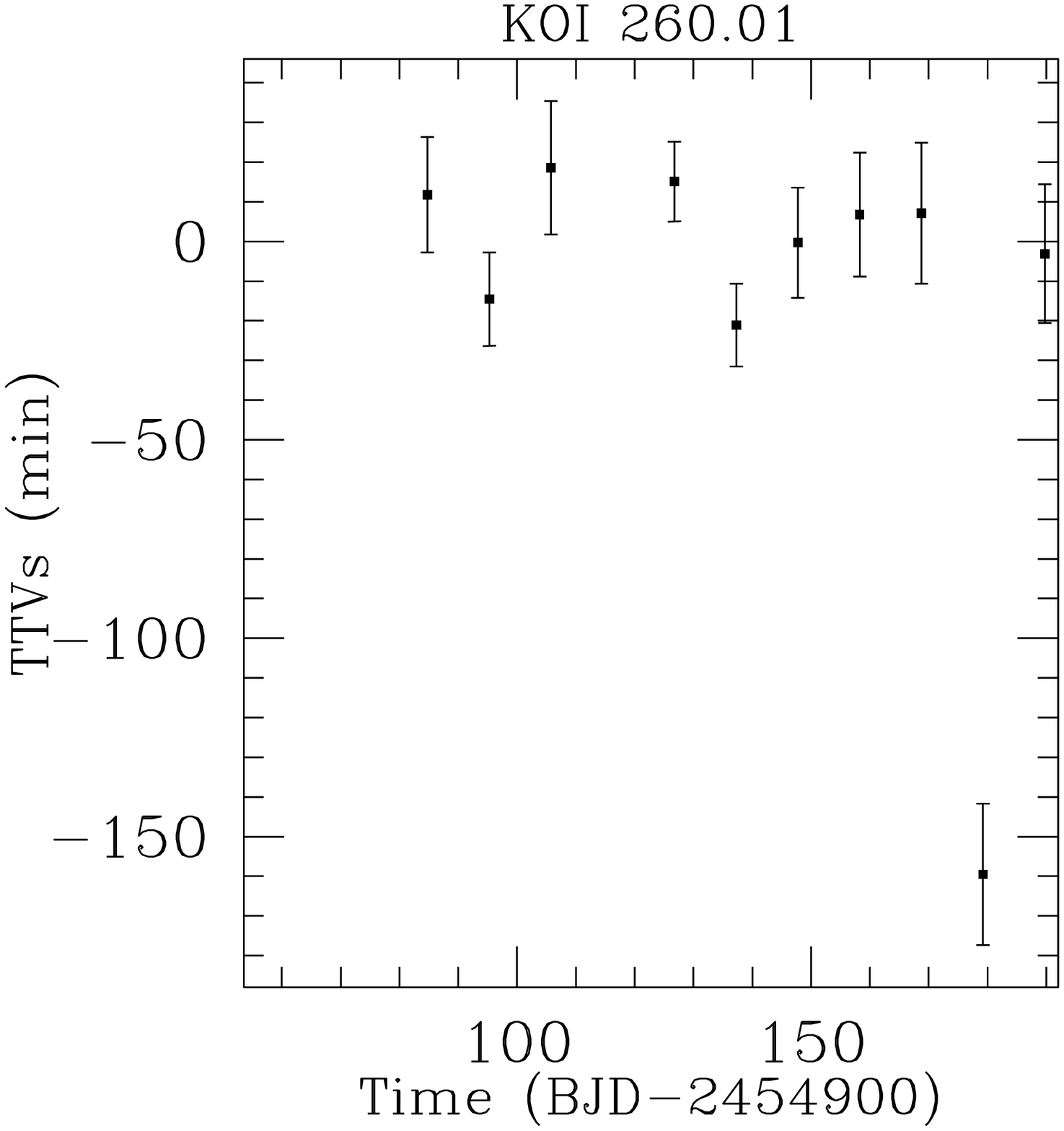}\end{figure*}\clearpage
\begin{figure*}\plottwo{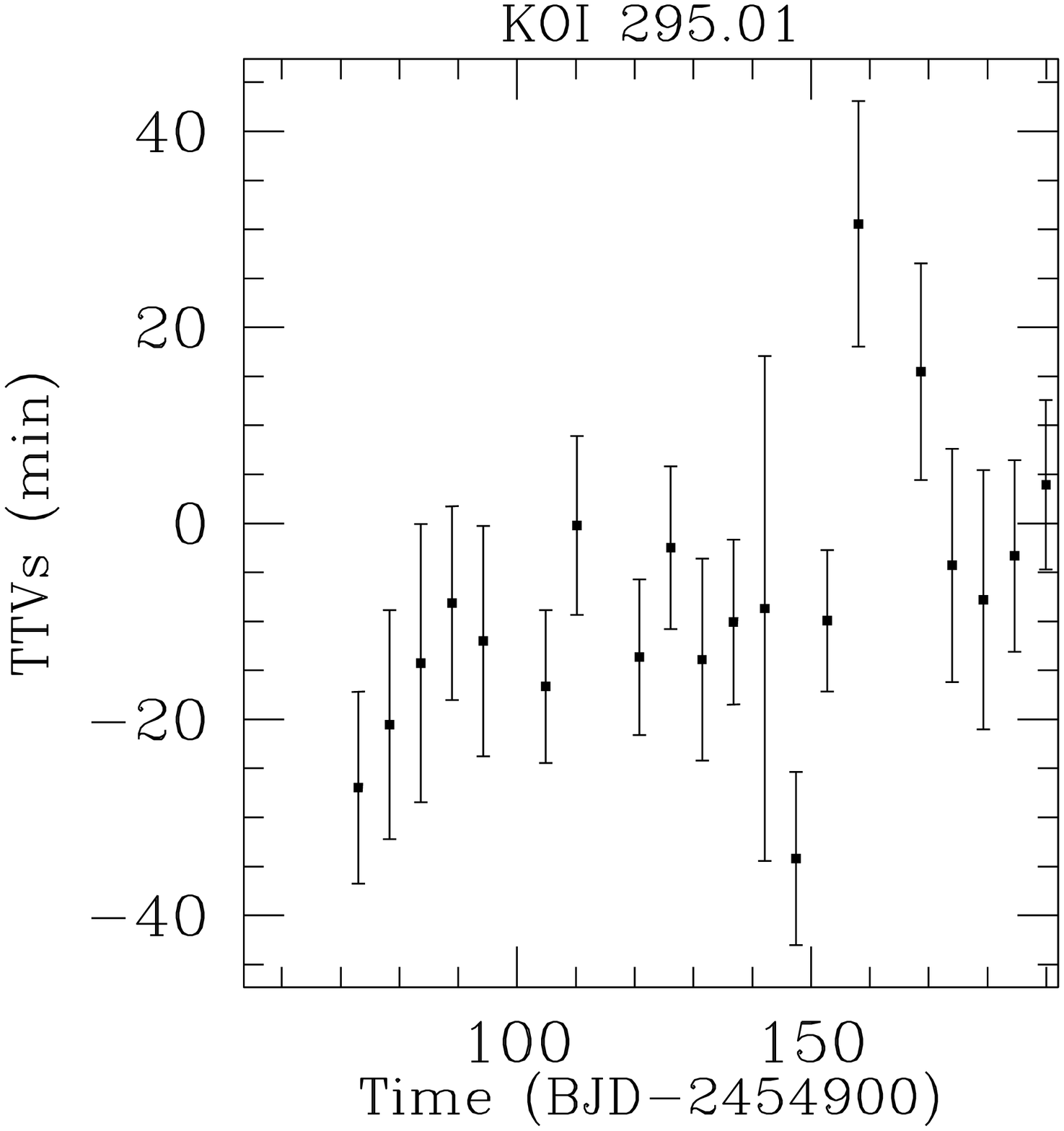}{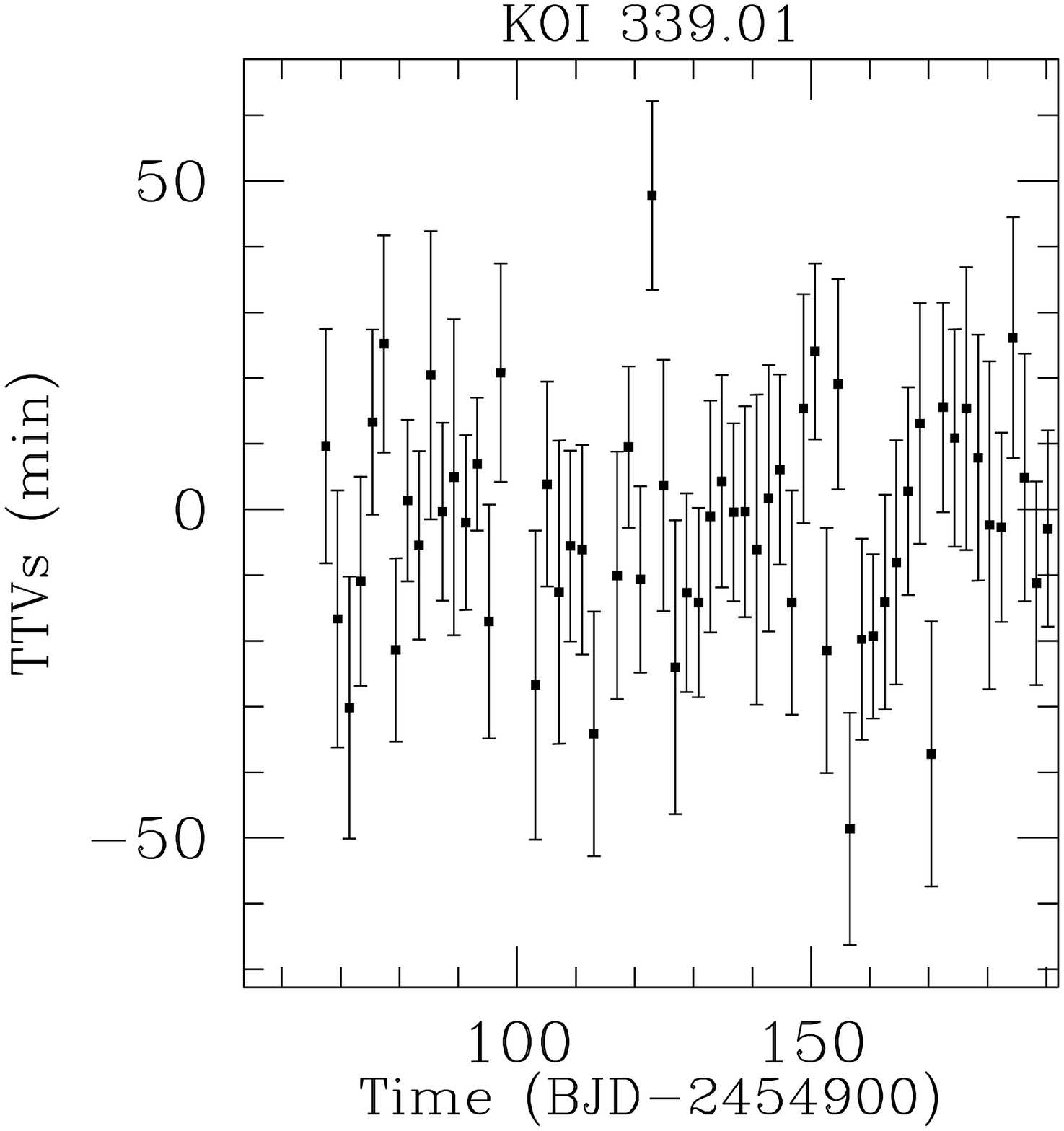}\end{figure*}
\begin{figure*}\plottwo{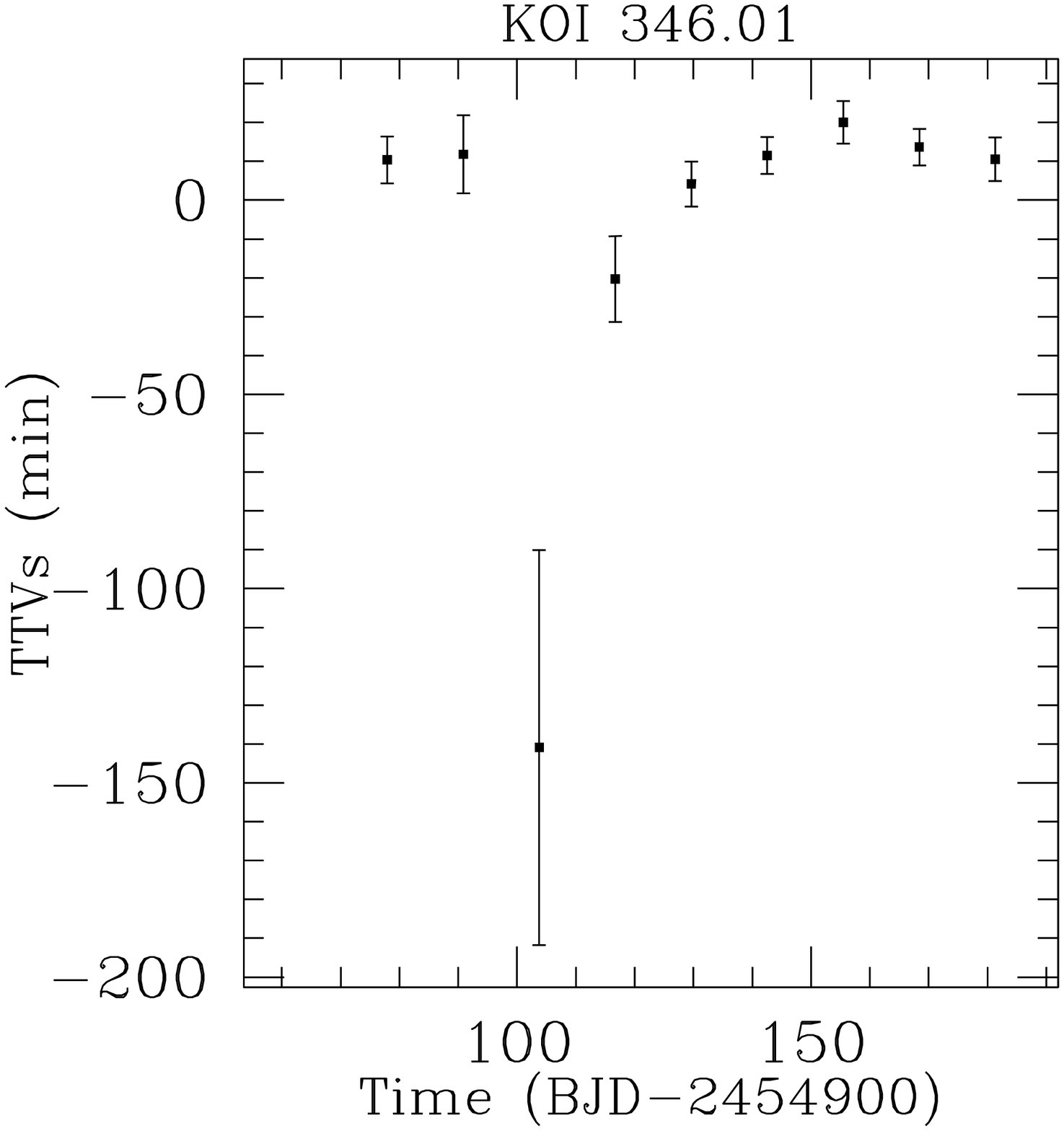}{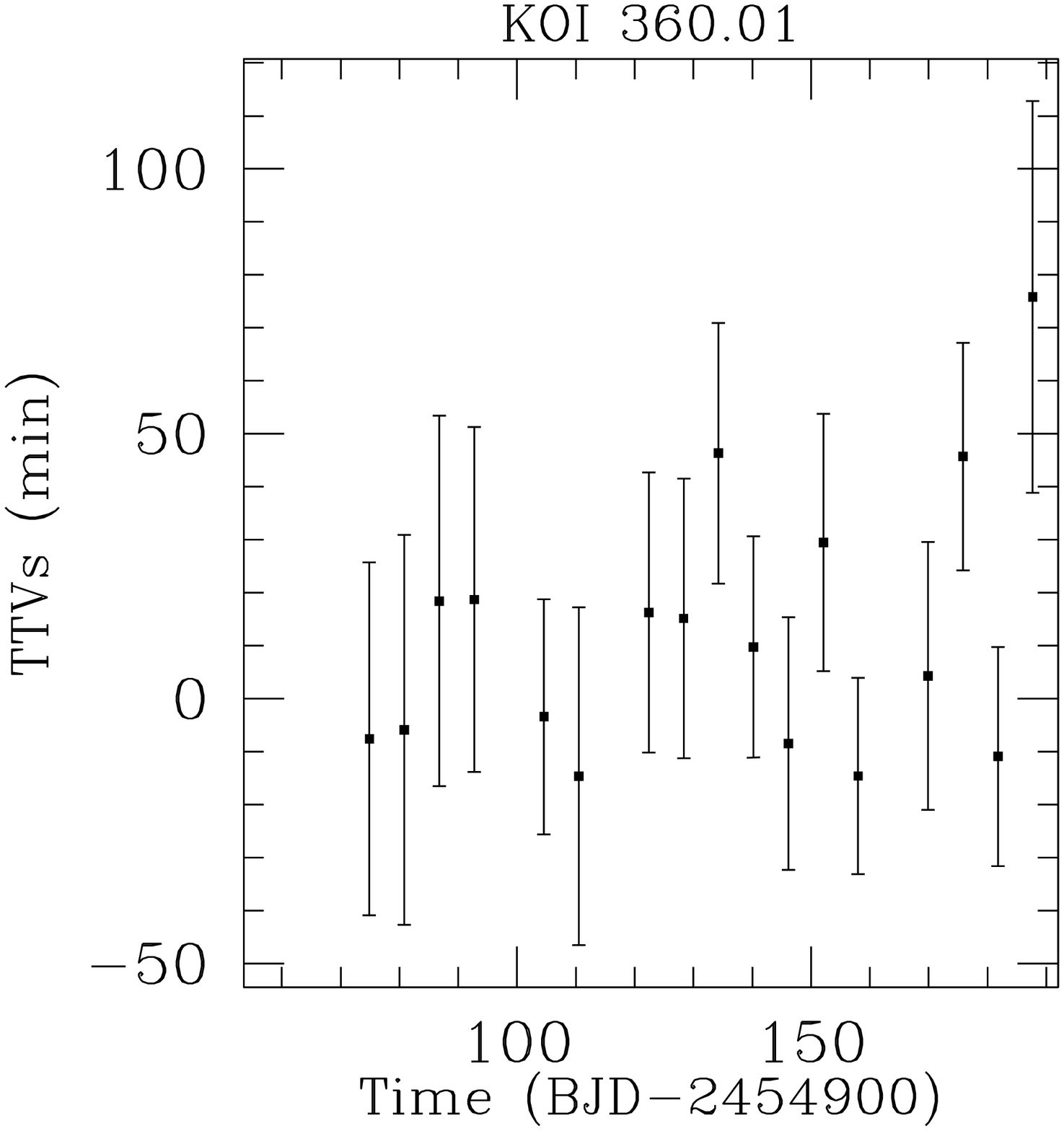}\end{figure*}\clearpage
\begin{figure*}\plottwo{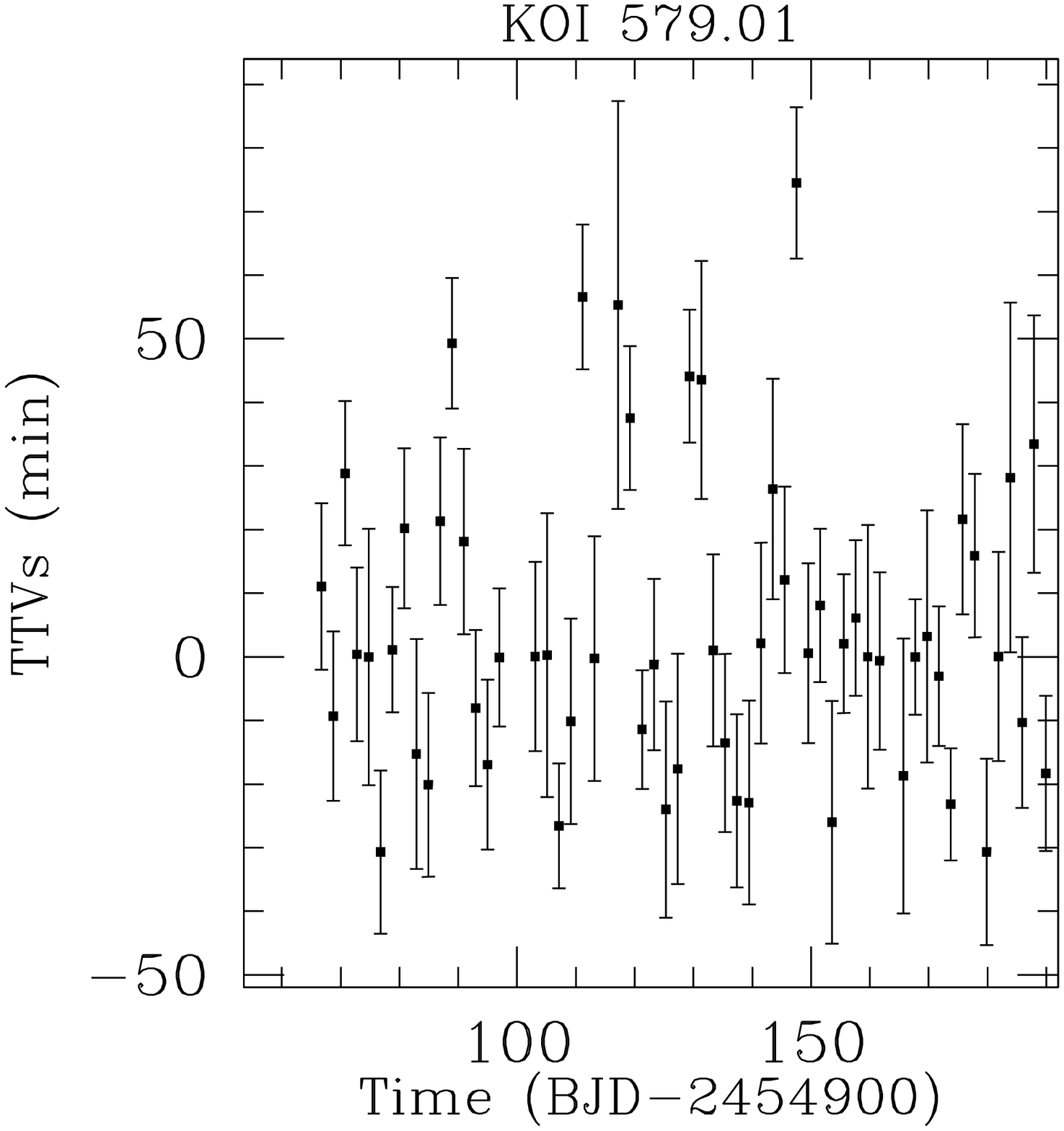}{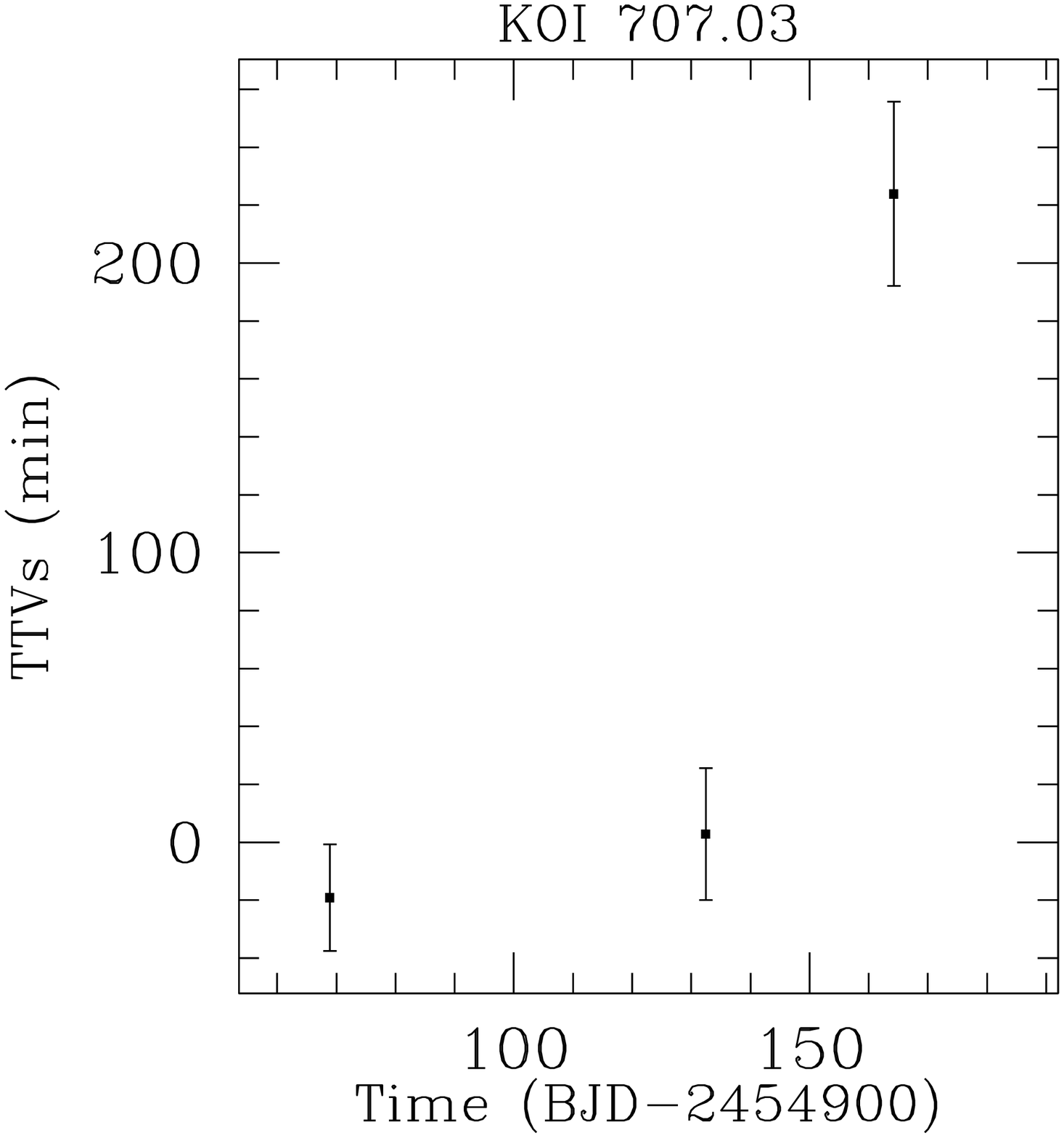}\end{figure*}
\begin{figure*}\plottwo{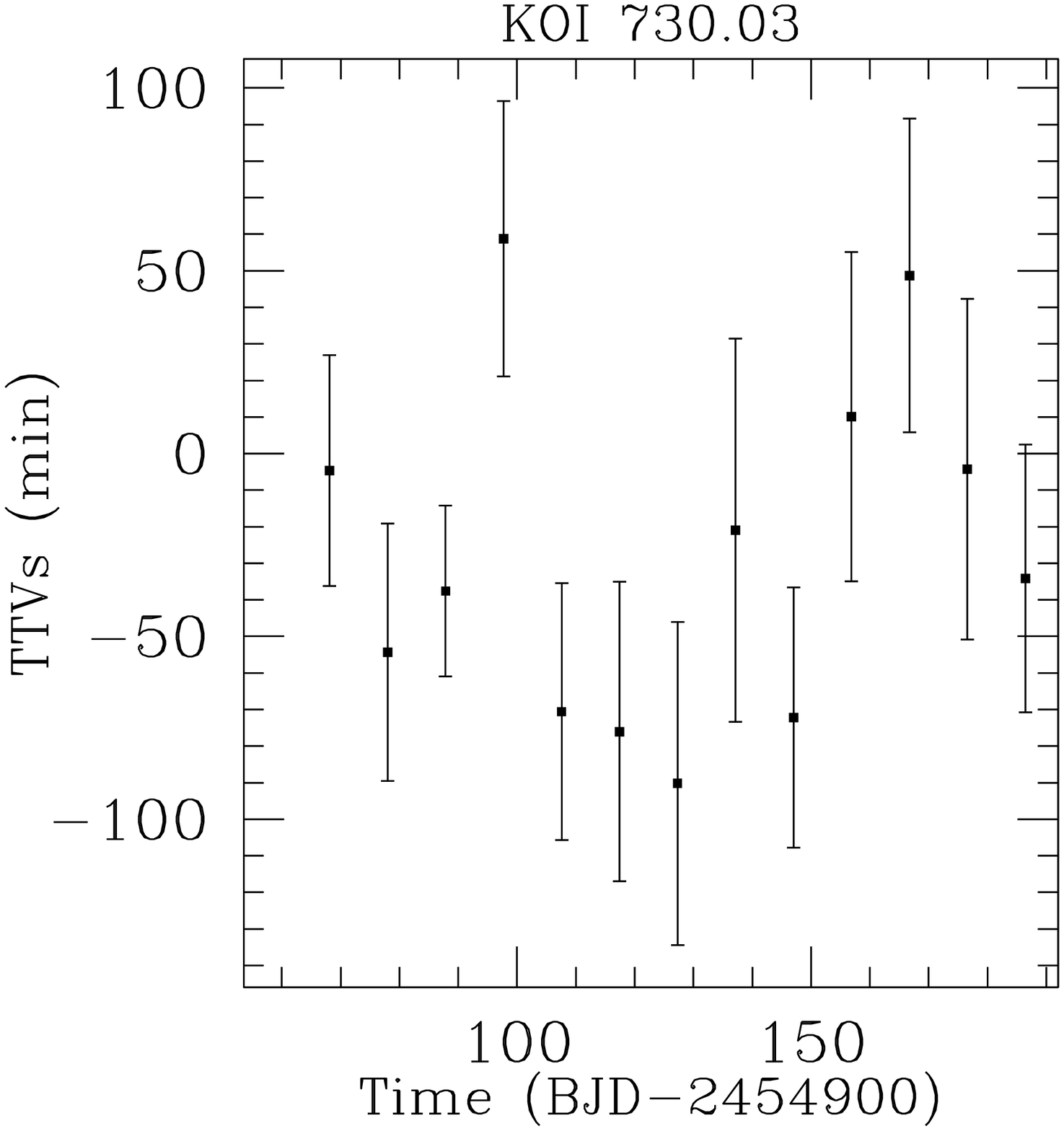}{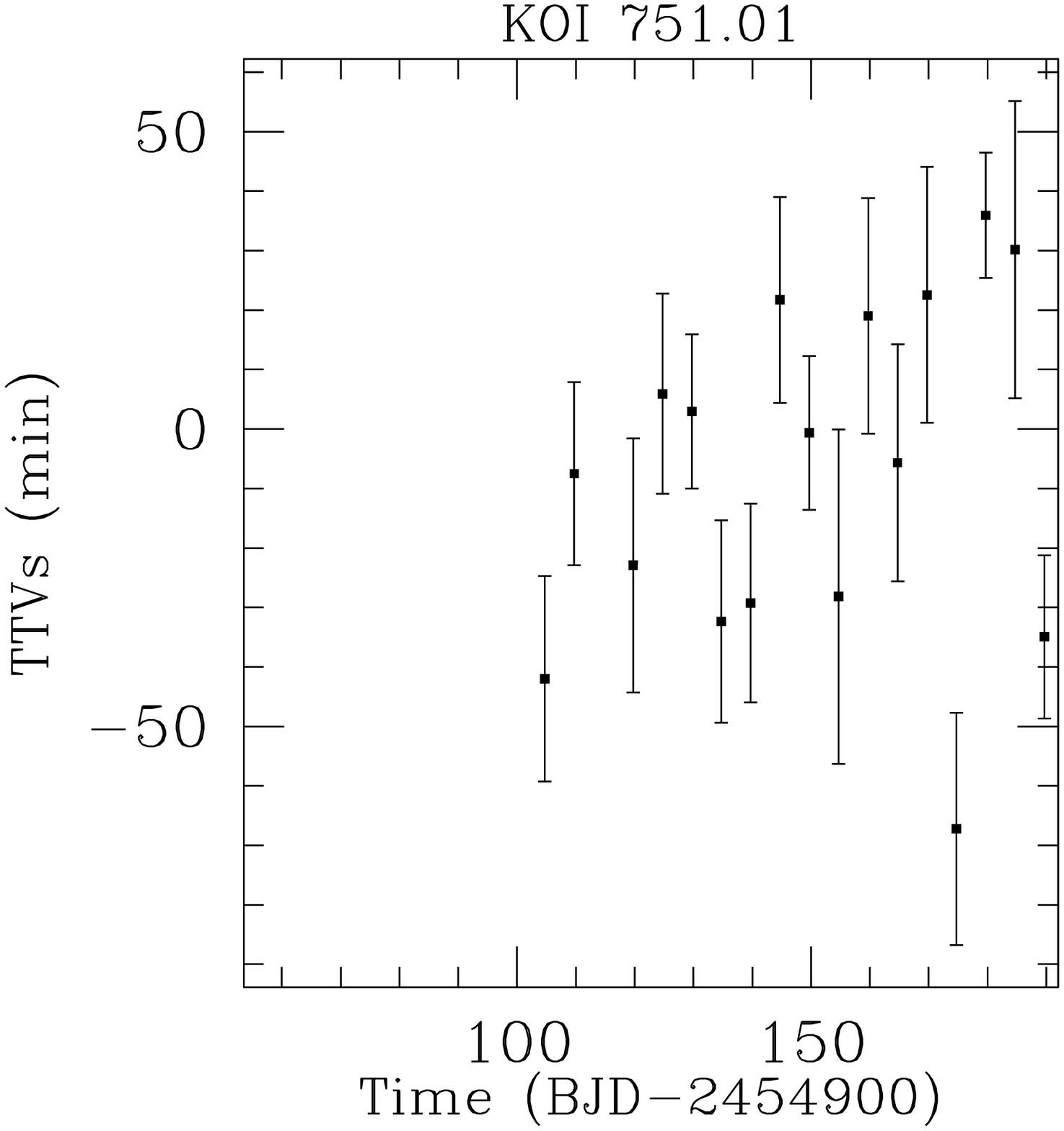}\end{figure*}\clearpage
\begin{figure*}\plottwo{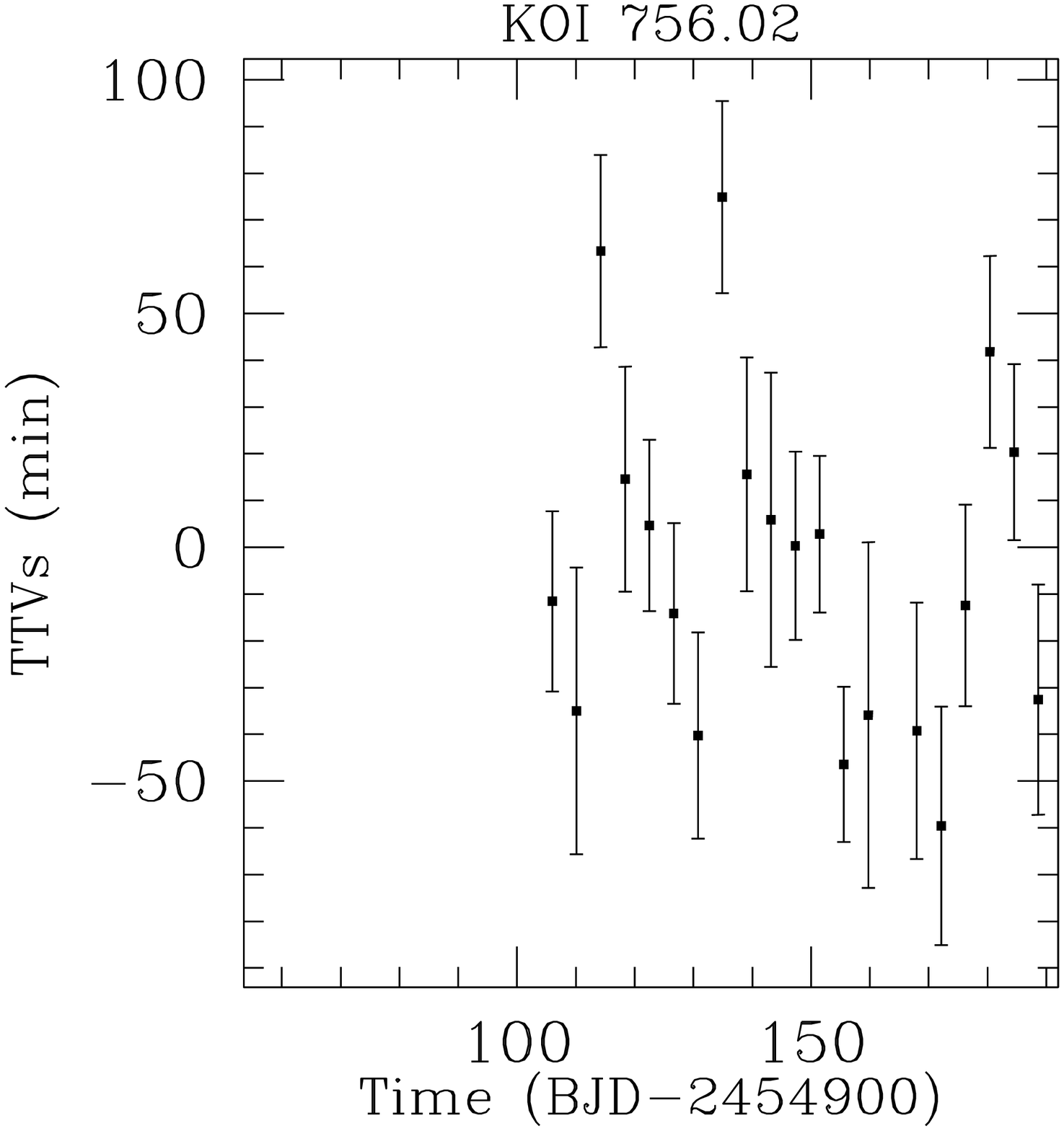}{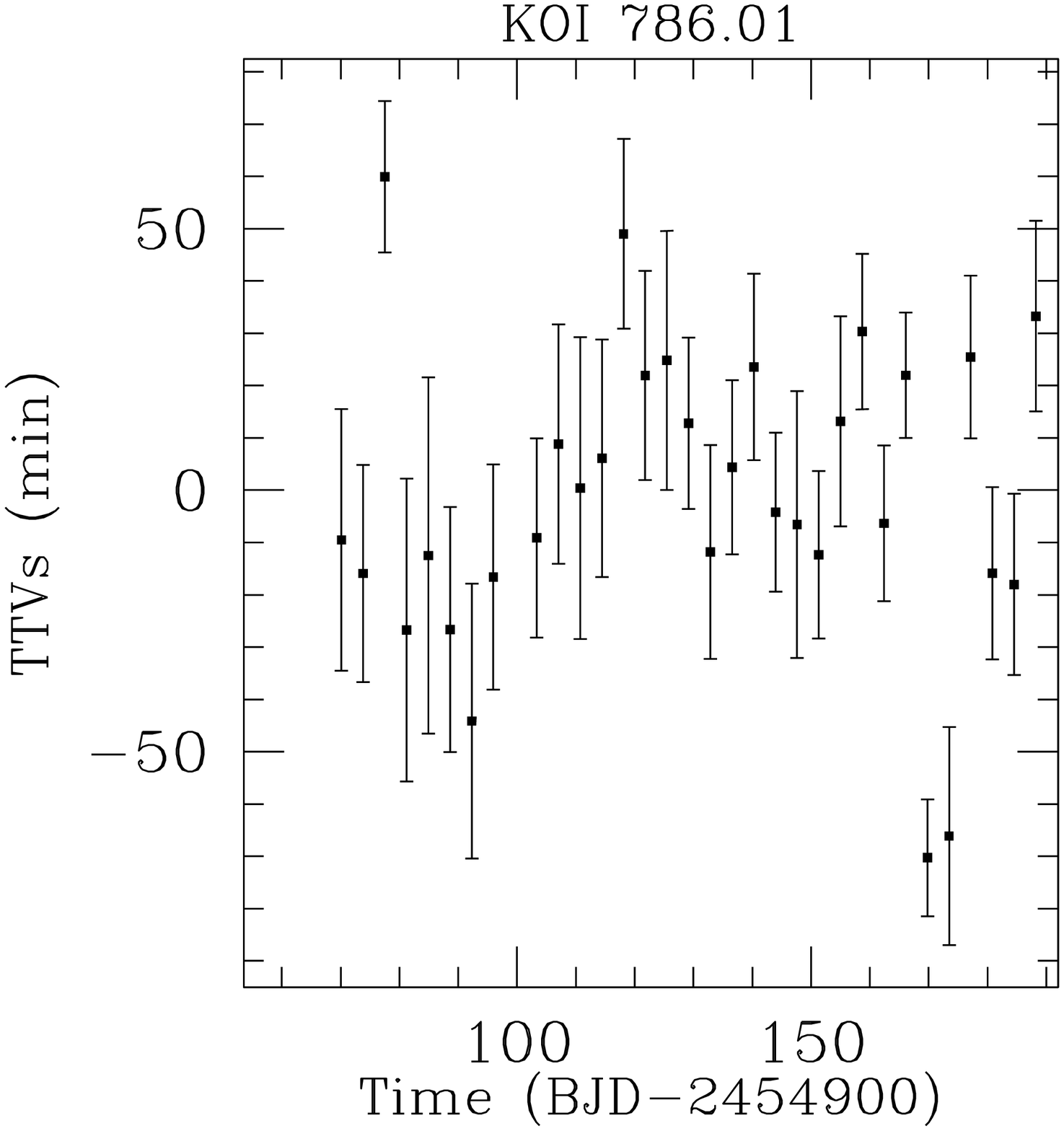}\end{figure*}
\begin{figure*}\plottwo{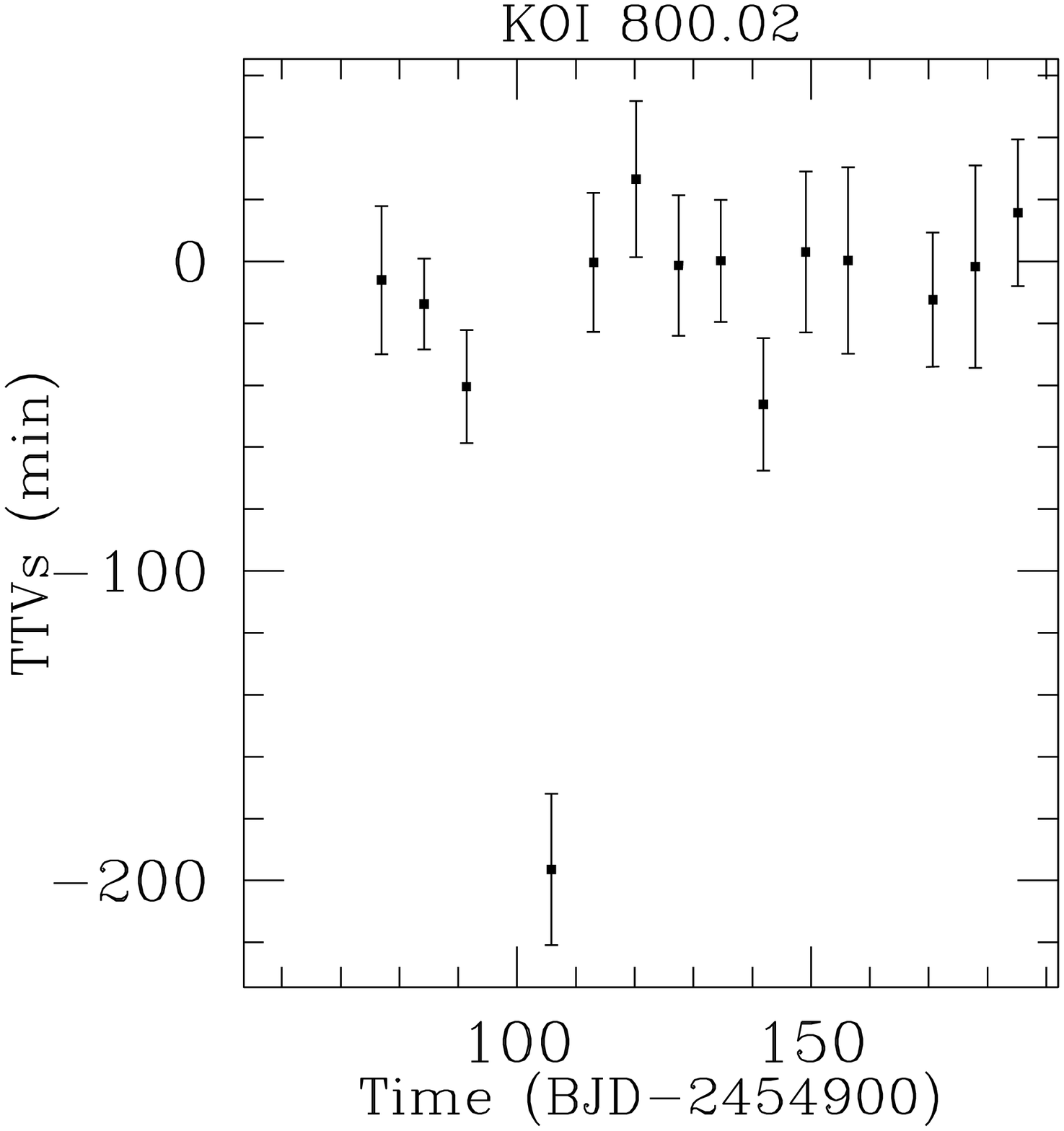}{koi834.03.eps}\end{figure*}\clearpage
\begin{figure*}\plottwo{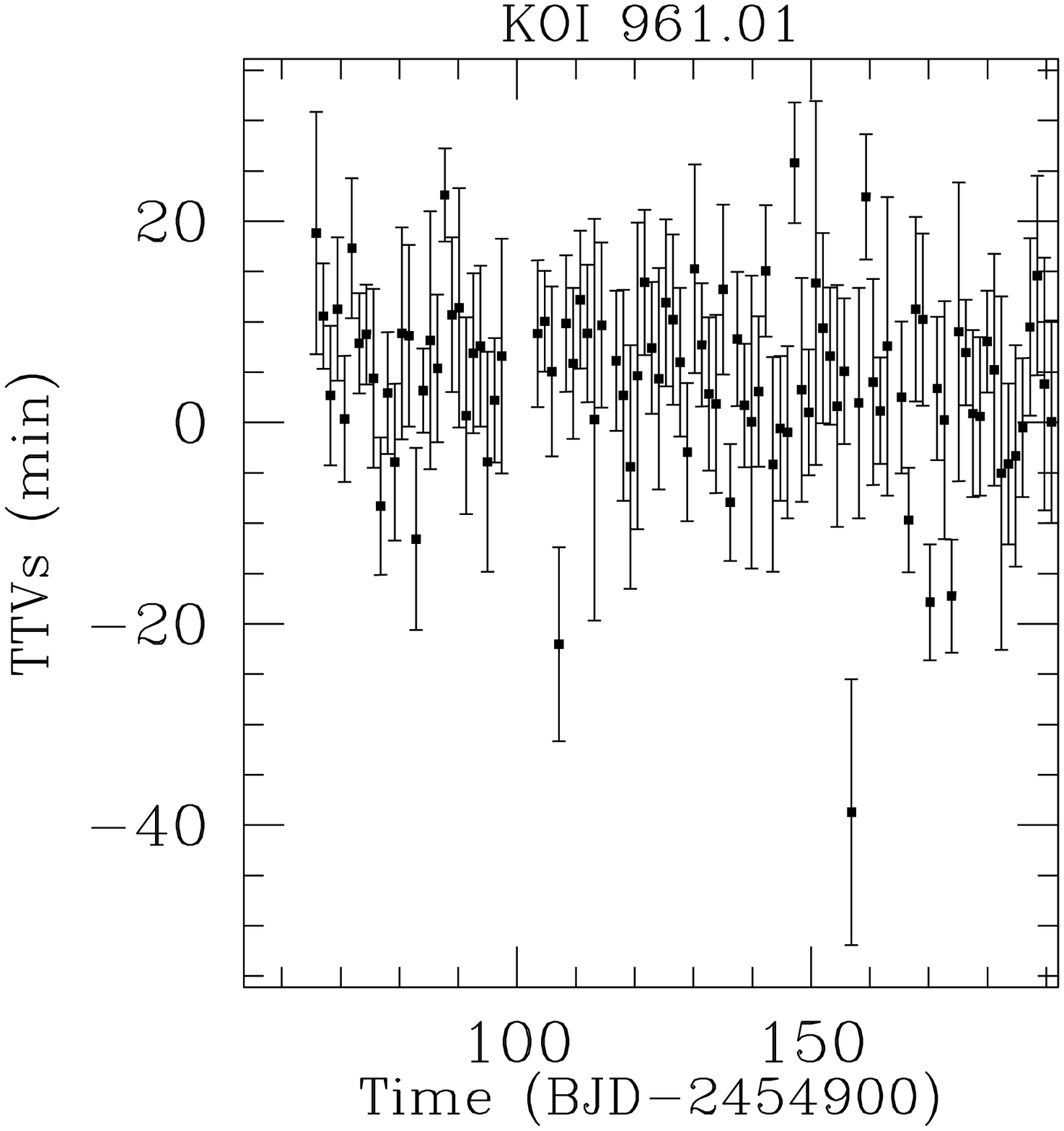}{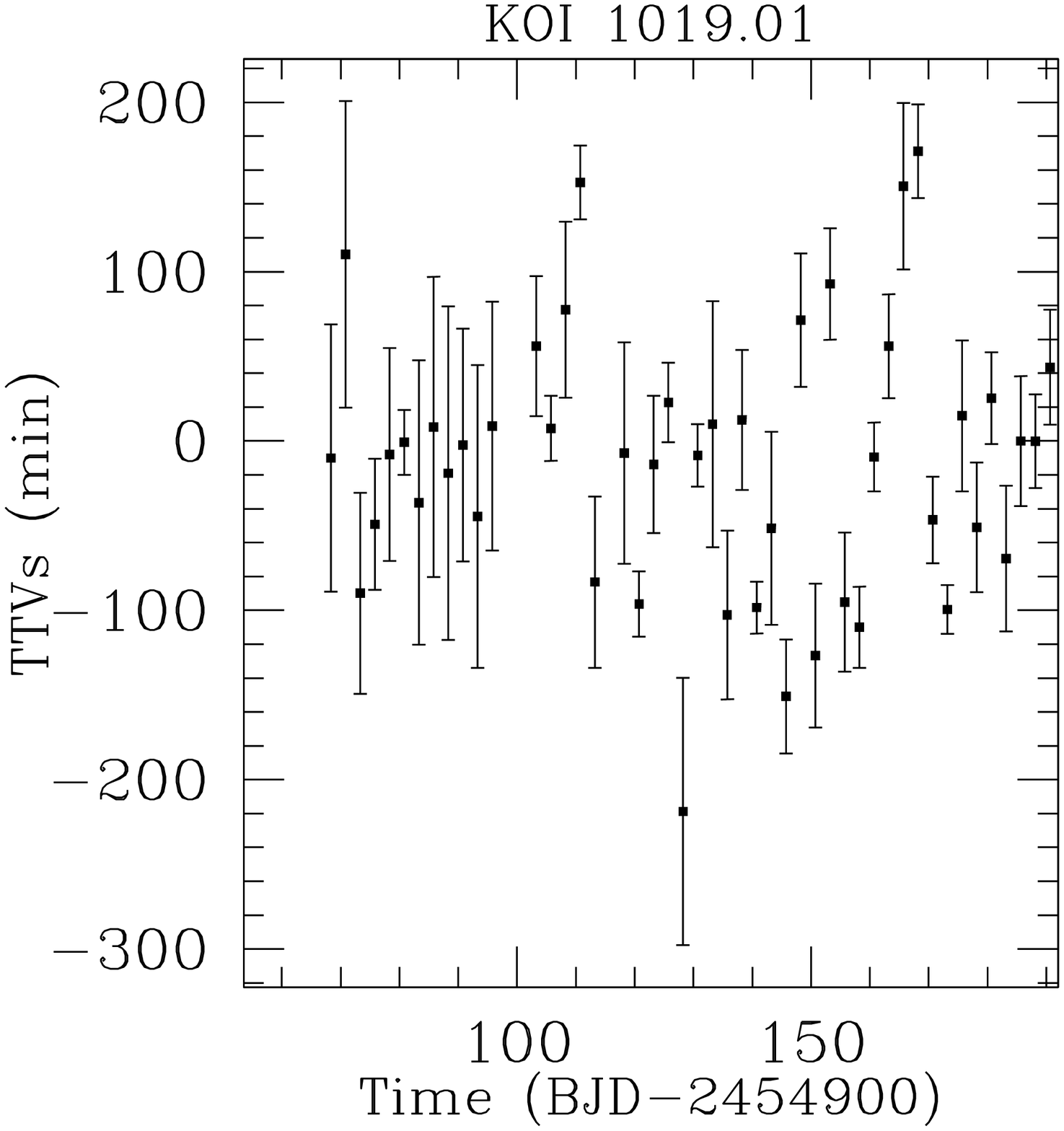}\end{figure*}
\begin{figure*}\plottwo{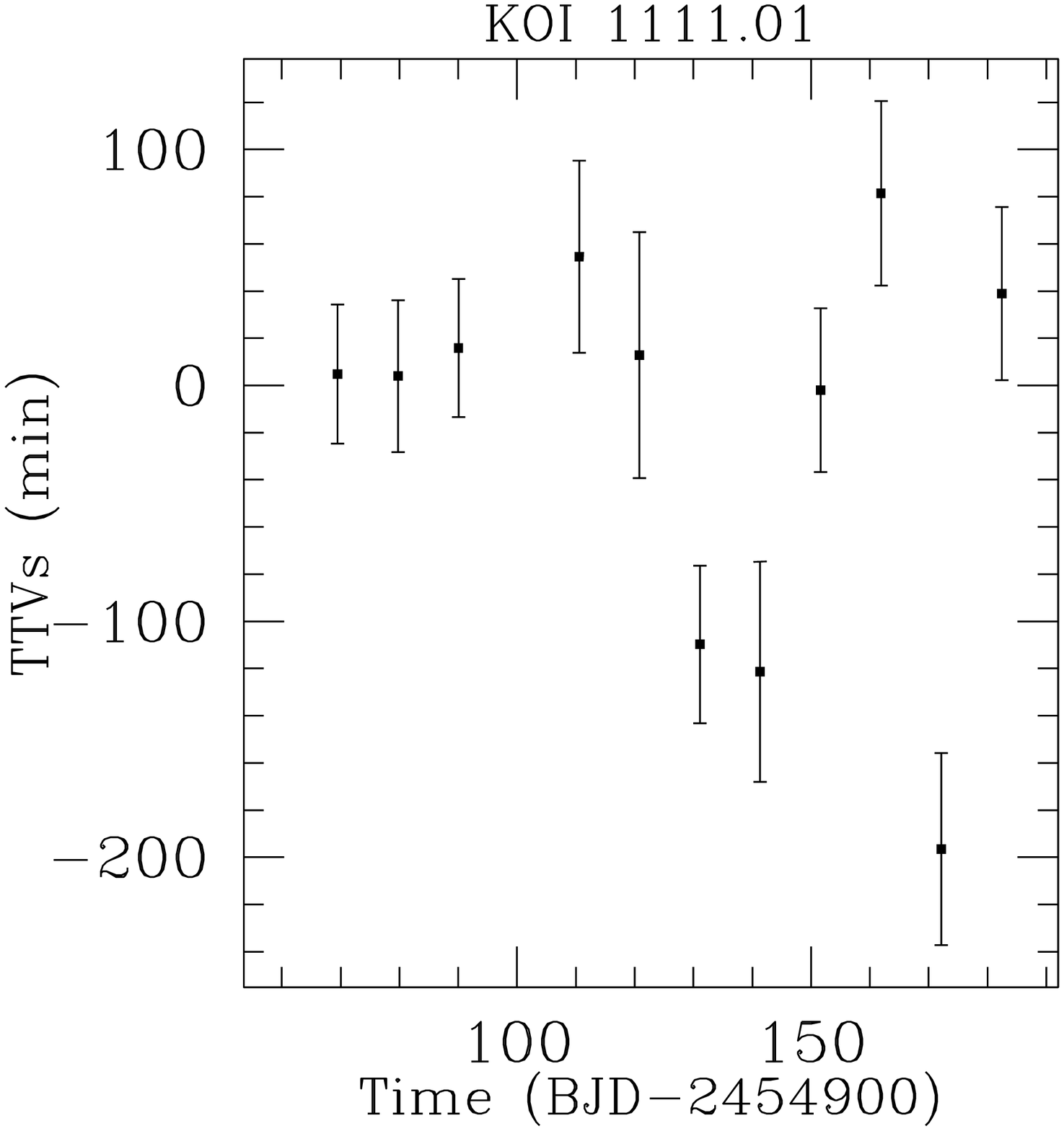}{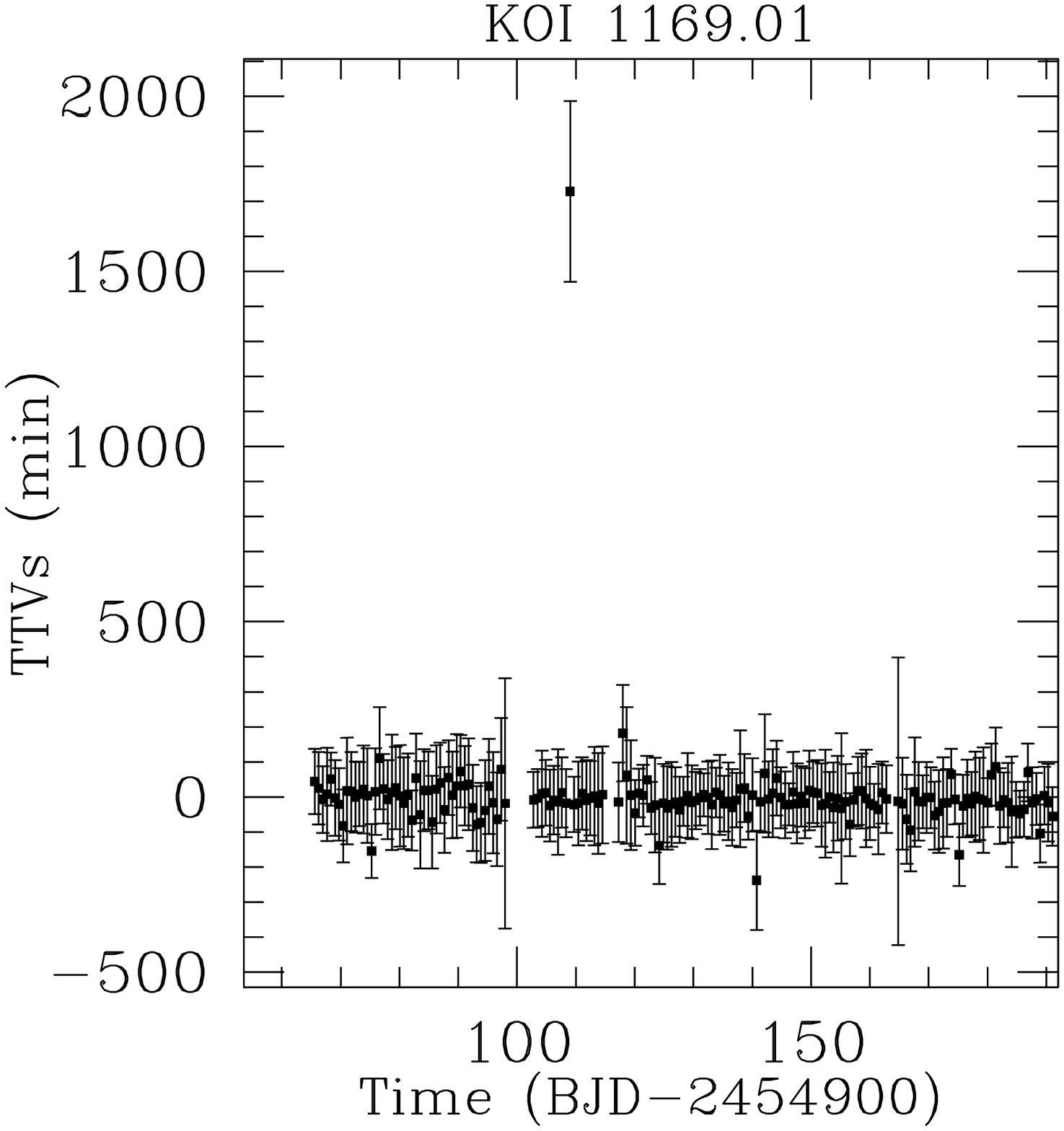}\end{figure*}\clearpage
\begin{figure*}\plottwo{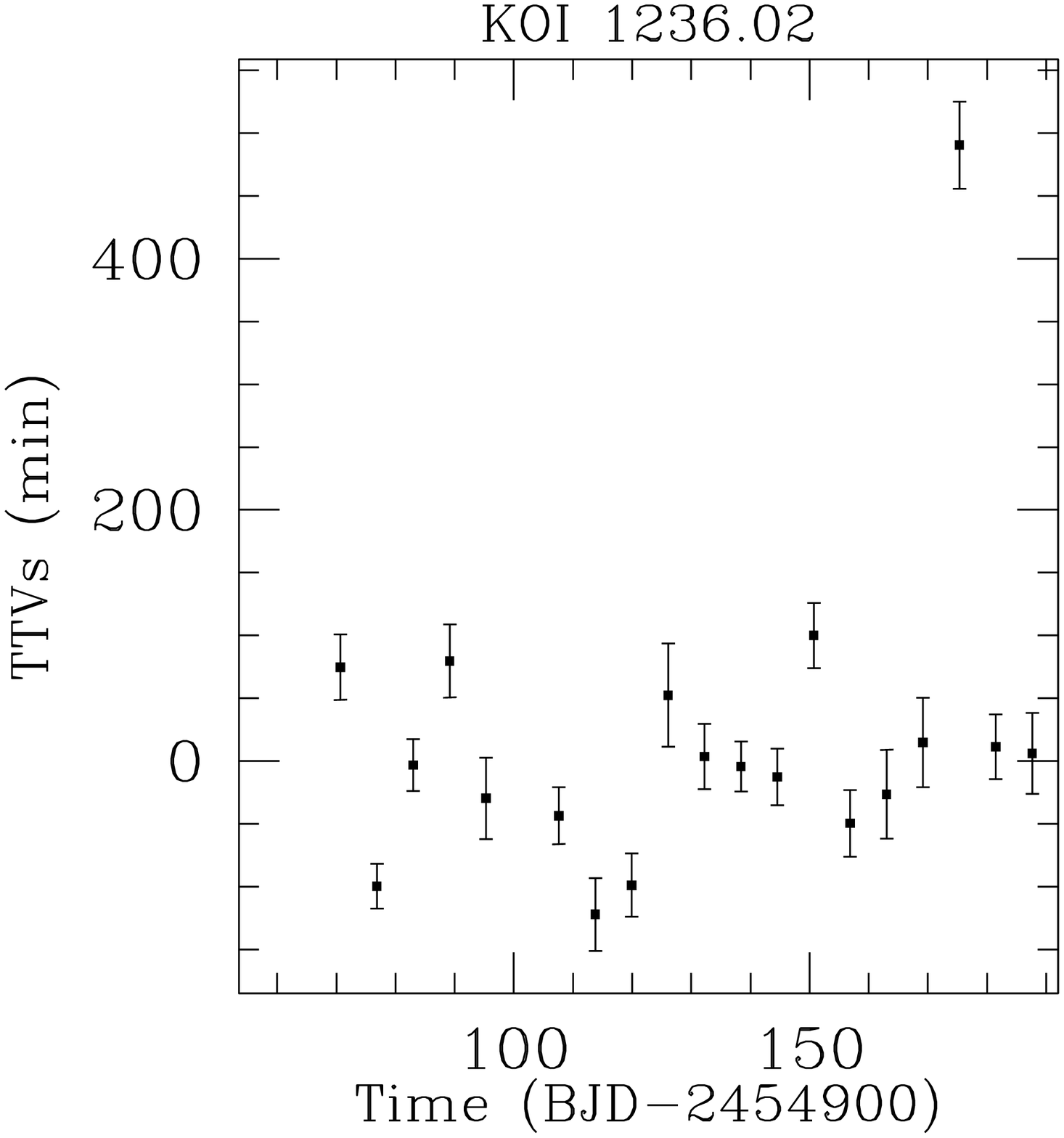}{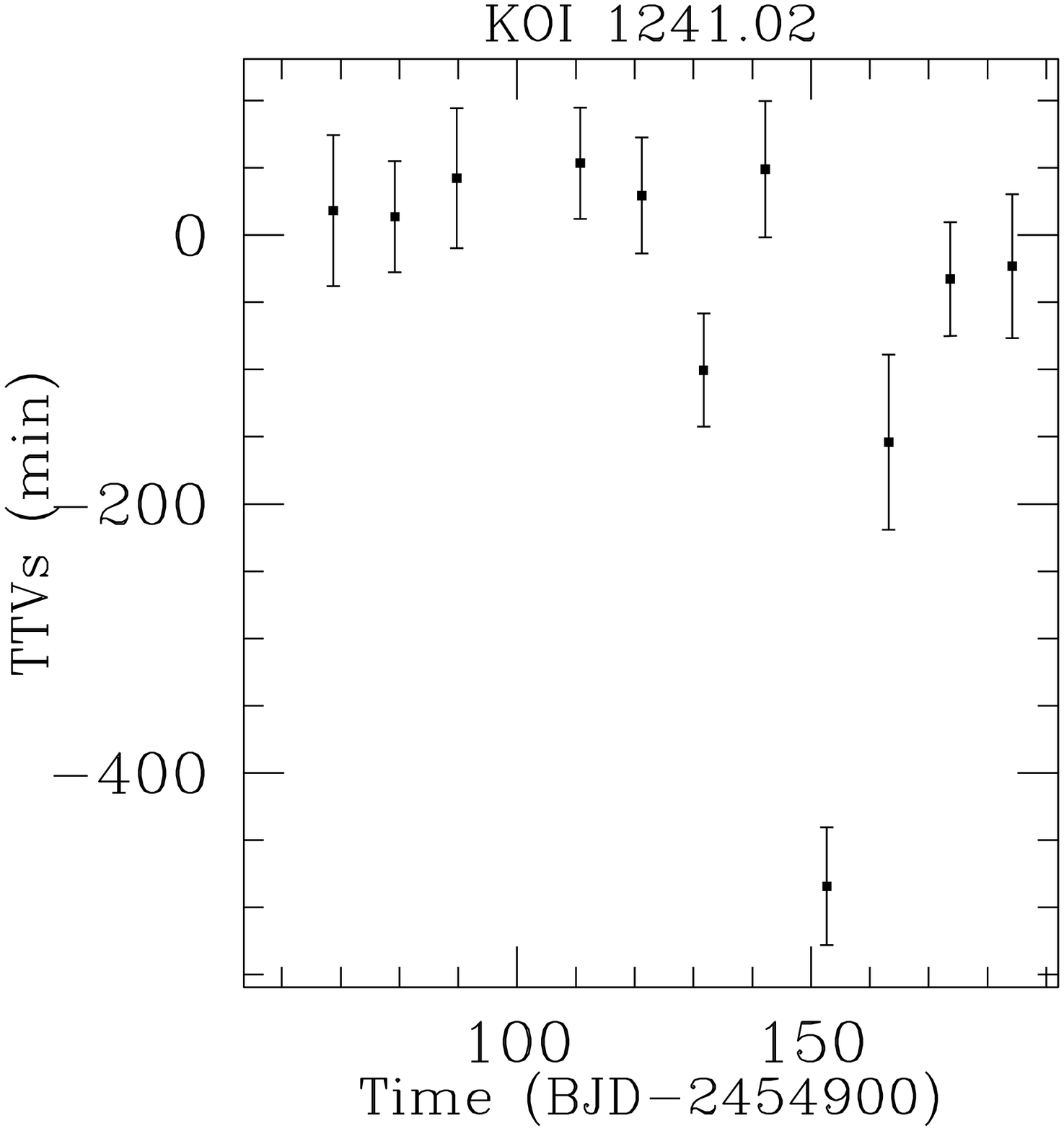}\end{figure*}
\begin{figure*}\plottwo{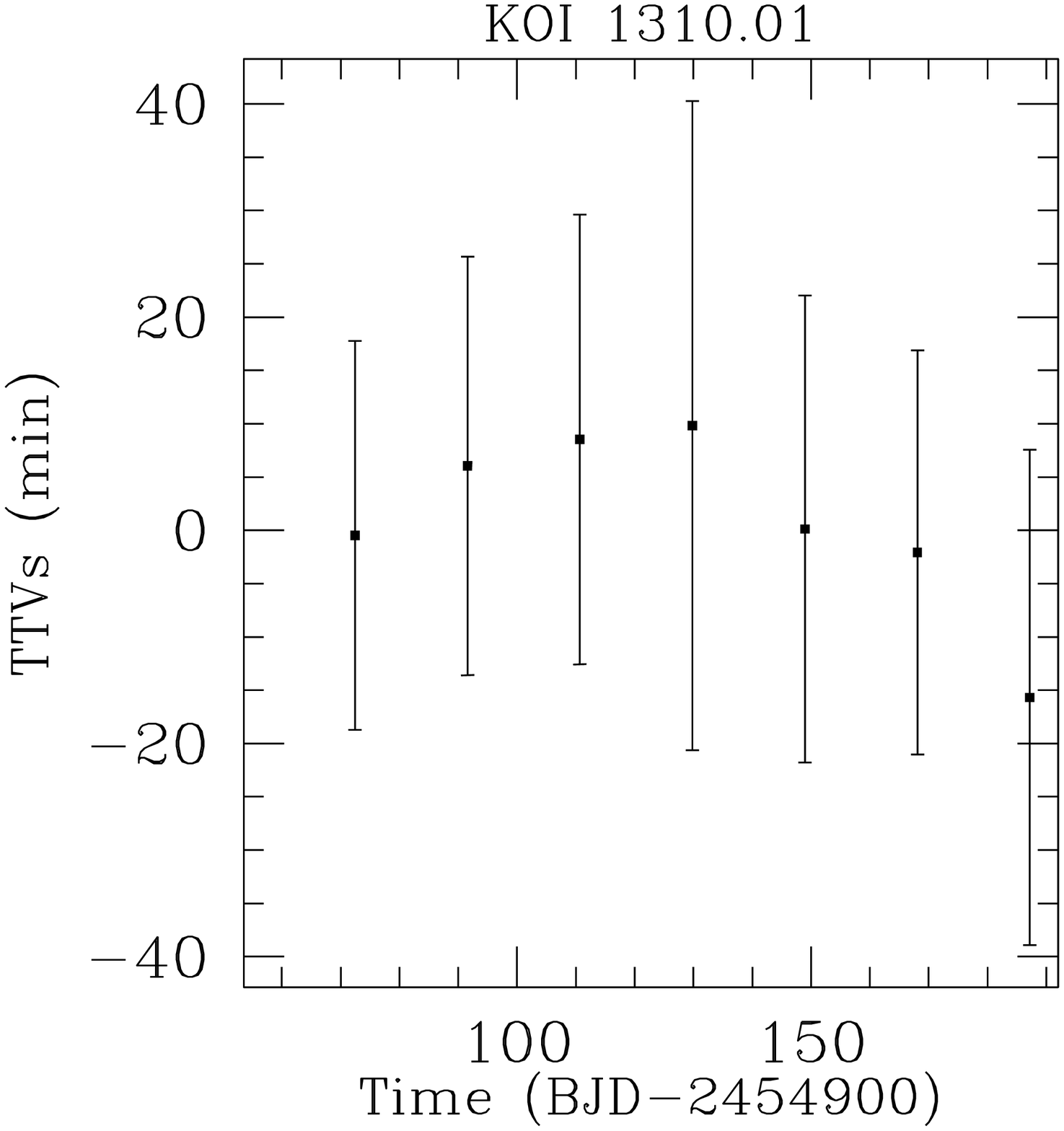}{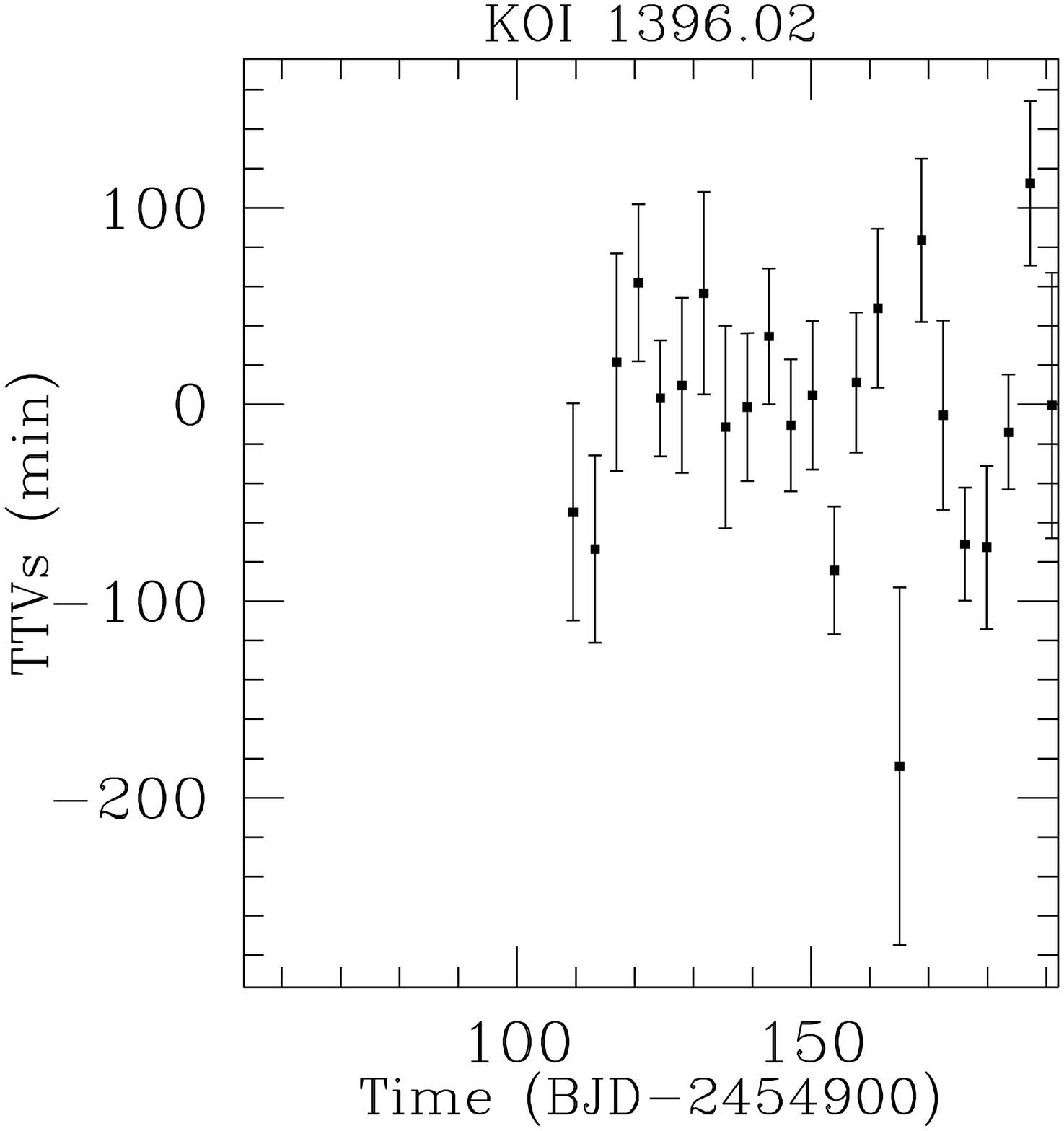}\end{figure*}\clearpage
\begin{figure*}\plottwo{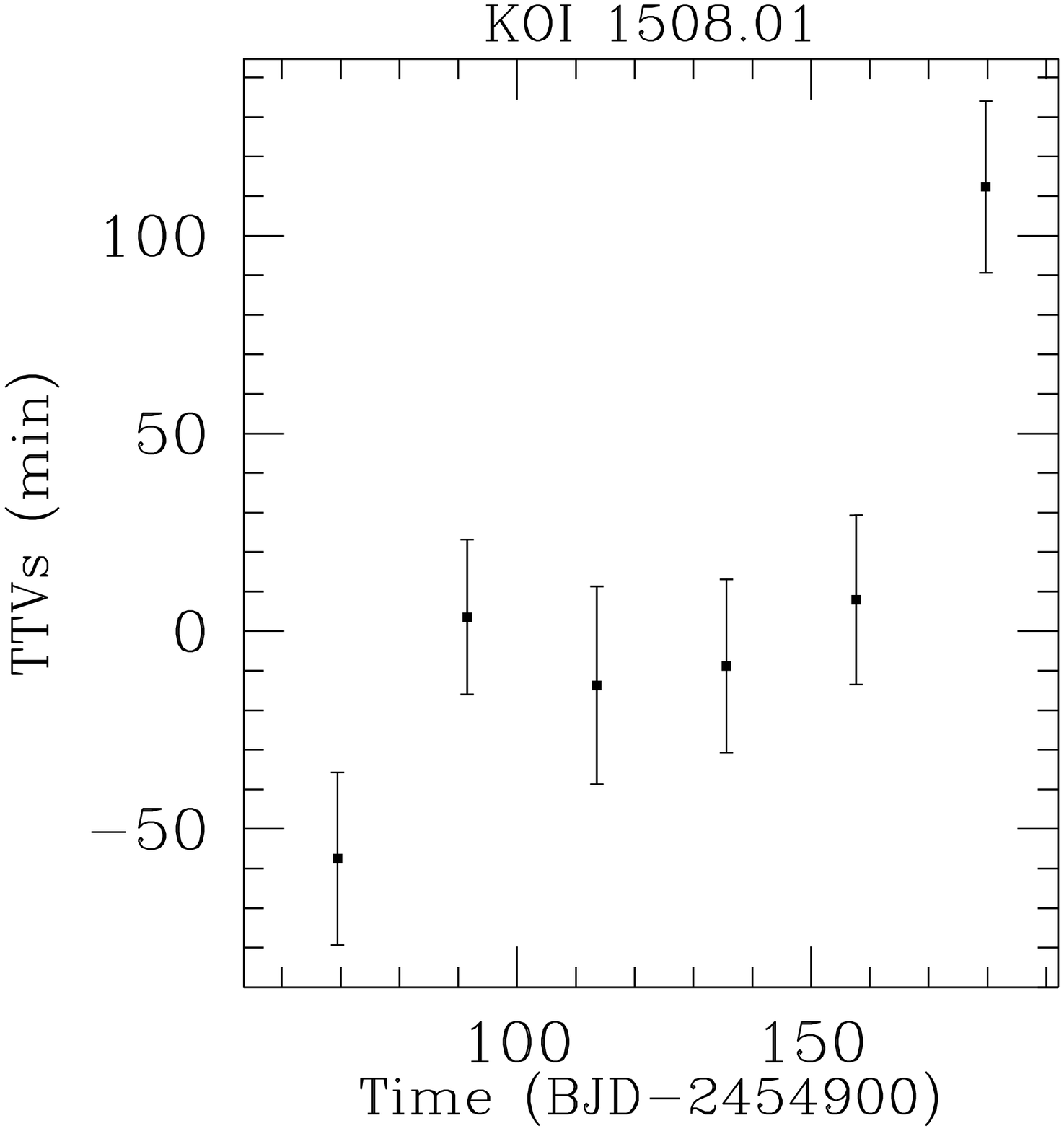}{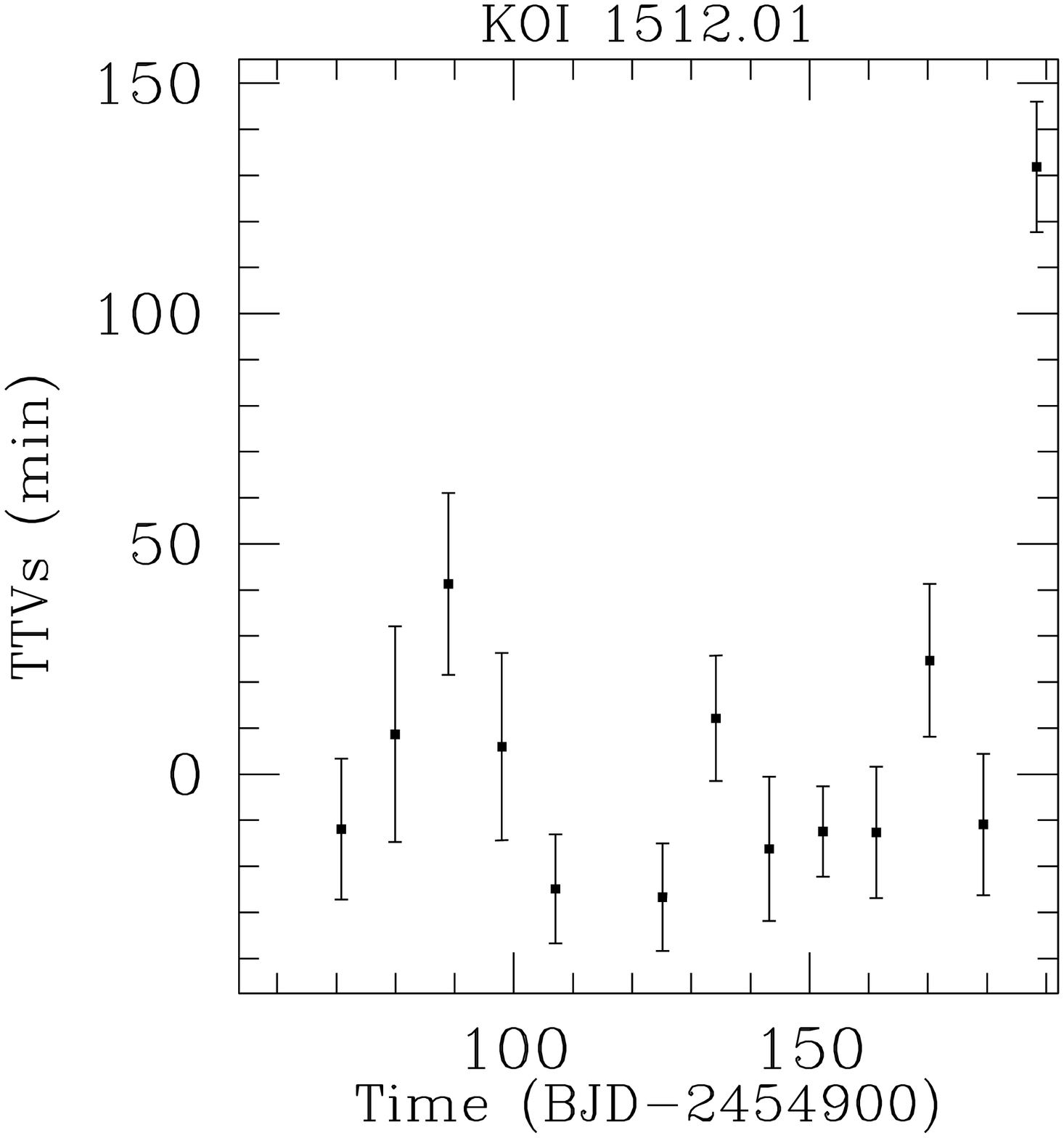}\end{figure*}

\clearpage

\section*{Non-Detections of TTVs}
\begin{figure*}\plottwo{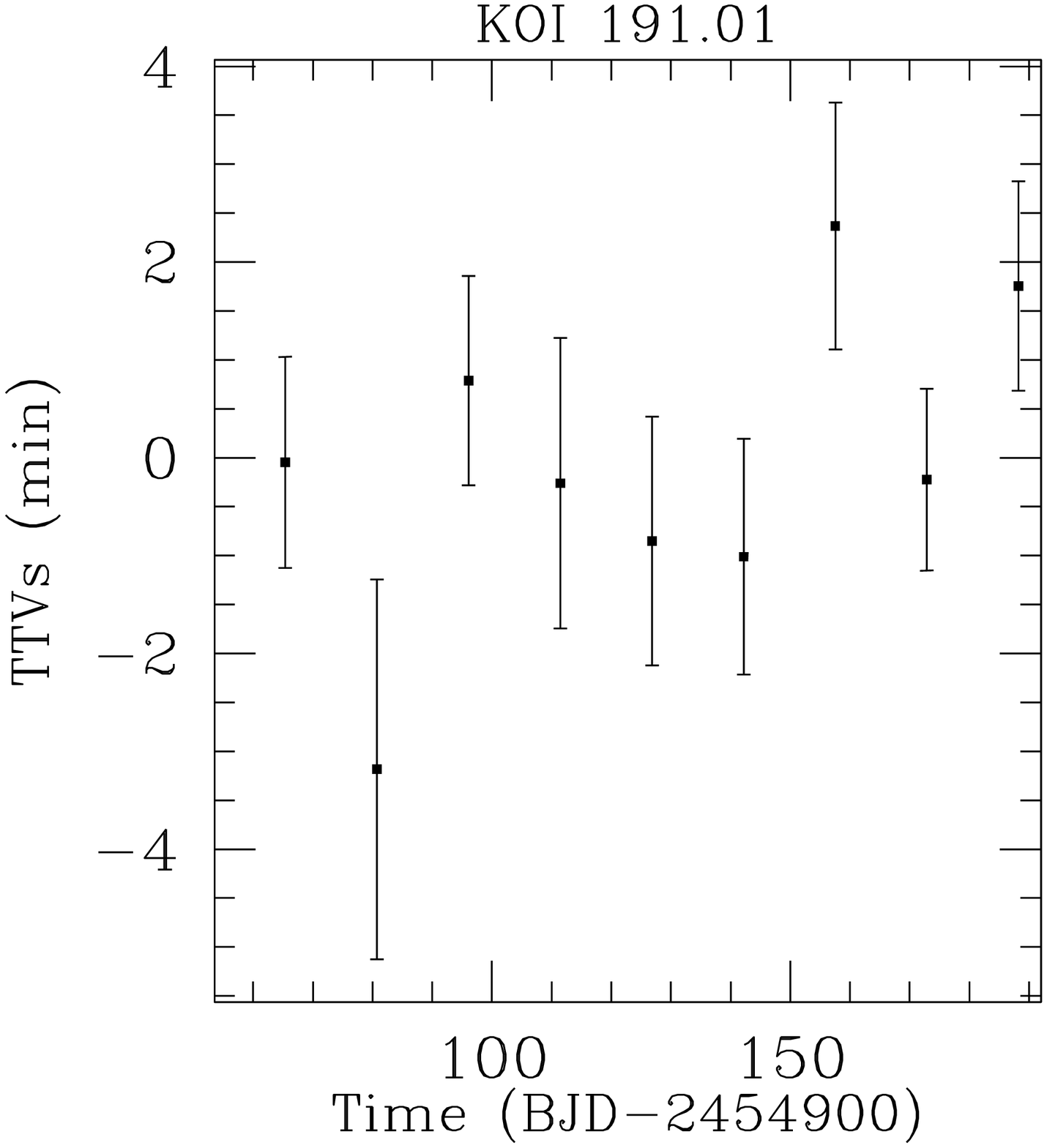}{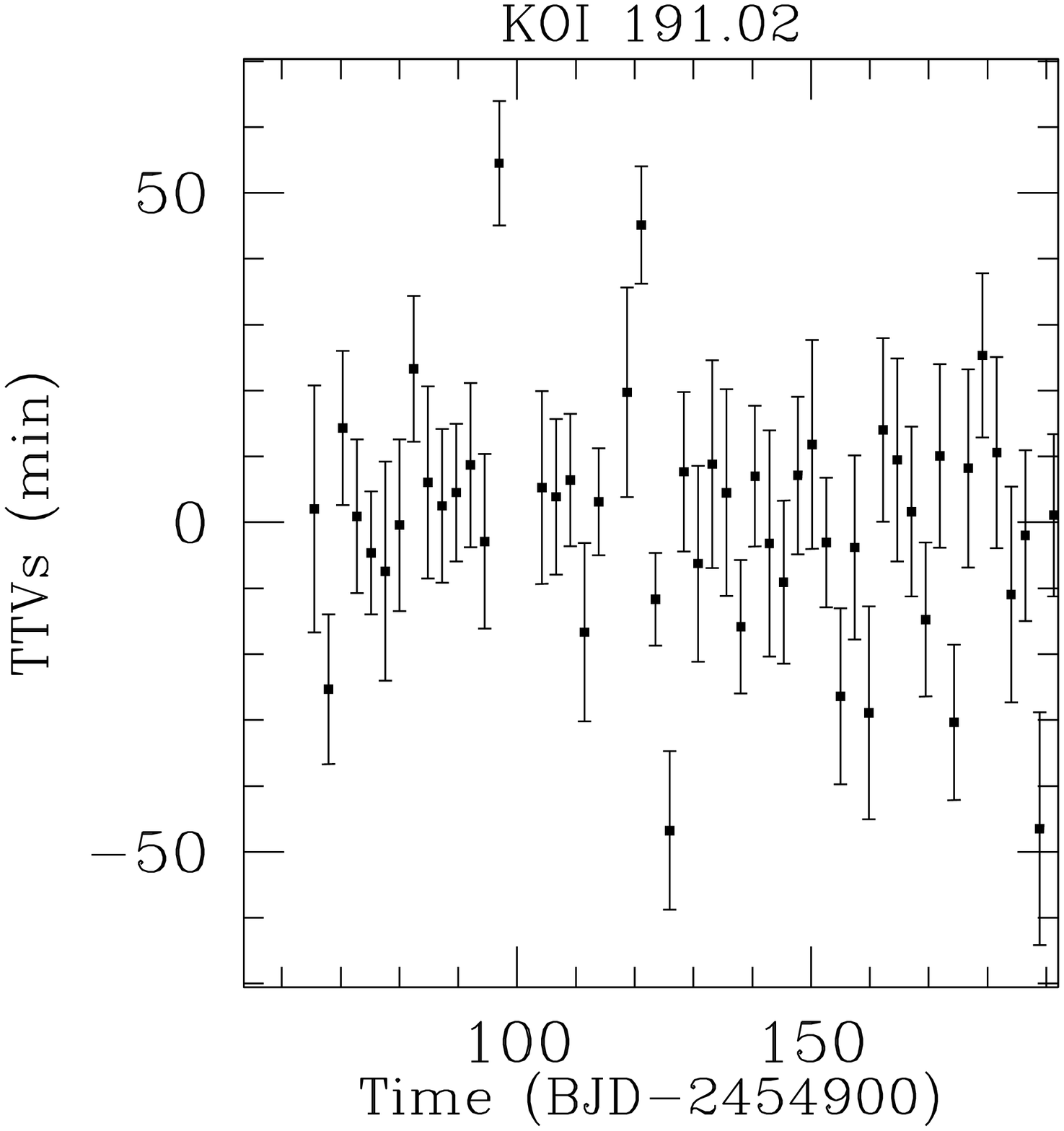}\end{figure*}
\begin{figure*}\plottwo{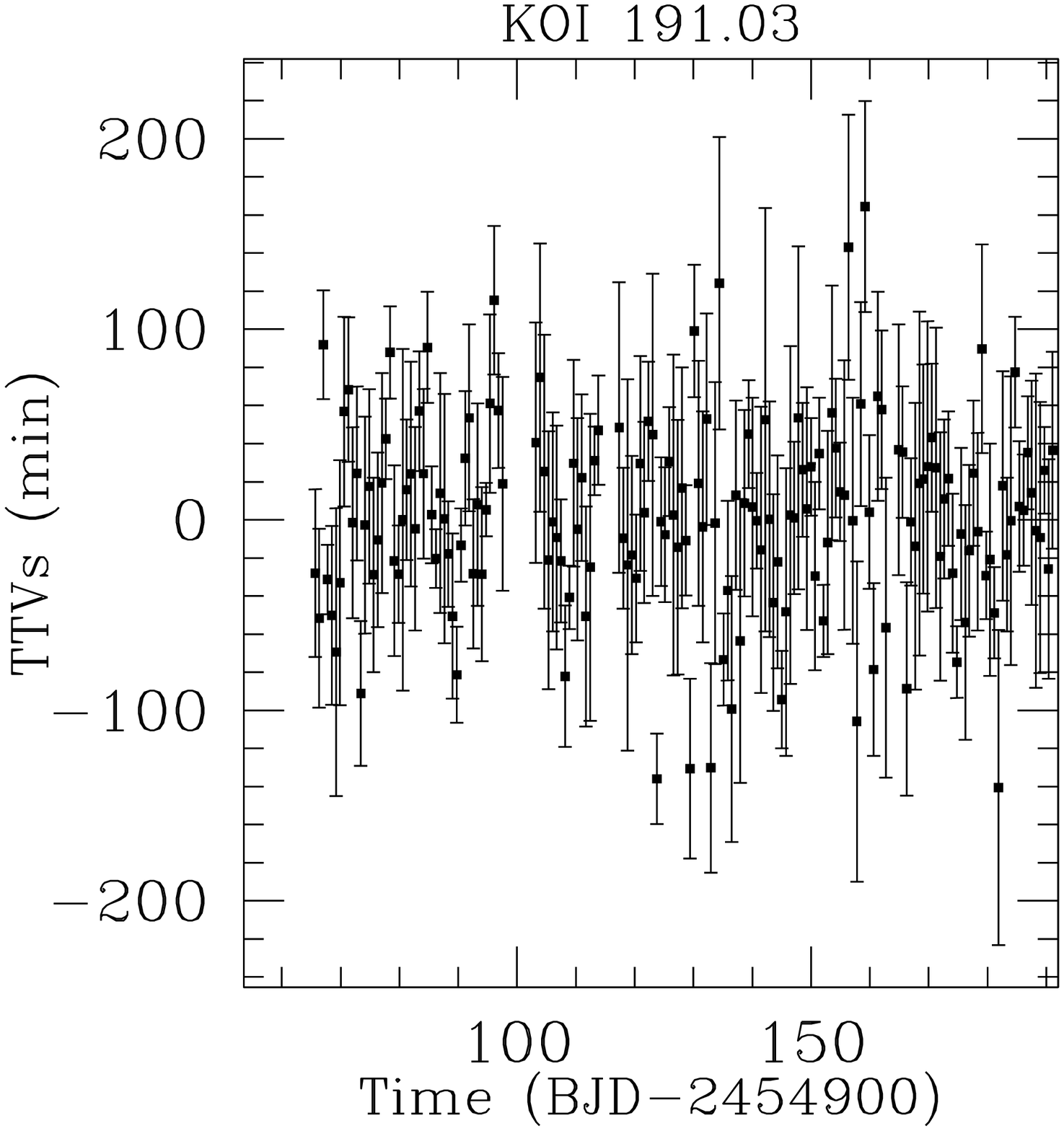}{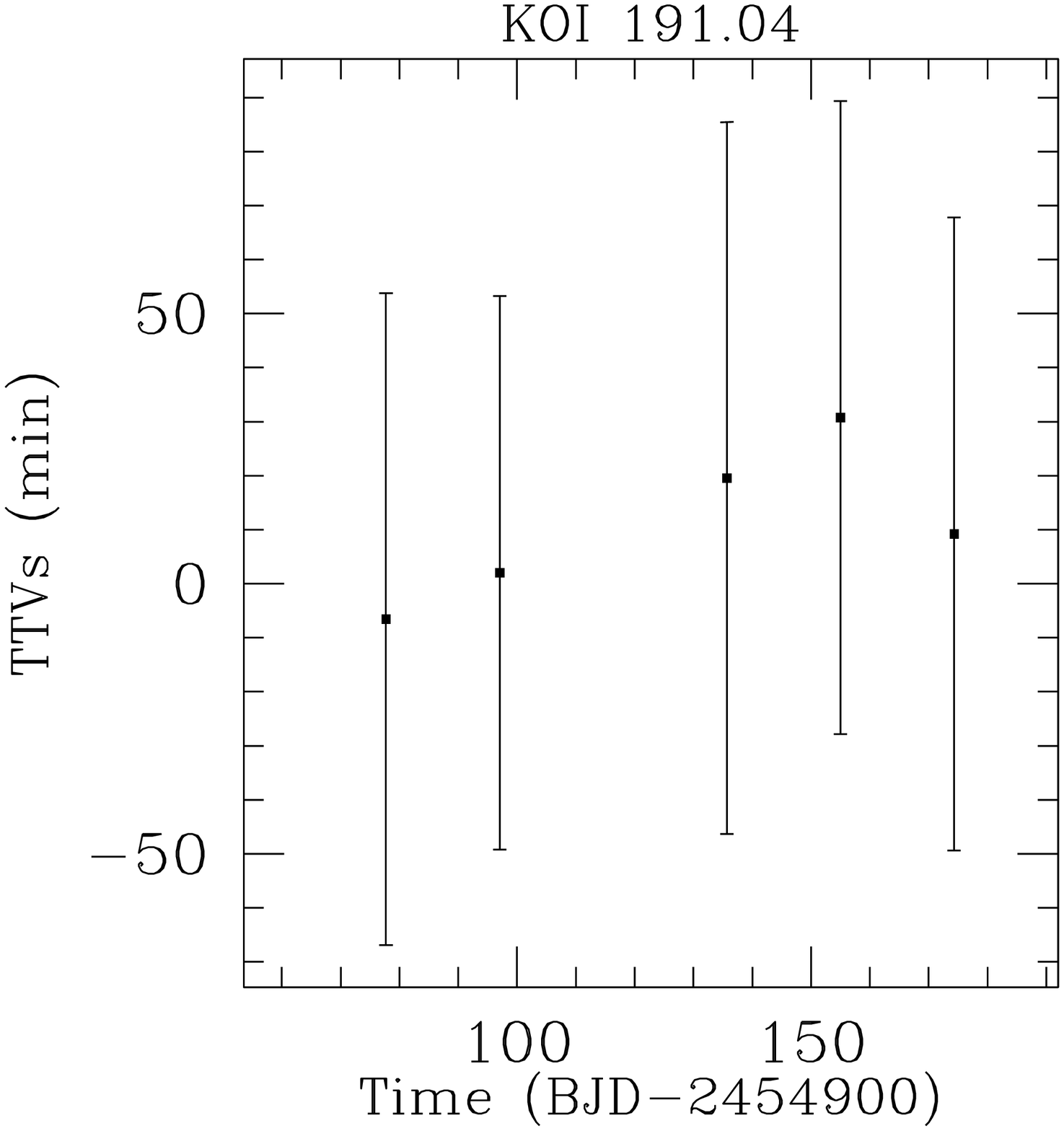}\end{figure*}
\clearpage
\begin{figure*}\plottwo{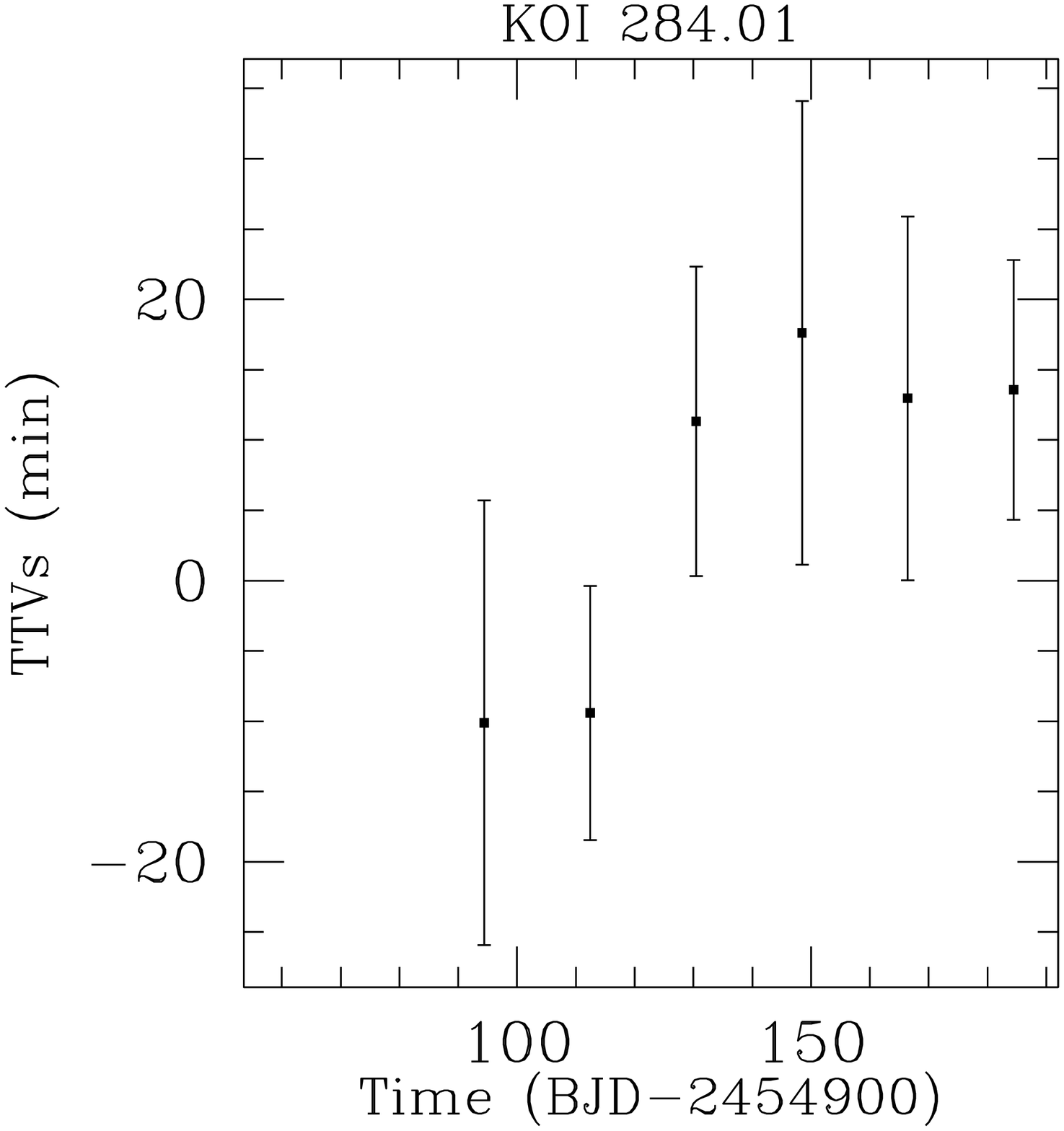}{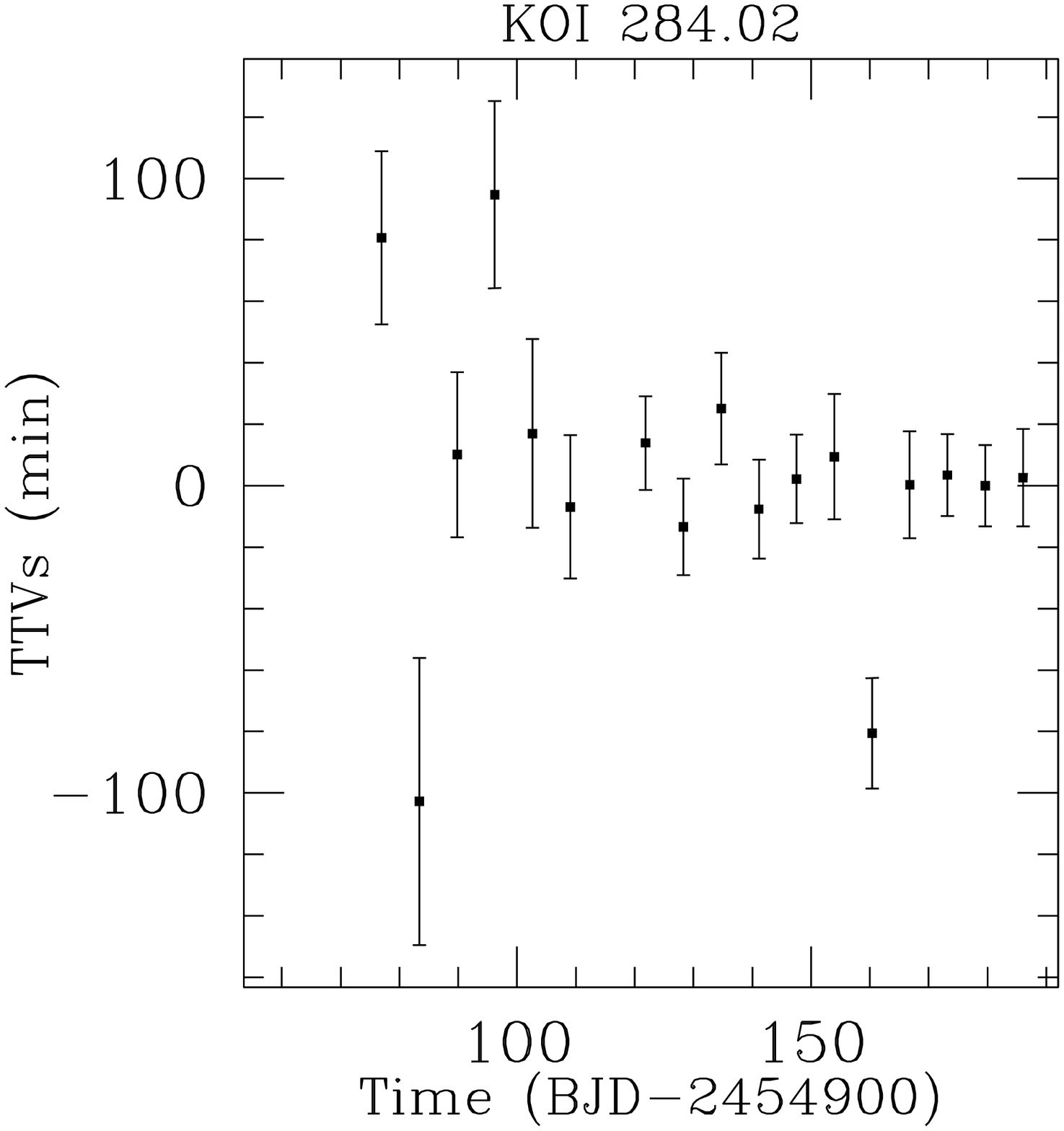}\end{figure*}
\begin{figure*}\plottwo{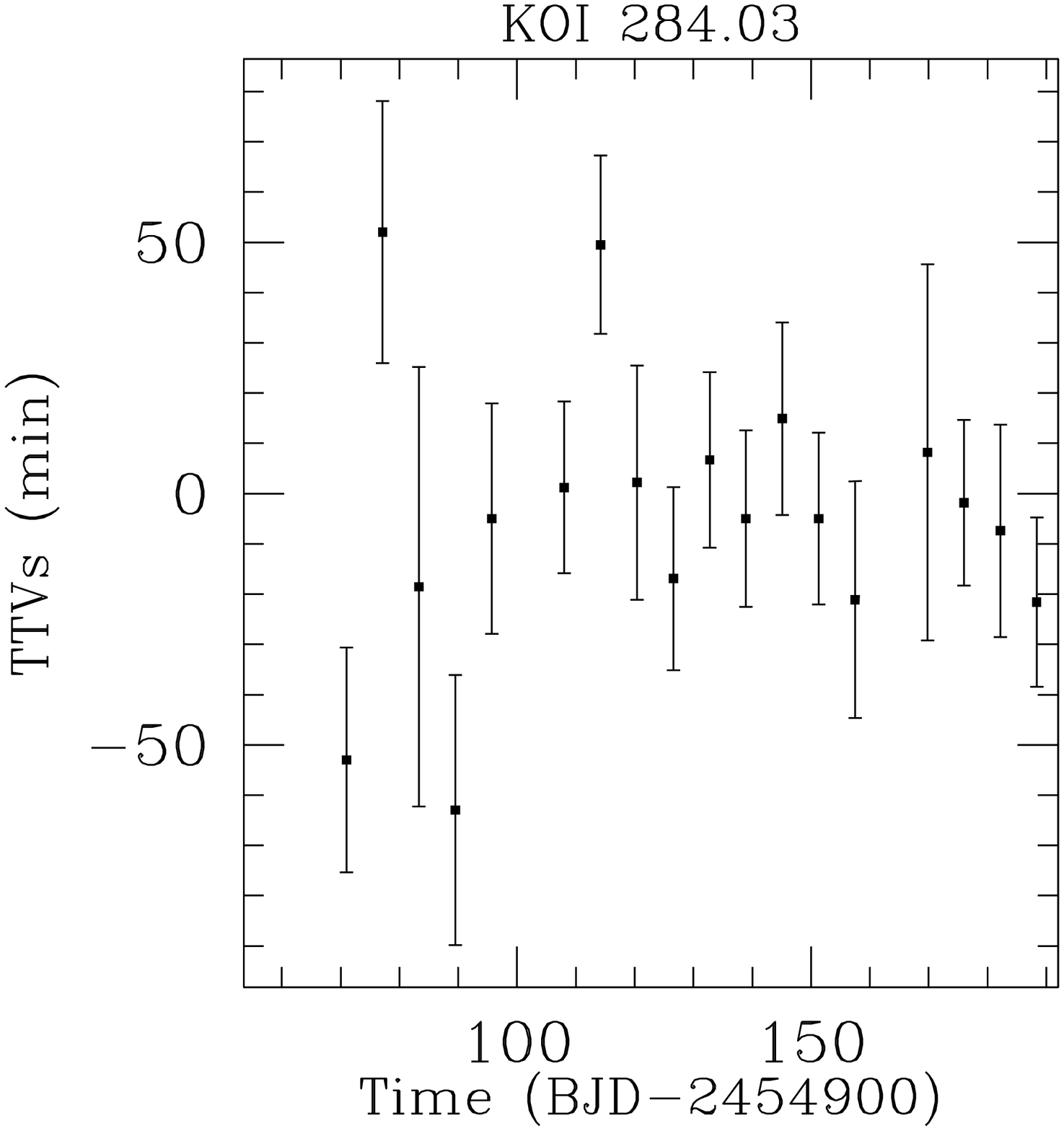}{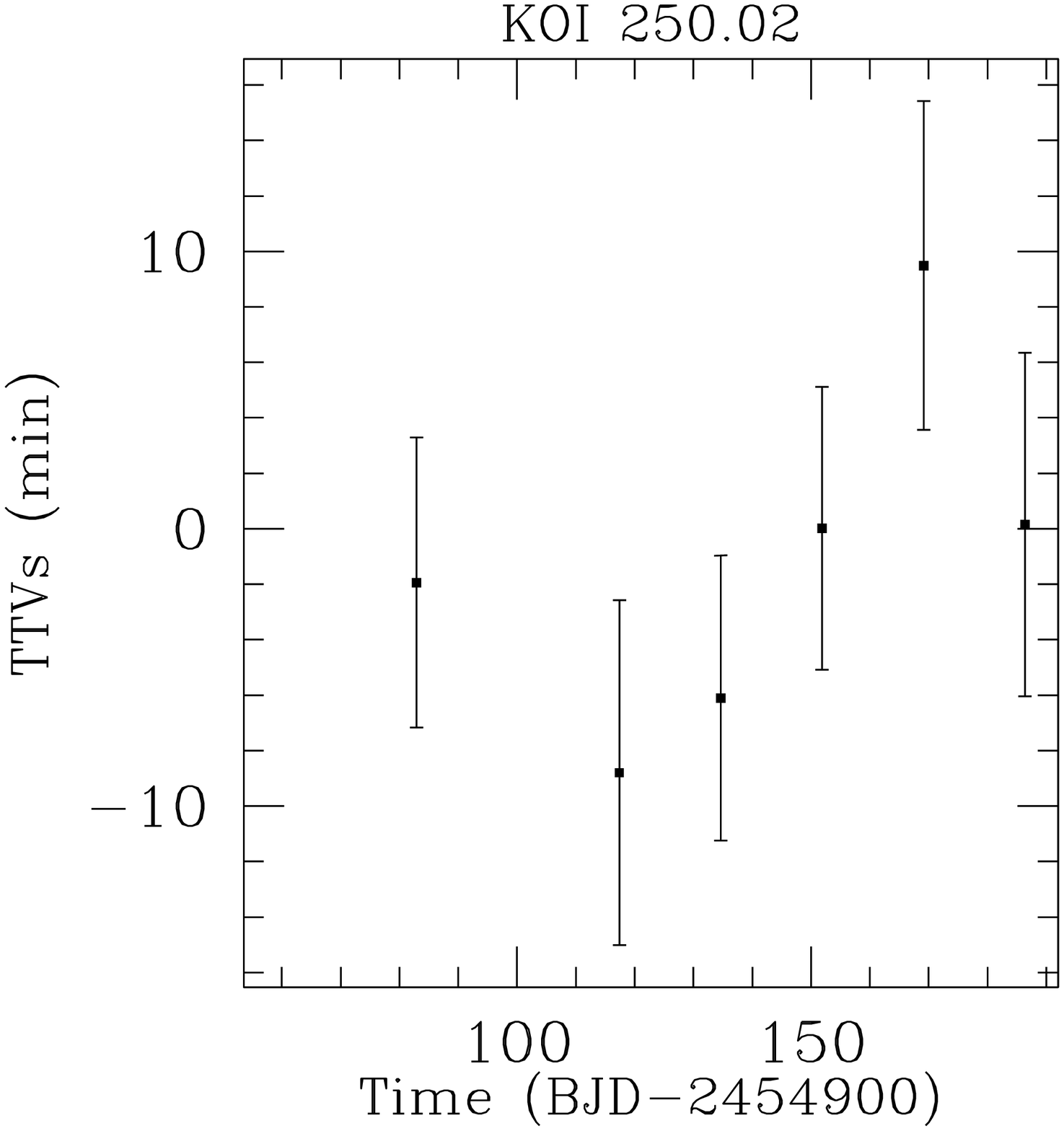}\end{figure*}
\clearpage

\end{document}